%% file: HenriqueVSouza_Tese.tex
\documentclass[12pt,a4paper,oneside,english,portuges,brazil]{memoir}
\usepackage[T1]{fontenc}
\usepackage[utf8]{inputenc}
\usepackage{graphicx}
\usepackage{caption}
\usepackage{subcaption}
\usepackage{amsmath,amsfonts,amssymb,amsthm,thmtools,epsfig,epstopdf,titling,url,array}
\usepackage{bbm}
\usepackage{cite}

\usepackage[paperwidth=210mm, paperheight=297mm, top=3cm, left=3cm, right=2cm, bottom=2cm]{geometry}
\usepackage{multirow}

\usepackage{ragged2e}
\usepackage{longtable}

\theoremstyle{plain}

\theoremstyle{definition}

\theoremstyle{remark}

\makeatother

\usepackage[version=3]{mhchem}

\usepackage{color}
\usepackage[table,xcdraw]{xcolor}
\usepackage[portuguese]{babel}
\usepackage{textcomp}
\usepackage{esint}

\usepackage{ucs}
\usepackage{makeidx}
\usepackage{lmodern}
\usepackage{fourier}
\usepackage{physics}
\usepackage{indentfirst}
\usepackage{blindtext}
\usepackage{enumitem}
\usepackage{float}
\usepackage{chemfig}
\usepackage{mathtools}

\renewcommand{\url}[1]{\href{#1}{Link}}

\usepackage{cancel}

\everymath{\displaystyle} 

\newcommand{\me}{\mathrm{e}}

\usepackage{url}

\usepackage{makecell}

\usepackage{footnote}
\makesavenoteenv{tabular}
\makesavenoteenv{table}

\usepackage[unicode=true,
bookmarks=true,bookmarksnumbered=true,bookmarksopen=true,bookmarksopenlevel=2,
breaklinks=true,pdfborder={0 0 0},backref=false,colorlinks=true]
{hyperref}

\usepackage{breakurl} 
\hypersetup{pdftitle={ARAPUCA, light trapping device for the DUNE experiment},
	pdfauthor={Henrique Vieira de Souza},
	pdfsubject={ARAPUCA, light trapping device for the DUNE experiment},
	pdfkeywords={Light detector, LArTPC, neutrinos},
	linkcolor=blue, citecolor=red, urlcolor=blue}
\usepackage{todonotes}
\makeatletter

\setsecnumdepth{subsection} 

\usepackage[square,numbers,compress]{natbib}

\usepackage{titlesec}
\titleformat{\chapter}[display]   
{\normalfont\huge\bfseries}{\chaptertitlename\ \thechapter}{1pt}{\Huge}   
\titlespacing*{\chapter}{0pt}{-70pt}{20pt}

\titleformat{\section}[block] 
{\Large\bfseries}
{\thesection. }{0em}{}

\titleformat{\subsection}[block] 
{\large\bfseries}
{\thesubsection. }{0em}{}

\titleformat{\subsubsection}[block] 
{\normalfont\bfseries}
{\thesubsubsection. }{0em}{}

\newcommand{\ptp}{PTP}
\newcommand{\xara}{X-ARAPUCA}
\newcommand{\ara}{ARAPUCA}

\newcommand{\sphe}{s.p.e.}
\newcommand{\phe}{p.e.}

\newcommand{\numu}{\nu_{\mu}}
\newcommand{\nue}{\nu_{e}}
\newcommand{\nutau}{\nu_{\tau}}
\newcommand{\nua}{\nu_{\alpha}}
\newcommand{\nub}{\nu_{\beta}}
\newcommand{\anu}{\bar{\nu}}
\newcommand{\anumu}{\bar{\nu}_{\mu}}
\newcommand{\anue}{\bar{\nu}_{e}}
\newcommand{\anutau}{\bar{\nu}_{\tau}}
\newcommand{\anua}{\bar{\nu}_{\alpha}}
\newcommand{\anub}{\bar{\nu}_{\beta}}

\newcommand{\eV}{\text{eV}}
\newcommand{\MeV}{\text{MeV}}
\newcommand{\GeV}{\text{GeV}}
\newcommand{\dm}{\Delta m^2}
\newcommand{\av}[1]{\left\langle{#1}\right\rangle}
\newcommand*\diff{\mathop{}\!\mathrm{d}}

\newcommand{\error}{~$\pm$~}
\newcommand{\fprompt}{$F_{\text{prompt}}$}
\newcommand{\slfrac}[2]{\left.#1\middle/#2\right.}
\newcommand{\dedx}{\mathrm{d}E/\mathrm{d}x}

\newcommand{\viku}{Vikuiti}

\newcommand{\noun}[1]{\textsc{#1}}

\usepackage{enumitem}

\makeatletter
\renewcommand\@biblabel[1]{#1.}
\makeatother

\AtBeginDocument{\addtocontents{toc}{\protect\thispagestyle{empty}}}

\begin{document}
\emergencystretch 3em
\textcolor{white}{\thispagestyle{empty}
\pagenumbering{gobble}}
\begin{flushleft}
\includegraphics[scale=0.3]{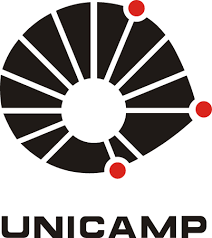}
\par\end{flushleft}

\ $ $ \ 

\begin{center}
\noun{\LARGE Universidade Estadual de Campinas}

\large{Instituto de Física ``Gleb Wataghin''}
\par\end{center}{\Large \par}

\vspace{3cm}

\begin{center}
\noun{\LARGE{}Henrique Vieira de Souza}\noun{\huge{} }
\par\end{center}{\par}

\vspace{2cm}

\begin{center}
\noun{\LARGE{} ARAPUCA, light trapping device for the DUNE experiment}

\vspace{2cm}
\noun{\LARGE{} ARAPUCA, dispositivo de coleta de luz para o experimento DUNE}
\par\end{center}{\LARGE \par}

\vspace{2cm}

\begin{center}
{\Large{}Campinas}
\par\end{center}{\Large \par}

\begin{center}
{\Large{}2021}
\par\end{center}{\Large \par}

\newpage
$ $

\begin{center}
\noun{\large{}Henrique Vieira de Souza}
\par\end{center}{\large \par}

\vspace{1cm}

\begin{center}
\noun{\LARGE{}ARAPUCA, light trapping device for the DUNE experiment}

\vspace{1.5cm}

\noun{\LARGE{}ARAPUCA, dispositivo de coleta de luz para o experimento DUNE}
\par\end{center}{\LARGE \par}

\vspace{1cm}
\begin{flushright}
\begin{minipage}[t]{10cm}
Tese apresentada ao Instituto de Física ``Gleb Wataghin'' da Universidade Estadual de Campinas como parte dos requisitos exigidos para a obtenção do título de Doutor em Ciências, na área de Física.

\ $ $ \

Thesis presented to the ``Gleb Wataghin'' Institute of Physics of the University of Campinas in partial fulfillment of the requirements for the degree of Doctor in Science the area of Physics. 
\end{minipage}
\vspace{1.cm}
\end{flushright}

\begin{flushleft}
\begin{minipage}[t]{8.5cm}
	
{\large Orientador: Ettore Segreto \vspace{0.5cm}}

{Este exemplar corresponde à versão final da tese defendida pelo aluno Henrique Vieira de Souza e orientada pelo Prof. Dr. Ettore Segreto.}
\end{minipage}
\end{flushleft}
\vspace{0.5cm}

\begin{center}
{\Large{}Campinas}
\par\end{center}{\Large \par}

\begin{center}
{\Large{}2021}
\par\end{center}{\Large \par}

\newpage
\newgeometry{top=0cm, left=0cm, right=0cm, bottom=0cm}
\includegraphics[scale=0.99]{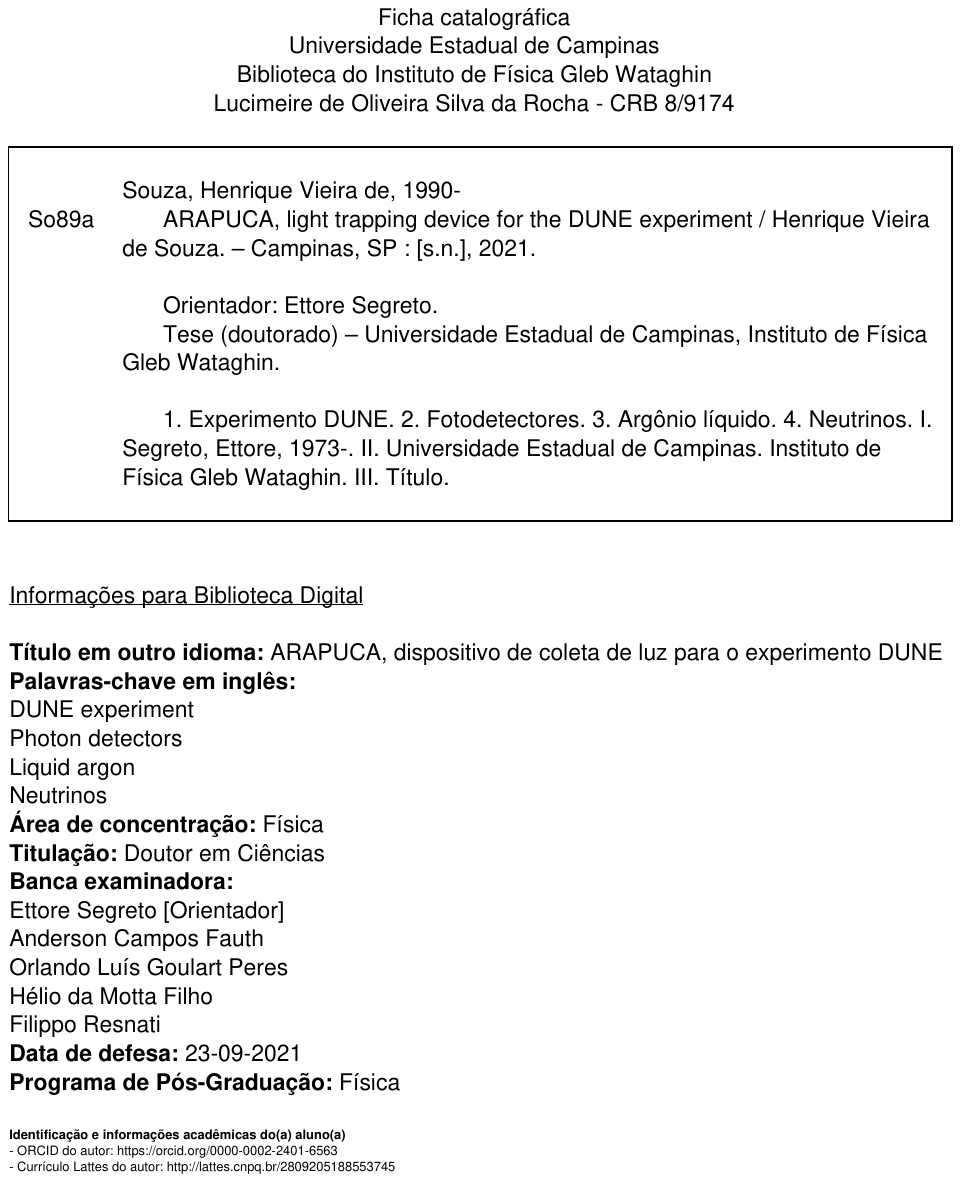}
\restoregeometry
\newpage \thispagestyle{empty}\textcolor{white}{\_}

\begin{flushleft}
\begin{picture}(0,0) \put(0,0){\includegraphics[width=1\textwidth]{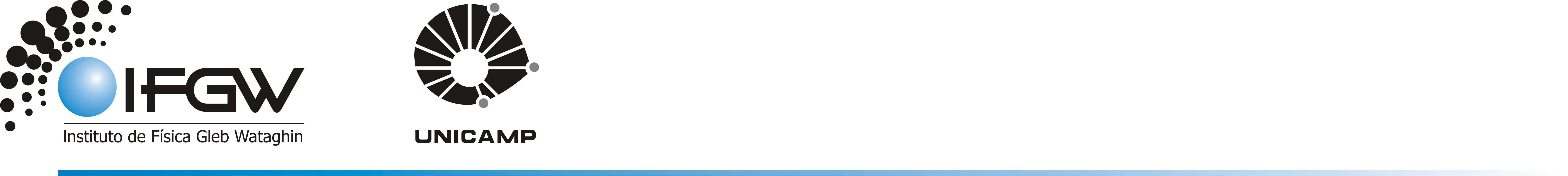}} \end{picture}
\end{flushleft}
\vspace{1.5cm}
\justify{
MEMBROS DA COMISSÃO JULGADORA DA DISSERTAÇÃO DE DOUTORADO  DE  \textbf{HENRIQUE VIEIRA DE SOUZA – RA: 119549} APRESENTADA E APROVADA AO INSTITUTO DE FÍSICA ``GLEB WATAGHIN'', DA UNIVERSIDADE ESTADUAL DE CAMPINAS, EM 23/09/2021.}
\par

\vspace{2cm}

\justify{\textbf{COMISSÃO JULGADORA:}}

\vspace{2cm}

\justify
Prof. Dr. Ettore Segreto - Orientador (IFGW/UNICAMP)
\justify
Prof. Dr. Anderson Campos Fauth (IFGW/UNICAMP)
\justify
Prof. Dr. Orlando Luis Goulart Peres (IFGW/UNICAMP)
\justify
Dr. Filippo Resnati (European Organization for Nuclear Research)
\justify
Dr. Hélio da Motta Filho (Centro Brasileiro de Pesquisas Físicas)

\vspace{1.5cm}

\justify
\textbf{OBS.:} Ata da defesa com as respectivas assinaturas dos membros encontra-se no SIGA/Sistema de Fluxo de Tese e na Secretaria do Programa da Unidade.

\vspace{1.5cm}
\begin{center}
{\Large{}Campinas}
\par\end{center}{\Large \par}

\begin{center}
{\Large{}2021}
\par\end{center}{\Large \par}

\selectlanguage{english}%
\chapter*[]{}

\selectlanguage{english}%
\vspace{11cm}

\begin{flushright}

\ $ $ \

\vspace{6cm}
\begin{minipage}[t]{8cm}

{A meus pais, que me incentivaram, apoiaram e me ensinaram que se Deus fecha uma porta Ele abre uma janela, dedico esta tese, com todo meu amor.}
\end{minipage}

\par\end{flushright}

\begin{flushright}

\par\end{flushright}

\newpage{}

\textbf{\LARGE{}Acknowledgements}{\LARGE \par}

\ $ $ \

First, I would like to thank Ettore Segreto, for the supervision and friendship along this years. It has been a pleasure being his student and I've learned so much and grew a lot professionally thanks to him. I would also like to thank Ana Amelia Machado and all the students and employees of the laboratory who made my days much better and helped me out in my research. 

I wish to thank Francesco Terranova and Carla Cattadori for the amazing experience in Italy. Thank you for the guidance and friendship. Thanks also to all students and employees of Milano Bicocca and to everyone from the residence. I cannot express how much I enjoyed my work and staying in Italy, and this is mostly because of you all.

I also thank all the professors who were part of my formation. Special thanks to prof. Anderson C. Fauth, who was my master advisor and has taught me so much, and professors Orlando Peres, Sandro Guedes, Edmilson José, Carola Chinellato, Marcelo Guzzo, Pascoal Pagliuso, Ernesto Kemp, Eduardo Granado, Antonio Riul and Fernando Iikawa. Thanks to the Secretaria de Pós-Graduação for all the help and to all the employees of the Universidade Estadual de Campinas (UNICAMP).

Thanks to Vinicius Pimentel, Laura Paulucci, Franciole Marinho, David Martinez, Roberto Acciarri, Vishvas Pandey, Vincent Basque and everybody else with whom I had the pleasure to collaborate. Thanks also to Hélio da Motta Filho and Filippo Resnati for accepting being part of my Ph.D. committee.

I want to thank specially my parents for all the love and support. You have provided me everything, thank you so much... I am glad the vasectomy failed. I want to thank Aline Souza for being with me for most of this journey giving a lot of support. To all my friends, in particular to João Diniz, Adnei Ercule, Marina Guzzo, Rafaela Ramos, Jully Nascimento, Maria Cecilia, Yago Silva, Marco Ayala, Esteban Cristaldo, Ariel, Bruno Gelli, Greg de Souza and many others that put up with me for so long. To all my friends that I made in Italy, specially to Madalina Munteanu, Antonio Branca, Claudia Brizzolari, Andrea Falcone, Marta Torti, Maura Spanu, Niccolò Gallice, Arenc Tukaj, Danny Giove, Aitesam Tahir, Wenxiang Guo, Hakam Abushanab, Federica Cecuta and many others, thank you for making my stay in Italy so amazing. 

Most importantly, I thank God because without Him I am nothing. Even though I do not deserve, He is always by my side and have put all this amazing people in my life. 

Thanks to the ``Ministry of Science, Technology and Innovation'' and the ``National Council for Scientific and Technological Development - CNPq'' for the scholarship. This study was financed in part by the Coordenação de Aperfeiçoamento de Pessoal de Nível Superior - Brasil (CAPES) - Finance Code 001.

\newpage{}

\textbf{\LARGE{}Resumo}{\LARGE \par}

\ $ $ \

O Deep Underground Neutrino Experiment (DUNE) será o primeiro mega programa científico em solo americano e irá esclarecer algumas questões em aberto na física de neutrinos. O experimento prevê a realização de um feixe de neutrinos intenso no Fermilab (Chicago - EUA), um detector próximo ao feixe para monitoramento e um detector distante em Sanford Underground Research Facility (SURF) a 1300~km de distância. O detector distante será composto por quatro módulos de câmeras de projeção temporal de argônio líquido (do inglês, LArTPC) de 10~kt de volume ativo cada que realizarão as medidas precisas exigidas para o experimento DUNE. A técnica experimental utiliza sinais de carga e luz de radiações ionizantes em argônio líquido para reconstruir interações de neutrinos com excelente resolução espacial, medidas de calorimetria e identificação de partículas.

A luz de cintilação do argônio líquido é produzida em torno de 127~nm dentro de poucos nanosegundos da passagem da radiação. A sua detecção pode melhorar a calorimetria da LArTPC além de fornecer o registro temporal de eventos não provenientes do feixe, o que é vital para a averiguação de eventos de decaimento de núcleos. O sistema de detecção de luz do DUNE (do inglês, PDS) é responsável por coletar a luz de forma eficiente, a fim de caracterizar o volume ativo do detector com eficiência superior a 99\%. O PDS será composto por 1.500 módulos de \xara, um dispositivo de coleta de luz que será o assunto principal dessa tese. 

Neste trabalho, a pesquisa e desenvolvimento do dispositivo \ara\ será apresentado. A caracterização completa dos protótipos foram realizadas em testes dedicados em argônio líquido no Brasil e na Itália, no qual eficiências comparáveis de (2.2\error0.4)\% e (1.9\error0.1)\% foram encontradas, respectivamente. O novo wavelength shifter desenvolvido na Itália em colaboração com a \textit{Università degli Studi di Milano-Bicocca} será apresentado. Componente que permitiu aumentar a eficiência de coleta de luz em cerca de 50\% e resultou em uma eficiência de (2.9\error0.1)\% para o protótipo de \xara.

Os dispositivos de \xara s foram também instalados no protótipo do DUNE (ProtoDUNE) para testes de dopagem de xenônio. Será apresentada nessa tese parte da análise de dados realizada que permitiu a colaboração do DUNE entender e estabelecer a dopagem de xenônio como alternativa para o detector distante. 

\vspace{5mm}

\noindent \textbf{Palavras-chave: Fotodetectores; Argônio líquido; Experimento DUNE; Neutrinos} 

\newpage{}

\textbf{\LARGE{}Abstract}{\LARGE \par}

\ $ $ \

The Deep Underground Neutrino Experiment (DUNE) will be the first mega-science program on the US soil and will shade light on some of the open questions in neutrino physics. The experiment foresees the realization of an intense neutrino beam at Fermilab (Chicago - USA), of a near detector to monitor the beam and a far detector installed in the Sanford Underground Research Facility (SURF), 1300~km far away. Four 10~kt Liquid Argon Time Projection Chambers (LArTPC) will compose the 40~kt far detector modules to perform the precise measurements required for DUNE. The experimental technique uses charge and light signal from ionizing radiations in liquid argon to fully reconstruct neutrino interactions with excellent spatial resolution, calorimetric measurements and particle identification.

The liquid argon scintillation light is produced around 127~nm within a few nanoseconds from the radiation passage. Its detection may improve the LArTPC calorimetric besides giving the time stamp of non-beam events, which is vital in fiducializing nucleon-decay events. The DUNE Photon Detection System (PDS) is responsible to efficiently collect this light in order to fiducialize the active volume of the detector with $>99\%$ efficiency. The PDS will be composed by 1,500 \xara\ modules, a light trapping device that was the main subject of this thesis.

In this work, the research and development of the \ara\ devices will be presented. The full characterization of the prototypes were performed in dedicated liquid argon tests, in Brazil and Italy, where comparable efficiencies of (2.2\error0.4)\% and (1.9\error0.1)\% were found. A new wavelength shifter developed in Italy in collaboration with the \textit{Università degli Studi di Milano-Bicocca} will be presented. It allowed to increase the light detection efficiency of about 50\% and resulted in an overall efficiency of (2.9\error0.1)\% of the \xara\ prototype.

The \xara\ devices were also deployed in the DUNE prototype (ProtoDUNE) for xenon doping tests. It will be shown in this thesis part of the analysis performed which aided the DUNE Collaboration to understand and establish the xenon doping as an alternative for the far detector modules.  

\vspace{5mm}

\selectlanguage{english}%
\noindent \textbf{Keywords: Photon detectors; Liquid argon; DUNE experiment; Neutrinos} 

\selectlanguage{english}%



\newpage
\chapter*{List of Abbreviations}
\pagenumbering{arabic}
\setcounter{page}{10}
\thispagestyle{empty}
\pagestyle{empty}

\newlist{abbrv}{description}{1}
\setlist[abbrv,1]{labelwidth=1in,align=parleft,noitemsep,leftmargin=!}
\begin{abbrv}

\item[ADC]{Analog-to-Digital Converter}
\item[APA]{Anode Plane Assemblies}
\item[APD]{Avalanche Photodiode}
\item[BBT]{2,5-Bis(5-tert-butyl-benzoxazol-2-yl)thiophene}
\item[BisMSB]{1,4-Bis(2-methylstyryl)benzol}
\item[CC]{Charged Current}
\item[CERN]{European Organisation for Nuclear Research}
\item[C.L.]{Confidence Level}
\item[CP]{Charge Parity}
\item[CPA]{Cnode Plane Assemblies}
\item[CPT]{Charge Parity Transformation}
\item[CT]{Charge Time}
\item[DAQ]{Data Acquisition}
\item[DUNE]{Deep Underground Neutrino Experiment}
\item[DUT]{Device Under Test}
\item[ES]{Elastic Scattering}
\item[eV]{Electron Volt}
\item[FD]{Far Detector}
\item[Fermilab]{Fermi National Accelerator Laboratory}
\item[GAr]{Gas Argon}
\item[HD]{Horizontal Drift}
\item[ICARUS]{Imaging Cosmic And Rare Underground Signals}
\item[IO]{Inverted Ordering}
\item[KamLAND]{Kamioka Liquid Scintillator Antineutrino Detector}
\item[LAr]{Liquid Argon}
\item[LArIAT]{Liquid Argon In A Test Beam}
\item[LArTPC]{Liquid Argon Time Projection Chamber}
\item[LBNE]{Long Baseline Neutrino Experiment}
\item[LET]{Linear Energy Transfe}
\item[MH]{Mass Hierarchy}
\item[MicroBooNE]{Micro Booster Neutrino Experiment}
\item[NC]{Neutral Current}
\item[ND]{Near Detector}
\item[NO]{Normal Ordering}
\item[NOvA]{NuMI Off-Axis $\nue$ Appearance}
\item[PD]{Photon Detection}
\item[PDE]{Photon Detection Efficiency}
\item[PDS]{Photon Detection System}
\item[P.E.]{Photo-electron}
\item[PMMA]{Poly(methyl methacrylate)}
\item[PMNS]{Pontecorvo-Maki-Nakagawa-Sakata}
\item[PMT]{Photomultiplier tube}
\item[POT]{Protons On Target}
\item[\ptp]{Para-Terphenyl}
\item[PVT]{Polivyniltoluene}
\item[RMS]{Root Mean Square}
\item[SBND]{Short Baseline Neutrino Detector}
\item[SiPM]{Silicon Photomultiplier}
\item[SNO]{Sudbury Neutrino Observatory}
\item[SP]{Single-Phase}
\item[S.P.E.]{Single photo-electron}
\item[SSM]{Standard Solar Model}
\item[SURF]{Sanford Underground Research Facility}
\item[T2K]{Tokai-to-Kamioka}
\item[TPB]{Tetra-phenyl butadiene}
\item[TPC]{Time Projection Chamber}
\item[VD]{Vertical Drift}
\item[VUV]{Vacuum Ultra Violet}
\item[WLS]{Wavelength shifter}

\end{abbrv}

\newpage

\pagestyle{empty}
\tableofcontents*
\cleardoublepage


\newpage
\input{Introduction}

\input{dune_nu}

\input{lar_tpc}

\input{arapuca_pd_system}
\input{r_n_d}

\input{lar_tests}

\input{protodune}

\input{Conclusion}

\end{document}

%% file: Introduction.tex
\chapter*{Introduction}
\addcontentsline{toc}{chapter}{Introduction}
\pagenumbering{arabic}
\setcounter{page}{16}
\thispagestyle{myheadings}
\pagestyle{myheadings}

In 1914, James Chadwick demonstrated that the electron emitted in $\beta$-decays has a continuous spectrum which seemed not to conserve energy. The idea of the neutrino came in 1930, when Wolfgang Pauli proposed the existence of a neutral and low mass particle that would also be emitted during the $\beta$-decay to preserve the principle of energy conservation. It was only 26 years later, in 1956, that the neutrino was first directly detected by Cowan and Reines in a nuclear reactor experiment~\cite{REINES1956}. 

Neutrinos are classified as neutral fermions, an elementary half-spin particle that interact only via weak interaction and gravity.  The Standard Model is a successful theory of elementary particles that has been profoundly probed along the years. It has, however, a few imperfections that still need further theoretical and experimental improvement to a proper understanding of fundamental physics. For example, the predominance of matter over antimatter in the early universe and the observation of Dark Matter and Dark Energy. More recently, the experimental observation of the neutrino oscillation implied  that neutrinos have mass, which was not included or predicted by the model, offering the most compelling evidence of physics beyond the standard model.

Although many advances have been made in the field of neutrino physics, there are still open questions such as the origin of neutrino mass, the precise parameters of neutrino mixing, if neutrinos are their own antiparticle or not and if there are more than three flavors of neutrinos. On top of it, the study of neutrinos may contribute to answer some of the open questions in astrophysics and cosmology.

The Deep Underground Neutrino Experiment (DUNE)~\cite{DUNE_Vol1_TDR,DUNE_vol2,DUNE_vol3,DUNE_vol4,near_detector} will be a world-class neutrino observatory and nucleon decay detector. The experiment aims to answer the matter and antimatter disparity, the dynamic of supernova neutrino bursts (SNBs) and whether protons eventually decay. The neutrino oscillation mechanism and the key features for experimental measurements in the field are given in Chapter~\ref{chap:dune_nu}, together with the SNB and proton decay signatures in the detector. The DUNE experiment and the technique of detection, the Liquid Argon Time Projection Chamber (LArTPC), are explored in Chapter~\ref{chap:dune}, highlighting the detector proprieties and specifications.

Chapter~\ref{chap:pd_system} is dedicated to the DUNE Photon Detection System (PDS), and the \ara\ is introduced. The PDS has the role to detect argon scintillation light at a wavelength of 127~nm, a challenging and important task to reach the DUNE ambitious physical goals. An efficient light collection not only improves DUNE's physics, but is crucial to measure proton decay efficiently. The work developed in this thesis is aimed to study, characterize and improve the \ara. In Chapter~\ref{chap:RnD}, the Research and Development~(R\&D) of the device is presented, with the main experiments performed and results achieved.

The two main experiments in liquid argon that fully characterized the \ara\ device are presented in Chapter~\ref{chap:lar_test}. The experiments were performed in Brazil and Italy with different prototypes of \ara s and retrieved comparable results. In Brazil, a dedicated cryogenic setup was built in the\textit{ ``Laboratório de Léptons}'' and three different ionizing radiations were used to retrieve the device photon detection efficiency (PDE). In Italy, an alpha source was used to further investigate the device efficiency and to validate the enhancement in the PDE previously investigated in Chap.~\ref{chap:RnD}.

The final chapter (Chapter~\ref{chap:protoDUNE}) presents the investigation of xenon doping in the DUNE prototype (namely ProtoDUNE), which resulted to be an excellent alternative for DUNE in case of nitrogen contamination. Data collected with two \ara\ devices helped the collaboration to analyze the effects of xenon doping in liquid argon, showing a recovery of light and increased light collection uniformity inside the detector.

As a disclaimer, the figures without reference in the caption were done by the author. The exceptions are the analysis plots for the \xara\ single-cell and double-cell (Chapters~\ref{chap:RnD} and \ref{chap:lar_test}), where the author was responsible for the analysis.

%% file: dune_nu.tex
\chapter{DUNE and neutrino physics}
\label{chap:dune_nu}
\thispagestyle{myheadings}

The scientific program of the DUNE experiment is extremely rich and includes: \textbf{(1)} measuring the charge-parity symmetry violation phase in the leptonic sector, \textbf{(2)} the determination of the octant of $\theta_{23}$ mixing angle and the determination of the neutrino mass hierarchy, \textbf{(3)} search for proton decay modes, providing a portal to Grand Unification of Theory (GUT) which predicts that, if there was an unified physical force at the birth of the universe, protons should decay and \textbf{(4)} Detection and measurements of the $\nu_e$ flux from a core-collapse supernova within our galaxy. This would help to understand the supernova phenomenon and would give fundamental information about the nature of neutrinos \cite{DUNE_Report_V1}.

\begin{figure}[h!]
	\centering
	\includegraphics[width=0.49\linewidth]{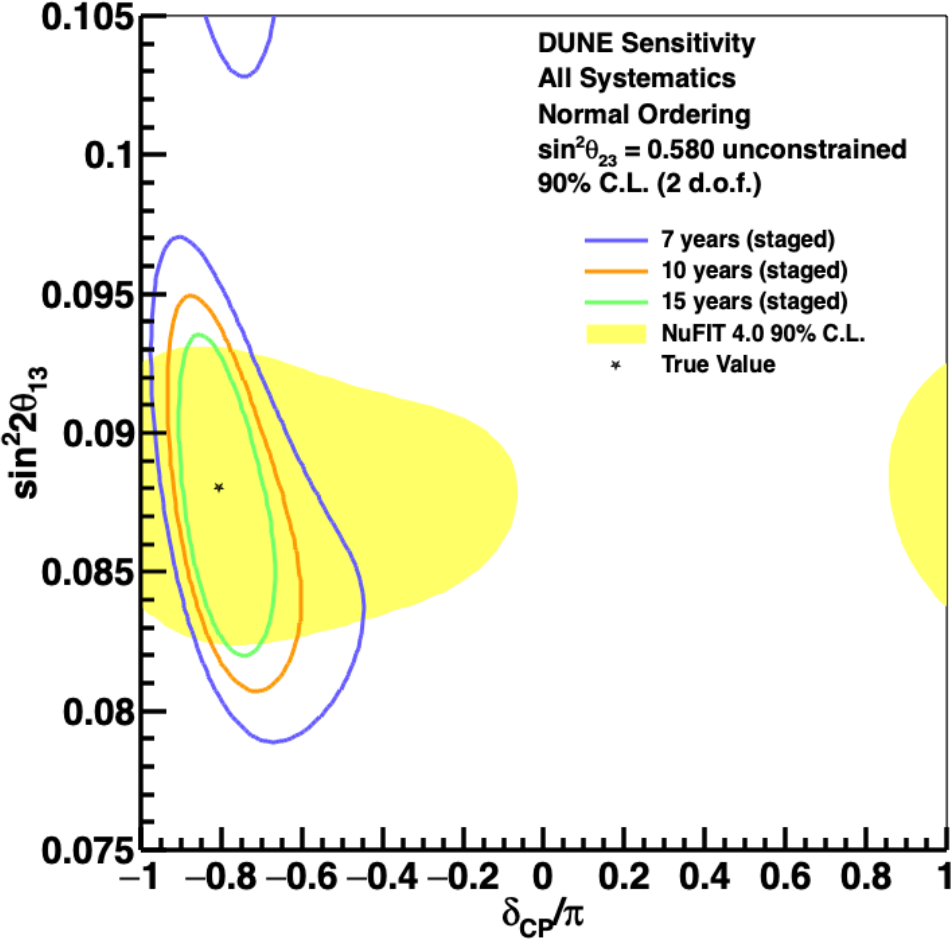}
	\includegraphics[width=0.49\linewidth]{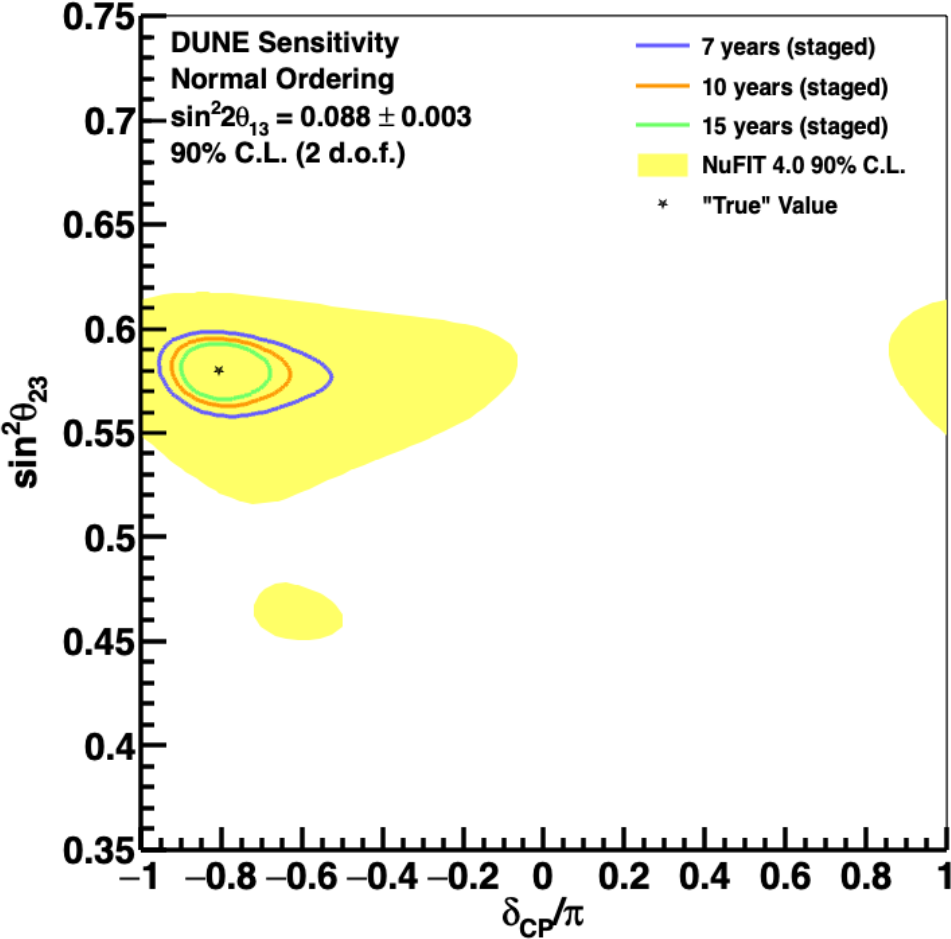}
	\caption{(Left) Allowed region within 90\% C.L. for $\sin[2](2\theta_{13})$ and $\delta_{CP}$ parameters for 7, 10 and 15~years of beam exposure. The global fit to neutrino oscillation data (NuFIT~4.0) 90\% C.L. region is represented by the yellow area and the black star is the ``true value'' assumed to be the central value of the global fit. (Right) Allowed region within 90\% C.L. for $\sin[2](\theta_{23})$ and $\delta_{CP}$ parameters~\cite{DUNE_vol2}.}
	\label{fig:dunesin_vs_cp}
\end{figure}

The charge-parity symmetry phase ($\delta_{CP}$) will be measured in DUNE due the fact that neutrinos travel through matter (Sec.~\ref{sec:oscillation_in_matter}), and it may provide insights into the origin of the matter-antimatter asymmetry. The same matter effect makes possible the determination of the neutrino mass hierarchy, due to the sensitivity to the sign of $\dm_{31}\equiv m^2_3 - m^2_1$. One of the DUNE primary goals is to measure the long-baseline oscillation parameters without constrains~\cite{DUNE_vol2}. Figure~\ref{fig:dunesin_vs_cp} shows the allowed region for $\sin[2](2\theta_{13})$ (Left) (and $\sin[2](\theta_{23})$ (Right)) and $\delta_{CP}$ within 90\% Confidence Level (C.L.) for 7, 10 and 15~years of operation. The black star represents the hypothetical ``true value'' assumed to be the central value of the  global fit to neutrino oscillation data (NuFIT~4.0). 

Furthermore, the intense neutrino beam, the massive 40 kt LArTPC far detector and the high-resolution DUNE near detector makes DUNE an outstanding tool for a rich ancillary science program. Many important researches are included, such as: Beyond Standard Model (BSM) measurements in order to search for sterile neutrinos and Dark Matter, measurements of tau neutrino appearance,  measurements of neutrino oscillation phenomena using atmospheric neutrinos, neutrino cross-section and others.

This chapter gives an introduction of the field of neutrino, with an overview of the current challenges and open questions, to provide a context for the subject of this thesis. We give an introduction of neutrino oscillation, with explanation of the usual two-neutrino oscillation mechanism and how this approach is good enough to understand the case of three neutrino mixing. An emphasis will be given to the DUNE experiment, showing why it is important and exploring the basics of DUNE's physical goals. 

\section{Historical context}

\subsection{Solar neutrino anomaly}

\begin{figure}[h!]
	\centering
	\includegraphics[width=0.55\linewidth]{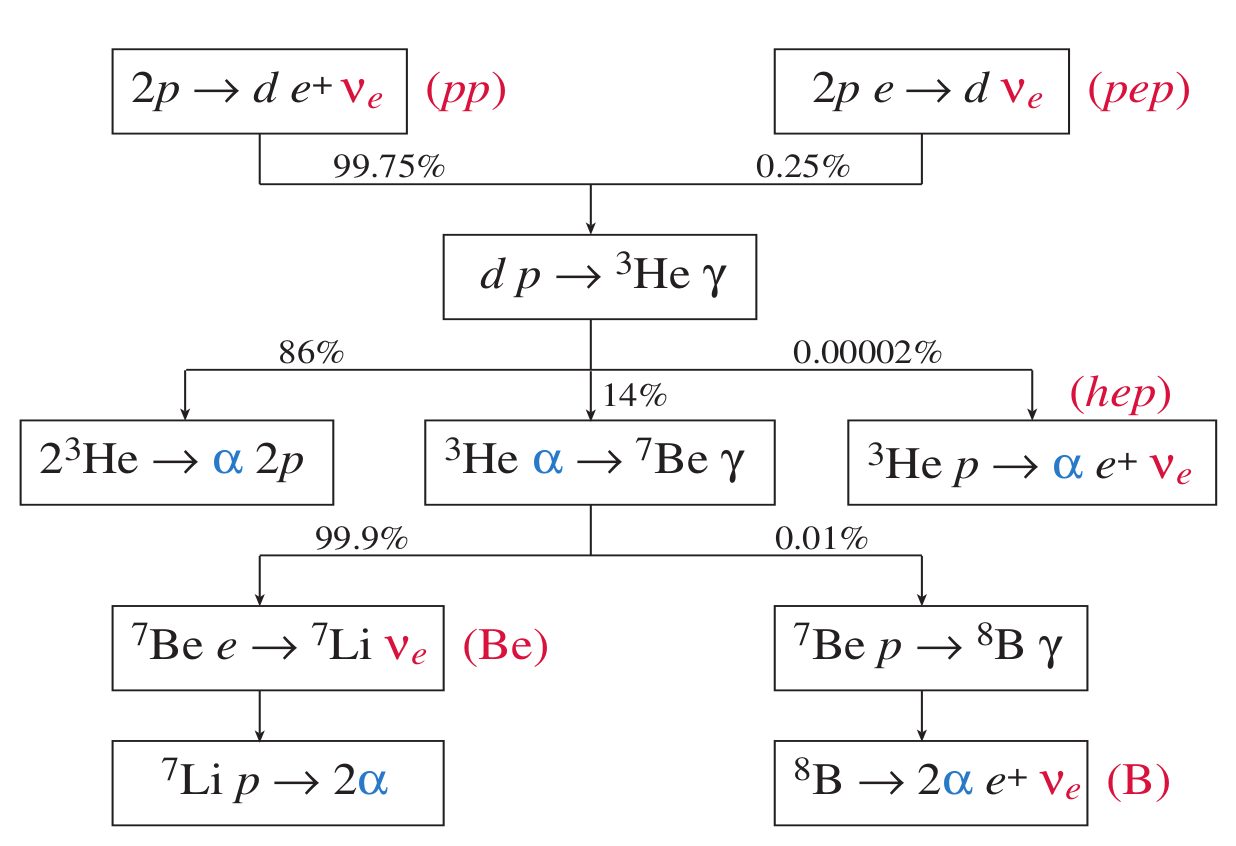}
	\caption{Solar chain for electron neutrino production~\cite{vissani}.}
	\label{fig:solar_chain}
\end{figure}

Earth receives a solar neutrino flux of about  6$\times10^{10}$~$\nu$/cm$^2$s with energy ranging from 0.1 to 20~MeV. The sun produces energy and neutrinos mostly through $^4$He reaction as:
\begin{equation}
	4p+2e^- \rightarrow {^4\text{He}} + 2\nu_{e},
\end{equation}
where 4 protons ($p$) and 2 electrons ($e^-$) will result in an alpha-particle ($^4$He) and two electron neutrinos ($\nu_e$) with total energy release of 26.7~MeV. The main solar chain is shown in Figure~\ref{fig:solar_chain}, where the highlighted reactions $pp$, $pep$, $hep$, Be and B always have the release of an alpha-particle ($\alpha$) and two neutrinos $\nu_e$, highlighted in blue and red, respectively. Heavier stars will follow the CNO chain, while it gives only a minor additional contribution to the solar neutrino flux~\cite{vissani}.

\begin{figure}[h!]
	\centering
	\includegraphics[width=0.7\linewidth]{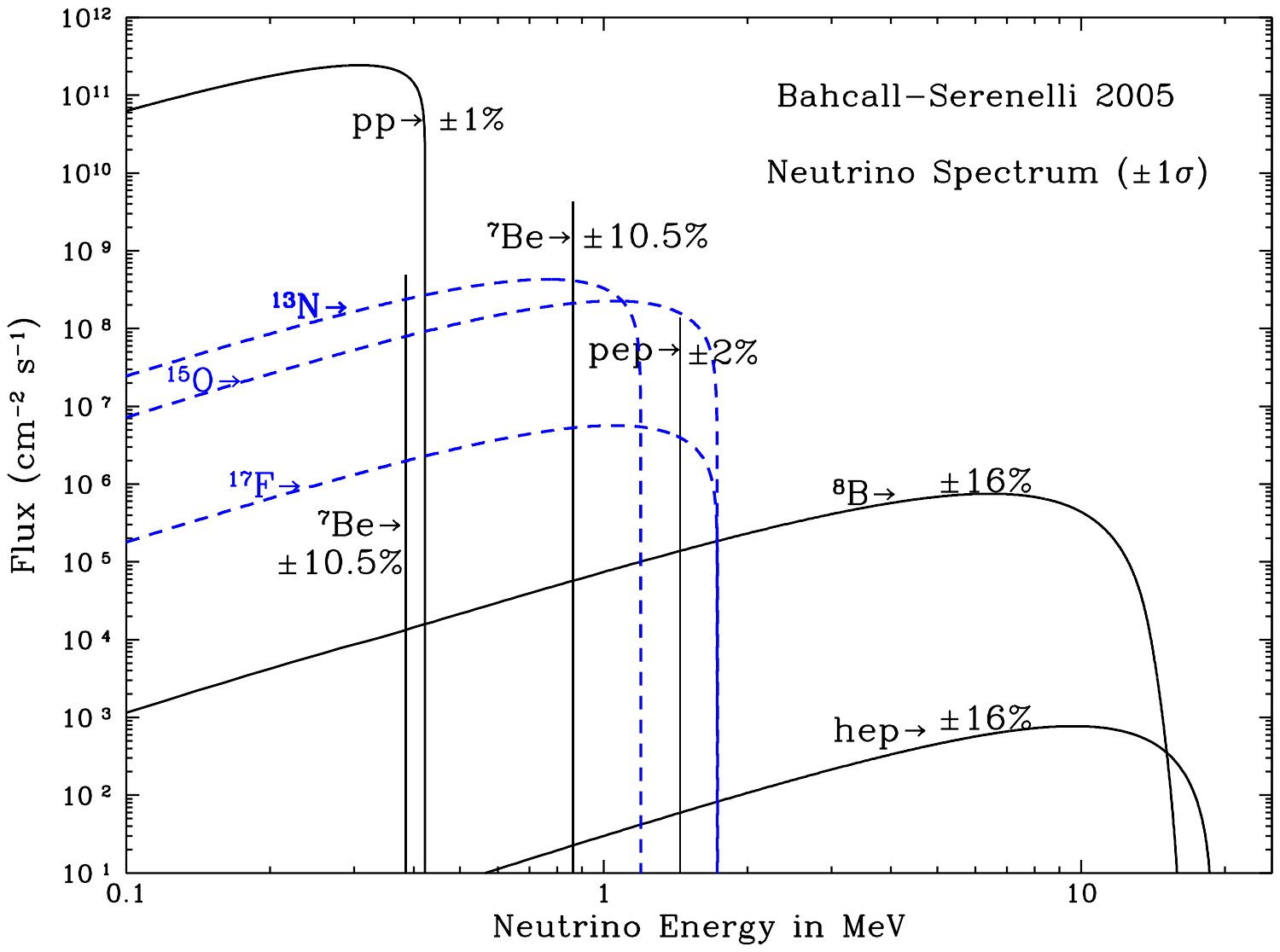}
	\caption{Predicted solar neutrino energy spectra. The solid lines represent neutrinos produced by the $pp$ and $pep$ chains and the dashed blue lines represent neutrinos from the CNO cycle. The theoretical fractional uncertainties are labeled and are shown for each source~\cite{solar_nu_flux_2005}.}
	\label{fig:solar_nu_spectrum}
\end{figure}

The expected neutrino energy spectrum is shown in Figure~\ref{fig:solar_nu_spectrum} for both chains ($pp$ and $pep$) as solid lines and for the CNO cycle as blue dashed lines.

The first experiment to investigate the neutrino flux was led by Ray David in 1968. The experiment, proposed by B. Pontecorvo, used Chlorine to detect neutrinos through inverse beta decay ($\nu+n\rightarrow p +e^-$). The Cl reaction has a threshold of $E_{\nue} > 0.814$~MeV, which leaves the experiment mostly sensitive to Boron (B) neutrinos, as can be noticed from Fig.~\ref{fig:solar_nu_spectrum}. The experiment measured, however, a neutrino flux three times lower than the predicted value, giving rise to the so called ``solar anomaly''.

Years later the Gallium based experiments, SAGE and GALLEX~\cite{solar_sage_experiment,solar_gallex_experiment}, measured the neutrino flux also through inverse beta decay but now sensitive to the main $pp$ chain as the $^{71}$Ga target had a lower threshold of 0.233~MeV. Both experiments also reported a disappearance of electron neutrinos, measuring a flux two times lower than the expected, while the predicted flux had an error of only 1\%~\cite{vissani}. Therefore, most probably, a new physics was present in the experimental data. 

\subsection{Atmospheric neutrino anomaly}

Cosmic rays collisions in the upper atmosphere produce pions (mostly) and kaons, the decay of these particles originates the so-called atmospheric neutrinos through the reactions~\cite{MuonsCecchini2012,pdg}:
\begin{subequations}
	\label{eq:atmospheric_nu}
	\begin{equation}
		\pi^- \rightarrow \mu^- + \anumu, \qquad\qquad \mu^- \rightarrow e^- + \anue + \numu
	\end{equation}
	\begin{equation}
		\pi^+ \rightarrow \mu^+ + \numu, \qquad\qquad \mu^+ \rightarrow e^+ + \nue + \anumu
	\end{equation}
\end{subequations}
where $\pi^{\mp}$ are negative and positive pions, $\mu^\mp$ are negative and positive muons, $e^\mp$ are electrons and positrons and $\nu_{\mu,e}$ ($\anu_{\mu,e}$) are muon and electron neutrinos (antineutrinos). The atmospheric neutrino flux is about 1 $\nu$/cm$^2$s with an uncertainty of 20\% due to, mostly, uncertainties in the cosmic rays flux and hadronic interactions~\cite{vissani}. 

\begin{figure}[h!]
	\centering
	\includegraphics[width=0.635\linewidth]{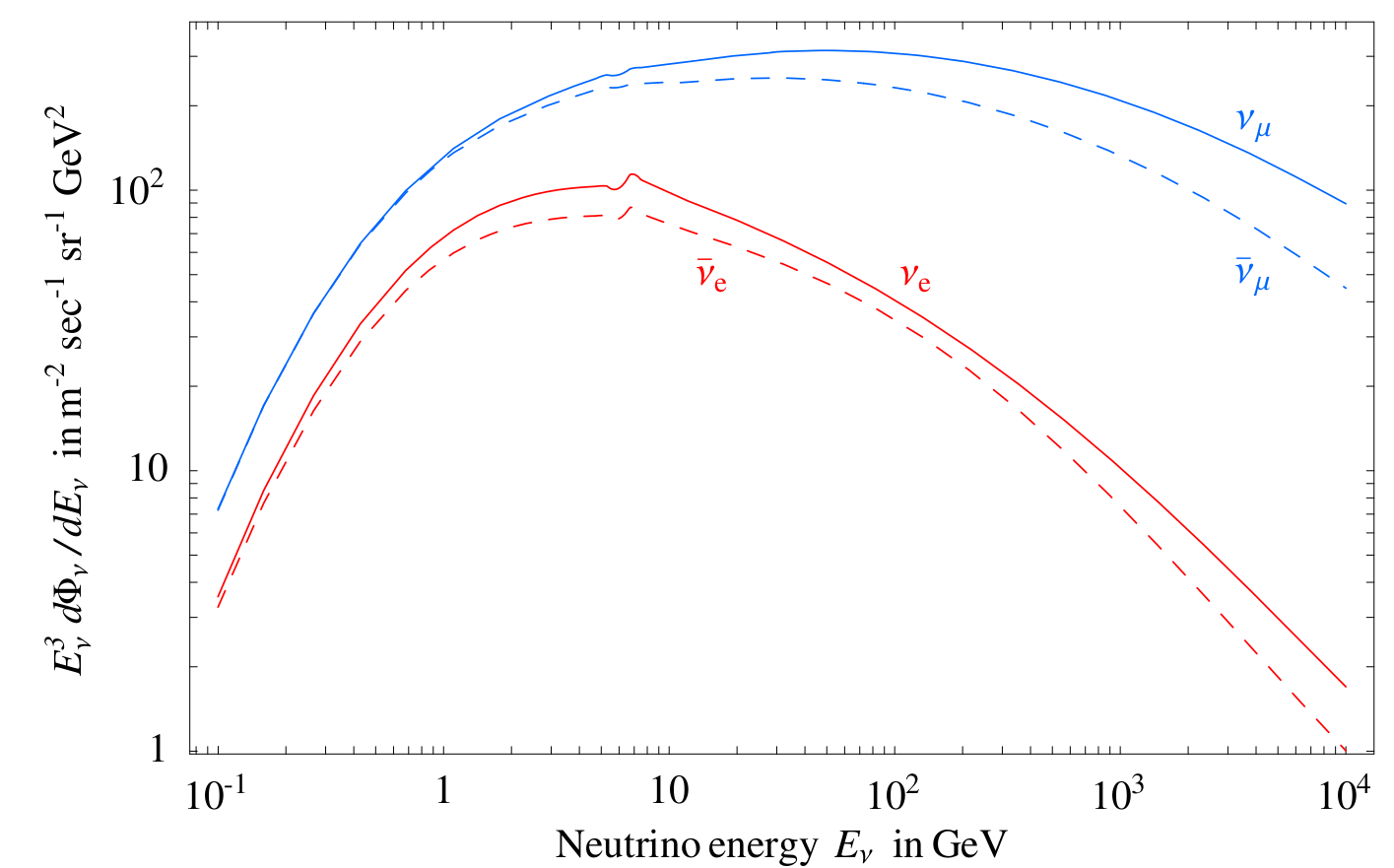}
	\includegraphics[width=0.355\linewidth]{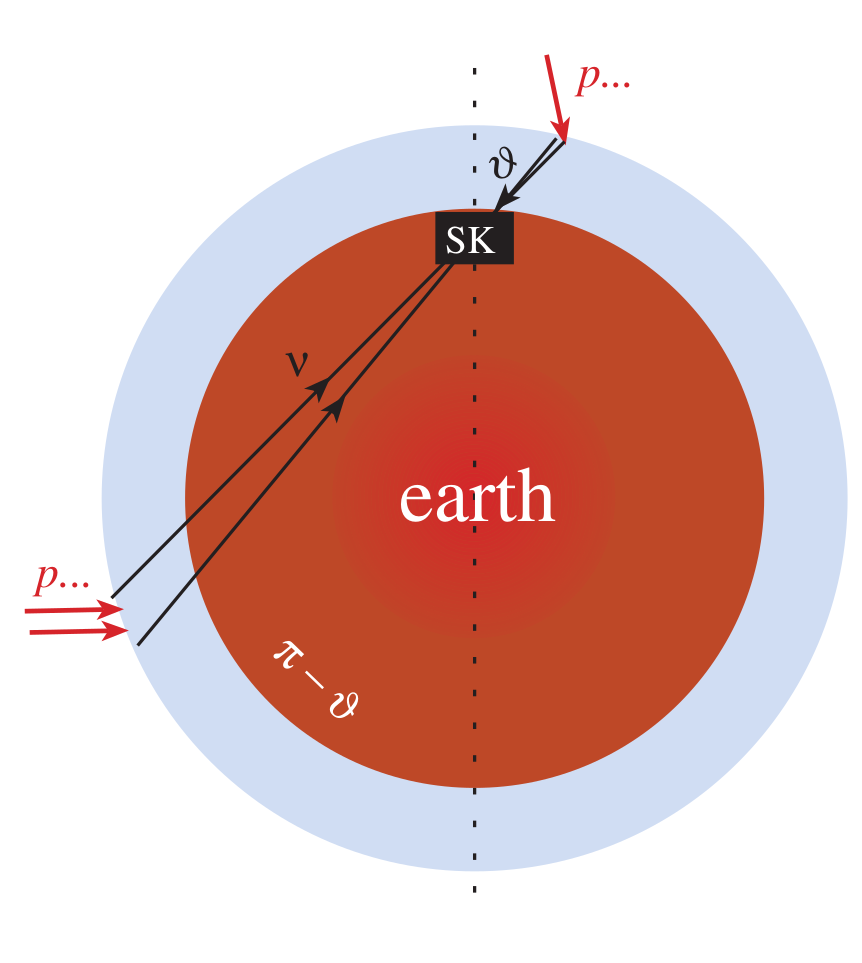}
	\caption{(Left) Predicted atmospheric neutrinos flux in absence of oscillation, average over the zenith angle was taken. (Right) Cartoon of the up and down symmetry on the neutrino flux~\cite{vissani}. Without considering oscillations, the flux of atmospheric neutrinos would be symmetric following the flux in the left panel.}
	\label{fig:atmospheric_nu_flux}
\end{figure}

The reactions of Eq.~\ref{eq:atmospheric_nu} imply that the ratio between the flux of muon-flavor and electron-flavor neutrinos is\footnote{Except for energies above a few~GeV, where the muons will reach the earth's surface before decaying, giving a $\numu$~:~$\nue$ ratio higher than 2.} 2~:~1. Taking into account the symmetry of the Earth, the flux should be the same by measuring neutrinos coming from the positive zenith angle (straight from the sky) or negative (down from Earth). Figure~\ref{fig:atmospheric_nu_flux} (Left) shows the expected flux for $\numu$ ($\anumu$) in solid (dashed) blue lines and for $\nue$ ($\anue$) in solid (dashed) red lines at the Super Kamiokande (SK) location, while a cartoon (Righ) shows that the atmospheric neutrino flux should be the same measured upward-going and or downward-going~\cite{vissani}.

The ``atmospheric anomaly'' was noticed by neutrino experiments and confirmed by the Kamiokande experiment~\cite{kamiokande} which measured a lower ratio of $\numu$~:~$\nue$ than expected. The (Super)Kamiokande is a Water Cherenkov (WC) detector that detects atmospheric neutrinos through charged current nucleon scattering ($\nu_\ell + n \rightarrow \ell + p$, $\ell = e,\mu,\tau$).  The charged particles which travel faster than light in water will emit Cherenkov radiation~\cite{Jelley}.  This was a hint that the ratio could depend on the distance traveled by the neutrino. 

\subsection{The neutrino oscillation evidence}

In the late 1900s, the Kamiokande and Super Kamiokande experiments in Japan~\cite{kamiokande,SuperK_plots} and the SNO experiment in Canada~\cite{SNO_first_phase_data_2007} produced unquestionable evidences of the neutrino oscillation together with explanations for the discrepancies observed. The SNO experiment found evidence of $\nu_{\mu,\tau}$ appearance with confirmation of the KamLAND experiment~\cite{KamLAND2002} discovering disappearance of $\anue$ from reactors~\cite{vissani}. The (Super)Kamiokande measured disappearance of atmospheric $\numu$ and $\anumu$ with confirmation of K2K~\cite{K2K2006} experiment with $\numu$ beam.

\subsubsection{Kamiokande and Super Kamiokande measurements}

The analysis of atmospheric neutrinos in (Super)Kamiokande was originally regarded as background for proton decay searches. The SK is an underground (1000 m) experiment consisting of a 50 kton cylindrical water tank surrounded by 13,142 photomultipliers (PMT). The experiment detected the rings of light produced by scattered leptons in the water. The detector can distinguish between $\numu$ and $\nue$ due to the topology of the rings~\cite{vissani,Thomson}, however, it could not differentiate particles from antiparticles.

\begin{figure}[h!]
	\centering
	\includegraphics[width=0.85\linewidth]{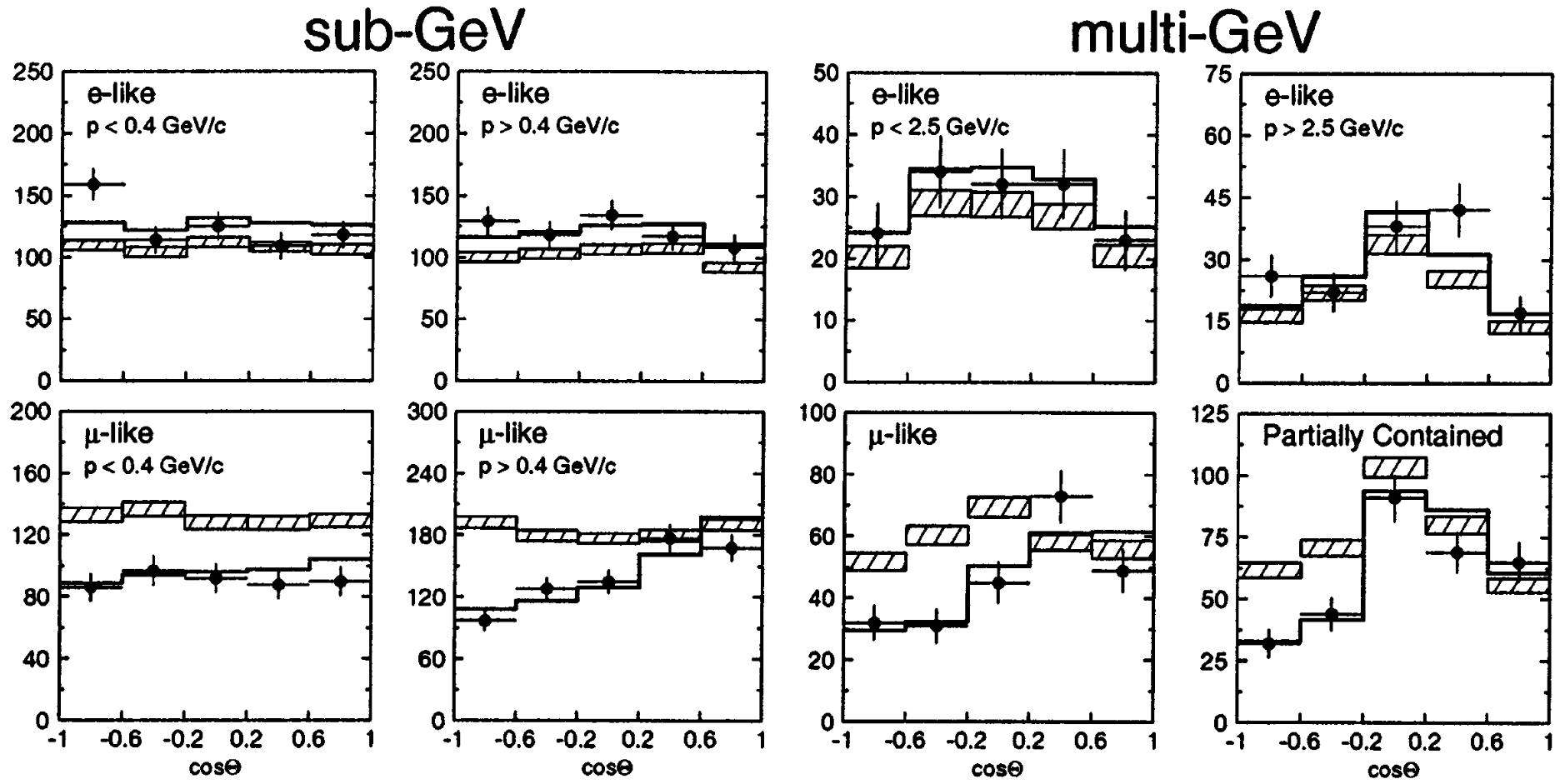}
	\caption{Zenith angle distribution of $\mu$-like (bottom panels) and $e$-like (top panels) events for sub-GeV and multi-GeV events in the Super Kamiokande. Upward-going and downward-going particles have $\cos\Theta<0$ and $\cos\Theta>0$, respectively. The dashed regions are Monte Carlo expectation for no oscillations and the bold line is the best-fit expectation for $\nu_\mu \rightarrow \nu_\tau$ oscillations~\cite{SuperK_plots}.}
	\label{fig:sk_plots_nu_oscillation}
\end{figure}

Figure~\ref{fig:sk_plots_nu_oscillation} shows the famous result from Super Kamiokande for the neutrino oscillation, where the zenith angle distribution is shown for muon-like and electron-like (bottom and top panels, respectively) events in two different set of data, sub-GeV and mult-GeV event. The $\mu$-like events with mult-GeV are separated in fully contained and partially contained\footnote{High energy muons will travel too further, escaping the detector.}, while the others have the momentum threshold displayed. Upward-going and downward-going particles\footnote{Vertically downward-going particles traveled 15~km in the atmosphere, while vertically upward-going particles have traveled about 13,000~km before interacting with the detector.} have $\cos\Theta<0$ and $\cos\Theta>0$, respectively. The dashed regions are Monte Carlo expectation for no oscillations, the bold line is the best-fit expectation assuming \mbox{$\nu_\mu\rightarrow\nu_\tau$} oscillation. 

It is noticeable that the $\mu$-like events show a disappearance of $\numu$ when comparing between upward-going ($\cos\Theta=-1$) and downward-going ($\cos\Theta=1$) muons, specially for the multi-GeV events. This means that muons crossing the Earth are missing and the data suggest that the oscillation $\numu\rightarrow \nutau$ is maximum. Looking at the data, the oscillation begins for horizontal neutrinos ($\cos\Theta\sim 0$), traveling a distance $L\sim1000$~km. Multi-GeV netrinos have energy of $E_\nu\sim 3~\GeV$, this means that the relation $E_\nu/L\sim\dm_{atm}\sim3\times10^{-3}~\eV^2$ (see Sec.~\ref{sec:averaged_probability}, Eq.~\ref{eq:matching_condition}) is well satisfied\footnote{The current value is $2.6\times10^{-3}~\eV^2$.}. While the zenith-angle distribution of $\mu$-like events is clearly asymmetric, $e$-like events show no asymmetry.

\subsubsection{SNO measurements}

The SNO solar experiment~\cite{SNO_first_phase_data_2007} is a Water Cherenkov detector with a major improvement: SNO employs 1 kton of salt heavy water (D$_2$O) rather than water, allowing to measure separately the $\nue$ and $\nu_{\mu,\tau}$ fluxes by different neutrino interactions. SNO is the first solar neutrino appearance experiment. The experiment used 12~m radius acrylic spheres shell at a depth of 6010~m of water equivalent with 9,456 PMTs to readout the signals.

The SNO was sensitive to charged-current (CC) and neutral-current, (NC) interactions and electron elastic scattering (ES), they can be written respectively as: 
\begin{align}
	\label{eq:sno_cc}
	\nue + d &\rightarrow p+p+e^- \hspace{70pt} \text{(CC)}\\
	\label{eq:sno_nc}
	\nu_\ell+d &\rightarrow p+n+\nu_\ell \hspace{70pt} \text({NC)} \\
	\label{eq:sno_es} 
	\nu_\ell+e^- &\rightarrow \nu_\ell+e^- \hspace{83pt} \text{(ES)}.
\end{align}

The CC interaction (Eq~\ref{eq:sno_cc}) is sensitive only to $\nue$ flux, while the NC and ES are sensitive to all flavors but cannot distinguish between them. The fragmentation of deuterons in the NC interaction (Eq.~\ref{eq:sno_nc}) has a threshold of $E_\nu>2.2~\MeV$ and it is detected by 6.25~MeV $\gamma$-ray emission after the neutron capture~\cite{vissani}.

\begin{figure}[h!]
	\centering
	\includegraphics[width=0.7\linewidth]{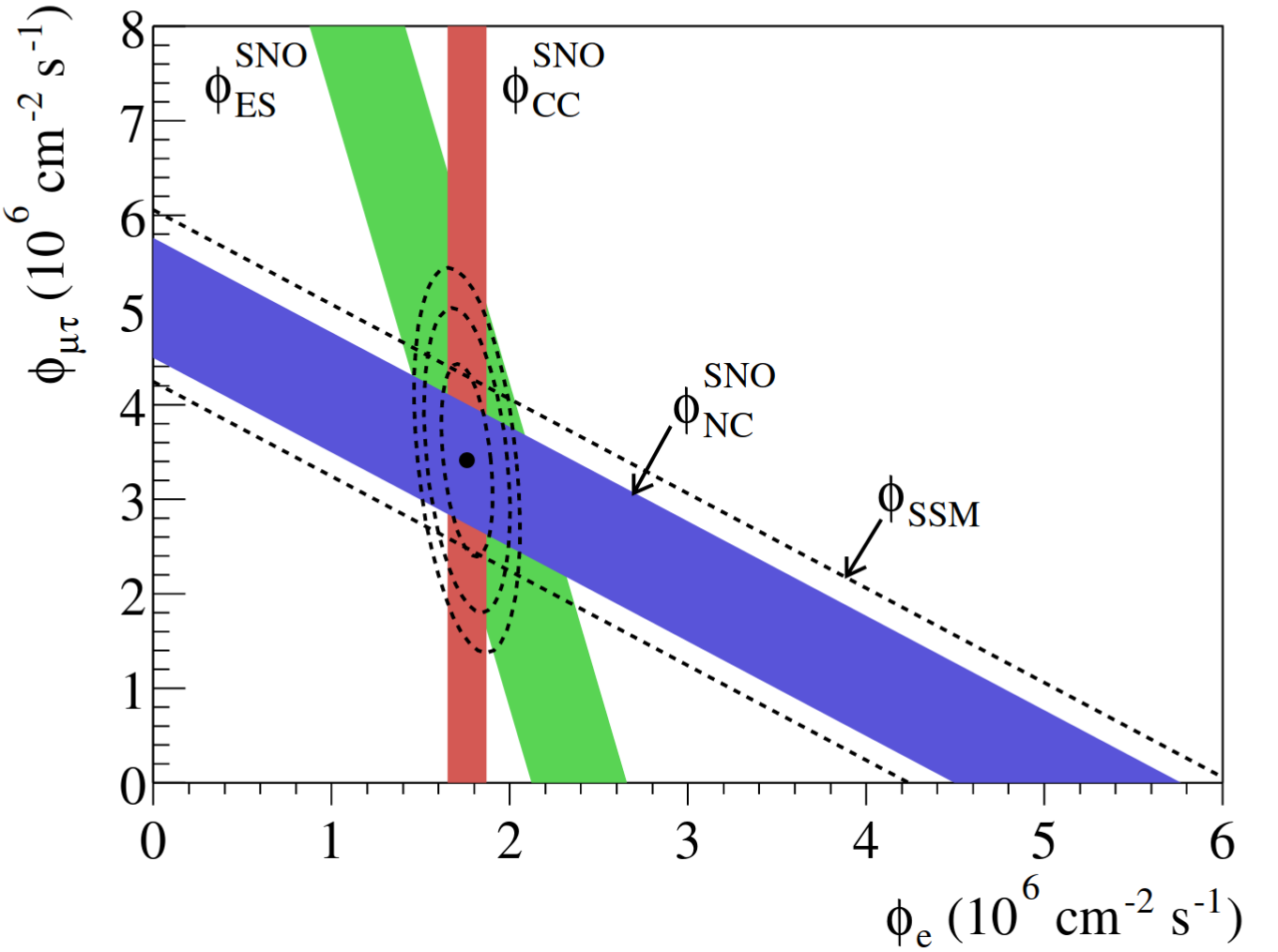}
	\caption{Flux of $^8$B solar neutrino for the SNO experiment measured via CC, NC and ES interactions ($\phi^{SNO}_{CC}$, $\phi^{SNO}_{NC}$ and $\phi^{SNO}_{ES}$). The axes are the inferred fluxes of $\numu+\nutau$ versus $\nue$. The sensitive of NC and ES interactions give the slope of the fluxes bands. The CC measurement is sensitive only to $\nue$, so the slope is infinity. The intersection of three bands give the best estimation of $\phi_{\mu,\tau}$ and $\phi_e$, with the 1$\sigma$, 2$\sigma$ and 3$\sigma$ confidence level expressed as dashed ellipses. The solar neutrino flux predicted by the SSM is indicated as $\phi_{SSM}$~\cite{SNO_first_phase_data_2007}.}
	\label{fig:sno_plots_nu_oscillation}
\end{figure}

Figure~\ref{fig:sno_plots_nu_oscillation} shows the famous SNO experiment plot that gives a 5$\sigma$ evidence for $\nue\rightarrow\nu_{\mu,\tau}$ appearance, proving that the Solar Standard Model (SSM) correctly predicted the neutrino flux and that electron neutrinos where not disappearing but rather changing flavor to $\mu$, $\tau$~neutrinos. The plot is made to show the flux of $\nu_{\mu,\tau}$ as function of the $\nue$ solar flux from the $^8$B process (see Fig.~\ref{fig:solar_chain} and~\ref{fig:solar_nu_spectrum}). The red area represents the neutrino flux measured with the CC interactions ($\phi^{SNO}_{CC}$), were the $\nu_{\mu,\tau}$ flux is unknown. The NC and ES measured neutrino flux in blue and green ($\phi^{SNO}_{NC}$ and $\phi^{SNO}_{ES}$) are composed by $\nue$ and/or $\nu_{\mu,\tau}$ fluxes, therefore they are represented as function of $\nue$ flux. The intersection of the three bands gives the best estimate of $\phi_e$ and $\phi_{\mu,\tau}$ with 1$\sigma$, 2$\sigma$ and 3$\sigma$ confidence level. The data clearly agrees with the SSM predicted flux ($\phi_{SSM}$)~\cite{SNO_first_phase_data_2007}.

\subsubsection{Oscillation parameters}

\begin{figure}[h!]
	\centering
	\includegraphics[width=0.99\linewidth]{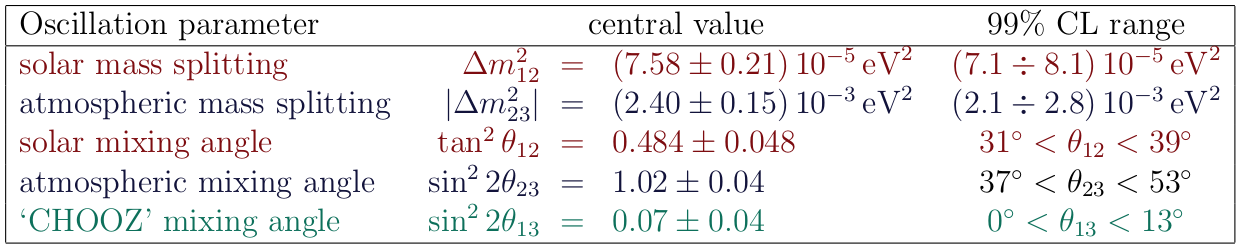}
	\caption{Summary of neutrino masses and mixing from oscillation data~\cite{vissani}.}
	\label{fig:nuparamsummary}
\end{figure}

The neutrino oscillation has been extensively studied and many experiments contributed to retrieve its parameters. A summary of the squared-mass difference and mixing angles, as the region where the confidence level (C.L.) is 99\%, is given in the table in Figure~\ref{fig:nuparamsummary} by~\cite{vissani}.

\section{Neutrino oscillation}

Neutrino oscillation was considered already in 1967 by Pontecorvo~\cite{Pontecorvo1967}, only 9 years after the first direct observation of neutrinos in 1956 by Cowan and Reines. However, only in the years 2000 the anomalies in different neutrino flavors were confirmed by the experiments KamLAND, (Super)Kamiokande and K2K~\cite{KamLAND2002,SuperK2004,K2K2006}. To understand the results from this experiments and the physical goal of DUNE, it is important to understand the process of oscillation and explore it's characteristics. 

It happens that, the evidence of neutrino oscillation implies massive neutrinos and flavor-mixing in the leptonic charged current (CC). This means that muon, electron and tau neutrinos ($\nue$, $\numu$ and $\nu_{\tau}$, respectively), which are the CC interaction eigenstates, can be expressed as a linear combination of the mass eigenstates $\nu_1$, $\nu_2$ and $\nu_3$ (with masses $m_1$, $m_2$ and $m_3$ respectively) in the form: 
\begin{equation}
\label{eq:nu_matrix}
	\begin{pmatrix}
	\nue \\
	\numu \\
	\nutau 
	\end{pmatrix}
=
	\begin{pmatrix}
	U_{e,1} & U_{e,2} &  U_{e,3} \\
	U_{\mu,1} & U_{\mu,2} &  U_{\mu,3} \\
	U_{\tau,1} & U_{\tau,2} &  U_{\tau,3} 
	\end{pmatrix}
	\begin{pmatrix}
	\nu_1 \\
	\nu_2 \\
	\nu_3 
	\end{pmatrix},	
\end{equation}
or in a more compact way:
\begin{equation}
\label{eq:nu_mixing}
	\ket{\nu_{\alpha}}  = \sum_{k = 1}^{3} U^*_{\alpha k} \ket{\nu_k} \qquad (\alpha = e, \mu, \tau),
\end{equation}
Where $U_{\alpha k}$ is the Pontecorvo-Maki-Nakagawa-Sakata (PMNS) matrix.

Reminding that $U$ must be unitary:
\begin{equation}
\label{eq:orthonormal}
U^\dagger U = 1 \qquad \Longleftrightarrow \qquad \sum_\alpha U_{\alpha k}^* U_{\alpha j}^{ } = \delta_{kj}.
\end{equation}

In this way, the massive neutrino states $\ket{\nu_k}$ are eigenstates of the Hamiltonian $\mathcal{H}$ in the form $\mathcal{H} \ket{\nu_k} = E_k \ket{\nu_k}$ where $E_k = \sqrt{p^2 + m_k^2}$ is the energy. With this, the Schrödinger equation will have the solution as: 
\begin{align}
i\frac{d}{dt} \ket{\nu_k(t)} &= \mathcal{H} \ket{\nu_k(t)} \\
\ket{\nu_k(t)} &= e^{-iE_k t} \ket{\nu_k}.
\end{align}

Which describes the time evolution of the massive states. By using Eq.~\ref{eq:nu_mixing} one can retrieve the time evolution of a neutrino with flavor state $\alpha$ created at $t=0$ as:
\begin{equation}
\label{eq:nu_evolution}
\ket{\nu_\alpha(t)} = \sum_{k} U^*_{\alpha k}\, e^{-iE_k t}\, \ket{\nu_k}.
\end{equation}

Equation~\ref{eq:nu_evolution} already gives the principle of neutrino oscillation, it described how the linear combination of the mass eigenstates changes in time for a flavor $\alpha$ neutrino. One can calculate the transition probability of an $\alpha$ neutrino to a $\beta$ neutrino, that is, what is the probability of finding the original $\nua$ as $\nub$ at any given time. This can be achieved by writing the state $\nua$ as function of $\nub$. We can invert Eq.~\ref{eq:nu_mixing} performing:
\begin{equation*}
 \left[ \sum_\alpha U_{\alpha j}^{} \right]\ket{\nu_{\alpha}}  = \sum_\alpha \sum_{k} U_{\alpha j}^{} U^*_{\alpha k} \ket{\nu_k} \qquad
\Rightarrow \qquad
\sum_\alpha U_{\alpha j}^{} \ket{\nu_{\alpha}}  = \sum_{k} \left(\sum_\alpha U_{\alpha j}^{*} U^{}_{\alpha k}\right)^* \ket{\nu_k},
\end{equation*} 
where, using Eq.~\ref{eq:orthonormal} and inverting the sides of the equality, leads to:
\begin{equation*}
\ket{\nu_{k}}  = \sum_{\alpha = e, \mu, \tau} U_{\alpha k} \ket{\nu_{\alpha}},
\end{equation*}
or, as we are going to use,
\begin{equation}
\label{eq:inverting}
\ket{\nu_{k}}  =  \sum_{\beta = e, \mu, \tau} U_{\beta k} \ket{\nu_{\beta}}.
\end{equation}

Substituting Eq.~\ref{eq:inverting} back to Eq.~\ref{eq:nu_evolution} we have:
\begin{equation}
\label{eq:nubeta}
\ket{\nu_\alpha(t)} = \sum_{\beta = e, \mu, \tau} \left( \sum_k U^*_{\alpha k} e^{-iE_kt}U^{}_{\beta k} \right) \ket{\nu_{\beta}}.
\end{equation}

One can notice that at $t=0$ the sum inside the parenthesis becomes simply the unitary property of Eq.~\ref{eq:orthonormal}. For $t>0$, the neutrino becomes a superposition of all the three different flavors, but only if the mixing matrix is not diagonal\footnote{To see this, one can choose the electronic flavor and open the terms of the sum, for instance. The only term that will not vanish is $U^*_{e1}e^{-iE_1t}U_{e1}$, which no longer mixes the flavors.}.

Therefore, the amplitude of the transition between $\nua$ to $\nub$ as function of time ($A_{\nua \rightarrow \nub}(t)$) is defined as
\begin{equation}
A_{\nua \rightarrow \nub}(t) \equiv \braket{\nub}{\nua(t)} = \sum_k U^*_{\alpha k}U^{}_{\beta k} e^{-iE_kt}.
\end{equation}
And the transition probability is defined as:
\begin{equation}
\label{eq:probdef}
P_{\nua \rightarrow \nub}(t) \equiv |A_{\nua \rightarrow \nub}(t)|^2 = \sum_{k,j} U^{*}_{\alpha k} U^{}_{\beta k} U^{}_{\alpha j} U^{*}_{\beta j} e^{-i(E_k-E_j)t}.
\end{equation}

To move further in any analysis, one should take here the assumption that the neutrinos are ultra-relativistic. Giving that the masses are expected to be lighter than about 0.2~eV~\cite{vissani}, this is a good approximation from an experimental point of view. In this case, the limit can be taken as: 
\begin{equation}
E_k = \sqrt{p^2+m_k^2} = p\left( 1+\frac{m_k^2}{p^2} \right)^{1/2} \xrightarrow{p\gg m} E\left(1+\frac{m_k^2}{2E^2}\right),
\end{equation}
which allow us to use,
\begin{equation}
\label{eq:diff_energy}
E_k - E_j \simeq \frac{\Delta m_{kj}^2}{2E},
\end{equation}
where 
\begin{equation}
\label{eq:squaremass}
\Delta m_{kj}^2 =  m_k^2 - m_j^2
\end{equation}
is the squared-mass difference.

One can notice that $p = p_k = p_j$ was assumed to retrieve Eq.~\ref{eq:diff_energy}. As a matter of fact, the proper treatment would consist in considering the neutrinos as wave-packets, allowing massive neutrinos to have different velocities and, therefore, being able to reach an observer with different times. This effect has interesting implications from the theoretical point of view which will not be explored in this thesis. However the result of Eq.~\ref{eq:diff_energy} still holds when using the wave-packet treatment~\cite{Thomson,fundamentals_nu}.

Finally, we take into account that neutrino oscillation experiments do not measure the time of flight but rather the distance $L$ traveled by the neutrino. For ultra-relativistic neutrinos we have $L \approx t$ ($c=1$, natural units) and the probability of transition as function of distance and energy becomes:
\begin{equation}
\label{eq:prob_E_L_simple}
P_{\nua \rightarrow \nub}(L,E) = \sum_{k,j} U^{*}_{\alpha k} U^{}_{\beta k} U^{}_{\alpha j} U^{*}_{\beta j} \exp\left(-i \frac{\Delta m^2_{kj}L}{2E} \right).
\end{equation} 

We will come back to this equation later, lets first introduce the two-neutrinos mixing and understand better the processes in this specific case. Since in the end it will be shown that the experiments are usually sensitive to the oscillation between two of the three neutrinos flavor, this approach results to be pretty realistic. This simplification of the problem makes the experimental results and discussions much easier to be understood.

\subsection{Two neutrino mixing}

Assume only 2 massive neutrinos are coupled with the flavor eigenstates. In this case we can only have two flavor neutrinos oscillating from one to another. Lets consider $\alpha$ and $\beta$ as the flavors of the neutrinos which are a linear superposition of the mass eigenstates $\nu_1$ and $\nu_2$:
\begin{equation}
\label{eq:matrix_two_flavors}
\begin{pmatrix}
\nua \\
\nub
\end{pmatrix}
=
\begin{pmatrix}
\cos\theta & \sin\theta\\
-\sin\theta & \cos\theta
\end{pmatrix}
\begin{pmatrix}
\nu_1 \\
\nu_2
\end{pmatrix},
\end{equation}
where $0 \le \theta \le \pi/2$ is the mixing angle. $U^\dagger U =1$ is satisfied. Lets assume that $\nu_1$ is the lightest massive neutrino and the squared-mass difference will be positive as:
\begin{equation}
\label{eq:twosquaremass}
\Delta m^2 \equiv \Delta m^2_{21} \equiv m^2_2-m^2_1.
\end{equation}

Now, from Eq.~\ref{eq:prob_E_L_simple}, follows that
\begin{align}
P_{\nua \rightarrow \nub}(L,E) &= U^{*}_{\alpha 1} U^{}_{\beta 1} U^{}_{\alpha 1} U^{*}_{\beta 1} \cancelto{1}{exp\left(-i \frac{\Delta m^2_{11}L}{2E}\right)} + U^{*}_{\alpha 2} U^{}_{\beta 2} U^{}_{\alpha 1} U^{*}_{\beta 1}exp\left(-i \frac{\Delta m^2_{21}L}{2E}\right) \nonumber\\
& + U^{*}_{\alpha 1} U^{}_{\beta 1} U^{}_{\alpha 2} U^{*}_{\beta 2} exp\left(+i \frac{\Delta m^2_{21}L}{2E}\right) + U^{*}_{\alpha 2} U^{}_{\beta 2} U^{}_{\alpha 2} U^{*}_{\beta 2} \cancelto{1}{exp\left(+i \frac{\Delta m^2_{22}L}{2E}\right)}.
\end{align}

Which, using the fact that $U_{ij} = U^*_{ij}$ and the fact that $e^{i\phi} + e^{-i \phi} = 2\cos{\phi}$, can be compressed as: 
\begin{equation}
P_{\nua \rightarrow \nub}(L,E) = |U_{\alpha 1}|^2 |U_{\beta 1}|^2 |U_{\alpha 2}|^2 |U_{\beta 2}|^2 + 2 U_{\alpha 1} U_{\alpha 2} U_{\beta 1} U_{\beta 2} \cos{\left(\frac{\Delta m^2 L}{2E}\right)}.
\end{equation}

From here, we can explore the transition ($\nua \rightarrow \nub$) or the survival probability ($\nua \rightarrow \nua$). Using the matrix from Eq.~\ref{eq:matrix_two_flavors} the equation above becomes
\begin{equation}
P_{\nua \rightarrow \nub}(L,E) = 2 \cos^2\theta \sin^2\theta \left(1-\cos\left(\frac{\Delta m^2 L}{2E}\right)\right),
\end{equation}
which, using the trigonometrical properties $\sin2\phi = 2\cos\phi \sin\phi$ and $\cos2\phi = 1 - \sin^2\phi$, gives the transition probability: 
\begin{equation}
\label{eq:transition_prob_two_flavor}
P_{\nua \rightarrow \nub}(L,E) = \sin^2 2\theta \sin[2](\frac{\Delta m^2 L}{4E})
\end{equation}
or, in another common form,
\begin{equation}
\label{eq:transition_prob_two_flavor_cos}
P_{\nua \rightarrow \nub}(L,E) = \frac{1}{2} \sin^2 2\theta \left[ 1-\cos(\frac{\Delta m^2 L}{2E})\right].
\end{equation}

The easiest way to retrieve the survival probabilities is using the fact that one of the two cases, $\nua \rightarrow \nub$ or $\nua \rightarrow \nua$ should always occur, resulting in: 
\begin{equation}
\label{eq:survival_prob_two_flavor}
P_{\nua \rightarrow \nua}(L,E) = 1 - P_{\nua \rightarrow \nub}(L,E) = 1 - \sin^2 2\theta \sin[2](\frac{\Delta m^2 L}{4E}),
\end{equation}
which is the survival probability. 
\subsubsection{Averaged probability}
\label{sec:averaged_probability}

Experiments that evaluate the transition probability (Eq.~\ref{eq:transition_prob_two_flavor}) between different neutrinos flavors are called \textit{Appearance experiments}. For this experiments, if the final neutrino state is not present in the source, the background is reduced. Differently, the ones that measure the survival probability given by Eq.~\ref{eq:survival_prob_two_flavor} are called \textit{Disappearance experiments}. The choice between the two type of experiments is given by the sensitivity of the detector (threshold of interactions, flavor distinguishing, uncertainties, etc.) and the values of mixing angles.

There are two factors that must be taken into consideration: experiments have different ranges of energy and distance and measurements have uncertainties in energy and distance. As a consequence of this, neutrino experiments measure an average oscillation probability. As a starting point, it is more convenient to write  Equations~\ref{eq:transition_prob_two_flavor} and~\ref{eq:survival_prob_two_flavor} with units in agreement with the experiment ranges of energy and distance. Experiments such as nuclear reactors will usually operate in the~MeV and meters range, while Long baselines experiments (LBL) such as DUNE will operate in the~GeV and kilometers range, for instance. Lets take the case of DUNE, with $\Delta m^2$~in~$\text{eV}^2$, $E$~in~$\text{GeV}$ and $L$~in~$\text{km}$, we can rewrite the term inside the sine of Eq.~\ref{eq:transition_prob_two_flavor} as:
\begin{align}
	\frac{\Delta m^2 L}{4E} &= \left(\frac{1}{4}\right)\frac{\left({\dm}/{\eV^2}\right) \left({L}/{\text{km}}\right)}{E/\GeV} \; \left\lbrace \frac{\eV^2 \;\text{km}}{\GeV} \right\rbrace = \frac{1}{4} \frac{\dm[\eV^2]\; L[\text{km}]}{E[\GeV]} \; \left\lbrace \frac{10^3}{10^9} \cdot \eV \; \text{m} \right\rbrace \nonumber\\
	&= \frac{1}{4} \frac{\dm[\eV^2]\; L[\text{km}]}{E[\GeV]} \; \left\lbrace 10^{-6} \cdot \frac{1}{1.97327\times 10^{-7}} \;  \right\rbrace \approx 1.27 \; \frac{\dm[\eV^2]\; L[\text{km}]}{E[\GeV]},
\end{align}
\newline
where $E[\GeV]$ means that the energy is expressed in~GeV and natural units were used. Exactly the same procedure can be used for $L[\text{m}]$ and $E[\eV]$, which gives a more convenient way to write Eq.~\ref{eq:transition_prob_two_flavor} as:
\begin{align}
P_{\nua \rightarrow \nub}(L,E) &= \sin^2 2\theta \sin[2](1.27 \; \frac{\dm[\eV^2]\; L[\text{m}]}{E[\eV]}) \\
&= \sin^2 2\theta \sin[2](1.27 \; \frac{\dm[\eV^2]\; L[\text{km}]}{E[\GeV]}).
\end{align}

For the averaged probability, the simplest case is to consider that the measurement L/E follows a Gaussian distribution, with average value $\av{L/E}$ and standard deviation $\sigma_{L/E}$ as
\begin{equation}
\label{eq:gaussian_cosine}
	\phi\left(\frac{L}{E}\right) = \frac{1}{\sqrt{2 \pi \sigma^2_{L/E}}} \, \exp\left[\frac{(L/E-\av{L/E})^2}{2\sigma^2_{L/E}}\right].
\end{equation}

In this case, the transition probability must be averaged in terms of the cosine for Eq.~\ref{eq:transition_prob_two_flavor_cos} (or sine for Eq.~\ref{eq:transition_prob_two_flavor}) as
\begin{equation}
\label{eq:averaging}
\left\langle P_{\nua \rightarrow \nub}(L,E)\right\rangle = \frac{1}{2} \sin^2 2\theta \left[ 1-\av{\cos(\frac{\Delta m^2 L}{2E})}\right],
\end{equation}
where 
\begin{equation}
\label{eq:integral_to_be_solved}
\av{\cos(\frac{\Delta m^2 L}{2E})} = I = \int \cos(\frac{\Delta m^2 L}{2E}) \phi \left(\frac{L}{E}\right) \diff\frac{L}{E}.
\end{equation}

To solve the integral $I$ of Eq.~\ref{eq:integral_to_be_solved}, one can rewrite the cosine as sum of exponentials and use Eq.~\ref{eq:gaussian_cosine} as the probability density. After simplifications, we obtain:
\begin{equation}
I = \frac{1}{\sqrt{2 \pi \sigma^2}} \int \cos(\alpha x) e^{-\frac{(x-\mu)^2}{2 \sigma^2}} \diff x =  \frac{1}{2\sqrt{2 \pi \sigma^2}} \left({\int e^{i\alpha x} e^{-\frac{(x-\mu)^2}{2 \sigma^2}} \diff x + \int e^{-i\alpha x} e^{-\frac{(x-\mu)^2}{2 \sigma^2}} \diff x}\right),
\end{equation}  
which using the substitution $x-\mu = \sigma\sqrt{2} u$ becomes
\begin{equation}
I = \frac{\sqrt{2}\sigma}{2\sqrt{2 \pi \sigma^2}}\left[e^{i\alpha\mu}\int e^{i\alpha \sigma \sqrt{2}u } e^{-u^2}\diff u + e^{-i\alpha\mu}\int e^{-i\alpha \sigma \sqrt{2}u } e^{-u^2}\diff u\right].
\end{equation}

Noticing that
\begin{equation*}
	-(u^2-i\alpha\sigma\sqrt{2}u) = -\left(u-i\alpha \frac{\sqrt{2\sigma}}{2}\right)^2 - \frac{\alpha^2 \sigma^2}{2}
\end{equation*}
we can rewrite the integral as:
\begin{equation}
I = \frac{1}{2\sqrt{\pi}} e^{- \frac{\alpha^2 \sigma^2}{2}}\left[e^{i\alpha\mu}\int_{-\infty}^{\infty} e^{-\left(u-i\alpha\sqrt{2\sigma}/2\right)^2}\diff u + e^{-i\alpha\mu}\int_{-\infty}^{\infty} e^{-\left(u+i\alpha\sqrt{2\sigma}/2\right)^2}\diff u \right].
\end{equation}

Using the result  
\begin{equation}
I' = \int_{-\infty}^{\infty} e^{-(x\pm ib)^2} \diff x = \sqrt{\pi}
\end{equation}
it follows that 
\begin{equation}
I = \frac{1}{2} e^{-\frac{\alpha^2 \sigma^2}{2}} \left[e^{i\alpha\mu} + e^{-i\alpha\mu}\right] = e^{-\frac{1}{2}(\alpha \sigma)^2} \cos(\alpha\mu)
\end{equation}
and we finally obtain the average over the cosine function of Eq.~\ref{eq:integral_to_be_solved}: 
\begin{equation}
\label{eq:integral_solved}
\av{\cos(\frac{\Delta m^2 L}{2E})} = \exp(-\frac{1}{2}\left(\frac{\dm}{2} \sigma_{L/E}\right)^2) \cos(\frac{\dm}{2}\av{\frac{L}{E}}).
\end{equation}

We then have the averaged transition probability as:
\begin{equation}
\label{eq:averaged_transition_prob}
\left\langle P_{\nua \rightarrow \nub}(L,E)\right\rangle = \frac{1}{2} \sin^2 2\theta \left[ 1-\exp(-\frac{1}{2}\left(\frac{\dm}{2} \sigma_{L/E}\right)^2) \cos(\frac{\dm}{2}\av{\frac{L}{E}})\right].
\end{equation}

It is reasonable to assume an uncertainty $\sigma_{L/E}$ proportional to $\av{L/E}$~\cite{fundamentals_nu}, and since the uncertainties of L and E are independent, the uncertainty of the ratio is:
\begin{equation}
\left(\frac{\sigma_{L/E}}{\av{L/E}}\right)^2 = \left(\frac{\sigma_L}{\av{L}}\right)^2 + \left(\frac{\sigma_E}{\av{E}}\right)^2,
\end{equation} 
where $\av{L}$ and $\sigma_L$ are the average distance and the distance uncertainty and $\av{E}$ and $\sigma_E$ are the average energy and energy uncertainty, respectively.

There are, of course, different uncertainties for different ranges of energies and distances. In particular, minimum ionizing particles will have a better defined energy, while low or high energy particles will be more tricky to have their  energy reconstructed. However, as a first order approximation, one can assume that both, energy and distances, have uncertainties proportional to their values. It is worth considering the case of $\sigma_{L/E} = 0.2\av{L/E}$~\cite{fundamentals_nu}, as an illustrative one.

\begin{figure}[H]
	\centering
	\includegraphics[width=0.95\linewidth]{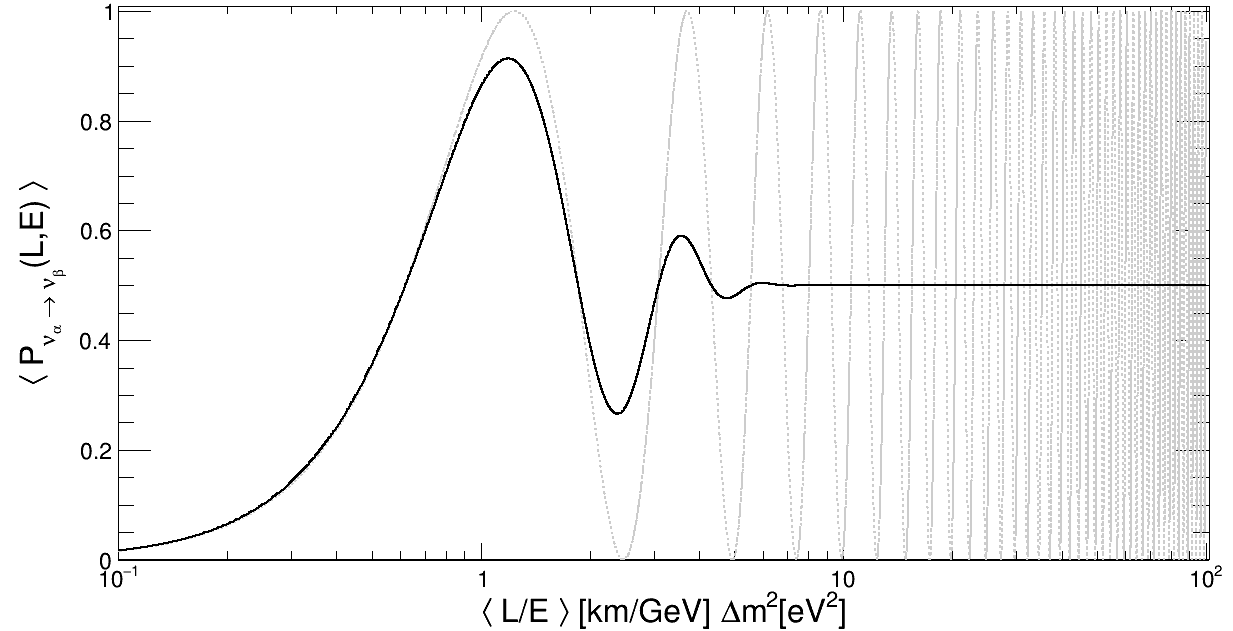}
	\caption{Transition probability of $\nua \rightarrow \nub$ for $\sin^22\theta=1$ as function of $\av{L/E}[\text{km/GeV}] \dm[\eV^2]$. The solid line is the averaged transition from Eq.~\ref{eq:averaged_transition_prob} with $\sigma_{L/E} = 0.2\av{L/E}$ and the dashed line comes from Eq.~\ref{eq:transition_prob_two_flavor_cos} without any averaging.}
	\label{fig:nu_osc_theorical}
\end{figure}

Figure~\ref{fig:nu_osc_theorical} shows the averaged and unaveraged (solid and dashed lines) transition probability from $\nua$ to $\nub$ as function of $\av{L/E}[\text{km/GeV}] \dm[\eV^2]$ with $\sin^22\theta=1$ and $\sigma_{L/E} = 0.2\av{L/E}$, using Eq.~\ref{eq:averaged_transition_prob} and~\ref{eq:transition_prob_two_flavor_cos} respectively. 

This result implies that experiments will have a range of sensitivity to neutrino oscillation, making it necessary to match the distance and energy to the squared-mass difference. Experiments will usually have a fixed energy with variable distance or vice-versa. Matching the ratio L/E of the experiment is the so called \textit{sensitivity} to $\dm$ of an experiment~\cite{fundamentals_nu}. The result from Fig.~\ref{fig:nu_osc_theorical} indicates the best matching will happen when
\begin{equation}
\label{eq:matching_condition}
\frac{\dm L}{E} \sim 1
\end{equation}
is satisfied. If $\frac{\dm L}{E} \ll 1$ the oscillation probability is very small and the transition between flavors cannot be easily measured, if $\frac{\dm L}{E} \gg 1$, only information about $\sin^22\theta$ can be retrieved.

The reason why the two-flavor mixing formalism is a good-enough approximation for the three-flavor one relies on the fact that the square mass-differences are very distant from each other, making the frequency of oscillation too apart. As consequence, if an experiment satisfies the condition~\ref{eq:matching_condition} for a given $\dm$, it will not satisfy for the other and the oscillation is either negligible or have average equal to zero. In the next section, this result will be shown for the DUNE, emphasizing the similarity of the two methods hoping to convince the reader. 

\subsubsection{Type of experiments}

Regarding the sensitivity condition of Eq.~\ref{eq:matching_condition} one can check the sensitivity of few experiments in order to understand better the implications of these results. A summary of the different types of experiments and sensitivities is given in~\cite{fundamentals_nu,pdg}. A \textbf{Short baseline experiment (SBL)} has a nuclear reactors or accelerators as source of neutrinos. The reactor experiments will be constrained by the energy $E \sim 1~\MeV$ of antineutrinos from beta decays and need to have a distance L~$\sim$~10~m, to be sensitive to $\dm \sim 0.1~\eV^2$. In this case, the antineutrino's energy is too low to produce muons or taus, making it only possible to measure the survival probability of $\anue$. Accelerator experiments, on the other hand, have different energy ranges depending on the type of neutrino production and must be built to match the distance. The sensitivity here goes from 1 to 100~$\eV^2$~\cite{fundamentals_nu}.

For instance, The PROSPECT detector is placed 7 - 9 meters from the reactor, with a sensitivity of $\dm \sim 1~\eV^2$~\cite{pdg}. The SBND~\cite{sbnd} experiment will use a beam with energy around 1~GeV and a baseline 0.1~km, which gives a sensitivity around $\dm\sim0.1~\eV^2$. This experiments search for oscillations evidence with models beyond 3 neutrinos oscillation, such as sterile neutrinos.

The \textbf{Long baseline experiments (LBL)} operate with much greater distance from the source, about 1~km away in the case of reactors and 100~to~1000~km away for beam experiments. LBL reactor and beam experiments are sensitive to \mbox{$\dm\gtrsim 10^{-3}~\eV^2$}, because of their range of energy ($L/E\lesssim 10^3$~m/MeV for reactors and $L/E\lesssim 10^3$~km/GeV for beam experiments). Examples of reactor and beam experiments are Double CHOOZ and ICARUS~\cite{fundamentals_nu,ICARUS_old}, respectively. The experiments are sensitive to the so-called atmospheric mass splitting (see Fig.~\ref{fig:nuparamsummary}).  

Experiments with a source-detector distance larger by one or two orders of magnitude than LBL are called \textbf{Very Long-Baseline experiments (VLB)}. Here the mass difference sensitivity will be around $\dm\gtrsim 10^{-5}~\eV^2$ for reactors, which have a distance on the order of 100~km from the source. Neutrino beam experiments, on the other hand, will have a sensitivity of $\dm\gtrsim 10^{-4}~\eV^2$ and a distance of several thousands of~km. At this same range, \textbf{Atmospheric neutrino experiments (ATM)} are able to take measurements. They measure the flux of atmospheric neutrinos\footnote{Cosmic rays interaction in the upper atmosphere generate pions and kaons that will decay into muons. This negative (or positive) muons may decay in flight into electron (positron), muon neutrino (antineutrino) and electron antineutrino (neutrino). These neutrinos are called atmospheric neutrinos.}, which have a wide range of energy (from 500~MeV to 100~GeV), in two different ranges: around 20~km for neutrinos coming from above and about $1.3 \times 10^4 \text{km}$ for neutrinos crossing the Earth.

The KamLAND experiment~\cite{KamLAND2002}, for instance, has the detector $\sim$180~km away from the reactor and energy between 1 to 10~eV. This gives a sensitivity from $\sim 2\times10^{-5}$ to $ 2\times10^{-4}~\eV^2$. The DUNE experiment have its optimum energy around 0.5 to 4~GeV and it the far detector is 1300~km away from the near detector~\cite{DUNE_Report_V1,DUNE_Vol1_TDR}, giving a sensitivity of $\dm\sim 10^{-4}~\eV^2$. The Super Kamiokande~\cite{SuperK2004} is one example of ATM experiment. 

Finally, there are the \textbf{Solar neutrino experiments (Sol)}, where the distance of $1.5\times10^{11} \text{ m}$ between the Sun and the Earth is used. The thermonuclear reactions in the core of the Sun can only produce neutrinos with energy between 0.5 to 14~MeV~\cite{solar_nu_energy}, giving a sensitivity around $\dm \sim 10^{-12}~\eV^2$. The so-called solar mass splitting come to the fact that experiments had measured a deficit in the electronic neutrinos flux coming from the Sun, which was later confirmed by SNO experiment~\cite{SNO_first_phase_data_2007} detecting solar neutrinos and also by KamLAND with reactor ones.

A summary of the different neutrino oscillation experiments is given in Table~\ref{tab:summary_exp}, with their respective distance, energy and $\dm$ sensitivity.

\begin{table}[tbph!]
	\centering
	
	\caption{Summary of experiments types and sensitivities.}
	\label{tab:summary_exp}
	\resizebox{\textwidth}{!}{%
	\begin{tabular}{|lcccr|}
		\hline
		Type of experiment & L               & E                & $\dm$ sensitivity     & Example      \\ \hline
		Reactor SBL        & $\sim$~10~m & $\sim$1~MeV & $\sim$0.1~eV$^2$ & PROSPECT \\ \hline
		Accelerator SBL    & $\sim$~10~m to 1~km            & $\sim$~0.01 to 10$^2$~GeV                & $\sim$~1 to 10$^2$~eV$^2$                    &    SBND          \\ \hline
		Reactor LBL    &    $\sim$~1~km           & $\sim$~1~MeV                & $\sim$~10$^{-3}$~eV$^2$                     & Double CHOOZ             \\ \hline
		Accelerator LBL    &  $\sim$~10$^3$~km              &  $\gtrsim 1$~GeV               & $\gtrsim 10^{-3}~\eV^2$                     & ICARUS             \\ \hline
		Reactor VLB    &     $\sim$~10$^2$~km           & $\sim$~1~MeV                & $\gtrsim 10^{-5}~\eV^2$                     & KamLAND             \\ \hline
		Accelerator VLB    &  $\sim$~$10^4$~km              & $\gtrsim~1~\GeV$                & $\gtrsim~10^4~\eV^2$                     & DUNE             \\ \hline
		ATM    &   20 to 10$^4$~km             & 0.5 to 10$^2$~GeV                & $\sim$~10$^{-4}$~eV$^2$                     & (Super)Kamiokande             \\ \hline
		Sol    &   $\sim$~10$^11$~km             & 0.2 to 15$^2$~MeV                & $\sim$~10$^{-12}$~eV$^2$                     & SNO             \\ \hline
		
	\end{tabular}}
\end{table}

Although this is the two flavors approach, the measurements in the three flavors formalism will be driven by massive neutrinos combination that matches the requirement of Eq.~\ref{eq:matching_condition}. Therefore, the mixing angle in the matrix of Eq.~\ref{eq:matrix_two_flavors} represents one the three mixing angles $\theta_{12}$, $\theta_{13}$ or $\theta_{23}$ that will be described at Sec.~\ref{sec:threenu}.

It is worth to present how experiments usually show their sensitivities. If an experiment does not measure any oscillation, the data suggest that the transition between flavors might be too small to be detected. In this case, the averaged transition probability must be below an upper limit determined by the experiment $\left(P^{\text{max}}_{\nua \rightarrow \nub}\right)$. This means that~\cite{fundamentals_nu}: 
\begin{equation}
\label{eq:prob_limit}
\av{P_{\nua \rightarrow \nub}(L,E)} \le P^{\text{max}}_{\nua \rightarrow \nub},
\end{equation}
which, using Eq.~\ref{eq:averaging}, leads to a limitation in $\sin[2](2\theta)$ and $\dm$ as
\begin{equation}
\label{eq:sin_limit}
\sin[2](2\theta) \le \frac{2P^{\text{max}}_{\nua \rightarrow \nub}}{1-\av{\cos(\frac{\dm L}{2E})}}.
\end{equation}

\begin{figure}[H]
	\centering
	\begin{subfigure}{0.45\textwidth}
		\includegraphics[width=0.99\textwidth]{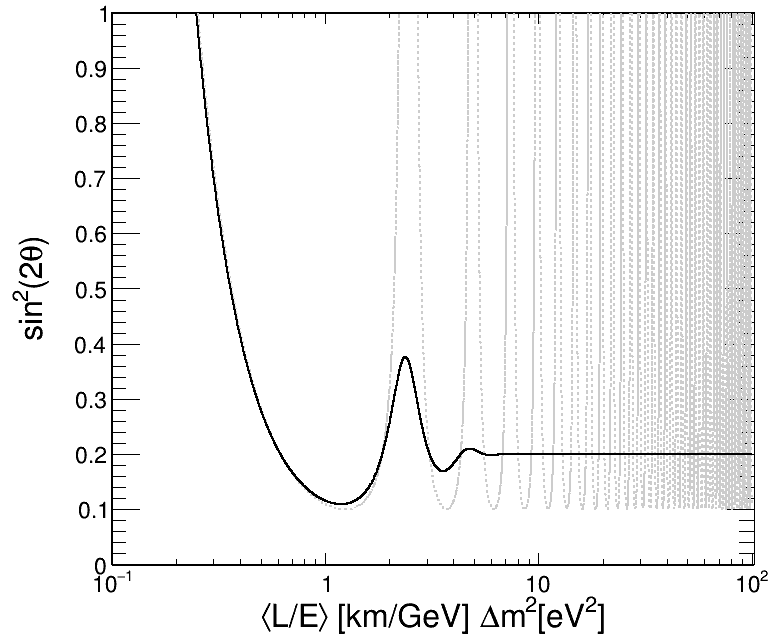}
		\caption{ }
		\label{fig:limits_sin_mass}
	\end{subfigure}
	\begin{subfigure}{0.45\textwidth}
		\includegraphics[width=0.99\textwidth]{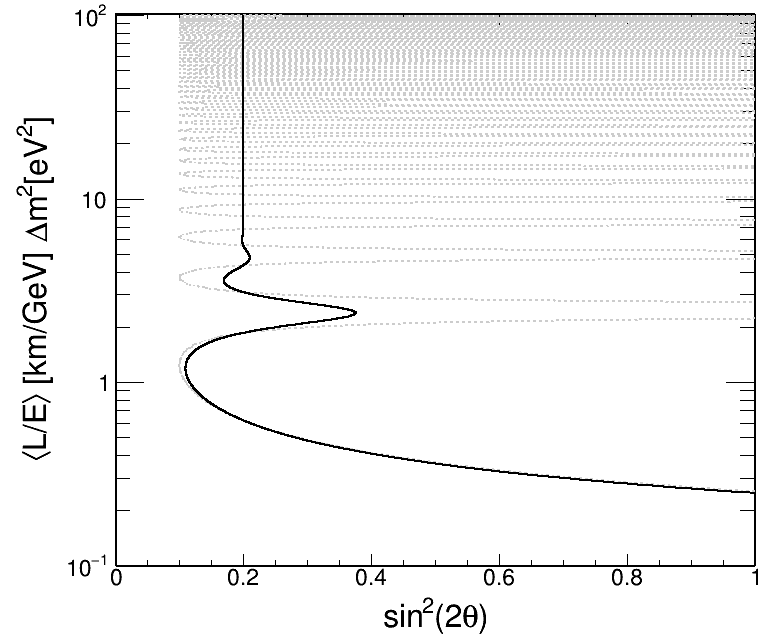}
		\caption{ }
		\label{fig:limits_sin_mass_inverted}
	\end{subfigure}
	\caption{\textbf{(a)} Limit between $\sin[2](2\theta)$ and $\av{E/L}[\GeV/\text{km}]\dm[\eV^2]$ given by Eq.~\ref{eq:sin_limit} with $P^{\text{max}}_{\nua\rightarrow\nub} = 0.1$. The black solid line is the evaluation with average given by Eq.~\ref{eq:integral_solved} with $\sigma_{L/E} = 0.2\av{L/E}$ and the gray dashed line is without any average. \textbf{(b)}~Inverting the axis as experiments will usually present their data.}
\end{figure}

Figure~\ref{fig:limits_sin_mass} shows the allowed region from Eq.~\ref{eq:sin_limit} not averaged (gray dashed lines) and averaged (black solid line) cosine term. The maximum probability was set to 0.1 and uncertainty used was $\sigma_{L/E} = 0.2\av{L/E}$. If the experiment did not detect any oscillation, it means that $\sin[2](2\theta)$ and $\dm$ lie in below the curve. Therefore the left region above the curve is then called excluded area. In the other hand, when the experiment actually detects oscillation, the same graph can be generated, but the excluded region will be on the opposite side of the curve, constraining the possible values of $\sin[2](2\theta)$ and $\dm$

Data results will be usually presented by inverting the axis as shown in Fig.~\ref{fig:limits_sin_mass_inverted}, only as a matter of preference. In the results from KamLAND~\cite{KamLAND2002} or K2K~\cite{K2K2006}, for instance, a graph similar to Fig.~\ref{fig:limits_sin_mass_inverted} can be found but with a fixed $\av{L/E}$ value, giving limits to the square mass-difference value instead.

The results presented in this section give the overview of oscillation experiments. Using this approximations one can understand where is the sensitivity and what are the results of the experiment. We can now come back to three flavors formalism and understand a little better the oscillations. 

\subsection{Three neutrinos mixing}
\label{sec:threenu}

One can represent the Pontecorvo-Maki-Nakagawa-Sakata (PMNS) matrix of Eq.~\ref{eq:nu_matrix} by the mixing angles in the same way done for Eq.~\ref{eq:matrix_two_flavors}. The PMNS mixing matrix $U$ is a 3 by 3 complex matrix that has 18 degrees of freedom. However, the unitary condition (Eq.~\ref{eq:orthonormal}) decreases it to 9, which can be expressed by 3 real angles and 6 complex phases\footnote{This is easily understood if you assume the matrix is real, in which case it can be parameterized  by the rotating matrix of Eq.~\ref{eq:rotating_matrix} (without the phase)}. One global phase can be absorbed and other 4 can be absorbed due to the invariance over global phases between the leptons and massive neutrinos (two for muons and taus and other two for $\nu_2$ and $\nu_3$, for example)~\cite{Thomson}. In this case, the PMNS matrix can be parameterized as:
\begin{equation}
	\label{eq:rotating_matrix}
	\small{
	U =
	\begin{pmatrix}
	1 & 0 & 0 \\
	0 & \cos\theta_{23} & \sin\theta_{23} \\
	0 & -\sin\theta_{23} & \cos\theta_{23} \\
	\end{pmatrix}
	\begin{pmatrix}
	\cos\theta_{13} & 0 & \sin\theta_{13}e^{-i\delta} \\
	0 & 1 & 0 \\
	-\sin\theta_{13}e^{i\delta} & 0 & \cos\theta_{13} \\
	\end{pmatrix}
	\begin{pmatrix}
	\cos\theta_{12} & \sin\theta_{12} & 0\\
	-\sin\theta_{12} & \cos\theta_{12} & 0\\
	0 & 0 & 1
	\end{pmatrix}.
}
\end{equation}

In order to be able to retrieve the individual elements, the expression can be rewritten using the definition $c_{ij} \equiv \cos{\theta_{ij}}$ and $s_{ij} \equiv \sin{\theta_{ij}}$:
\begin{equation}
\label{eq:pmns_matrix}
\small{
	U_{PMNS} = 
	\begin{pmatrix}
	U_{e,1} & U_{e,2} &  U_{e,3} \\
	U_{\mu,1} & U_{\mu,2} &  U_{\mu,3} \\
	U_{\tau,1} & U_{\tau,2} &  U_{\tau,3} 
	\end{pmatrix}
	=
	\begin{pmatrix}
	c_{12}c_{13} & s_{12}c_{13} & s_{13}e^{-i\delta} \\
	-s_{12}c_{23}-c_{12}s_{23}s_{13}e^{i\delta} & c_{12}c_{23}-s_{12}s_{23}s_{13}e^{i\delta} & s_{23}c_{13} \\
	s_{12}s_{23}-c_{12}c_{23}s_{13}e^{i\delta} & -c_{12}s_{23}-s_{12}c_{23}s_{13}e^{i\delta} & c_{23}c_{13}
	\end{pmatrix}.
}
\end{equation}

All the results and plots for experiments will be in terms of sines and cosines of the mixing angles. However, it is healthier to keep developing the probabilities in terms of the generic form of Eq.~\ref{eq:prob_E_L_simple} than with the (very) complex form presented above. Getting back to Eq.~\ref{eq:prob_E_L_simple}, one can separate the sum into three parts, when $k=j$, $k>j$ and $k<j$:
\begin{align}
\label{eq:opened_in_terms}
P_{\nua \rightarrow \nub}(L,E) = \sum_k |U_{\alpha k}|^2 |U_{\beta k}|^2 &+ \sum_{k>j,j} U^{*}_{\alpha k} U^{}_{\beta k} U^{}_{\alpha j} U^{*}_{\beta j} \exp\left(-i \frac{\Delta m^2_{kj}L}{2E} \right)\nonumber\\
& + \sum_{k<j,j} U^{*}_{\alpha k} U^{}_{\beta k} U^{}_{\alpha j} U^{*}_{\beta j} \exp\left(-i \frac{\Delta m^2_{kj}L}{2E} \right).
\end{align}

One should notice that for any of the two cases (k>j or k<j) it is true that $\dm_{kj} = -\dm_{jk}$ (Eq.~\ref{eq:squaremass}) and that $U^{*}_{\alpha k} U^{}_{\beta k} U^{}_{\alpha j} U^{*}_{\beta j}$ = $[U^{*}_{\alpha j} U^{}_{\beta j} U^{}_{\alpha k} U^{*}_{\beta k}]^*$. This makes clear that one can switch the indexes j and k in the last term, summing up for j$>$k, in Eq.~\ref{eq:opened_in_terms} and use the properties just mentioned to obtain:
\begin{align*}
\label{eq:opened_in_terms2}
P_{\nua \rightarrow \nub}(L,E) = \sum_k |U_{\alpha k}|^2 |U_{\beta k}|^2 &+ \sum_{k>j,j} U^{*}_{\alpha k} U^{}_{\beta k} U^{}_{\alpha j} U^{*}_{\beta j} \exp\left(-i \frac{\Delta m^2_{kj}L}{2E} \right)\nonumber\\
& + \sum_{k>j,j} \left[U^{*}_{\alpha k} U^{}_{\beta k} U^{}_{\alpha j} U^{*}_{\beta j} \exp\left(-i \frac{\Delta m^2_{kj}L}{2E} \right)\right]^*,
\end{align*}
which, using the definition of $\Re{Z} = (Z+Z^*)/2$, can be rewritten as
\begin{equation}
\label{eq:prob_L_E_opened}
P_{\nua \rightarrow \nub}(L,E) = \sum_k |U_{\alpha k}|^2 |U_{\beta k}|^2 + 2 \Re \sum_{k>j,j} U^{*}_{\alpha k} U^{}_{\beta k} U^{}_{\alpha j} U^{*}_{\beta j} \exp\left(-i \frac{\Delta m^2_{kj}L}{2E} \right).
\end{equation}

Now, using $P_{\nua \rightarrow \nub}(L=0,E)= \delta_{\alpha\beta}$, which means that there is no oscillation at the source, the first term in the right-hand side of Eq.~\ref{eq:prob_L_E_opened} can be written as:
\begin{equation}
\sum_k |U_{\alpha k}|^2 |U_{\beta k}|^2 = \delta_{\alpha\beta} - 2 \Re \sum_{k>j,j} U^{*}_{\alpha k} U^{}_{\beta k} U^{}_{\alpha j} U^{*}_{\beta j}.
\end{equation}

Substituting back in Eq.~\ref{eq:prob_L_E_opened} we have\footnote{Here we use the fact that $e^{-ix} = \cos{x} -i\sin x$ and that $\Re{iZ} = -\Im{Z}$, because if $Z = a+ib$, then $\Re{iZ} = \Re{ia-b} = -b$. This means that $\Re{Ze^{-ix}} = \Re{Z}\cos(x) + \Im{Z}\sin(x)$.}:
\begin{align}
\label{eq:prob_E_L_cos}
P_{\nua \rightarrow \nub}(L,E) = \delta_{\alpha\beta} &- 2 \sum_{k>j,j} \Re{U^{*}_{\alpha k} U^{}_{\beta k} U^{}_{\alpha j} U^{*}_{\beta j}}\left[1-\cos\left(\frac{\Delta m^2_{kj}L}{2E} \right)\right]\nonumber\\
& + 2 \sum_{k>j,j} \Im{U^{*}_{\alpha k} U^{}_{\beta k} U^{}_{\alpha j} U^{*}_{\beta j}}\sin\left(\frac{\Delta m^2_{kj}L}{2E} \right),
\end{align}
or in the form
\begin{align}
\label{eq:prob_E_L_sin}
P_{\nua \rightarrow \nub}(L,E) = \delta_{\alpha\beta} &- 4 \sum_{k>j,j} \Re{U^{*}_{\alpha k} U^{}_{\beta k} U^{}_{\alpha j} U^{*}_{\beta j}}\sin[2](\frac{\Delta m^2_{kj}L}{4E})\nonumber\\
& + 2 \sum_{k>j,j} \Im{U^{*}_{\alpha k} U^{}_{\beta k} U^{}_{\alpha j} U^{*}_{\beta j}}\sin\left(\frac{\Delta m^2_{kj}L}{2E} \right).
\end{align}


The transition probability, where $\beta\neq\alpha$, is straightforward in Eq.~\ref{eq:prob_E_L_cos} with only the Kronecker delta vanishing. The survival probability, on the other hand, can be taken noticing that the product between the mixing matrix elements inside the real and imaginary parts is real in the case where $\beta=\alpha$, giving the form
\begin{equation}
\label{eq:survival_prob_sin}
P_{\nua \rightarrow \nua}(L,E) = 1 - 4 \sum_{k>j,j} |U^{}_{\alpha k}|^2 |U^{}_{\alpha j}|^2 \sin[2](\frac{\Delta m^2_{kj}L}{4E}).
\end{equation}

It is straightforward to retrieve the survival probability of Eq.~\ref{eq:survival_prob_two_flavor} using the two mixing matrix (Eq.~\ref{eq:matrix_two_flavors}) in the equation above.  

\subsubsection{CPT, CP and T violations}

The oscillation for antineutrinos can be taken in the same way as for neutrinos, noticing that instead of the mixing from Eq.~\ref{eq:nu_mixing} we need to start from
\begin{equation}
\ket{\anua} = \sum_k U_{\alpha k} \ket{\anu_k}  \qquad (\alpha = e, \mu, \tau).
\end{equation}
for the antineutrinos. Without much effort, the transition probability can be found as
\begin{equation}
\label{eq:prob_E_L_sin_anti_simple}
P_{\anua \rightarrow \anub}(L,E) = \sum_{k,j} U^{}_{\alpha k} U^{*}_{\beta k} U^{*}_{\alpha j} U^{}_{\beta j} \exp(-i \frac{\dm_{kj} L}{2E}).
\end{equation}
Which is very similar to equation~\ref{eq:prob_E_L_simple}. Doing the same computations as for the neutrinos, the antineutrinos transition probability can be written as:
\begin{align}
\label{eq:prob_E_L_sin_anti}
P_{\anua \rightarrow \anub}(L,E) = \delta_{\alpha\beta} &- 4 \sum_{k>j,j} \Re{U^{*}_{\alpha k} U^{}_{\beta k} U^{}_{\alpha j} U^{*}_{\beta j}}\sin[2](\frac{\Delta m^2_{kj}L}{4E})\nonumber\\
& - 2 \sum_{k>j,j} \Im{U^{*}_{\alpha k} U^{}_{\beta k} U^{}_{\alpha j} U^{*}_{\beta j}}\sin\left(\frac{\Delta m^2_{kj}L}{2E} \right).
\end{align}

One can immediately notice the similarity between Eq.~\ref{eq:prob_E_L_sin} and~\ref{eq:prob_E_L_sin_anti} and the fact that they only differ by a negative sign in the imaginary part. This means that, if the mixing matrix from Eq.~\ref{eq:nu_matrix} have complex elements, the transition probability of neutrinos have a different amplitude than the antineutrinos one, i. e., neutrino oscillation violates charge and parity transformation (CP) by it self. The phase present in Eq.~\ref{eq:rotating_matrix}, if different than 0, $\pi$ or $2\pi$, will be responsible for the CP violation, because of that it is usually named CP phase ($\delta_{CP}$).  

\begin{figure}[h!]
	\centering
	\includegraphics[width=0.8\linewidth]{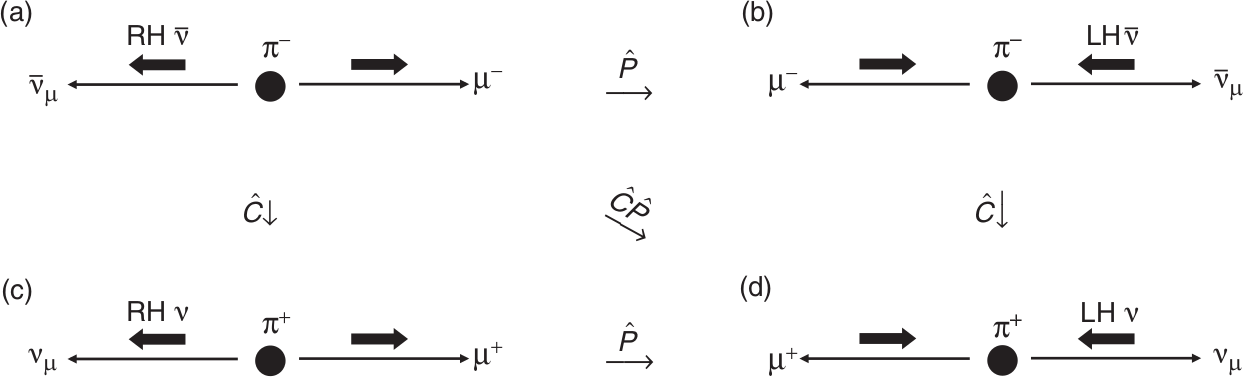}
	\caption{Pion decay (a), with the parity (b), charge (c) and CP (d) transformations, from~\cite{Thomson}.}
	\label{fig:piondecay}
\end{figure}

Particles and antiparticles are related by a Charge and Parity\footnote{The parity transformation is the spatial inversion only, this means $x\rightarrow-x$, $y\rightarrow-y$, $z\rightarrow-z$ and $t\rightarrow t$.} (CP) transformation, the example in Figure~\ref{fig:piondecay} by Thomson~\cite{Thomson} illustrates this relation very well. In (a) it shows a negative pion decaying into a right-handed (RH) muon and right-handed antineutrino. Performing a parity transformation the situation expressed in (b) is achieved with a left-handed (LH) antineutrino and muon. In this case, the weak-charged current matrix element is zero\footnote{As described in~\cite{Thomson} Chapter 6.4.2, ultra-relativistic particles, such as neutrinos, will have Chirality equivalent to Helicity. In this case, only left-handed particles and right-handed antiparticles will have a non-zero contribution in the weak-charged current.}. The same can be said of (c), where from (a) we performed only a charge transformation. The result is that the antiparticles will have helicity changed and the right-handed neutrino will cancel the weak current. In the end, transforming parity and charge (or charge and parity) the situation on (d) will give a non-zero current, as we have a left-handed neutrino and a left-handed anti-muon.

This is called the CP symmetry and, in principle, the same transition amplitude for the process in situation (a) and (d) should be found. The CP violation in the neutrino sector could contribute to explain better the baryon-antibaryon asymmetry of the universe~\cite{DUNE_Vol1_TDR,Thomson}. The CP transformation can be represented by $\nua \xleftrightarrow{\hat{CP}} \anua$.

\begin{figure}[h!]
	\centering
	\includegraphics[width=0.5\linewidth]{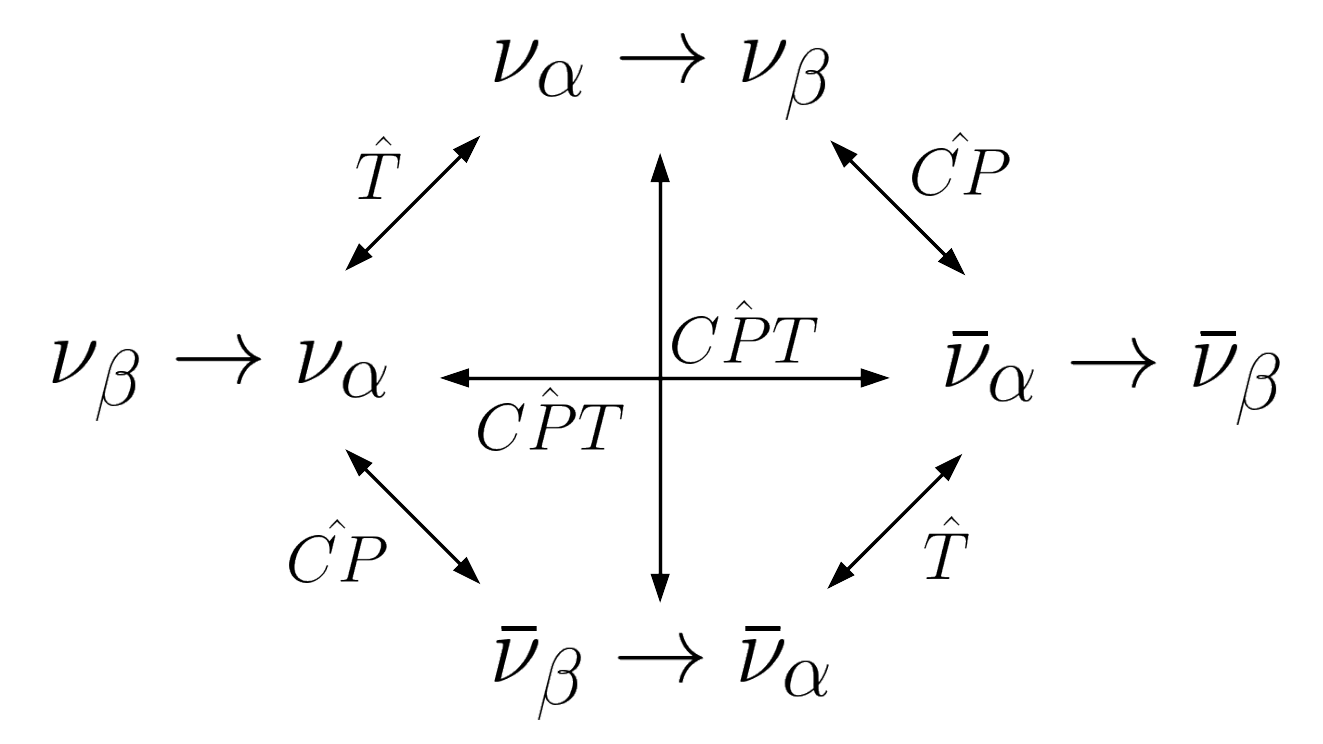}
	\caption{Schematic of CPT, CP and T transformations for neutrinos flavors transitions~\cite{fundamentals_nu}.}
	\label{fig:schematiccpt}
\end{figure}

The time transformation consists in looking the same event backwards, in this case a certain  $\nua \rightarrow \nub$ transition will change through a time transformation as \mbox{$\nua \rightarrow \nub \xleftrightarrow{\hat{T}} \nub \rightarrow \nua$}. And finally, if you combine CP and T transformation you have the CPT transformation that can be written as \mbox{$\nua \rightarrow \nub \xleftrightarrow{\hat{CPT}} \anub \rightarrow \anua$}. Figure~\ref{fig:schematiccpt} shows a simple scheme for all the three transformations for the transitions between two different neutrino's flavor. 

The asymmetry of any transformation can be evaluated by the difference in the transitions probabilities of the same event before and after transformations. Taking CPT as example, we have:
\begin{equation}
\label{eq:cpt_violation}
A^{\hat{CPT}}_{\alpha \beta} = P_{\nua \rightarrow \nub} - P_{\anub \rightarrow \anua}.
\end{equation}
And the same can be done for CP and T transformations following the schematic of \mbox{Fig.~\ref{fig:schematiccpt}}.

We can now easily evaluate the transition probability of $\anua \rightarrow \anub$, by simply changing the index of flavors in Eq. $\ref{eq:prob_E_L_sin_anti_simple}$. This leads to 
\begin{equation}
P_{\anub \rightarrow \anub}(L,E) = \sum_{k,j} U^{}_{\beta k} U^{*}_{\alpha k} U^{*}_{\beta j} U^{}_{\alpha j} \exp(-i \frac{\dm_{kj} L}{2E}),
\end{equation}
which is precisely the same as Eq.~\ref{eq:prob_E_L_simple}. This means that, in principle, there is no CPT violation in neutrinos oscillation.

The CP asymmetry, on the other hand, can be obtained by comparing Eq.~\ref{eq:prob_E_L_sin} and Eq.~\ref{eq:prob_E_L_sin_anti}. More precisely 
\begin{equation}
A^{\hat{CP}}_{\alpha \beta} = P_{\nua \rightarrow \nub} - P_{\anua \rightarrow \anub},
\end{equation}
which can be evaluated as:
\begin{equation}
A^{\hat{CP}}_{\alpha \beta} = 4 \sum_{k>j,j} \Im{U^{*}_{\alpha k} U^{}_{\beta k} U^{}_{\alpha j} U^{*}_{\beta j}}\sin\left(\frac{\Delta m^2_{kj}L}{2E} \right).
\end{equation}

If CPT is symmetric and the mixing matrix is complex, then the CP violation previously discussed implies T violation. The CPT symmetry implies, through Eq.~\ref{eq:cpt_violation}, that $P_{\nua \rightarrow \nub} = P_{\anub \rightarrow \anua}$, which causes the time asymmetry to be equal to the charge-parity asymmetry, because:
\begin{equation}
A^{\hat{T}}_{\alpha \beta} = P_{\nua \rightarrow \nub} - P_{\nub \rightarrow \nua} = P_{\nua \rightarrow \nub} - P_{\anua \rightarrow \anub} = A^{\hat{CP}}_{\alpha \beta},
\end{equation}
where we used a simply change of index in the equality given by the CPT symmetry of the example.

In such a case, one could measure the asymmetry in neutrinos oscillation by the T or CP transformed channels. That is, if the experiment measures the probability $P_{\nua \rightarrow \nub}$, it can measure either $P_{\nub \rightarrow \nua}$ or $P_{\anua \rightarrow \anub}$ to fully measure the asymmetry. This however is not a simple task because of the averaging effect described in the two flavor mixing, which is also present here.

The CPT symmetry is not a coincidence. In fact, CPT transformation is a symmetry of any local quantum field theory~\cite{fundamentals_nu} such as the one used for neutrino oscillations. However, Quantum Field Theory describes neutrino oscillations approximately, which can led to small violations of CPT.  


\subsubsection{Neutrino Mass Hierarchy}
\label{sec:nuMH}
\begin{figure}[h!]
	\centering
	\includegraphics[width=0.8\linewidth]{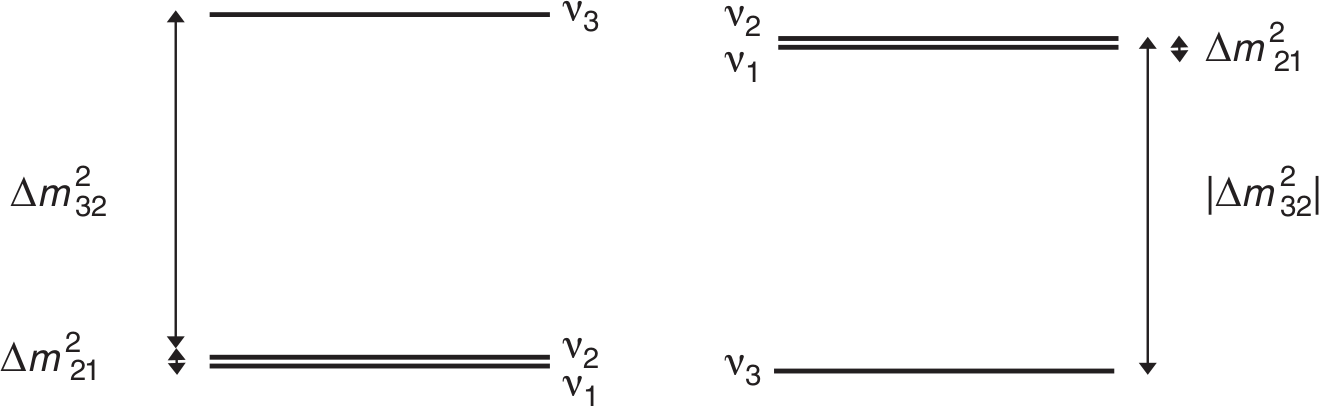}
	\caption{(Left) Normal neutrino mass hierarchy representation ($m_3>m_2$) and (Right) inverted mass hierarchy ($m_1>m_3$)~\cite{Thomson}.}
	\label{fig:masshierarchy}
\end{figure}

Current data allow to understand the modulus of the squared-mass difference, but not the sign. Because of this, it is not known in which order the massive neutrinos are distributed and what is the mass of the lightest neutrino.  Figure~\ref{fig:masshierarchy} shows the two possible mass hierarchies for the neutrinos choosing $m_2>m_1$. The left side is the so-called Normal Mass Hierarchy and the right side is the Inverted Mass Hierarchy.

The DUNE experiment will be able to determine the mass ordering with a significance of 5$\sigma$ level in 2-3 years for all possible values of $\delta_{CP}$ and, basically, regardless of all the other parameters constrains~\cite{DUNE_Vol1_TDR}.

\section{DUNE Context}
\label{sec:dune_oscillation}

The main goal of this section is to give an introduction and background for neutrinos oscillations in the framework of the DUNE experiment. We will use one of the main channels of detection $\numu\rightarrow\nue$ instead of a generic one. The distance from the near to far detector is defined as $L=1300$~km and will operate within 0.1 to 10~GeV, with the optimum energy around 0.5 to 5~GeV~\cite{DUNE_Vol1_TDR}. With this, we already know that DUNE will be sensitive to squared-mass difference around $10^{-3}~\eV^2$.

\begin{figure}[h!]
	\centering
	\includegraphics[width=0.8\linewidth]{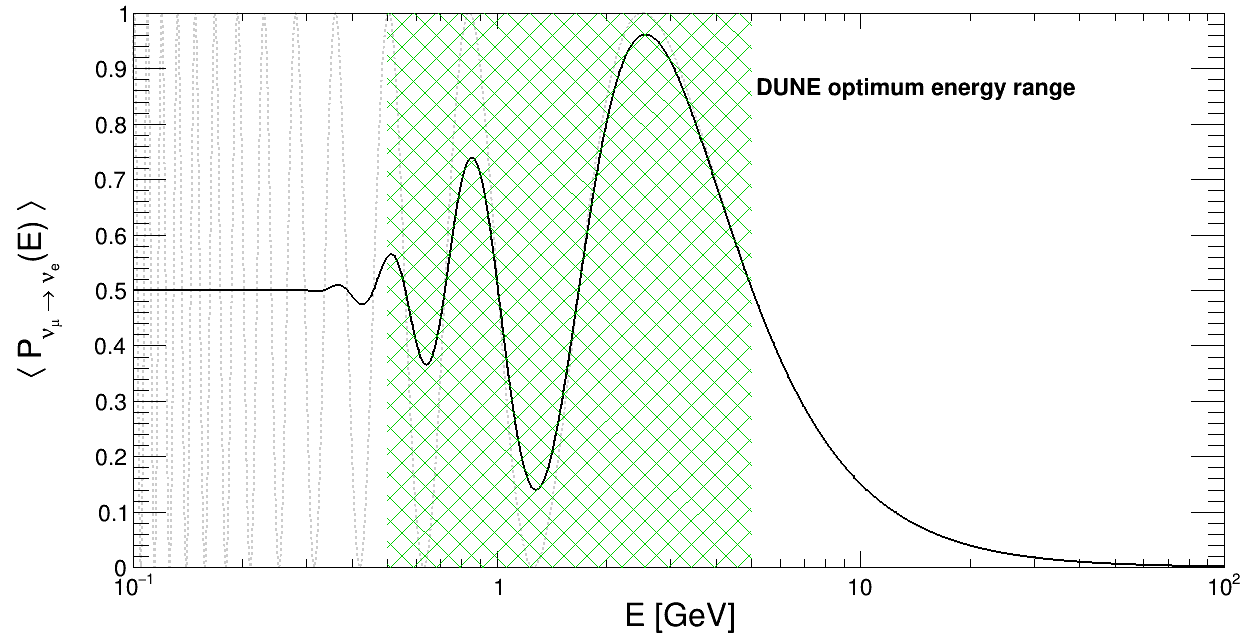}
	\caption{Averaged transition probability for $\numu$ to $\nue$ as function of energy from Eq.~\ref{eq:averaged_transition_prob} (solid line). The unaveraged transition probability from Eq.~\ref{eq:transition_prob_two_flavor_cos} (dashed line) is also displayed. The green crosshatched region represents the DUNE optimum energy.}
	\label{fig:duneoscbasic}
\end{figure}

Figure~\ref{fig:duneoscbasic} shows the transition probability $P_{\numu\rightarrow\nue}$ as function of energy from Eq.~\ref{eq:transition_prob_two_flavor_cos} (dashed line) and Eq.~\ref{eq:averaged_transition_prob} (solid line). The squared-mass difference $\dm_{32} = 2.4\cdot10^{-3}~\eV^2$ and $\sin[2](2\theta) = 1$ were chosen and the uncertainty selected is $\sigma_{L/E} = 0.13\av{L/E}$~\cite{DUNE_vol2}. The green area represents the optimum energy region. One can see that, in this ideal case, DUNE can measure three oscillations peaks and this explains why this distance and energy were chosen.

Lets make  a comparison between the two and three neutrinos formalism. The transition probability $P_{\numu\rightarrow\nue}$ can be extracted from Eq.~\ref{eq:prob_E_L_sin} using the PMNS matrix~\ref{eq:pmns_matrix}, it follows that~\cite{nu_ocs_prob_better,nu_osc_prob_vaccum}:
\begin{equation}
\label{eq:numu_nue_vacuum}
	P_{\numu\rightarrow\nue} \approx P_{atm} + 2\sqrt{P_{atm}}\sqrt{P_{sol}}\cos(\frac{\dm_{32}L}{4E}+\delta)+P_{sol},
\end{equation}
where
\begin{align}
	\sqrt{P_{atm}}\quad &\equiv \quad\sin\theta_{23}\sin2\theta_{13}\sin(\frac{\dm_{31}L}{4E}) \nonumber \\
	\sqrt{P_{sol}}\quad &\equiv \quad\cos\theta_{23}\cos\theta_{13}\sin2\theta_{12}\sin(\frac{\dm_{21}L}{4E}).
\end{align}
The reference $P_{atm}$ and $P_{sol}$ are due to the fact that the only squared-mass difference contributions are $\dm_{31}$ and $\dm_{21}$, respectively.

\begin{figure}[h!]
	\centering
	\includegraphics[width=0.8\linewidth]{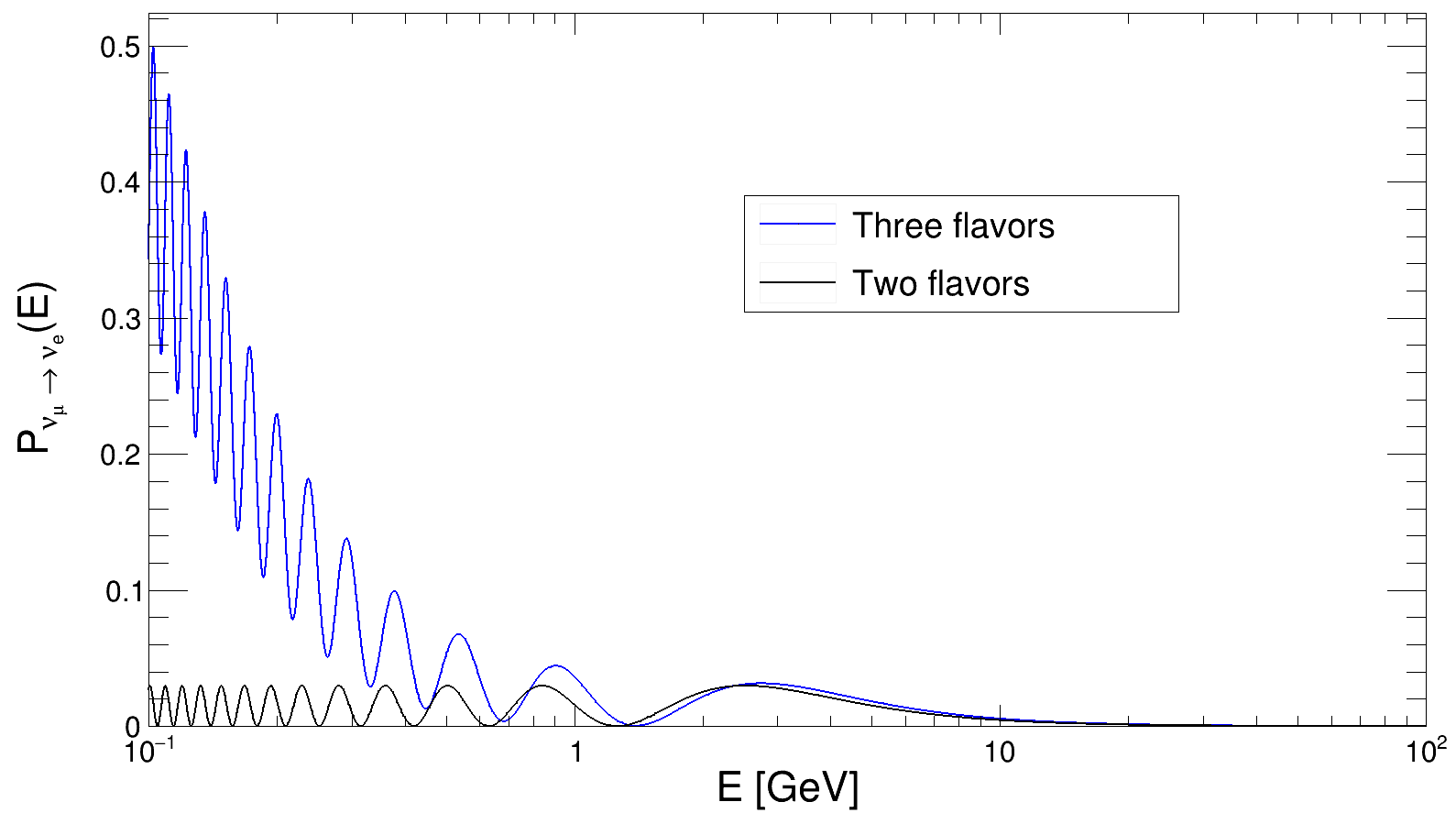}
	\caption{Three flavor (blue) and two flavor transition probability $P_{\numu\rightarrow\nue}$ as function of energy.}
	\label{fig:duneosccompare}
\end{figure}

Figure~\ref{fig:duneosccompare} shows the two different results of transitions probabilities (without averaging) for two neutrinos flavor (black, Eq.~\ref{eq:transition_prob_two_flavor}) and for three flavors (blue, Eq.~\ref{eq:numu_nue_vacuum}). All the parameters where set using the table of Figure~\ref{fig:nuparamsummary} and the phase was chosen $\delta = 0$. It is very clear that, although the two flavors approximation fails to fully describe the oscillation pattern, it can predicts quite well the oscillation peaks giving a hint of how well will the experiment perform in the energy (or mass) region.

By comparing Fig.~\ref{fig:duneoscbasic} and Fig.~\ref{fig:duneosccompare} one can notice that the oscillation due to the $\dm_{21}$ (solar) contribution only grows for lower energy, where the initial first peak (right to left) starts to increase as the oscillation due to $\dm_{31}$ (atmospheric) is already too fast. This is why the two flavors formalism manages to give a reasonable approach.

Now, one should notice from Fig.~\ref{fig:duneosccompare} that the transition probability for muon neutrinos to electron neutrinos is quite small. This small chance of transition between the two flavors causes the flux of electron neutrinos to be much smaller decreasing the statistics and making this channel not the ideal to measure $\numu$ oscillation. There are still two channels: the survival probability $\numu\rightarrow\numu$ or the transition to tau neutrino $\numu\rightarrow\nutau$.

To make it simple, but still with a realistic approach, we can use Eq.~\ref{eq:prob_E_L_sin} for all the channels and then use the approximation $\dm_{32} \cong \dm_{31} \cong \dm_{atm}$ and $\dm_{21} \cong \dm_{sol} \cong 0$. This is nicely done by~\cite{nu_osc_prob_vaccum} (with $\delta = 0$) which results\footnote{Note that this is the same approach as the two-mixing neutrinos, however now the amplitudes do not need to be guessed.} in:
\begin{align}
\label{eq:threeflavors}
P_{\numu\rightarrow\nue}(L,E) &= \sin[2](2\theta_{13})\sin[2](\theta_{23})\sin(\frac{\dm_{atm}L}{4E}) \\
\label{eq:threeflavors_tau}
P_{\numu\rightarrow\nutau}(L,E) &= \sin[2](2\theta_{23})\cos[4](\theta_{13}) \sin(\frac{\dm_{atm}L}{4E}) \\
\label{eq:threeflavors_mu}
P_{\numu\rightarrow\numu}(L,E) &= 1-\left[ \sin[2](2\theta_{13})\sin[2](\theta_{23}) + \sin[2](2\theta_{23})\cos[4](\theta_{13})\right] \sin(\frac{\dm_{atm}L}{4E})
\end{align}

\begin{figure}[h!]
	\centering
	\includegraphics[width=0.95\linewidth]{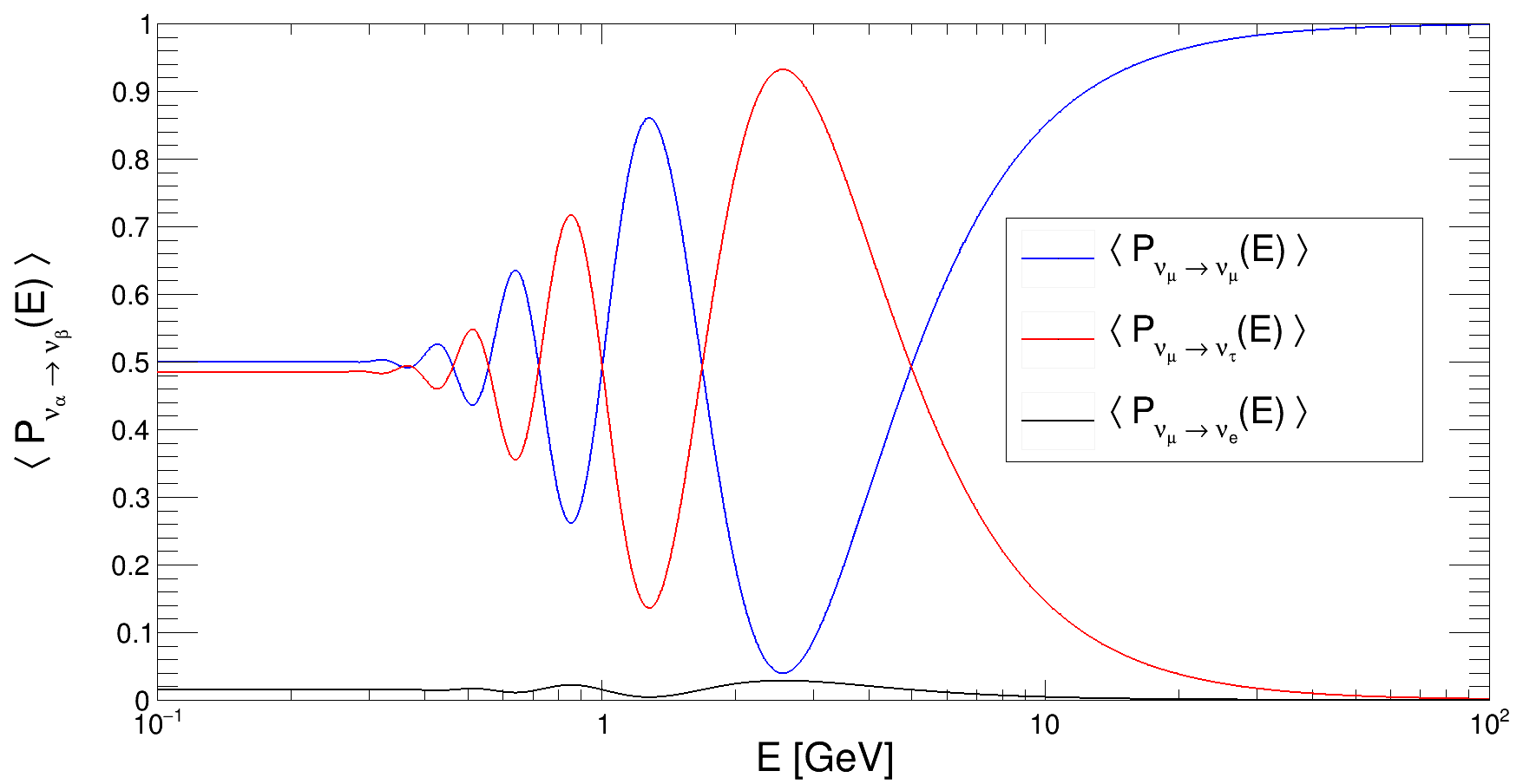}
	\caption{Averaged survival probability for $\numu$ (blue) and averaged transition probability $\numu\rightarrow\nutau$ (red) and $\numu\rightarrow\nue$ (black) as function of energy at the DUNE baseline $L=1300$~km.}
	\label{fig:nuoscthreeflavors}
\end{figure}

Figure~\ref{fig:nuoscthreeflavors} is the averaged transitions probabilities for equations~\ref{eq:threeflavors} to~\ref{eq:threeflavors_mu} with $\sigma_{L/E} = 0.13L/E$ and the constants from the table of Fig.~\ref{fig:nuparamsummary}. It can be easily seen that the transition probability to tau neutrinos or the survival probability are much higher than the transition probability to electron neutrinos. So the DUNE experiment should measure the oscillation in one of these two channels. To understand better this choice, one can notice that the first peak of oscillation (right to left) happens with an energy around 3~GeV, but the energy threshold to produce a $\tau$ lepton is $E_{\nu_{\tau}}>3.5~\GeV$~\cite{Thomson} (see Sec~\ref{sec:nu_detection}). This makes impossible to use this channel for the DUNE range of energy.

\subsection{Oscillation in matter}
\label{sec:oscillation_in_matter}
Before moving forward to DUNE scientific program, we need to go through neutrinos oscillation in matter. Up to here all the expressions where found for vacuum oscillations. However, the fact that the neutrinos cross matter instead of vacuum must be taken into account as it can induce a fake CP violation effect and allows the DUNE experiment to measure the mass ordering (Sec.~\ref{sec:nuMH})~\cite{DUNE_Vol1_TDR,nu_ocs_prob_better}. 

When traveling through matter, neutrinos can interact with it as incoherent inelastic scattering or coherent scattering. The first one is the usual process that causes the detection of the neutrinos in DUNE and have a very small cross section and, therefore, a small probability to happen. The latter is defined as an interaction that does not change the initial conditions of the medium after the interaction but interferes with the neutrino propagation. This sort of interaction will not affect the beam intensity, but the phase velocity of the neutrinos~\cite{pdg}. A good analogy to this effect is how the refraction index of mediums affects the phase velocity of light without absorbing it~\cite{vissani}.

Since Earth is composed by electrons instead of muons and taus, the interaction of $\nu_{e}$ will be different than $\numu$ and $\nutau$. All the three flavors will interact in the same way with electrons, protons and neutrons through neutral-current (NC), but only $\nue$ will interact with electrons mediated by $W$ boson in a charged-current (CC)~\cite{vissani,pdg} (see Sec.~\ref{sec:nu_detection}). Besides, a matter asymmetry can be expected due to the presence of electrons and absence of positrons in the Earth~\cite{DUNE_Vol1_TDR}. 

In this scenario, the transition probability $\numu\rightarrow\nue$ through a constant density matter will be given by~\cite{nu_ocs_prob_better}:
\begin{align}
\label{eq:osc_matter}
P(\numu\rightarrow\nue) &\cong \sin[2](\theta_{23})\sin[2](2\theta_{13})\frac{\sin[2](\Delta_{31}-aL)}{(\Delta_{31}-aL)^2}\Delta^2_{31} \nonumber \\
&+ \sin(2\theta_{23})\sin(2\theta_{13})\sin(2\theta_{12})\frac{\sin(\Delta_{31}-aL)}{(\Delta_{31}-aL)}\Delta_{31}\frac{\sin(aL)}{aL}\Delta_{21}\cos(\Delta_{31}+\delta_{CP})\nonumber \\
&+\cos[2](\theta_{23})\sin[2](2\theta_{12})\frac{\sin[2](aL)}{(aL)^2}\Delta_{21}^2,
\end{align}
where $\Delta_{ij} = \dm_{ij}L/4E$, $a = G_F N_e/\sqrt{2}$, $G_F$ is the Fermi constant, $N_e$ is the number density of electrons in the Earth. For the anti-neutrino transition $\anumu\rightarrow\anue$, both the $\delta_{CP}$ and $a$ change sign giving rise to the neutrino-antineutrino asymmetry due to CPV and matter effect~\cite{DUNE_Vol1_TDR}. 

\begin{figure}[h!]
	\centering
	\includegraphics[width=0.99\linewidth]{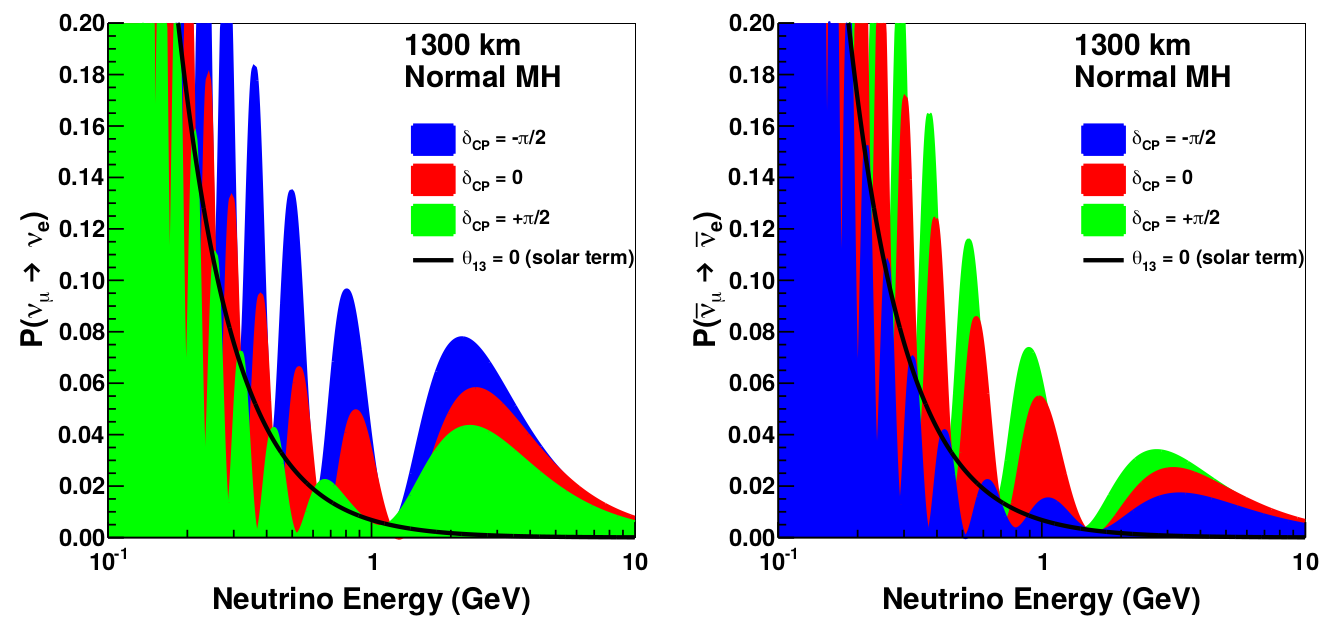}
	\caption{Electron neutrino appearance probability at the DUNE baseline $L=1300$~km as function of neutrino energy. Three different values of CP phase are used, $\delta_{CP}=-\pi/2$ (blue), 0 (red) and $\pi/2$ (green) for neutrinos (Left) and antineutrinos (right). Normal hierarchy is used and the black line indicates the oscillation probability if $\theta_{13}$ were equal to zero~\cite{DUNE_vol2}.}
	\label{fig:duneoscmatter}
\end{figure}

Figure~\ref{fig:duneoscmatter} shows the probabilities for neutrinos (left) and anti-neutrinos (right) for three values of $\delta_{CP}$ at the baseline distance $L=1300\;\text{km}$. The black line represents the oscillation only due to the solar term (making $\theta_{13}=0$) and ``Normal MH'' means Normal Mass Hierarchy. Present data corroborate with negative values of $\sin\delta_{CP}$, but poorly constrained, so all the values of $\delta_{CP}$ from $\pi$ to $2\pi$, including the CP-conservation values 0 and $\pi$, are consistent with neutrino data~\cite{DUNE_Vol1_TDR}.

Another new feature appears from Eq.~\ref{eq:osc_matter}: the transition is now sensitive to the sign of $\dm_{31}$. Comparing to equations~\ref{eq:numu_nue_vacuum} or~\ref{eq:threeflavors}, for instance, the sign of the squared-mass difference is suppressed by the sine square, allowing to retrieve only the modulus of $\dm_{ij}$. This new dependence on the signal is crucial to determine the neutrinos Mass Hierarchy (MH). One of the reasons why DUNE have a baseline of 1300~km is because the ability of distinguishing the CP asymmetry from matter effects starts around $\sim$1200~km and increases with the baseline~\cite{DUNE_Vol1_TDR,dune_baseline}. 

\begin{figure}[h!]
	\centering
	\includegraphics[width=0.9\linewidth]{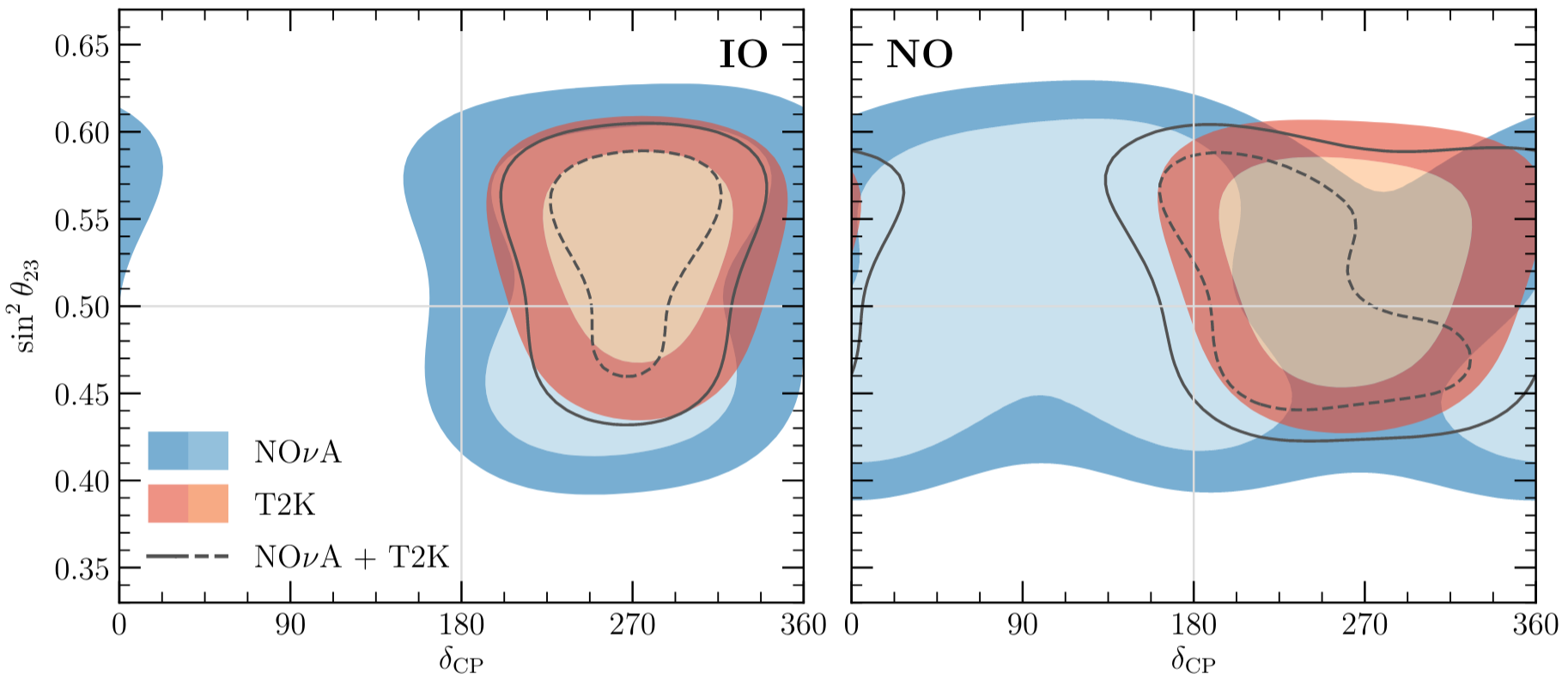}
	\caption{Allowed region inside 1$\sigma$ and 2$\sigma$ level (lighter and darker shading) for NOvA (blue shading), T2K (red shading) and their combination (black curves). Inverted ordering (IO) and normal ordering (NO) are taken into account (left and right panels, respectively)~\cite{nova_and_t2k_results}.}
	\label{fig:novaandt2kresults}
\end{figure}

Recent results from T2K~\cite{t2k_2018} and NOvA~\cite{nova_2019} showed a preference for maximum CP violation $\delta_{CP} = 3\pi/2$ when inverted neutrino mass ordering (see Sec.~\ref{sec:nuMH}) is assumed. However, the results are in tension when considering normal ordering, with $\delta_{CP}$ preferred as $3\pi/2$ for T2K and $\pi/2$ for NOvA. The combined result is shown by the $\delta_{CP}$ and $\sin[2](\theta_{23})$ two-dimensional plane in Figure~\ref{fig:novaandt2kresults}~\cite{nova_and_t2k_results}, where the allowed regions inside 1$\sigma$ and 2$\sigma$ are represented for NOvA (blue shading), T2K (red shading) and their combined result (black curves). The left panel represents the preferred inverted ordering (IO) while the right panel the normal ordering (NO).

Currently global fit analyses prefer the normal ordering, however, it went from 1.6$\sigma$ level to 2.7$\sigma$ due to the IO preferred data Super-Kamiokande~\cite{SuperK2004} presented in 2020~\cite{nova_and_t2k_results}. Nevertheless, there is still statistical agreement between the two experiment data and a joint analysis of the experiments will further improve the results. Both experiment have a moderated preference for  the upper octant of $\theta_{23}$ ($\sin[2](\theta_{23})>0.5$). 

\subsection{Parameters measurements}
\label{sec:parameters_dune}

The oscillation patterns of $\numu$ and $\anumu$ are going to be exhaustively studied by reconstruction of the energy spectra of $\numu$, $\anumu$, $\nue$ and $\anue$ as described above. To collect enough statistics the DUNE will have a beam of 1.2 MW minimum (upgradable to 2.4 MW after 6 years~\cite{DUNE_vol2}) and corresponding proton-on-target per year assumed as $1.1\times10^{21}$ POT~\cite{DUNE_Vol1_TDR}. This will result in an increase on $\nue$ and $\anue$ event yield  of one order-of-magnitude with respect to the current NOvA~\cite{nova_2019} and T2K~\cite{t2k_2018} data sample and even a bigger increase in the survival probabilities channels.

As result, besides being able to measure the mass hierarchy ordering and the CP violation (as described in Sections~\ref{sec:oscillation_in_matter} and~\ref{sec:nuMH}), DUNE will provide more precise measurements of some of the key parameters of neutrinos oscillations, like $\delta_{CP}$, $\sin[2](2\theta_{13})$, $\dm_{31}$ and $\sin[2](\theta_{23})$. Moreover, the experiment will be able to determine the octant of $\theta_{23}$, as it is known with the present data that $\theta_{23}$ is close to the maximal-mixing value of $\pi/4$ (see Fig.~\ref{fig:nuparamsummary}), but it is not known in which octant it lies, that is, if $\theta_{23}>\pi/4$ or $\theta_{23}<\pi/4$. 

\begin{figure}[!h]
	\centering
	\begin{subfigure}{0.49\textwidth}
		\includegraphics[width=0.99\textwidth]{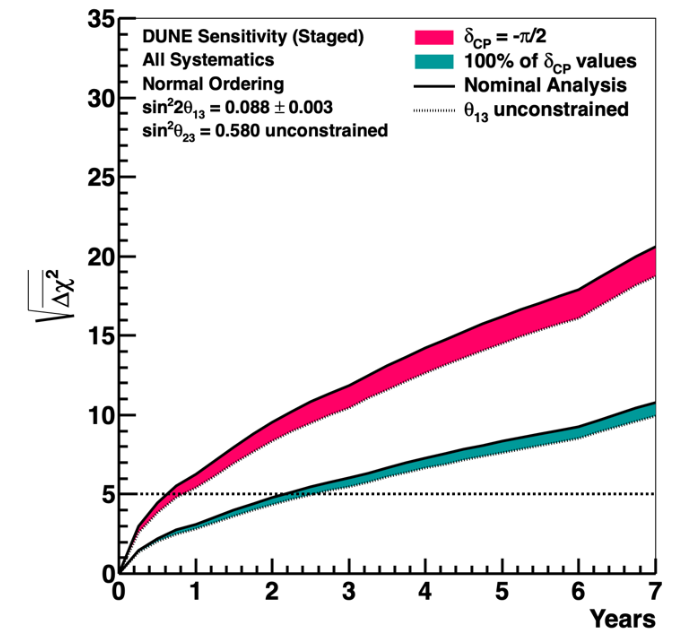}
		\caption{ }
		\label{fig:masshierarchysensitivity}
	\end{subfigure}
	\begin{subfigure}{0.49\textwidth}
		\includegraphics[width=0.99\textwidth]{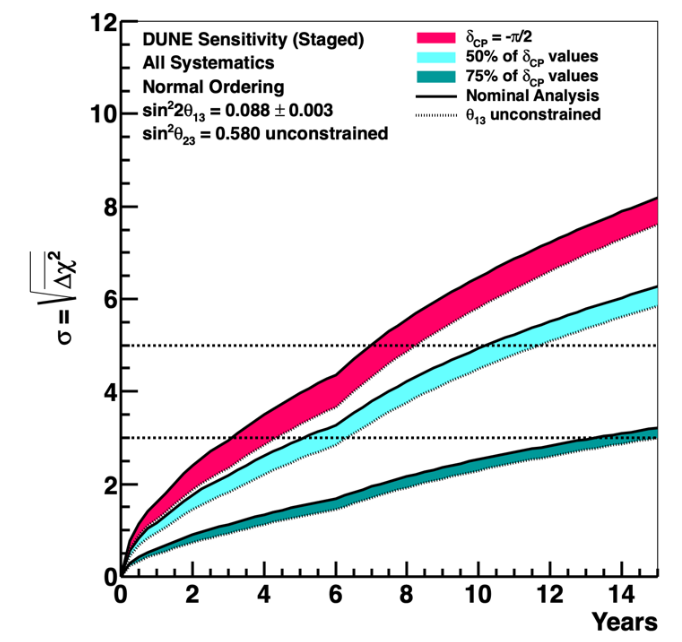}
		\caption{ }
		\label{fig:cpsensitivity}
	\end{subfigure}
	\caption{\textbf{(a)} Confidence level of the DUNE determination of the neutrino mass ordering in the case where $\delta_{CP}=-\pi/2$ (red area) and for all possible values of $\delta_{CP}$ (green area) as a function of years of operation. Normal ordering is assumed~\cite{DUNE_Vol1_TDR}. \textbf{(b)}~Confidence level for the determination of CP-violation (if any) for the case where $\delta_{CP}=-\pi/2$ (red area) and for 50\% (light green) and 75\% (green) of the possible values in normal ordering situation, as function of years of operation~\cite{DUNE_Vol1_TDR}.}
\end{figure}

Figures~\ref{fig:masshierarchysensitivity} and~\ref{fig:cpsensitivity} show the DUNE sensitivity for the mass hierarchy and CP violation, respectively, over the years of operation~\cite{DUNE_vol2}.  The colored areas represent different assumptions for $\delta_{CP}$, as a fixed value of $-\pi/4$ (red) and for every possible value (dark blue) for mass hierarchy for instance. The same is done for CP violation but for a fixed value, 50\% and 75\% of the values. The solid line delimits the region with nominal values reported in the figures and the dashed line is the limit for unconstrained $\theta_{13}$.

It can be seen that DUNE will be able to determine the mass ordering between 2-3 years of operation with a 5$\sigma$ confidence level as previously cited. The CP violation, on the other hand, can take up to 14 years operation to have a 3$\sigma$ sensitivity or about 6 years for 5$\sigma$ if $\delta_{CP} = -\pi/2$.
It is also reported in~\cite{DUNE_Vol1_TDR} the resolution of each parameter as function of the time of exposure\footnote{The exposure is given by kt$\cdot$MW$\cdot$year, this means a detector of 1 kt exposed to 1 MW beam for one year. To get the correct amount of years exposed one should remember that DUNE will have 4 phases of operation as described previously.}, which can change depending on its own values and on others parameters. For instance, $\delta_{CP}$ after 15 years of operation can have a resolution of 17 degrees (for $\delta_{CP}=\pm\pi/2$) or up to 7 degrees (for $\delta_{CP}=-\pi,0,\pi$).

\section{Beyond the neutrino beam}
Besides the measurements with the neutrino beam from the Long Baseline Neutrino Facility (LBNF), DUNE will be able to perform measurements with non-beam neutrinos. The photon detection (PD) system is crucial to measure the events not related to the neutrino beam, being responsible for the trigger system of supernova neutrinos and proton decay (if any). The PD will also enable the precise measurement of the $t_o$ for every event, time resolution for neutrinos bursts and reconstruction of events energy and topology in the drift coordinate. 

\subsection{Supernova neutrinos}
\label{sec:supernova_nu}
In 1987, the supernova SN 1987A~\cite{supernova1987} occurred in the Large Magellanic Cloud, about 50 kpc from Earth, and opened a new era of extragalactic neutrino astronomy. During this event, the 25 antineutrino events detected along three detectors have confirmed the basic physical picture of core collapse supernovas and gave new inputs for a wide range of new physics~\cite{DUNE_vol2}. 

It is expected that a core collapse supernova will occur in the Milk Way (within about 20 kpc) once every 10 to 50 years and also about the same rate for Andromeda Galaxy (the closest galaxy to Earth) about 780 kpc away~\cite{dune_supernova}. Although these events are rare, there is a reasonable chance of a supernova occurring in the few-decade long lifetime of a large neutrino detector. Because of that, it is important that at least one module of the DUNE far detector be online at all time to observe this unpredictable event, if and when it occurs~\cite{DUNE_Vol1_TDR}.

The Supernova burst (SNB) signature from the core collapse is expected to last few tens of seconds and produce from a few hundreds to several thousands interactions\footnote{This number depends on the distance of the supernova and the number of operating modules.}. The neutrino spectrum retrieved from these interactions will provide information about the progenitor, the collapse, the explosion and the remnant of the supernova~\cite{DUNE_vol2}. The neutrinos energy spectrum from SNB peaks around 10~MeV with appreciable flux to about 30~MeV. Such low energies are below the charged-current threshold for $\numu$, $\nutau$, $\anumu$ and $\anutau$ causing, together with the fact that neutral-current interactions have low cross sections and does not carry flavor information, the $\nue$ and $\anue$ flux measurements the most rich in information~\cite{DUNE_Vol1_TDR,DUNE_vol2}. 

Getting back to the importance of the photon detection system, it is expected that during the first 50 ms of a SNB 10 kpc away the mean interval between successive neutrino interactions is 0.5 - 1.7~ms depending on the model. Moreover, events closer to Earth can have a rate tens or hundreds of times higher~\cite{DUNE_vol2}. The TPC provides a resolution of 0.6 ms at 500~V/cm which is not enough for theses events. The PD system provides a resolution smaller than $10$~$\mu$s which allows DUNE to measure the flux without a time resolution constrain~\cite{DUNE_vol2}.

\begin{figure}[!h]
	\centering
	\includegraphics[width=0.7\linewidth]{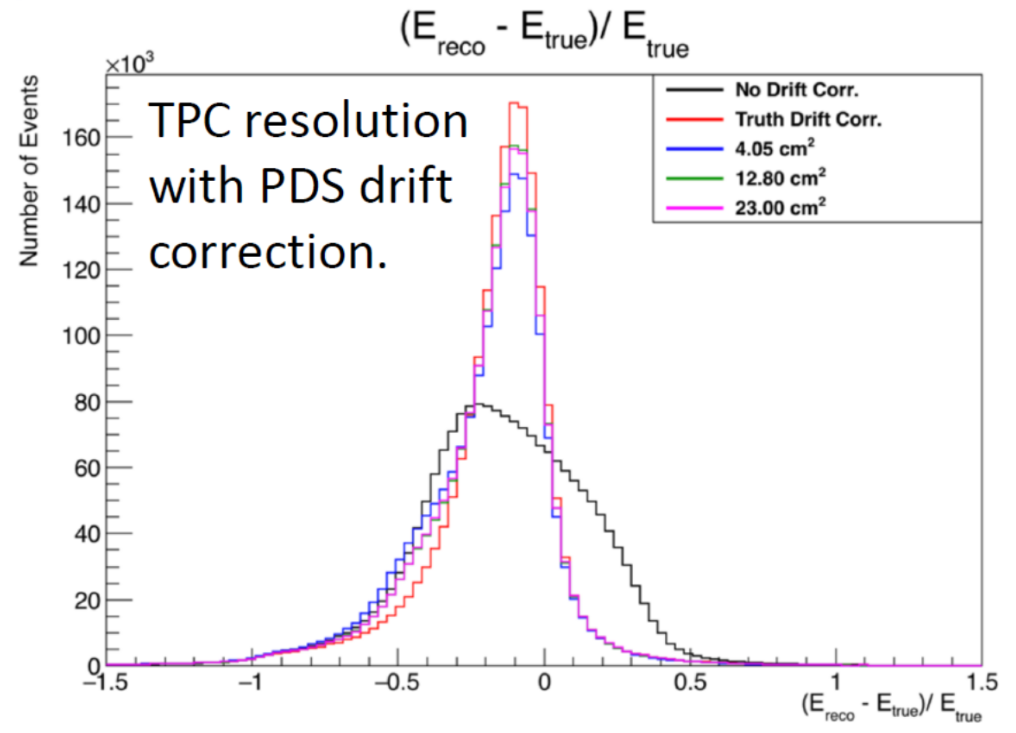}
	\caption{Residual between reconstructed and true energy for supernova neutrinos events without (black) and with (color) a drift time correction with the photon detection system to the reconstructed energy. The red histogram assumed the event vertex is known perfectly, while the others corresponds to effective area of the light collection. The 23~cm$^2$ histogram correspond roughly to the 0.5~\phe/MeV requirement~\cite{DUNE_vol2}.}
	\label{fig:energyres}
\end{figure}

Specially in the case of SN neutrinos, the PD system allows to improve the resolution of the reconstructed energy. By associating events on the TPC and the light output, one can determine where the event happened along the drift direction. This allows to define the detector fiducial volume and to correct the data for electron quenching. Figure~\ref{fig:energyres} shows the energy residuals for the supernova neutrinos events without (black) and with (colored) drift correction to the reconstructed energy. Blue, green and pink are effective areas (light collector area~$\times$~efficiency) of light collection and the red histogram assumes that event vertex is perfectly known. It can be easily noticed the improvement in resolution when the drift correction is applied. However, one should know that the effective are of 23~cm$^2$ is the minimum specification for DUNE.  

\subsection{Proton decay}

Due to the DUNE large fiducial mass, low background (because of its deep underground location), excellent event imaging, particle identification and calorimetry, proton and neutron decays signatures can be searched to probe baryon-number violation (BNV)~\cite{proton_decay_orlando}. It is unknown up to now if protons are stable particles and, therefore, if baryon-number is conserved or not. In the current Standard Model framework, baryon-number is defined as conserved, although no symmetry requires so and grand unified theories foresee nucleon decay. Because of that, nucleon decays would be a clear evidence of physics beyond Standard Model~\cite{DUNE_vol2}. 

DUNE will be sensitive to many possible proton decay channels but, in general, the favored channels are $p\rightarrow K^+ \anu$ or $p\rightarrow e^+ \pi^0$. In both cases, the detector needs to have a good particle identification capability, being able to distinguish all the particles in this kind of events, and of rejecting the background induced by atmospheric muons. Preliminary results from ProtoDUNE demonstrate a good particle identification capability for DUNE~\cite{DUNE_Vol1_TDR}. The fiducialization of the detector volume is crucial for the background removal and will be performed with the assistance of the PD system.

%% file: lar_tpc.tex
\chapter{Deep Underground Neutrino Experiment}
\label{chap:dune}
\thispagestyle{myheadings}

The Deep Underground Neutrino Experiment (DUNE)~\cite{DUNE_Vol1_TDR, DUNE_vol2,DUNE_vol3,DUNE_vol4} will be a 40 kt modular liquid argon time-projection chamber (LArTPC) placed 1.5 km underground exposed to the world's most intense neutrino beam, and  will answer some of the open questions in elementary particle physics. The experiment will take place in the Long-Baseline Neutrino Facility (LBNF) hosted by Fermilab, which will provide the infrastructure for DUNE, the deep-underground site and the neutrino beam.

The far detector (FD) will be installed in the Sanford Underground Research Facility (SURF) in South Dakota, 1,300 km away from Fermilab. The final configuration of the detector foresees the construction of four liquid argon time-projection chambers (LArTPC) with an active mass of 10~kt each. The total active mass will be of 40~kt.

\begin{figure}[h!]
	\centering
	\includegraphics[width=0.99\linewidth]{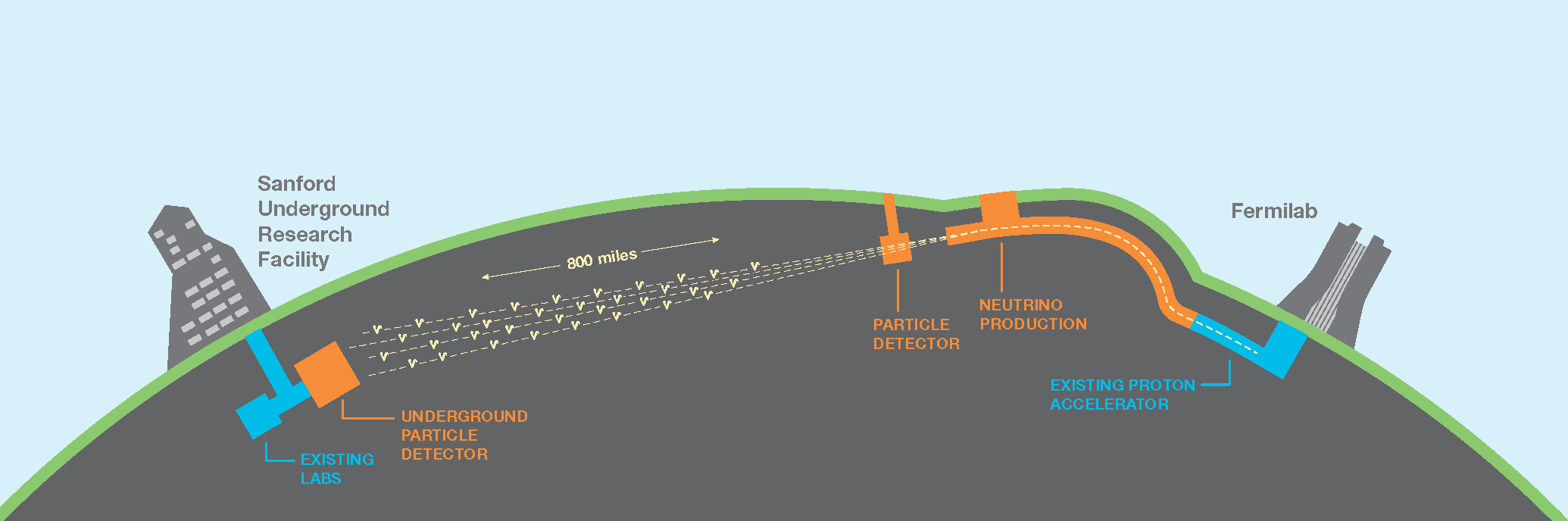}
	\caption{Schematic concept of neutrinos beam directed from Fermilab to SURF~\cite{DUNE_Vol1_TDR}.}
	\label{fig:fermilab}
\end{figure}

In this chapter, the DUNE experiment will be discussed starting with the description of the LArTPC technology, highlighting the features of this detector and the motivations of its use. There will be than an overview of the experiment, pointing out the sensitivities and requirements for potential discoveries. 

\section{Liquid Argon Time Projection Chamber (LArTPC)}
\label{sec:lartpc}

\subsection{Principle of operation}
The Liquid Argon Time Projection Chamber (LArTPC) follows the idea of a bubble chamber, with the concept of a readable trail left by the particle in a high quality three-dimensional image. It was first proposed by Carlo Rubbia in 1977~\cite{rubbia1977}, who pointed out that with simple techniques a desirable argon purity could be reach and ``\textit{In this case a multi-hundred-ton neutrino detector with good vertex detection capabilities could be realized}''.

The basic concept is that charged particles cross the active liquid argon (LAr) volume of the TPC leaving a track of ionized atoms and electrons. A strong and very uniform electric field is applied to the LAr volume, which allows to separate the electrons and Ar ions. A typical ionizing particle through the LAr produces around 60 thousand ionization electrons per centimeter. Electrons are drifted toward the anode grid, which is made by three parallel wire planes. Reading-out the signal induced by the drifting electrons on the wires allows to obtain a 2-dim representation of the ionizing event for each wire plane. Combining the three different 2-dim view of the same events allows to perform a complete 3-dim reconstruction of the same event.

Two different designs of LArTPC will be implemented for the DUNE experiment, the single-phase (SP) horizontal drift~(HD) and vertical drift\footnote{The original option was the dual-phase (DP) technology, which employs a layer of GAr above the liquid, through which the drift electrons are extracted. This grants a signal amplification due to electron avalanches in the gas phase with the electric field, decreasing the detection threshold allowing for a larger fiducial volume.}~(VD). In the SP-HD design, electrons are drifted horizontally in the liquid argon volume (whenever only ``SP'' is mentioned, it is understood to refer to the HD). In the vertical drift the electrons are drifted vertically in the LArTPC. The main subject of this chapter will be the SP design, but in order to contextualize the reader: the vertical drift has a simplified field cage design, which permits the LArTPC to be divided in two volumes while there are four volumes in the SP. The photon detection system can be placed behind the field cage and inside the cathode plane, but not behind the anode plane as in the HD design, since it is not transparent. This technology is still under development and a full description will be published soon. 

\begin{figure}[h]
	\centering
	\includegraphics[width=0.8\linewidth]{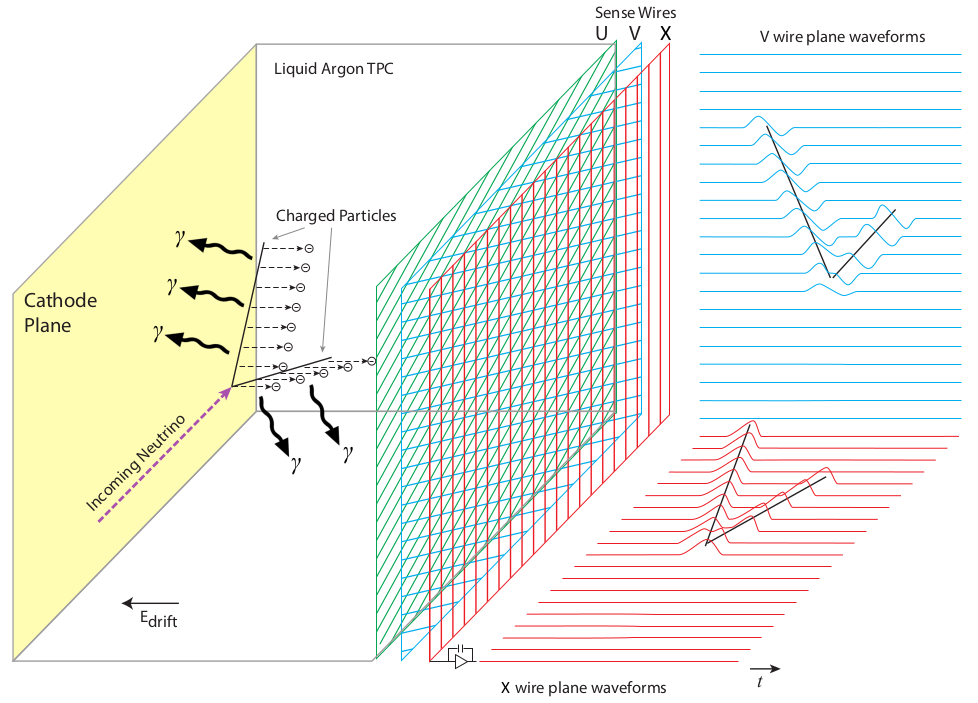}
	\caption{Diagram of the single-phase LArTPC basic operation~\cite{DUNE_vol4}.}
	\label{fig:schematic_LArTPC}
\end{figure}

Figure~\ref{fig:schematic_LArTPC} represents the interaction of a neutrino resulting in two charged particles in a single-phase LArTPC module. An electric field of 500 V/cm is applied between the cathode and the three anode wire grids to drift the electrons. The relative voltage between the layers is chosen
to ensure all but the final layer are transparent to the drifting electrons, and these first layers (U and V) produce bipolar induction signals (right side of the figure) as the electrons pass through them. The final layer (X) collects the drifting electrons, resulting in a monopolar signal~\cite{DUNE_vol4}.

Besides the charge signals from the ionized drifting electrons, liquid argon is also an excellent scintillator, emitting VUV light at a wavelength of 127~nm after de-excitation of argon molecules~\cite{LAr_fund_properties,DUNE_vol4}. Scintillation light in LAr is produced when ionizing radiation produces excitation and ionization of the argon atoms. The mechanism can be described by:
\begin{align}
	\label{eq:argon_excitation}
	\text{Ar}  \rightarrow \text{Ar}^* \rightarrow &\text{Ar}^*_2 \rightarrow 2 \text{Ar} + \gamma \\
	\label{eq:argon_ionization}
	\text{Ar} \rightarrow \text{Ar}^+ + e^- \rightarrow \text{Ar}^+_2 + e^- \xrightarrow{\text{recomb.}} &\text{Ar}^*_2 \rightarrow 2 \text{Ar} + \gamma.
\end{align}

In Eq.~\ref{eq:argon_excitation}, argon was excited to a higher energy state and in Eq.~\ref{eq:argon_ionization} ionized. In the latter case, the electron can recombine with the argon ionized molecule formed. The recombination will result in the same argon excited molecule of Eq.~\ref{eq:argon_excitation}, which can lose energy by emitting scintillation photons. The TPC signal is accompanied by a flash of scintillation light which may be detected and provide the time $T_0$ at which the ionization electrons start drifting toward the anode.

The ionizing electron has a certain probability of recombining with the argon ion or being drifted by the electric field. The electric field of 500~V/cm is chosen to balance between the charge from ionization and scintillation light which are competing processes (see Sec.~\ref{sec:lar_scintillation}).

A higher E field results in: a higher electron drift velocity, which reduces drift time and minimizes the losses due to impurities (see Sec.~\ref{sec:purity}); a decrease in the recombination probability and a lower free electron diffusion (proportional to the square root of the drift time). Therefore, improving the signal-to-noise, the calorimetry capabilities of the detector and spacial resolution~\cite{DUNE_vol4}.

However, without scintillation light, it is impossible to place an absolute time scale on the event detected, not allowing to solve the coordinate along the drift direction. The E field should not be raised beyond certain limits to not affect the electron recombination to a point where there is not enough scintillation photons to be detected and to avoid electric discharges in the LArTPC.

It was experimentally established~\cite{ICARUS} that a 500~V/cm field is an appropriate trade-off between charge and light signal. At this level, about 60\% of the electrons can recombine and a drift velocity of $\sim$1.6~mm/$\mu$s~\cite{protoDUNE_first_results} is achieved, taking around 2.2~ms to travel from the cathode to anode. Furthermore, the design and construction has already been tested to be cost-effective for a large scale experiment. 

\subsection{Why using liquid argon}
\label{sec:why_lar}
The use of a noble liquid rather than gas is necessary in neutrino experiments to provide a high enough target mass for increased neutrino interactions probability. Noble liquids have a high electron mobility for being electropositive; they are dielectrics, which allows high voltages; and they have high radiation length\footnote{Radiation length is defined as the mean distance in which a high energy electron loses all but 1/$e$ of its energy by bremsstrahlung and the 7/9 of the mean free path for pair production by high energy photons.}. Besides, LAr also has a high density, which makes it a good target for neutrino interactions, and abundantly produces scintillation light. Table~\ref{tab:noble_liquids} summarizes some of the properties of noble gases for a possible use in a TPC with water only for reference. Stopping power ($\slfrac{\diff E}{\diff x}$) is defined as the average energy lost by ionizing radiation in a medium per unit path length traveled. 

\begin{table}[htbp]
	\centering
	\caption{Properties of noble liquids~\cite{electron_mobility,electron_diffusion_mobility,helium_electron_mob,table_noble_liquids,neon_electron_mobility}.}
	\label{tab:noble_liquids}
	\begin{tabular}{lcccccc}
		\hline
		& Water         & He    & Ne   & {\color[HTML]{FE0000} Ar}   & Kr   & Xe   \\ \hline
		Boiling point {[}K{]} @ 1 atm            & 373           & 4.2   & 27.1 & {\color[HTML]{FE0000} 87.3} & 120  & 165  \\
		Density {[}g/cm$^3${]}                   & 1             & 0.125 & 1.2  & {\color[HTML]{FE0000} 1.4}  & 2.4  & 3.0  \\
		Radiation length {[}cm{]}                & 36.1          & 755.2 & 24   & {\color[HTML]{FE0000} 14}   & 4.9  & 2.8  \\
		Scintillation Yield {[}$\gamma$/keV{]}          & -             & 19    & 30   & {\color[HTML]{FE0000} 40}   & 25   & 42   \\
		Scintillation $\lambda$ {[}nm{]}         & -             & 80    & 78   & {\color[HTML]{FE0000} 127}  & 150  & 175  \\
		$\slfrac{\diff E}{\diff x}$ {[}MeV/cm{]}                     & 1.9           & 0.24  & 1.4  & {\color[HTML]{FE0000} 2.1}  & 3.0  & 3.8  \\
		Abundance (Earth atm) {[}ppm{]}          & 25$\times10^3$ & 5.2   & 18.2 & {\color[HTML]{FE0000} 9300} & 1.1  & 0.09 \\
		Electron mobility {[}cm$^2$/V$\cdot$s{]} & -           & $<0.3$   & $<0.01$  & {\color[HTML]{FE0000} $\sim$500}  & $\sim$1800 & $\sim$2200 \\ \hline
	\end{tabular}
\end{table}
Helium and Neon can be discarded because of the very low electron mobility, making it impossible to build a large TPC (they also have a very low boiling temperature, making very challenging to keep them at liquid state). Besides that, their scintillation light is very energetic, being easily absorbed and very challenging to be detected. Krypton, on the other hand, has a good electron mobility, density and stopping power ($\slfrac{\diff E}{\diff x}$). However, the low scintillation yield summed-up with its low abundance in nature makes challenging its use in a large TPC.

Even though liquid xenon (LXe) has better features, besides the short radiation length, its low abundance makes quite impossible to build a 10 kt LXeTPC. Liquid argon satisfies all the requirements for a high quality TPC with the additional advantage of a low ionization threshold of $W_l = 23.6 \pm 0.3$~eV~\cite{lar_for_low_en_part,lar_ion_pair_energy} which is the average energy expected per ion pair. This makes LAr the preferred choice for a massive detector such as the DUNE experiment.

\subsection{Neutrino detection in a LArTPC}
\label{sec:nu_detection}

Neutrinos of one of the three flavors ($\nu_\ell$) will interact in matter through charged current (CC) and neutral current (NC) following the Feynman diagrams~\cite{Thomson}:

\begin{figure}[h!]
	\centering
	\includegraphics[width=0.9\linewidth]{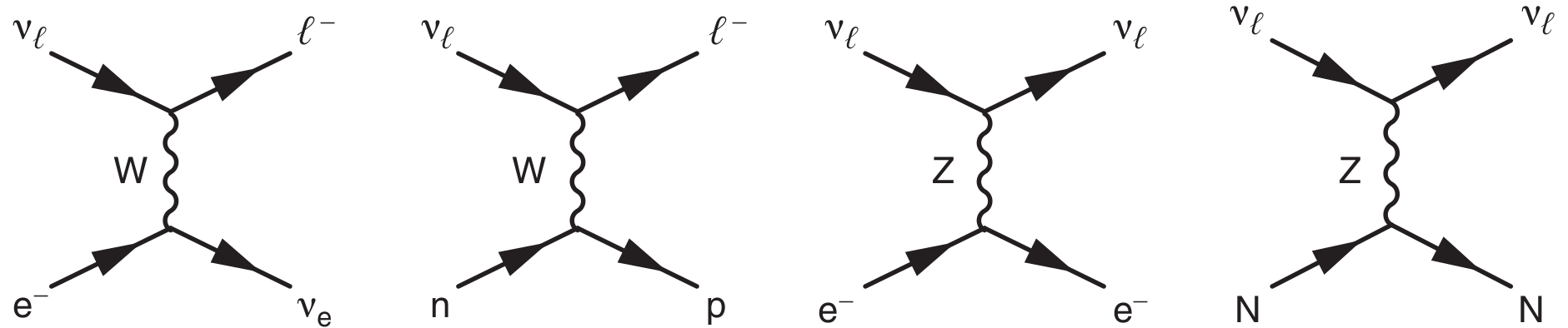}
	\caption{Feynman diagrams for CC and NC neutrino interactions in matter~\cite{Thomson}. For $\anue$ there is an extra $\anue+e^- \rightarrow \anue +e ^-$ CC and NC scattering.}
	\label{fig:feynman_diagrams}
\end{figure}

The two first diagrams are CC interactions and the last two are NC. In Sec.~\ref{sec:oscillation_in_matter} the difference between $\nue$ interactions and $\nu_{\mu,\tau}$ was mentioned, notice that in the first diagram the only neutrino flavor that can interact without changing flavor is $\nue$, this charged current scattering channel is not accessible to $\nu_{\mu,\tau}$ (without changing flavor). The neutral current scattering (third diagram) is accessible to all flavors. The fourth diagram is mostly not detectable in a LArTPC. 

While NC interactions happen for any neutrino energy, CC interactions need to have sufficient centre-of-mass energy to produce a charged lepton and the final stage hadronic system or $\nue$ (second and first diagrams, respectively)~\cite{Thomson}. Therefore, the CC absorption $\nu_\ell+n\rightarrow\ell^- +p$ will have the centre-of-mass energy squared\footnote{Centre-of-mass energy squared is a Lorentz invariant, formed by the total energy and momentum of the initial or final-state particles. For further explanation Ref.~\cite{Thomson} is exceptional.} in the laboratory frame as:
\begin{equation}
	\label{eq:s_square_lepton_p}
	s = (p_\nu +p_n)^2 = (E_\nu+E_n)^2-(\vec{p}_\nu+\vec{p}_n)^2,
\end{equation}  
where $p$ here is the four vector ($\vec{p},E$) of momentum and energy. The neutrino is considered massless here and the neutron without motion, so, considering the momentum in the x-axis:
\begin{align}
	p_\nu  &= (E_\nu,0,0,E_\nu) \\
	p_n &= (m_n,0,0,0),
\end{align}
which makes Eq.~\ref{eq:s_square_lepton_p} to become:
\begin{equation}
	s = (E_\nu+m_n)^2 - E^2_\nu = 2E_\nu m_n + m^2_n.
\end{equation}

The interaction $\nu_\ell+n\rightarrow\ell^- +p$ is only kinematically allowed if the lepton and proton are created in rest, that is $s>(m_\ell+m_p)^2$. Therefore, a threshold can be established~\cite{Thomson}:
\begin{equation}
	\label{eq:nu_threshold_interaction}
	E_\nu > \frac{(m^2_p-m^2_n)+m^2_\ell +2m_p m_\ell}{2m_n}.
\end{equation}
The resulting neutrino energy threshold for charged current interactions with a nucleon is:
\begin{equation}
	E_{\nue} > 0, \qquad E_{\numu} > 110~\MeV \qquad \text{and} \qquad E_{\nutau} > 3.5~\GeV.
\end{equation}

The threshold here is to produce the primary particles; nevertheless, the secondary particles need to be detected. Although there is no precise threshold for the detection of electrons, a few~MeV particle is needed to produce signal well above the electrical signal in the wire. Supernova muon and, evidently, tau neutrinos (see Sec.~\ref{sec:supernova_nu}) are above the threshold to produce signals in a LArTPC. The charged current absorption for supernova neutrinos in the LArTPC has the form:
\begin{equation}
	\label{eq:cc_scattering_lar}
	\nue + \text{$^{40}$Ar} \rightarrow e^- + \text{$^{40}$K}^*,
\end{equation}
which are likely to also give gamma rays signals from the de-excitation of K$^*$~\cite{DUNE_vol2}.

For CC interactions with atomic electron $\nu_\ell +e^- \rightarrow \nue+\ell^-$ (first diagram), the same approach can be done with the condition that the centre-of-mass energy must be high enough to produce a lepton at rest, that is, $s>m^2_\ell$ in the laboratory frame. Therefore:
\begin{equation}
	s = (p_\nu+p_e)^2 = (E_\nu + m_e)^2 - E^2_\nu = 2E_\nu m_e+ m^2_e,
\end{equation}
resulting in the condition: 
\begin{equation}
	E_\nu > \frac{m^2_\ell-m^2_e}{2m_e}.
\end{equation}
Therefore the neutrino electron charged current scattering has the thresholds: 
\begin{equation}
	E_{\nue} > 0, \qquad E_{\numu} > 11~\GeV \qquad \text{and} \qquad E_{\nutau} > 3090~\GeV.
\end{equation}

Considering the threshold energy for muon neutrino interaction through electron scattering, it is evident that this channel does not contain any peak of oscillation (by comparing with Fig.~\ref{fig:nuoscthreeflavors}).

In DUNE, even though the electron neutrino appearance is quite low compared to muon neutrino survival or tau neutrino appearance (see Fig.~\ref{fig:nuoscthreeflavors}), the matching sensibility of the detector for $\nue$ interactions, the fact that $\nu_{\mu,\tau}$ are background majorly in the NC interactions, in which the flavor cannot be distinguished, and the matter effect in $\nue$ make the measurement of electron neutrino appearance an obvious choice.

For supernova neutrinos, only three channels will be accessible: the CC nucleon absorption for electron neutrino, the CC absorption for antineutrino and the neutrino electron scattering for all flavors~\cite{DUNE_vol2}:
\begin{align}
	\nue + \text{$^{40}$Ar} &\rightarrow e^- + \text{$^{40}$K}^* \\
	\anue + \text{$^{40}$Ar} &\rightarrow e^+ + \text{$^{40}$Cl}^* \\
	\nu_\ell+e^- &\rightarrow \nu_\ell + e^-.
\end{align}

\subsection{LAr purity}
\label{sec:purity}

The purity of the LAr is of great concern for the DUNE experiment. As discussed in Sec.~\ref{sec:why_lar}, LAr has a high electron mobility, for being electropositive; however, electronegative contaminants such as oxygen or water will absorb ionized electrons in the drift. Moreover, nitrogen contaminants quench the scintillation of LAr and decreases the quality of light signals.

The DUNE FD has a requirement of electronegative contaminants in LAr lower than 100 ppt (part per trillion) 0$_2$-equivalent, which is enough to ensure a lifetime greater than 3~ms for the ionized electron~\cite{DUNE_vol4}. One can assume that the rate of electrons absorbed is proportional to the number of electrons, that is:
\begin{equation}
	\frac{\diff Q}{\diff t} = -\lambda_e Q.
\end{equation}
where $\lambda_e$ is the absorption constant and $Q$ is the total charge. Defining $\tau_e = 1/\lambda_e$ as the electron lifetime and solving the differential equation, it results that:
\begin{equation}
	Q(t) = Q_0\me^{-t/\tau_e},
\end{equation}
where $Q(t)$ is the charge as function of the drift time $t$ and $Q_0$ is the initial charge. For the DUNE SP module the maximum drift length will be 3.5~m (full distance from the cathode to anode planes) and, at nominal 500~V/cm drift voltage, this correspond to a drift time of $t_\text{max} = 2.2$~ms. In the target electron lifetime of 3~ms, an attenuation of about 48\% is expected for ionizing radiation near the cathode plane, ensuring a S/N ratio greater than 5 in the induction planes and S/N~>~10 for the collection planes, which are necessary to perform pattern recognition and two-track separation~\cite{DUNE_vol4}. 

In ProtoDUNE-SP run I, dedicated measurements were performed to determinate the electron lifetime. At the end of the Run, it reached approximately 89\error22~ms, which corresponds\footnote{Assuming a that the concentration is given by $N_{\text{O}_2}=(k_a \tau)^{-1}$, were $k_a$ = (300 ppt ms)$^{-1}$ being the attachment constant for oxygen~\cite{protoDUNE_first_results}.}to a concentration in liquid argon of 3.4\error0.7~ppt O$_2$-equivalent, much lower than the DUNE requirement of <100~ppm O$_2$ equivalent~\cite{protoDUNE_first_results}. During the run I, electron lifetime was stable above 20~ms.

\section{Experimental details}
\label{sec:dune_experiment_description}
The DUNE experiment has been carefully designed to reach its ambitious  physics goals. In this section we highlight some of the key features of the experiment. A complete description can be found in Refs.~\cite{DUNE_Vol1_TDR,DUNE_vol2,DUNE_vol3}. The near detector (ND) conceptual design report can be found at Ref.~\cite{near_detector}, but will not be explored in this thesis.

\subsection{Neutrino beam}
\begin{figure}[h!]
	\centering
	\includegraphics[width=0.99\linewidth]{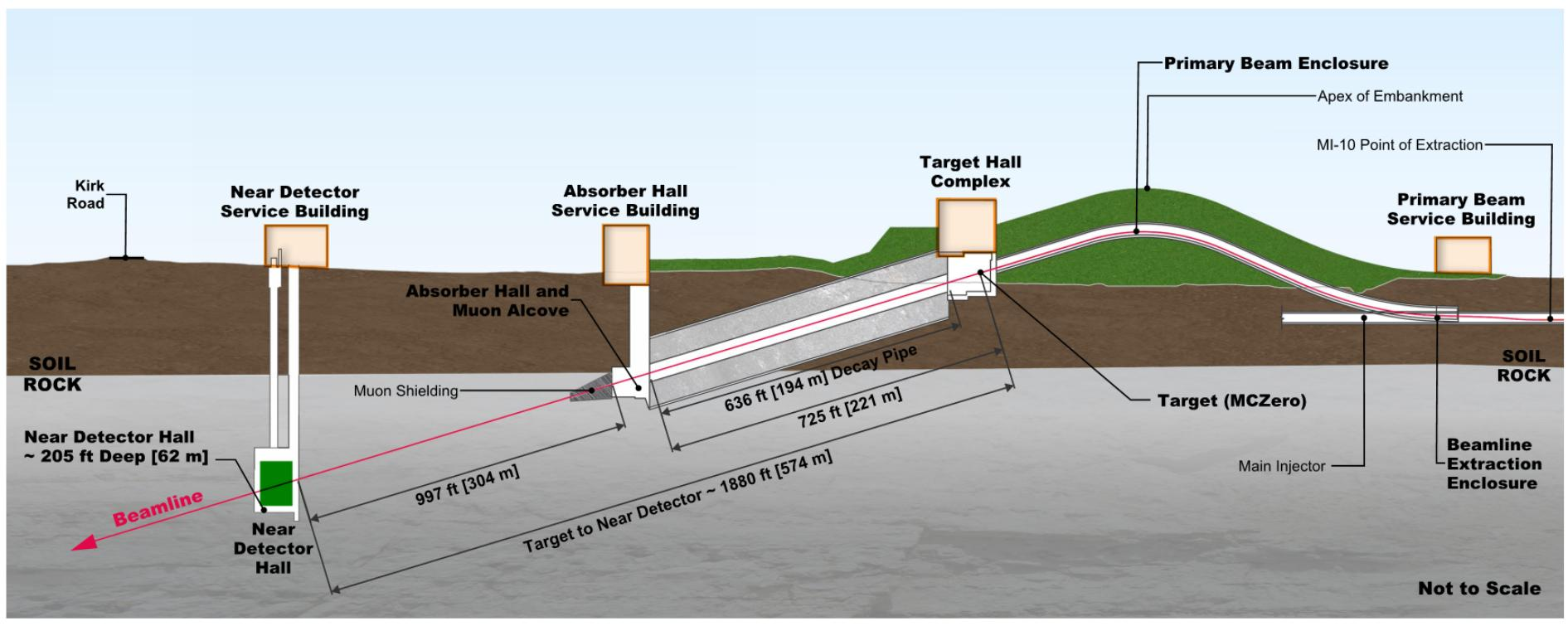}
	\caption{Longitudinal section of the Long Baseline Neutrino Facility beamline at Fermilab~\cite{DUNE_Vol1_TDR}.}
	\label{fig:schematicnubeam}
\end{figure}

As described in Sec.~\ref{sec:dune_oscillation}, the DUNE aims to study the two oscillation maximum for neutrinos around 1~GeV at 1300~km. A proton beam, from the Long Baseline Neutrino Facility (LBNF) is generated at Fermilab with an energy raging from 60 to 120~GeV and bent downwards towards the far detector as shown in Fig.~\ref{fig:schematicnubeam}~\cite{DUNE_Vol1_TDR}.

\begin{figure}[h!]
	\centering
	\includegraphics[width=0.99\linewidth]{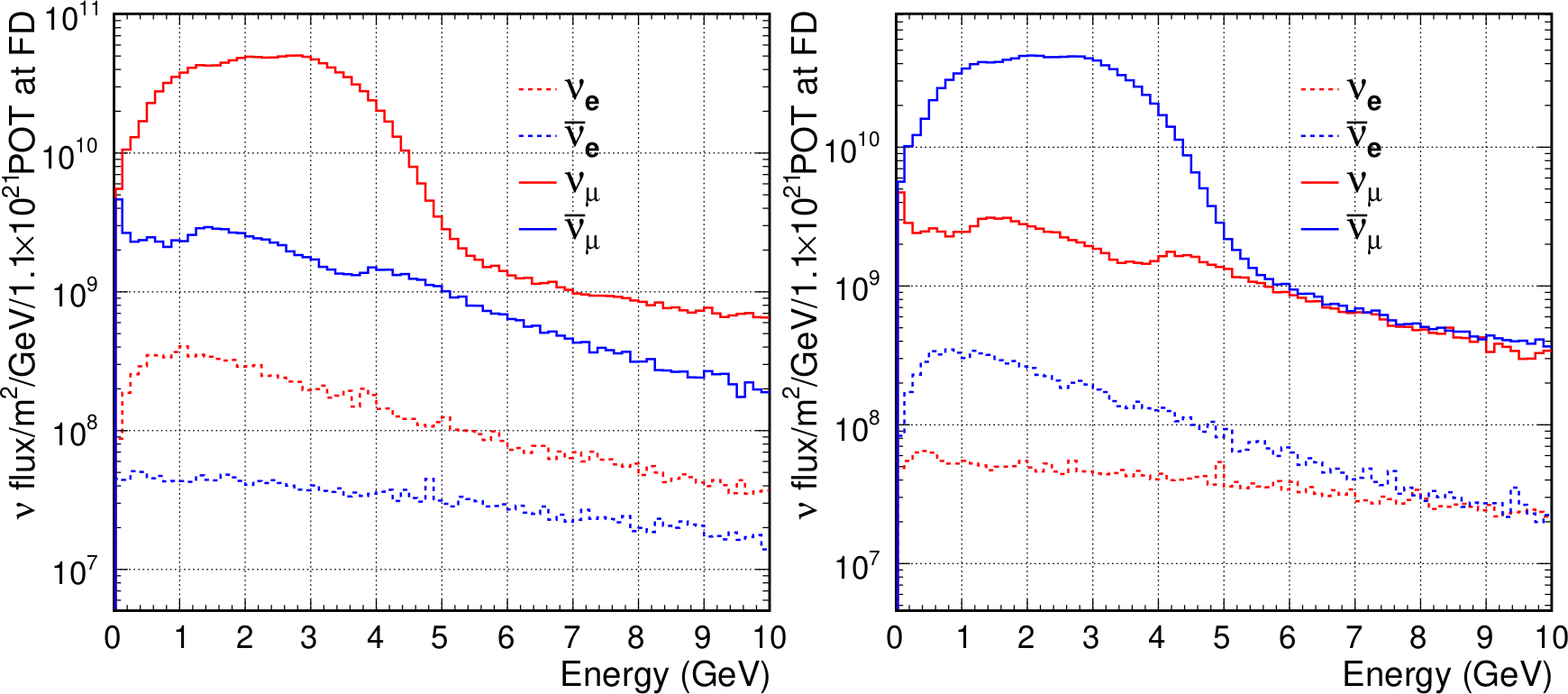}
	\caption{Predicted neutrino fluxes at the far detector (FD) when operating as neutrino mode (left) and antineutrino
		mode (right)~\cite{DUNE_vol2}.}
	\label{fig:nu_mu_bean_flux_energy}
\end{figure}

The beam will hit a target and produce secondary mesons, that will be focused in a decay pipe by magnetic horns and will successively decay into muons and neutrinos. Neutrino or antineutrino beam is chosen by switching the horn polarity. Muons produced and any residual hadron are stopped, producing the neutrino beam with tuned energy between 0.5 and 5~GeV~\cite{DUNE_Vol1_TDR}. Figure~\ref{fig:nu_mu_bean_flux_energy} shows the predicted neutrinos and antineutrinos fluxes at the Far Detector (FD) at 1300~km when operating as neutrino (left) and antineutrino mode (right). The flux is given as neutrinos per GeV per (1$\times$20$^{21}$) proton-on-target (POT) per m$^2$. It is assumed $1.1\times10^{21}$~POT per year for the 1.2~MW neutrino beam~\cite{DUNE_Vol1_TDR}. This numbers are given by the assumption that the uptime and efficiency of the Fermilab accelerator complex and the LBNF beamline are 56\% combined. An upgrade to 2.4~MW of beam power is foreseen by 2030.

\subsection{Far Detector}
\label{sec:far_detector_dune}

\begin{figure}[h!]
	\centering
	\includegraphics[width=0.72\linewidth]{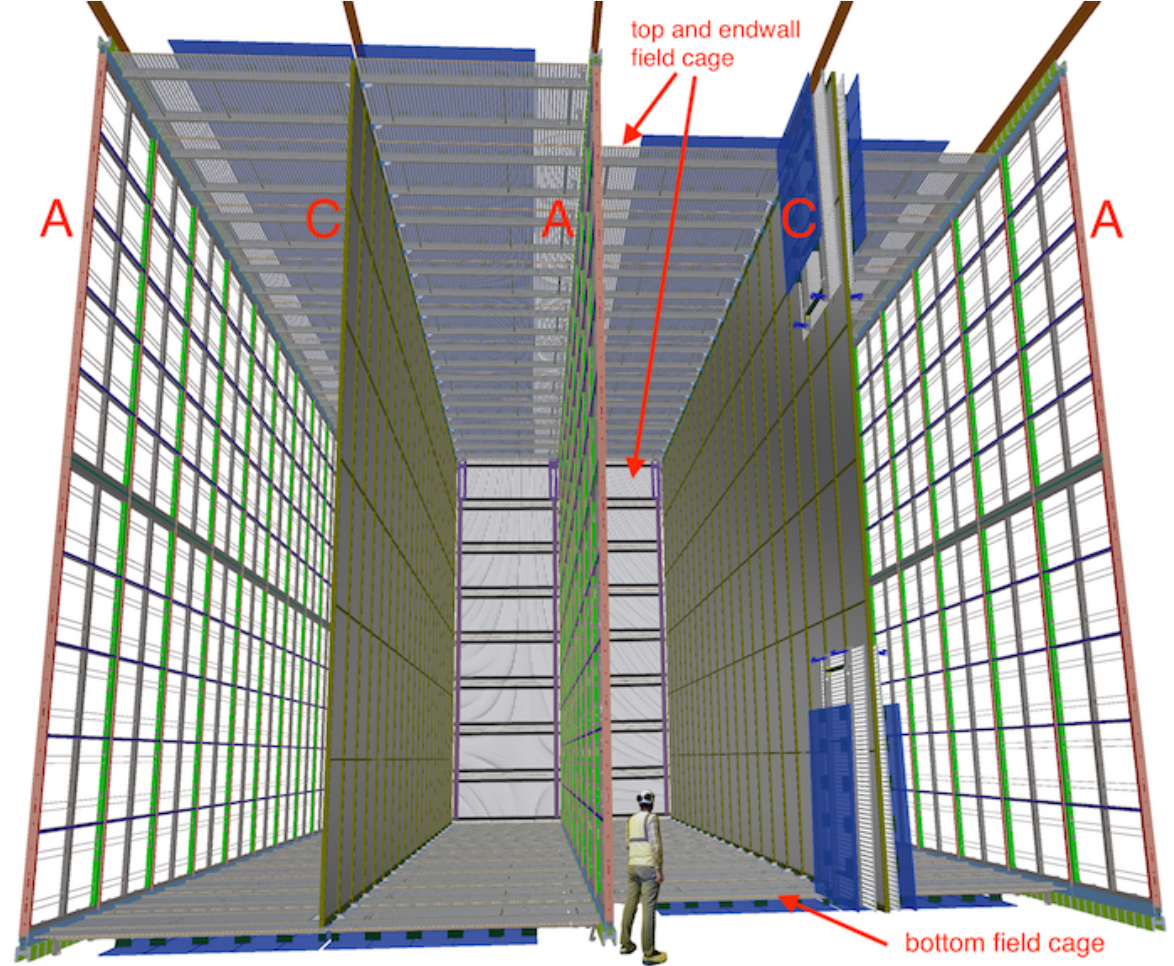}
	\caption{Basic design of the DUNE Far Detector (FD) SP module of 10~kt. The Anode (A) and Cathode (C) are separated by 3.5~m and are 12~m high~\cite{DUNE_vol4}.}
	\label{fig:fdmoduledesign}
\end{figure}

The DUNE Far Detector Single-Phase will consist of four modules of 10~kt LArTPCs. Figure~\ref{fig:fdmoduledesign} shows the 10~kt module (17.5~kt of total liquid argon mass). The LArTPC is composed by four drift volumes of 3.5~m between anode and cathode walls, with dimensions of 58~m~$\times$~12~m with the beam arriving into the 58~m dimension direction.

The two cathode walls are composed of cathode plane assembly (CPA) panels. The CPA is a 1.2~m~$\times$~4~m panel that form the CPA arrays of Fig~~\ref{fig:fdmoduledesign}. Each CPA array contains 150 CPAs (300 in total). The CPA arrays are set to $-180$~kV, which results in an uniform electric field of 500~V/cm across the drift volume to the grounded anode planes. The anodes are each composed by 50 anode plane assemblies (APA) units (150 in total) with 6~m~$\times$~2.3~m dimension placed vertically.

\begin{figure}[h!]
	\centering
	\includegraphics[width=0.8\linewidth]{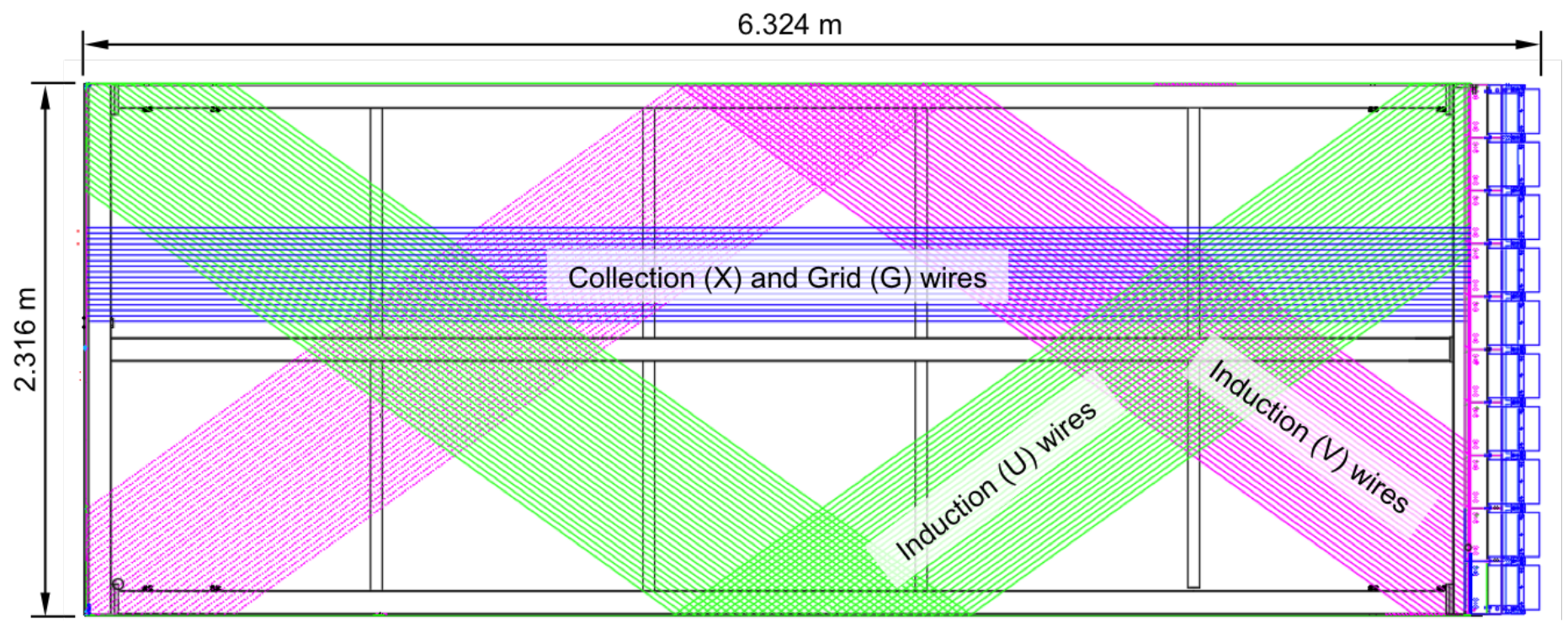}
	\caption{Illustration of the DUNE APA scheme showing small portions of the wires from the three signal planes (U, V, X) and fourth wire plane (G) present to improve the pulse shape on the U plane signals. The TPC electronics boxes, shown in blue on the right, mount directly to the frame and process signals from both the collection and induction channels~\cite{DUNE_vol4}.}
	\label{fig:apa_design}
\end{figure}

The APAs are two-sided, with three wire layers and an additional grid layer for shielding\footnote{The grid improves the signal shapes on the U induction channels.} wrapped around them. Figure~\ref{fig:apa_design} shows the APA schematic, the wire spacing on the layers is 4.75~mm. The collection layer X and the grid G are placed horizontally (in the figure), with wires spaced (pitch) 4.790~mm from each other. The induction U- and V-layers are wrapped around the APA at $\pm$35.7$^{\circ}$ from the vertical and pitch of 4.669~mm. A grounding mesh is mounted  on both sides of the APA frame. The blue boxes in Fig.~\ref{fig:apa_design} represent the read-out electronic of the TPC.

The $\pm$35.7$^{\circ}$  angle is chosen to ensure that each induction wire only crosses a given collection wire once\footnote{This condition is achieved if an angle $\theta<36.2^\circ$ is set.}, reducing the ambiguities on reconstruction, while keeping a performance close to the optimum 45$^\circ$ wire angle configuration~\cite{DUNE_vol4}.

The wires pitch is chosen so the volume can be fiducialize to a precision of 1\%, in which the vertex position resolution is of 1.5~cm along each coordinate. Besides, the radiation length for photons in LAr is of 14~cm, the spacial resolution given by the pitch of $\sim$0.47~cm allows the gap between a neutrino interaction vertex and a photon conversion point to be easily visible. The tolerance of the wire pitch is $\pm$0.5~mm to maintain the reconstruction precision required~\cite{DUNE_vol4}. 

Each wire will collect about 20k to 30k electrons from a minimum ionizing particle traveling parallel to the wire plane and perpendicular to the wire orientation. Therefore, a noise below 1k~$e^-$ is required.

The ProtoDUNE-SP experiment (see Chap.~\ref{chap:protoDUNE}) has set a bias voltage for the grids of -370~V, 0~V and +820~V for U, V and X planes respectively, and it has reported a signal-to-noise (S/N) of at least 12.1 for the U plane and up to 48.7 for the X plane~\cite{protoDUNE_first_results}. Results that safely satisfy the DUNE specifications of S/N~>~5 for the U and V and S/N~>~10 for the X plane~\cite{DUNE_vol4}.

\begin{figure}[h!]
	\centering
	\includegraphics[width=0.632\linewidth]{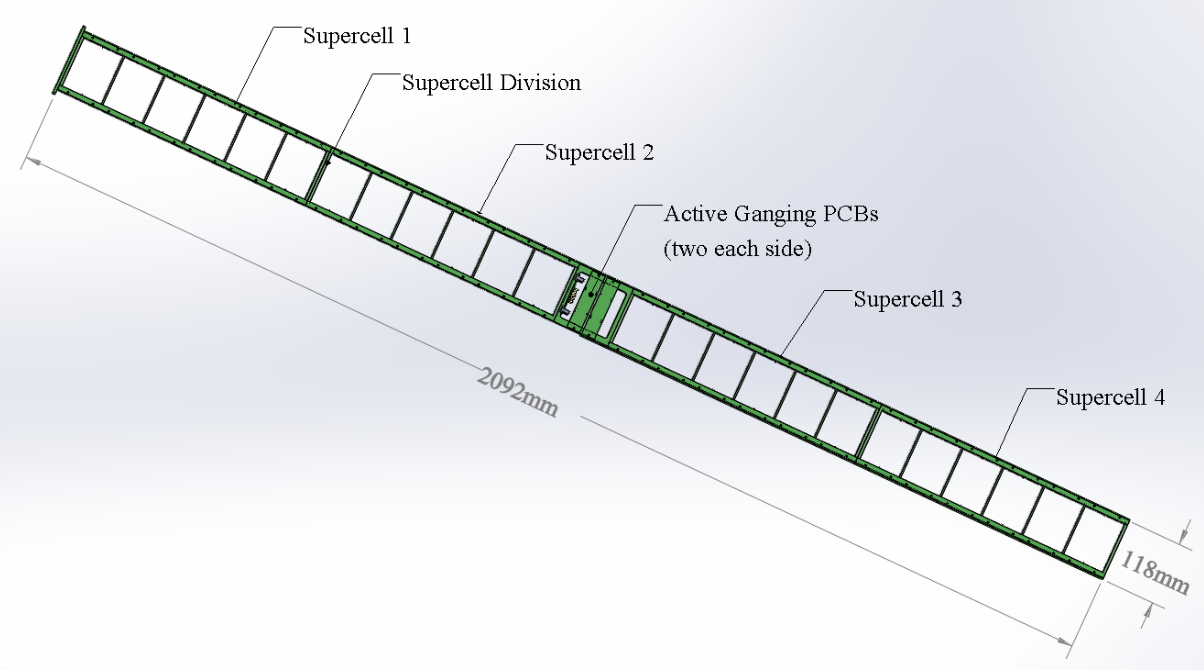}
	\includegraphics[width=0.358\linewidth]{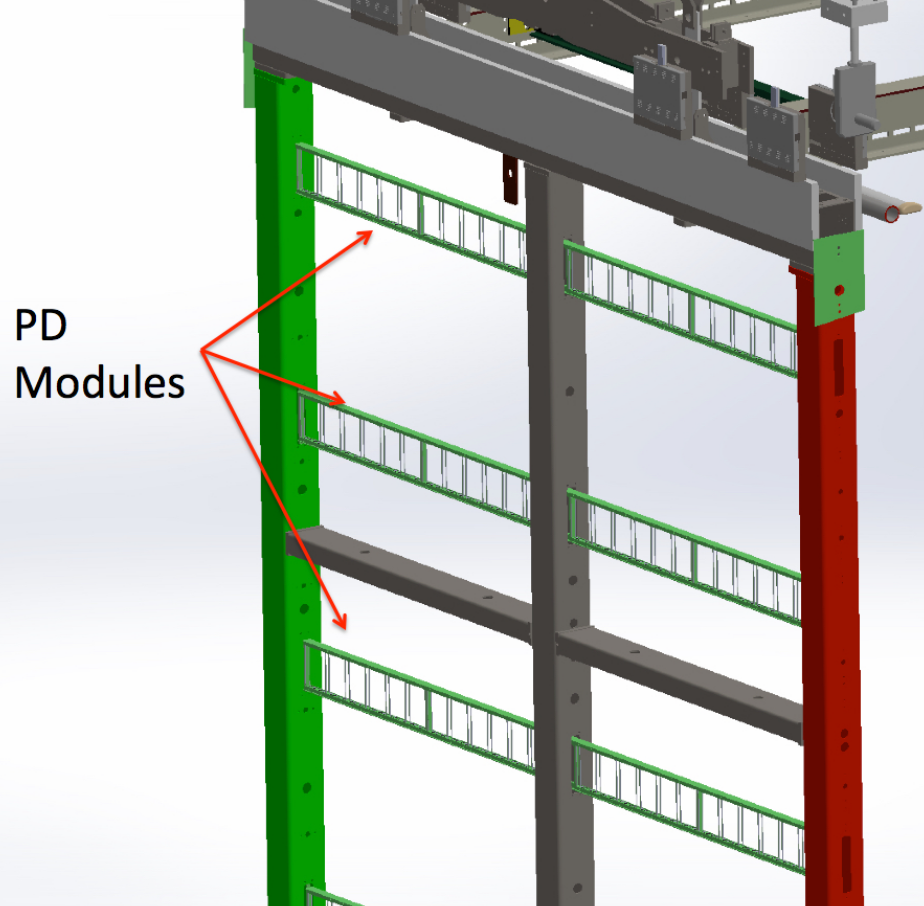}
	\caption{(Left) The \xara\ module schematic, composed by 24 \xara\ cells grouped into a set of four supercells. The active ganging PCBs, which read-out the signals, is displayed at the center~\cite{DUNE_vol4}. (Right) 3D model of the \xara\ module assembled horizontally in the APA.}
	\label{fig:super_cell}
\end{figure}

Each APA will host 10 Photon Detection (PD) modules shown in Figure~\ref{fig:super_cell} (Left). The \xara\ modules are 209~cm~$\times$~12~cm~$\times$~2~cm bars placed between the wire layers of the APA (Right). Each module is composed by 24 \xara\ cells assembled in groups of six to form four \xara\ supercell. In total, 1,500~\xara\ modules will be installed in one DUNE FD module, which gives a total of 6,000~supercells installed in the 10 kt LArTPC.

The \xara\ device is described in Sec.~\ref{sec:x_arapuca_description}.


\begin{figure}[h!]
	\centering
	\includegraphics[width=0.6\linewidth]{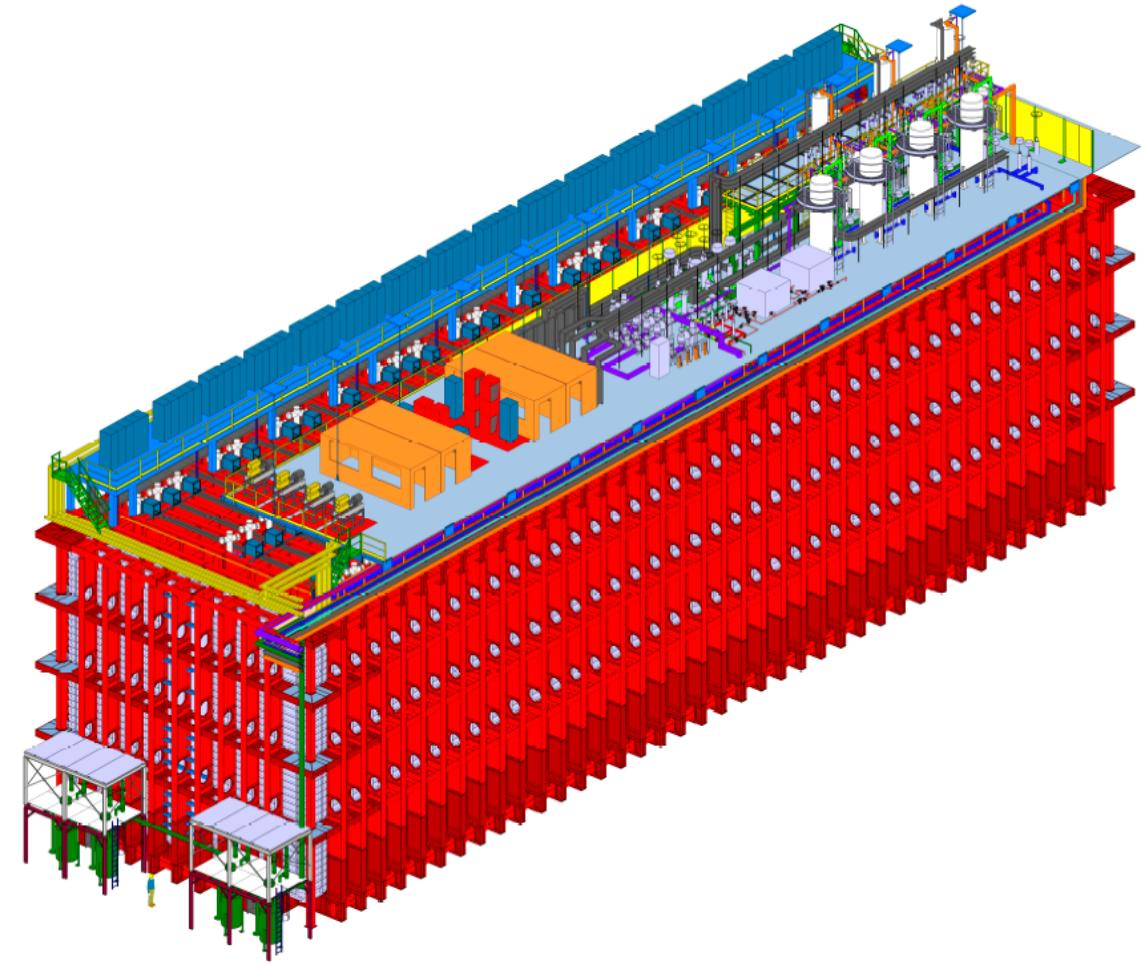}
	\caption{The cryostat that houses a 10 kt FD module. A mezzanine (light blue) installed 2.3 m above the cryostat supports both detector and	cryogenics instrumentation. At lower left, between the LAr recirculation pumps (green) installed on the cavern floor, the figure of a person indicates the scale~\cite{DUNE_vol4}.}
	\label{fig:dunecryostat}
\end{figure}

The 10 kt LArTPC will be hosted inside the cryostat of Figure~\ref{fig:dunecryostat}, with outer dimensions of 65.8~m~$\times$~17.8~m~$\times$~18.9~m.

\subsection{The DUNE specifications for the LArTPC}
\label{sec:dune_spec}
Here we report the high level specification of the far SP detector. The specifications shown in Table~\ref{tab:tpc_spec} concern the TPC and Table~\ref{tab:pds_spec} concern  the Photon Detection System. 

\begin{table}[h!]
	\centering
	\caption{TPC specifications~\cite{DUNE_vol4}.}
	\label{tab:tpc_spec}
	\begin{tabular}{|>{\centering\arraybackslash}m{2.1cm}|>{\centering\arraybackslash}m{4.6cm}|m{7.5cm}|}
		\hline
		\rowcolor[HTML]{C0C0C0} 
		Description &
		Specification (Goal) &
		\multicolumn{1}{|c|}{Justification} \\
		APA wire angles &
		0$^\circ$ for the collection wires; $\pm$35.7$^\circ$ for induction wires &
		Minimize inter-APA dead space. \\
		\rowcolor[HTML]{EFEFEF} 
		APA wire spacing &
		4.669~mm for U, V; 4.790~mm for X, G &
		Enable 100\% efficiency MIP detection and a 1.5~cm yz vertex resolution. Spatial resolution lower than the radiation length for photons allow a visible gap between the neutrino interaction vertex and photon conversion point. \\
		Minimum drift field & >~250 V/cm (>~500 V/cm) & Reduces recombination, reduces impact of $e^-$ lifetime and diffusion, and space charge. \\
		\rowcolor[HTML]{EFEFEF}
		System noise &
		< 1000 $e^-$     &
		Ensures S/N~>~5 on induction planes for pattern recognition and two-tracks separation.  \\
		Front-end peaking time (readout) &
		1~$\mu$s   & 
		Ensures vertex resolution; optimized for 5~mm wire spacing.\\ \hline 
	\end{tabular}
\end{table}

\begin{table}[h!]
	\centering
	\caption{PDS specifications~\cite{DUNE_vol4}.}
	\label{tab:pds_spec}
	\begin{tabular}{|>{\centering\arraybackslash}m{2.1cm}|>{\centering\arraybackslash}m{4.6cm}|m{7.5cm}|}
		\hline
		\rowcolor[HTML]{C0C0C0} 
		Description &
		Specification (Goal) &
		\multicolumn{1}{|c|}{Justification} \\
		Light Yield &
		>~20~\phe/MeV (average); >~0.5~\phe/MeV (minimum) &
		PDS energy resolution comparable with TPC for 5 to 7 neutrinos from Supernova. Allows tagging of >~99\% of nucleon decay backgrounds with light at all points of the detector. \\
		\rowcolor[HTML]{EFEFEF} 
		Time resolution &
		< 1 $\mu$s (< 100 ns) &  
		Position resolution of 1~mm for 10~MeV SNB candidate events in a rate lower than 1 event ~m$^{-3}$ms$^{-1}$.\\
		LAr nitrogen contaminant& 
		< 25 ppm&  
		Necessary to reach the 0.5~\phe/MeV PDS sensitivity for triggering proton decay near the cathode.\\
		\rowcolor[HTML]{EFEFEF}
		Signal-to-noise &
		>~4 &
		Keep data rate within electronic bandwidth limit. \\
		Dark noise rate. &
		< 1 kHz &
		Keep data rate within electronic bandwidth limit. \\ \hline
	\end{tabular}
\end{table}

%% file: arapuca_pd_system.tex
\chapter{Light collectors}
\label{chap:pd_system}
\thispagestyle{myheadings}

In DUNE, the Photon Detection System (PDS) is based on the \ara\footnote{\textit{Arapuca}, from the original \textit{ara'puka}, comes from tupi-guarani and refers to a handcraft bird trap.} technology (Sec.~\ref{sec:far_detector_dune}). The \ara\ device needs to efficiently detect liquid argon (LAr) scintillation light with a wavelength of 127~nm as it improves the determination of the neutrino interaction vertex for beam events and will be crucial to determinate the vertex for non-beam events. It will also help to trigger SNBs and to perform calorimetric measurements. A review of the LAr scintillation is given in the first section. In the following sections the \ara\ light trapping device is introduced with the description of the two main prototypes, the Standard \ara\ (S-\ara) and the \xara. The device components are described with more details in the succeeding sections.

\section{Scintillation light in LAr}
\label{sec:lar_scintillation}

\subsection{Process and photons production}
\begin{figure}[h!]
	\centering
	\includegraphics[width=0.7\linewidth]{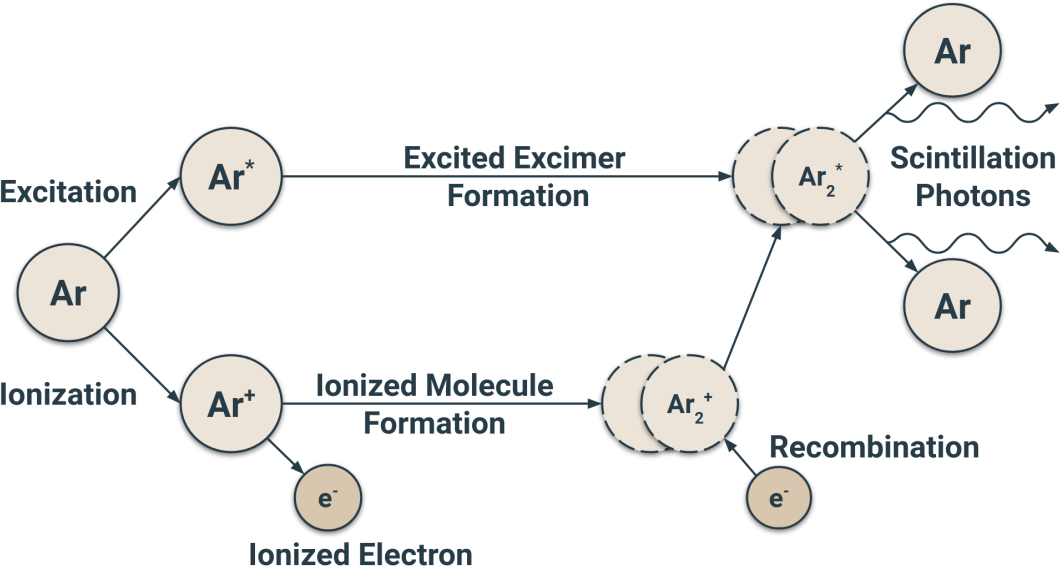}
	\caption{Schematic of scintillation light production in argon~\cite{DUNE_vol4}.}
	\label{fig:scintillationdiagram}
\end{figure}

The passage of ionizing radiation in LAr produces excitation and ionization of the argon atoms that will ultimately produce an excited dimer state~(Ar$_2^*$), as shown in the diagram of Fig.~\ref{fig:scintillationdiagram}. In the case of ionization (bottom part of the diagram), argon ions will gather together and form an ionized molecule (Ar$_2^+$) and, if the electric field is not high enough to completely separate the electron, the recombination process will generate an excited molecule (Ar$_2^*$). The excitation of argon atom (top part of the diagram) will result in the same excited molecule, this molecule will lose energy through the emission of a scintillation photon decaying to the fundamental state remaining with two separate argon atoms~\cite{LAr_fund_properties}.

One can estimate the maximum number of photons $N_{ph}$ produced in the scintillation. The number of photons originated from ionized atoms will be proportional to the energy deposited divided by the average energy expected per ion pair, that is $N_i = E_0/W_l$~\cite{thesi_ettore}. The ratio of the number of excited atoms to the number of ionized ones ($N_{ex}/N_i$) produced by the passage of an ionizing particle is assumed to be independent of the particle type and energy and is given by~\cite{absolute_scint,Segreto_2021,lar_for_low_en_part}:
\begin{equation}
\label{eq:ratio_nex_ni}
\frac{N_{ex}}{N_i} = 0.21
\end{equation}

Therefore, we can consider that:
\begin{equation}
N_{ph} = N_i+N_{ex} = N_i\cdot(1+N_{ex}/N_i) = E_0/W_l \cdot (1+N_{ex}/N_i),
\end{equation}
which can assume the form:
\begin{equation}
N_{ph} = \frac{E_0}{W_{ph}^{\text{min}}}
\end{equation}
with
\begin{equation}
W_{ph}^{\text{min}} = \frac{W_l}{1+N_{ex}/N_i} = 19.5 \pm 1.0~\eV
\end{equation}
as the minimum (or average) energy\footnote{In Ref.~\cite{absolute_scint} $W_{ph}^{\text{min}}$ is reported as $W_{ph}(\text{max})$ to emphasize that this is the maximum absolute scintillation.} to produce one scintillation photon (at maximum yield)~\cite{lar_for_low_en_part,absolute_scint}. Therefore, the maximum number of photons emitted per MeV deposited is taken by assuming $E_0 = 1~\MeV$:
\begin{equation}
\label{eq:light_yield}
Y_{ph}^\text{ideal} \sim 51.3\times10^3 \text{photons}/\MeV.
\end{equation}

The light yield of Eq.~\ref{eq:light_yield} assumes that all the ionized or excited atoms produce scintillation light, thus neglecting quenching processes (see Sec.~\ref{sec:purity}) and electrons escaping recombination (see Fig.~\ref{fig:electron_recomb_figs}). In Chapters~\ref{chap:lar_test} and~\ref{chap:protoDUNE} a maximum light yield of 51,000~photons/MeV was assumed as input of Monte Carlo simulations.

However, the light yield of LAr is affected by the type of particle and the Linear Energy Transfer\footnote{The main difference between LET and stopping power is the fact that LET does not take into account the production of high energetic electrons production~\cite{stoppingpowe_vs_let}, being a locally energy loss described as restricted stopping power. If there is no production of large energy electrons, LET becomes the unrestricted linear energy transfer which equals to the stopping power} (LET). Light particles, such as electrons and muons, will have a decreased light yield due to escaping electrons from recombination because of the low density of ionized atoms. Non-relativistic heavy particles, such as protons, alpha particles and fission fragments, will cause a high density of excited states, that can diffuse and suffer a quenching process due to biexcitonic interactions~\cite{Segreto_2021}.

\begin{figure}[h!]
	\centering
	\includegraphics[width=0.75\linewidth]{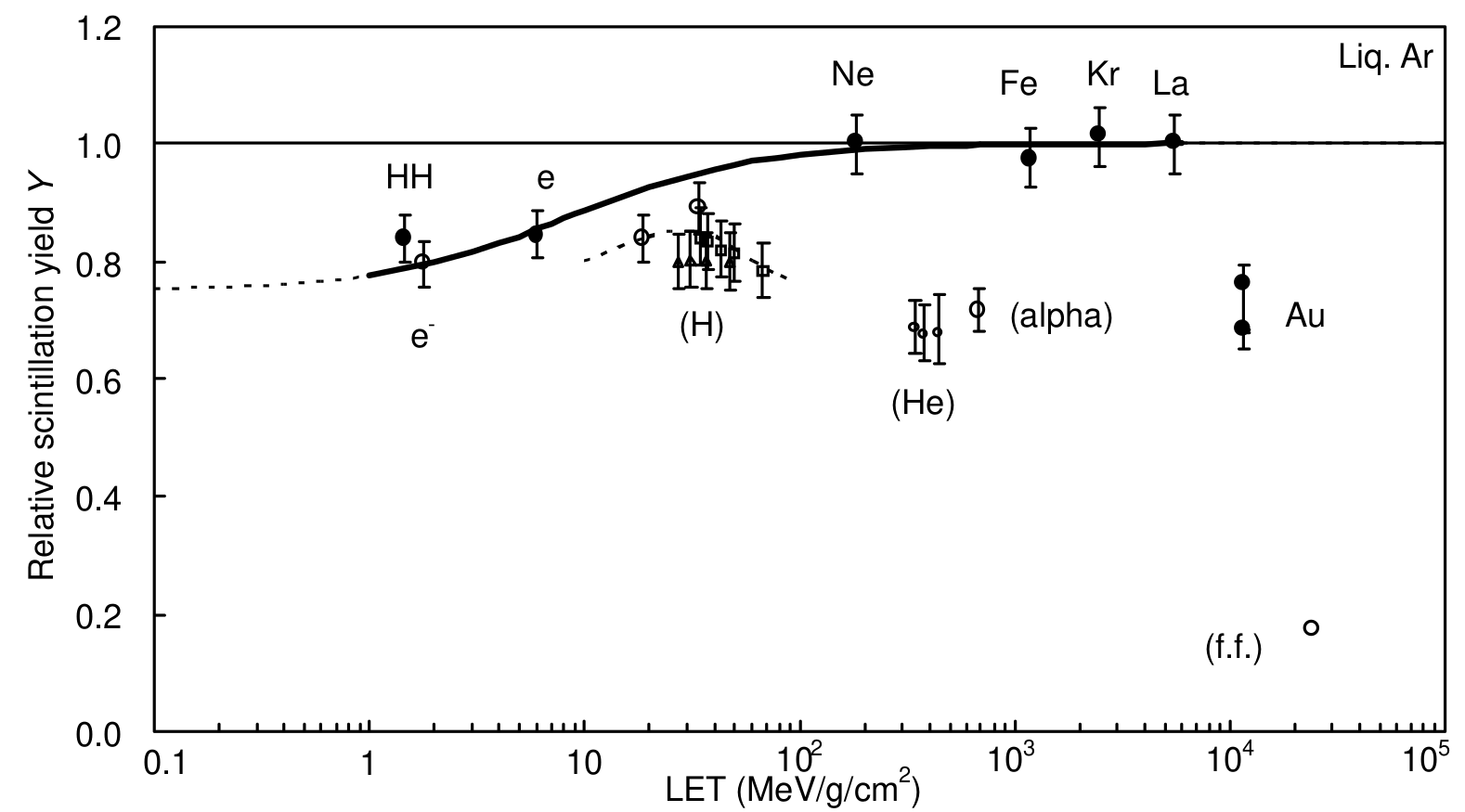}
	\caption{Relative light yield as function of LET for different particles in liquid argon. Solid circles are measurements for relativistic particles, while open circles are non-relativistic. Squares and small circles represent non-relativistic protons and helium, respectively~\cite{absolute_scint}.}
	\label{fig:reallightyield}
\end{figure}

Figure~\ref{fig:reallightyield} shows the relative light yield with respect to the maximum yield (Eq.~\ref{eq:light_yield}) as function of LET for different particles~\cite{absolute_scint}. Therefore, it can be expected from a minimum ionizing particle (mip), which LET in LAr is about 1.5~MeV/(g/cm$^2$), a relative light yield of $Y_{ph}^\text{mip} = 0.8\cdot Y_{ph}^\text{ideal}$. Alpha particles, on the other hand, will have a relative light yield of 0.71. These values are used for the Monte Carlo simulation of Chapters~\ref{chap:lar_test} and~\ref{chap:protoDUNE}.

\subsection{Electron-ion recombination}
\label{sec:recombination}

\begin{figure}[h!]
	\centering
	\begin{subfigure}{0.48\textwidth}
		\includegraphics[width=0.98\linewidth]{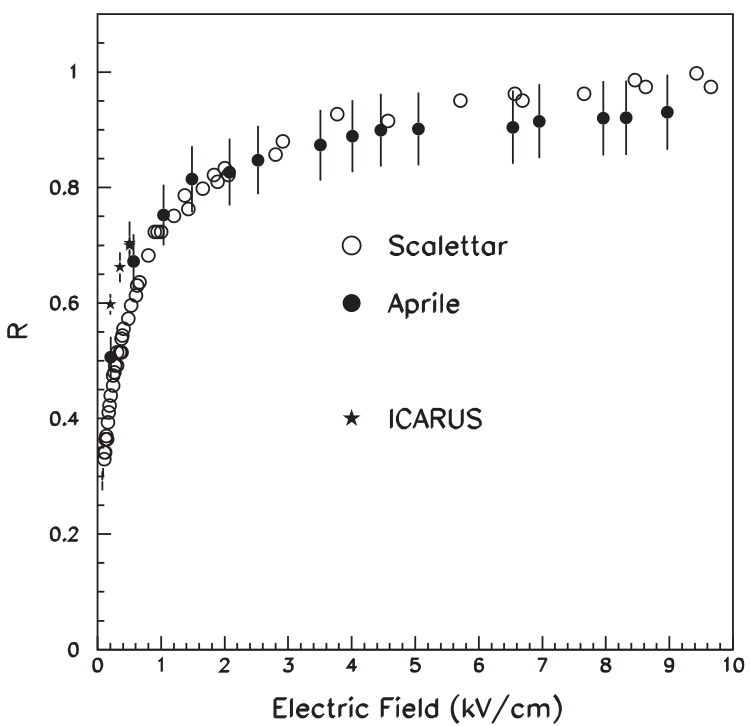}
		\caption{ }
		\label{fig:recombvsefield}
	\end{subfigure}
	\begin{subfigure}{0.483\textwidth}
		\includegraphics[width=0.99\linewidth]{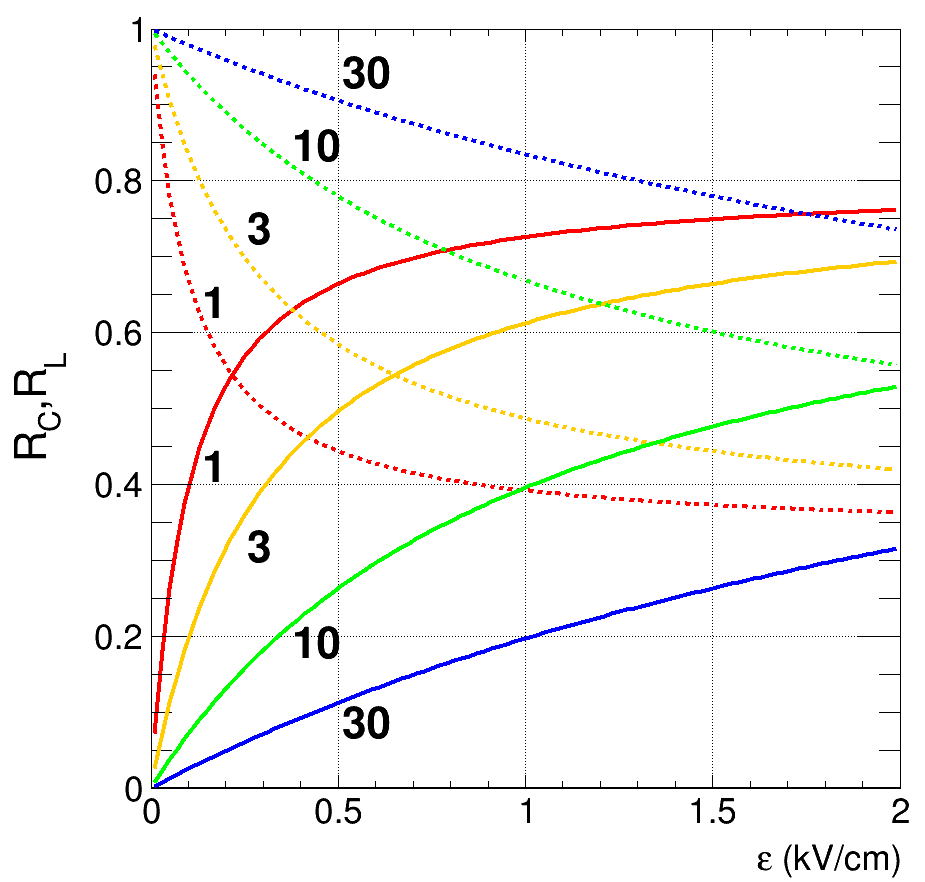}
		\caption{ }
		\label{fig:recombination_plot}
	\end{subfigure}
	\caption{\textbf{(a)}~Charge recombination of the electron-ion pairs with ICARUS~\cite{electron_recomb_icarus}, Scalettar et al. experiment~\cite{recombination_scalettar} and Aprile at al.~\cite{recombination_aprile} versus electric field applied. \textbf{(b)}~Recombination factors versus the electric field applied. The solid lines represent the charge recombination ($R_C$) and the dashed lines represent the light recombination ($R_L$) for different values of energy loss ($\dedx$) in units of minimum ionizing particles~\cite{electron_recomb_old,electron_recomb_icarus,recombination_old}. This plot is based on the image created by Craig Thorn in the LBNE DocDB 4482-v1.}
	\label{fig:electron_recomb_figs}
\end{figure}

The light yield is also affected by the electric field applied. Figure~\ref{fig:recombvsefield} shows the charge recombination probability versus the applied electric field, where  the ICARUS experiment~\cite{ICARUS} has measured a recombination of the electron ion pair of 60\%~\cite{electron_recomb_old,electron_recomb_icarus} for a minimum ionizing particle and an electric field of 500~V/cm. The charge recombination factor $R_C$ can be modeled as~\cite{electron_recomb_icarus}:
\begin{equation}
R_C = \frac{Q}{Q_0} = \frac{A}{1+(k/\varepsilon)\dedx}, 
\end{equation}
where $Q_0$ and $Q$ are the initial and collected charge, respectively. $\varepsilon$ is the electric field in kV/cm, $\dedx$ is the particle stopping power, $A = 0.800\pm0.003$ is a normalization parameter and $k = 0.0486\pm0.0006$~(kV/cm)(g/cm$^2$)/MeV is a Birks parameter~\cite{electron_recomb_icarus}. By assuming that only light from self excitation is produced when the charge collection is maximum, it was measured~\cite{electron_recomb_old} that about 67\% of the light of the LAr scintillation is produced by ionization process and only 33\% is caused by the self excitation process\footnote{The light yield (luminosity) and charge collected were measured as function of an electric field up to 10~kV/cm. At the highest field, the collected charge reached a maximum of 95\% of the produced and the luminosity decreased 67\%. At this electric field, only the self excitation process contributes with the light yield.}. The light recombination factor $R_L$ will be the complementary of $R_C$ as:
\begin{equation}
R_L = 1-\alpha R_C,
\end{equation}
where $\alpha$ is an attachment factor~\cite{lar_proprierties_calc}. Figure \ref{fig:recombination_plot} shows the charge and light recombination factors as function of the electric field~\cite{electron_recomb_old,electron_recomb_icarus,recombination_old}. The electric field of 500 V/cm is a good compromise between light production and charge collection. 

The DUNE FD expects about 24,000~photons/MeV~\cite{DUNE_vol4}, because the light yield of 41,000~photons/MeV for minimum ionizing particles will have 60\% reduction in the recombination due to the 500~V/cm electric field.

\subsection{Emission spectrum and time profile}

\begin{figure}[h!]
	\centering
	\includegraphics[width=0.6\linewidth]{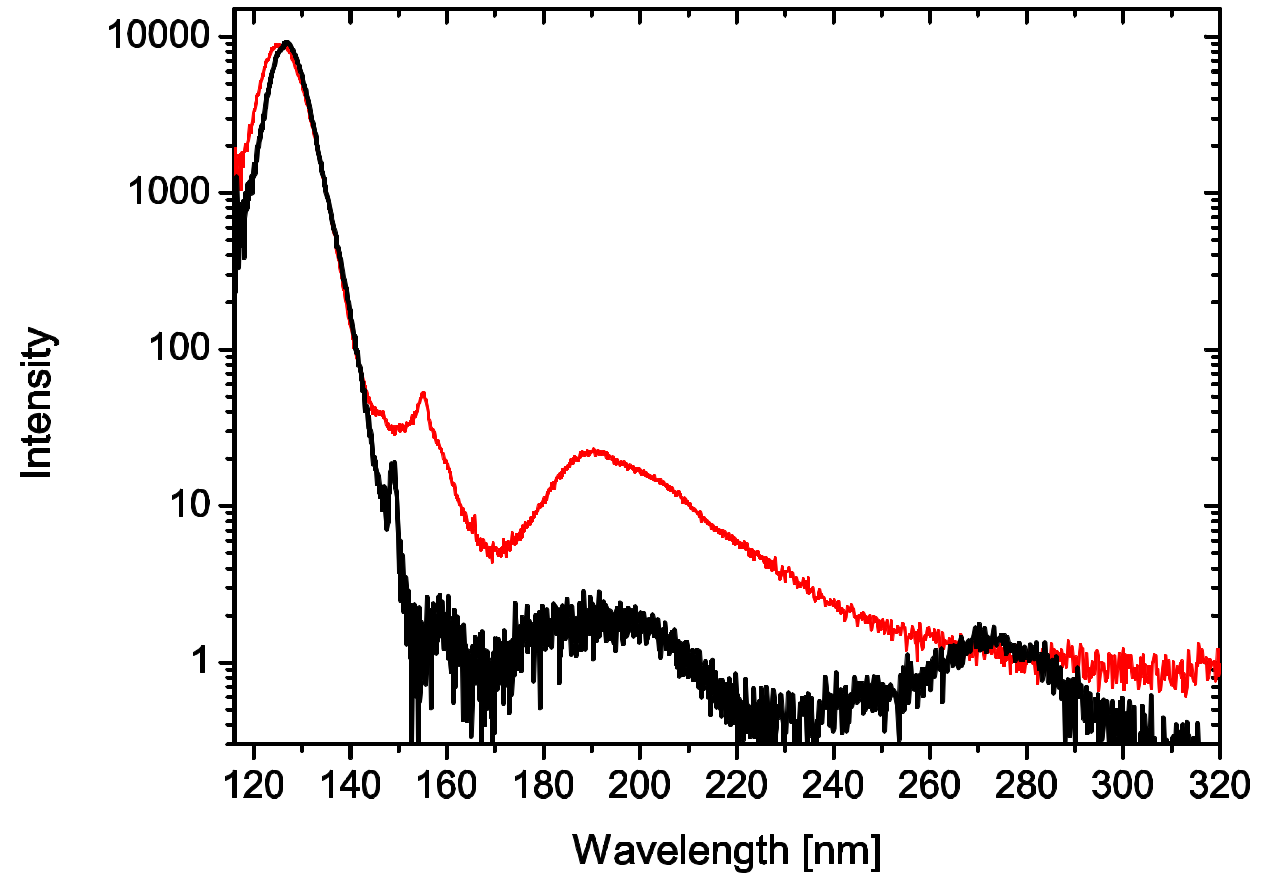}
	\caption{Scintillation light emission spectrum for LAr (Black line) and GAr (red line)~\cite{lar_scint_spec}.}
	\label{fig:scintillationspectrum}
\end{figure}

The scintillation light is emitted in the Vacuum Ultra Violet (VUV) in a ten nm band centered around 127 nm~\cite{LAr_arapuca_test} as shown in Figure~\ref{fig:scintillationspectrum}~\cite{lar_scint_spec} for liquid and gaseous argon, black and red lines respectively. 

Excited argon molecules have two possible excimer states, singlet and triplet ($^1\Sigma_u$ and $^3\Sigma_u$). The de-excitation of the $^1\Sigma_u$ state is very fast with a characteristic time of $\sim$6~ns, while the $^3\Sigma_u$ state, for been a forbidden transition, de-excites with a characteristic time of $\sim$1600~ns~\cite{LAr_arapuca_test,LAr_fund_properties}.

The prompt light response of the LAr to the passage of ionizing radiation is the key to point the $T_0$, the time when the electrons started drifting, for non-beam events such as proton decays or supernovas since for beam events the $T_0$ is calculated on the basis of the protons spill time at Fermilab. The drift time is taken by the difference between $T_0$ and the arrival of the ionized electrons on the anode wires and it is used to reconstruct the coordinate along the drift direction with a resolution better than 1 mm. The coordinate along the drift direction is important to correct for the ionized charge absorption along the drift and to fiducialize the active volume of the detector~\cite{DUNE_vol4}.

The time evolution of the LAr scintillation signal $I(t)$ can be described as the sum of two exponentially decaying distributions~\cite{LAr_fund_properties,nitrogen_contamination_roberto}:
\begin{equation}
\label{eq:_lar_fast_and_slow_fit}
I(t) = A_S \exp(-\frac{t}{\tau_S}) + A_T \exp(-\frac{t}{\tau_T}),
\end{equation}
where $A_S$ and $A_T$ are the relative amplitudes and $\tau_S$ and $\tau_T$ are the time constants of the singlet and of the triplet dimer states\footnote{The components are often labeled as fast and slow to emphasize the time profile of each emission.}, respectively, and the normalization $A_S + A_T = 1$ is typically applied. The relative intensity of the singlet and triplet components ($A_S/A_T$ is related to the ionization density of LAr and depends on the ionizing particle: 0.3 for electrons, 1.3 for alpha particles and 3 for neutrons~\cite{LAr_fund_properties,nitrogen_contamination_roberto,abudance_dependence}.

The dependence of singlet and triplet (fast and slow) contribution with the type of particle is useful feature for particle identification (PID). In Chapter~\ref{chap:lar_test}, a Pulse-shape Discrimination (PSD) method is applied to separate $\alpha$-particle signals from cosmic muons. For the DUNE experiment, the PID is mainly done by the TPC signal, because in such a large detector, the chance of having single particle events is pretty small~\cite{DUNE_vol4}.

\subsection{Nitrogen contamination and quenching of light}
\label{sec:quenching_nitrogen}

The inelastic collisions of Ar excimers with Nitrogen molecules can cause the non-radiative de-excitation of the excimer, with consequent loss of the scintillation photons. The process can be summarized as follows:
\begin{equation}
\label{eq:n2_contaminat}
\text{Ar}_2^* + \text{N}_2 \rightarrow 2 \text{Ar} + \text{N}_2.
\end{equation}

\begin{figure}[h!]
	\centering
	\begin{subfigure}{0.443\textwidth}
		\includegraphics[width=0.99\linewidth]{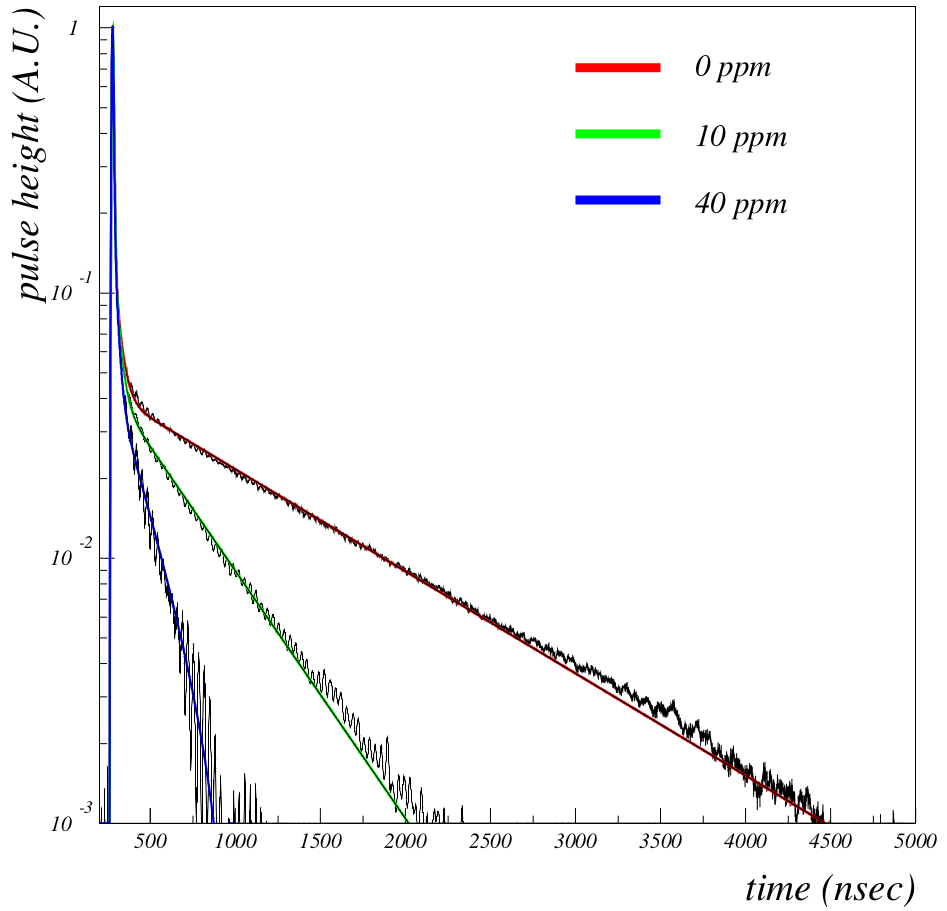}
		\caption{ }
		\label{fig:lar_scint_time_profile}
	\end{subfigure}
	\begin{subfigure}{0.545\textwidth}
		\includegraphics[width=0.99\linewidth]{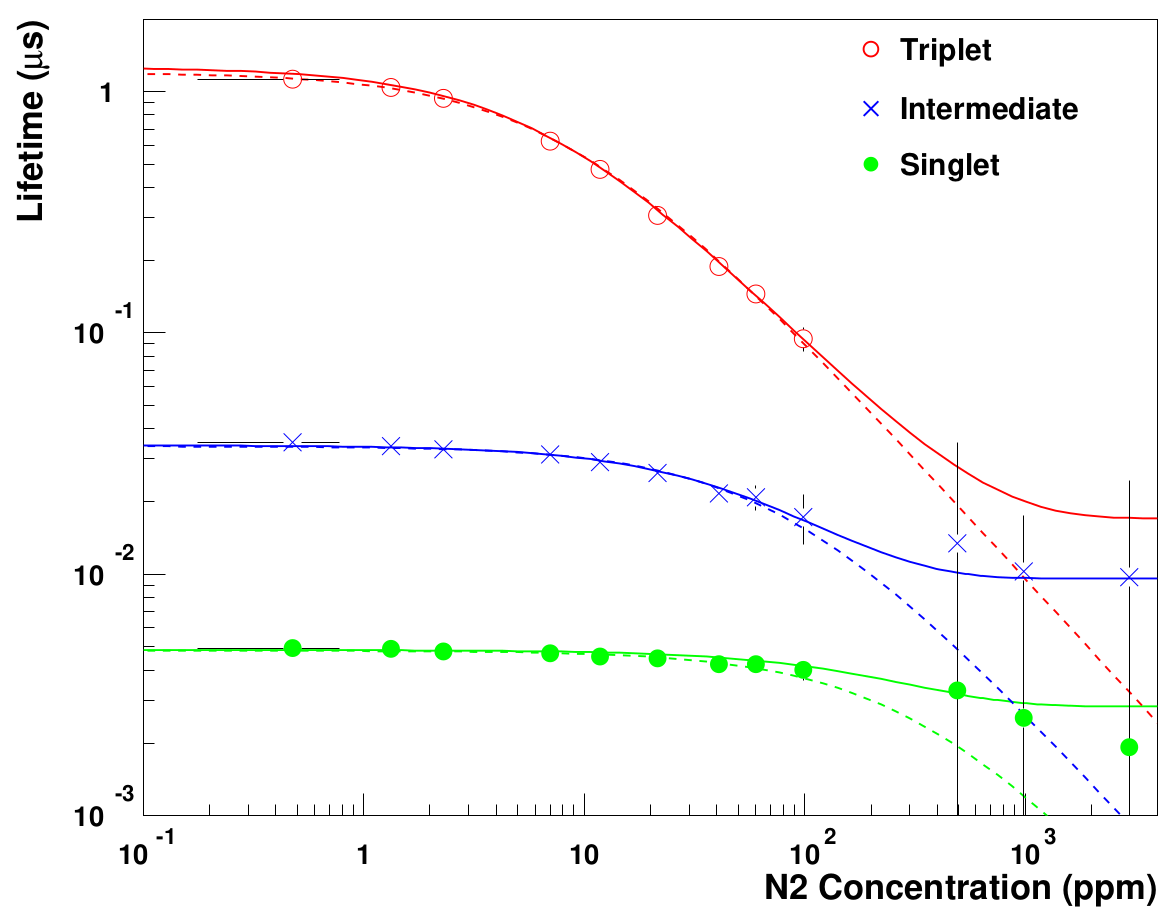}
		\caption{ }
		\label{fig:nitrogen_contamination_tau}
	\end{subfigure}
	\caption{\textbf{(a)}~LAr scintillation signal shape for 0~ppm, 10~ppm and 40~ppm of N$_2$ contamination~\cite{nitrogen_contamination_roberto}. The fits are performed with an intermediate component to better adjust the data. \textbf{(b)}~Lifetime of singlet, triplet and intermediate components versus $N_2$ contaminant in ppm~\cite{nitrogen_contamination_roberto}. The dashed lines are fits considering Eq.~\ref{eq:n2_contaminat}, while the solid line is the fit considering a possible saturation at increasing concentration.}
	\label{fig:nitrogen_contamination}
\end{figure}

In this situation, the quenching process decreases the excimer concentration $[\text{Ar}_2^*]$, while keeping constant the contaminant concentration [N$_2$]. As a first approximation, the first-order rate can be assumed as~\cite{nitrogen_contamination_roberto}:
\begin{equation}
\frac{\diff [\text{Ar}_2^*]}{\diff t} = - k_Q [\text{N}_2][\text{Ar}_2^*] \qquad \xrightarrow{\qquad} \qquad  [\text{Ar}_2^*](t) = [\text{Ar}_2^*](t=0)\me^{-k_Q[\text{N}_2]t},
\end{equation}
where $k_Q = 0.11$~$\mu$s$^{-1}$ppm$^{-1}$ is the quenching rate constant. The N$_2$ quenching process affects mainly the triplet component, due to its longer lifetime. The effective decay time constant can be written as:
\begin{equation}
\frac{1}{\tau^\text{eff}_j} = \frac{1}{\tau_j} + k_Q [\text{N}_2],
\end{equation}
where $j = [S,T]$ for singlet and triplet. Also the amplitude of each component is affected. Figure~\ref{fig:lar_scint_time_profile} shows the signal shapes of the LAr scintillation light for three different concentration of N$_2$~\cite{nitrogen_contamination_roberto}. One can see the two pure exponential components (Eq.~\ref{eq:_lar_fast_and_slow_fit}) for 0~ppm concentration and the triplet time constant decreases with increasing N$_2$ contaminant. To better illustrate this effect, Figure~\ref{fig:nitrogen_contamination_tau} brings the fitted values of singlet, triplet and intermediate\footnote{The intermediate component was introduced to better fit the data and, as reported at Ref~\cite{nitrogen_contamination_roberto} it is not well understood.} components. The singlet lifetime is affected only with high contaminant concentrations, otherwise, the triplet lifetime is the most affected.

The DUNE experiment requires that Nitrogen contamination must be below 10~ppm~\cite{DUNE_vol4}. 

\subsection{Summary}

Table~\ref{tab:lar_properties} summarizes the LAr characteristics that are relevant for this work. The three first rows are parameters relevant for the TPC and scintillation, while the others concern only the light emission properties of LAr.

Rayleigh scattering is the elastic scattering of light by argon atoms. The Rayleigh scattering length of a photon strongly depends on its wavelength ($\propto \lambda^{-4}$) and on the optical properties of the medium. Recent measurements\footnote{The Rayleigh scattering length was previously calculated to be about 90~cm which was in disagreement with the measured value of 66\error3~cm~\cite{rayleigh_ar_old,rayleigh_ar_xe}. Recently calculations~\cite{rayleigh_ar_new} have pointed to a length of 55\error5~cm, but with a higher refraction index.} point to a Rayleigh scattering length of 99.1\error2.3~cm for $\lambda = 127$~nm~\cite{rayleigh_most_new_and_awesome}. The absorption length is related to the absorption of LAr scintillation light from impurities such as dissolved nitrogen.
\begin{table}[h!]
	\centering
	\caption{Summary of liquid argon properties.}
	\label{tab:lar_properties}
	\begin{tabular}{lr}
		\hline
		\rowcolor[HTML]{EFEFEF} 
		Mean energy loss (mip)                                       & $\left\langle{\mathrm{d}E_\text{mip}/\mathrm{d}x}\right\rangle$ = 1.519 MeV/(g/cm$^2$)~\textsuperscript{\cite{pdg}} \\
		Average energy for pair production ($\me^-$, $Ar^+$)         & $W_l = 23.6 \pm 0.3$~\textsuperscript{\cite{lar_for_low_en_part,lar_ion_pair_energy}}                                                                   \\
		\rowcolor[HTML]{EFEFEF} 
		Excited to ionized atoms ratio                               & $N_{ex}/N_i=0.21$~\textsuperscript{\cite{absolute_scint,Segreto_2021,lar_for_low_en_part}}                                                                      \\ \hline
		$\gamma$ emission spectrum                                   & $\av{\lambda_{\text{scint}}} = 127$~nm; $\sigma_\text{scint} \approx 3$ nm~\textsuperscript{\cite{LAr_arapuca_test}}             \\
		\rowcolor[HTML]{EFEFEF} 
		Decay time consntats                                         & $\tau_S \sim 6$ ns; $\tau_T \sim 1600$ ns~\textsuperscript{\cite{LAr_arapuca_test,LAr_fund_properties}}                                              \\
		Relative intensity                                           & \multicolumn{1}{l}{$A_S/A_T = 0.3$ for electrons and muons}                            \\
		& \multicolumn{1}{l}{\hspace{3.4em}$= 1.3$ for alpha particles}                          \\
		& \multicolumn{1}{l}{\hspace{3.4em}$= 3.0$ for neutrons~\textsuperscript{\cite{LAr_fund_properties,nitrogen_contamination_roberto,abudance_dependence}}}                                 \\
		\rowcolor[HTML]{EFEFEF} 
		Average energy for $\gamma$ production                       & $W_{ph}^{\text{min}} = 19.5 \pm 1.0~\eV$~\textsuperscript{\cite{absolute_scint,Segreto_2021,lar_for_low_en_part}}                                               \\
		\begin{tabular}[c]{@{}l@{}}Light Yield [$\epsilon = 0$ V/cm] (ideal)\\ \hspace{5.3em}[$\epsilon = 0$ V/cm] (mip)\\ \hspace{5.3em}[$\epsilon = 500$ V/cm] (mip)\end{tabular} &
		\begin{tabular}[c]{@{}r@{}}$Y_{ph}^\text{ideal} = 5.1\times 10^4~\gamma/\MeV$\\ $Y_{ph}^\text{mip} = 4.1\times 10^4~\gamma/\MeV$\\ $Y_{ph}^\text{mip} = 2.4\times 10^4~\gamma/\MeV$~\textsuperscript{\cite{absolute_scint,DUNE_vol4}}\end{tabular} \\
		\rowcolor[HTML]{EFEFEF} 
		Rayleigh scattering length ($\lambda_\text{scint} = 127$ nm) & 99.1\error2.3~cm\textsuperscript{\cite{rayleigh_most_new_and_awesome}} \\ 
		Absorption length (for N$_2$ concentration < 5 ppm)          & $L_A > 20$ m~\textsuperscript{\cite{lar_atten_length}}                                                                           \\
		\rowcolor[HTML]{EFEFEF} 
		Refractive index                                             & $n_\text{LAr} = 1.38$~\textsuperscript{\cite{rayleigh_ar_xe}}                                                                  \\ \hline
	\end{tabular}
\end{table}


In the DUNE experiment, Rayleigh scattering and absorption length must be taken into account to properly fiducialize the detector and to perform calorimetric measurements.

\section{The Standard \ara}
\label{sec:s_arapuca}
\begin{figure}[h!]
	\centering
	\begin{subfigure}{0.515\textwidth}
		\includegraphics[width=0.99\linewidth]{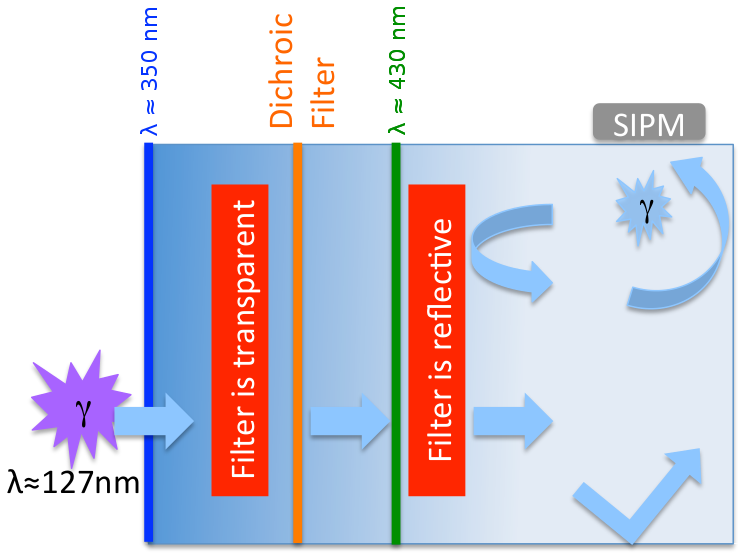}
		\caption{ }
		\label{fig:arapuca_schematic}
	\end{subfigure}
	\begin{subfigure}{0.465\textwidth}
		\includegraphics[width=0.99\linewidth]{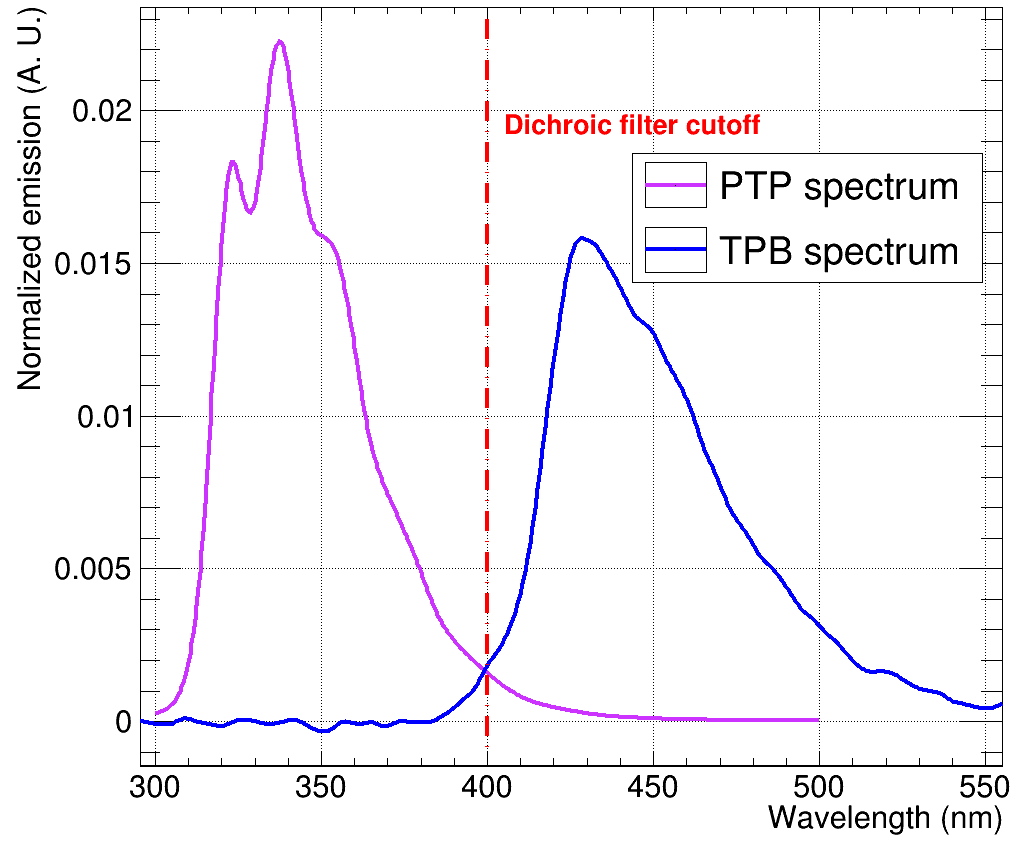}
		\caption{ }
		\label{fig:arapuca_wavelength_shifters}
	\end{subfigure}
	\caption{\textbf{(a)}~\ara\ principle of work~\cite{propostaARA}. \textbf{(b)}~The dichroic filter cutoff (red dashed line), the \ptp\ (purple) and TPB emission spectra.}
	\label{fig:standard_arapuca}
\end{figure}

The \ara~\cite{propostaARA} is a light trap device that captures wavelength shifted photons inside a reflective box until they are either detected by Silicon Photomultipliers (SiPMs) or lost. The working principle of the Standard \ara\ (S-\ara) is the following: the wavelength shifter (WLS) para-Terphenyl (\ptp)~\cite{pTP} downshifts\footnote{Downshift here refers to the photon energy: as the wavelength increases, the photon energy decreases.} the 127~nm light from LAr scintillation to around 350~nm. A dichroic filter (Sec.~\ref{sec:dichroic}) has high transparency for wavelengths below a certain cutoff and high reflectance for wavelengths above it. Light of the \ptp\ can go through the dichroic filter and enters the device to find a layer of wavelength shifter tetra-phenyl butadiene (TPB)~\cite{TPB_LAr} coated in highly reflective foil (\viku~\cite{vikuiti}). The TPB downshifts again the light to around 430~nm,  which can now be reflected by the dichroic filter (and \viku\ foils around the device walls) until detected by the SiPMs or lost as schematized in Fig.~\ref{fig:arapuca_schematic}.

Figure~\ref{fig:arapuca_wavelength_shifters} shows the emission spectrums of the two WLS: the \ptp\ spectrum (in purple) have wavelength below the dichroic filter cutoff of 400~nm (red dashed line) and enters the \ara. The TPB spectrum (in blue) is above the cutoff so its downshifted light is reflected inside the device. 

The S-\ara\ prototype was initially proposed as the light collector for the DUNE experiment. It was deployed for tests at LArIAT~\cite{lariat} and ProtoDUNE-SP~\cite{protoDUNE_TDR}. In 2016 a liquid argon test was performed at the Brazilian Synchrotron Light Laboratory (LNLS) were the efficiency of (1.0\error0.2)\% and (1.2\error0.2)\% was found for alphas and muons~\cite{LAr_arapuca_test}, respectively. During the start of this Ph.D thesis, the \ara\ trapping effect was verified, and studies were conducted trying to establish the best WLS to be used in the device.

The S-\ara\ has a disadvantage that both WLS need to be vacuum evaporated onto \viku\ foils and on the dichroic filters. The process of vacuum evaporation is time consuming and TPB films can be mechanically unstable during the cool down process to LAr temperature.

\section{The \xara}
\label{sec:x_arapuca_description}

\begin{figure}[h!]
	\centering
	\begin{subfigure}{0.5265\textwidth}
		\includegraphics[width=0.99\linewidth]{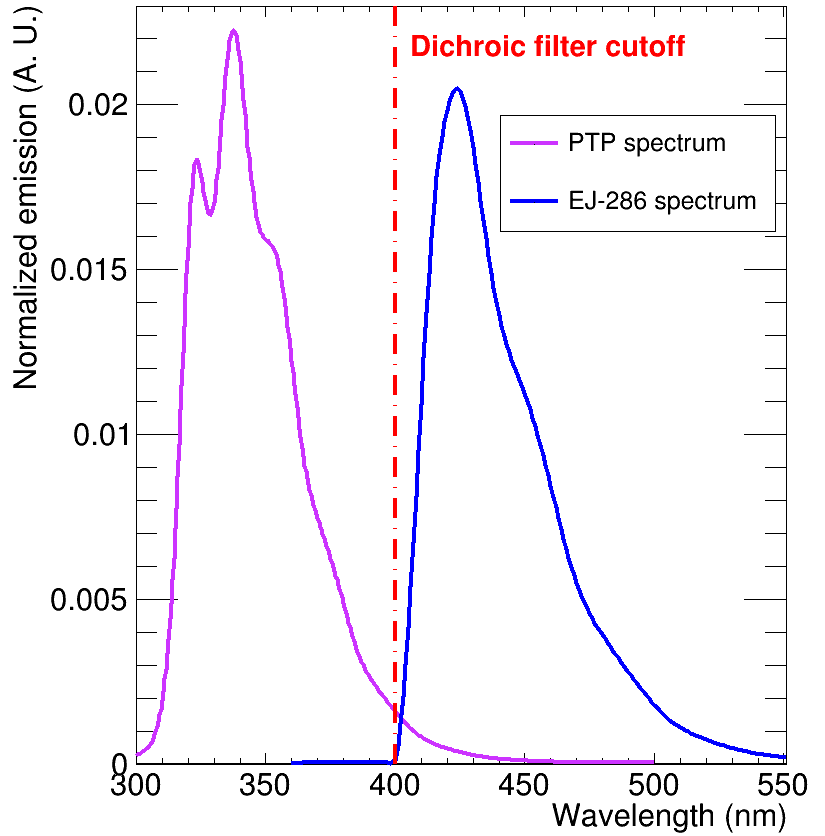}
		\caption{}
		\label{fig:ptp_ej286_spectrum}
	\end{subfigure}
	\begin{subfigure}{0.3465\textwidth}
		\includegraphics[width=0.99\linewidth]{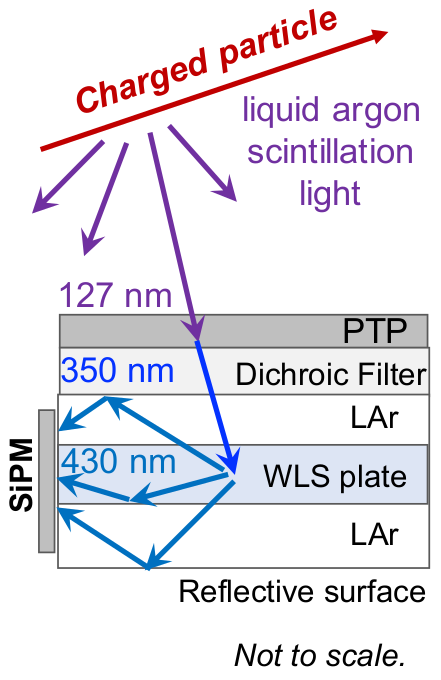}
		\caption{}
		\label{fig:xa_concept}
	\end{subfigure}
	\caption{\textbf{(a)}~The dichroic filter cutoff (red dashed line), the \ptp\ (purple) and the EJ-286 emission spectra. \textbf{(b)}~\xara\ principle of work, with total internal reflection and the reflective cavity trapping the photons~\cite{DUNE_vol4}.}
	\label{fig:xa_concept_spectrum}
\end{figure}

The \xara~\cite{x_arapuca} is an improvement of the S-\ara. It uses the principle of total internal reflection onto a wavelength shifter slab instead of WLS coated on a reflective foil. The outer \ptp\ layer of WLS coating is still part of the device, and the matching between wavelengths and dichroic filter cutoff is still used as shown in Figure~\ref{fig:ptp_ej286_spectrum} with the emission of the Eljen EJ-286 slab~\cite{eljen_286}. However, a portion of light that is downshifted inside the WLS slab cannot exit due to total internal reflection (see Appendix~\ref{chap:total_internal_reflection}).

Figure~\ref{fig:xa_concept} shows the schematic of the \xara, where the light re-emitted by the WLS can be trapped by total internal reflection or escape the bar and be reflected on the dichroic filter or on the \viku\ foils.

\begin{figure}[h!]
	\centering
	\includegraphics[width=0.488\linewidth]{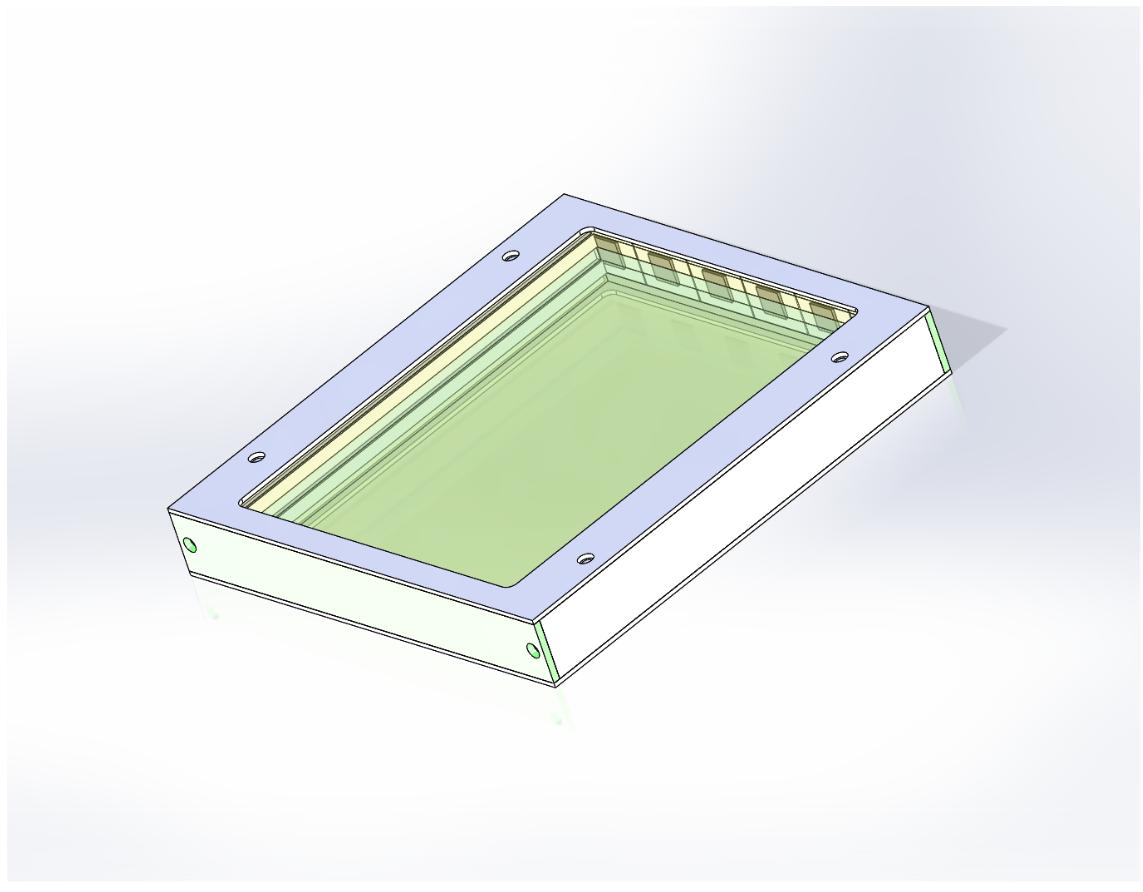}
	\includegraphics[width=0.49\linewidth]{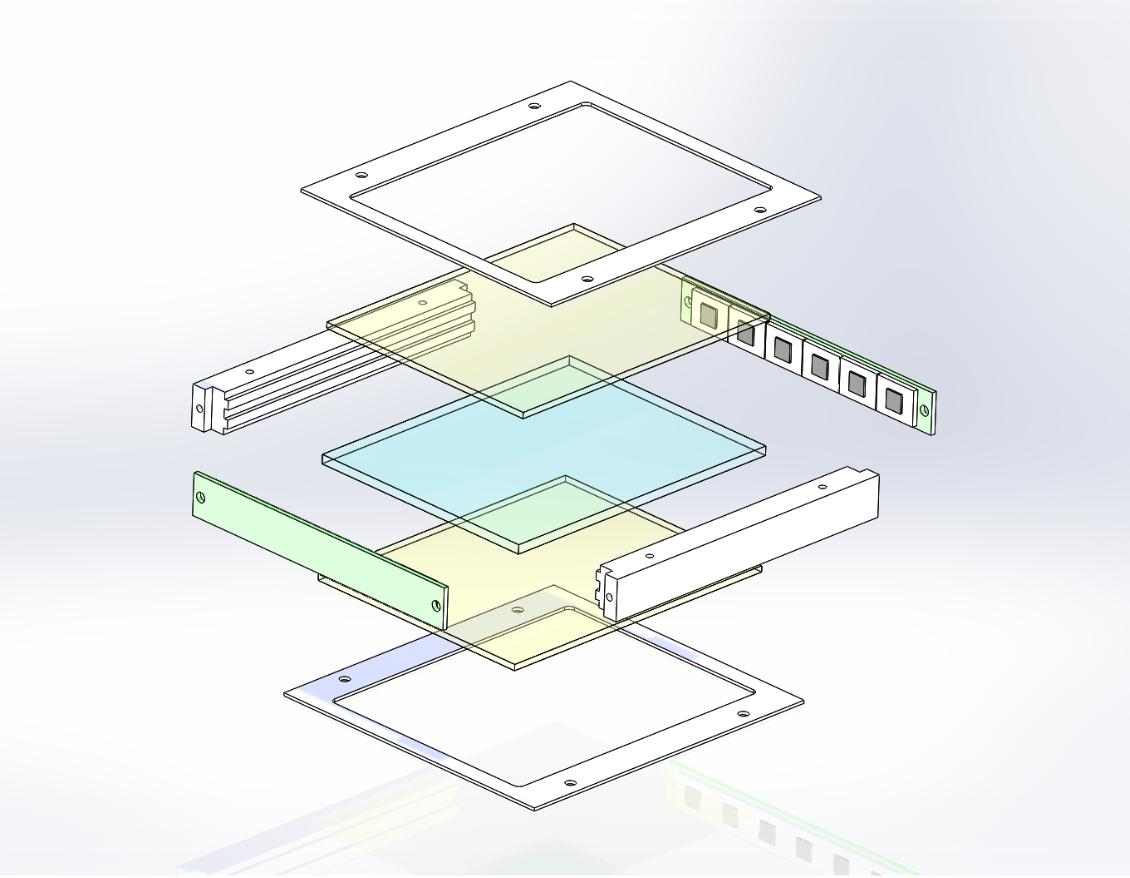}
	\caption{Simplified schematic of a double-sided \xara ~\cite{DUNE_vol4} assembled (left) and exploded view (right). The yellow plates are dichroic filters coated with \ptp\ externally to the \xara\ and the blue plate is the WLS slab.}
	\label{fig:double_sided}
\end{figure}

Besides the fact that the internal WLS is no longer coated onto \viku\ foil, the \xara\ can be deployed as either single-sided or double-sided by using either an opaque reflector plate (single) or a second dichroic filter window (double) on the second face (bottom part of Fig.~\ref{fig:xa_concept}). This is an interesting feature of the device, allowing to place the double-sided \xara s in the central APA of DUNE (see Fig.~\ref{fig:fdmoduledesign}), collecting light from both sides. The schematic of a double-sided \xara\ is displayed in Fig.~\ref{fig:double_sided}, the two yellow transparent plates are the dichroic filters and the blue one is the wavelength shifter.

The structure of the \xara\ is made of FR-4 G-10 (Garolite$^\text{\textregistered}$), a material with good adhesion to \viku\ foils, cryogenic resistant, low thermal contraction, isolating and with a low cost and easy to manufacture.   

\subsection{The device challenges}

Different factors can impact the \xara\ efficiency: the matching of spectrums and dichroic filter cutoff, the transmittance and reflectance of the dichroic filter, the thickness of the WLS slab, the thickness of the \ptp\ deposition, the quality of the deposition, the efficiency of both WLS deployed, the amount of SiPMs to detect the light, the gap between the WLS and the SiPMs, the SiPMs quantum efficiency. Some of these characteristics are quite easy to control and do not play a major role in the efficiency, while others can be more difficult to investigate and may significantly contribute to the light collection efficiency.


The gap between the WLS and the SiPMs is caused due to the higher thermal contraction coefficient of the WLS slab compared to the FR-4, in which the SiPMs are attached. The Shrinkage Factor for a 206~$^\circ$C temperature drop was measured\footnote{The units given is a shrink of one meter for each one meter of material.} to be $1.4\times~10^{-2}$~m/m for Polystyrene (WLS Bars) and $2.1\times10^{-3}$~m/m for the FR-4 G10~\cite{DUNE_vol4}. Simulations performed for the \xara\ prototype~\cite{simulation_xara_Paulucci_2020} showed a drop of $\sim$30\% in the total efficiency of the device for a separation of 0.5~mm, while keeping almost constant up to 1.5~mm. New designs are under investigation to achieve a better optical coupling between the SiPMs and the WLS bars.

The minimum value of the \ptp\ layer must be between 100 to 200~$\mu$g/cm$^2$ to ensure the VUV is fully absorbed. For the DUNE FD module a thickness of 400~$\mu$g/cm$^2$ was chosen, to ensure that the minimum thickness is reached everywhere on the filter, even in the case of exceptional fluctuations~\cite{DUNE_vol4}. The WLS slab thickness was chosen as 3.5~mm to allow almost complete absorption of the photons emitted by the \ptp\, to ensure the nominal conversion efficiency and to keep a 2~mm gap in which LAr can enter in both sides preventing physical contact with the surface. 

\subsection{The Short Baseline Neutrino Detector (SNBD)}
\label{sec:sbnd}

The \xara\ with two dichroic filters deployed, namely \xara\ double-cell, will also be part of the Short Baseline Neutrino Detector~(SBND) light detection system. The Short Baseline Neutrino (SBN) program at Fermilab~\cite{sbnd,sbnd_at_fermilab,SBND_construction} consists of three LArTPCs: the SBND, the ICARUS~T600~\cite{ICARUS} and MicroBooNE~\cite{MicroBooNE_TDR}, and aims to resolve the short baseline $\nue$ and $\anue$ appearance measured by MiniBooNE and LSND experiments. The SBND will be a 112~t single phase LArTPC 110~meters away from a neutrino beam with energy up to 3~GeV. Besides the primary scientific goal of measuring the $\nue$ appearance and $\numu$ disappearance at the same time, the experiment will also perform high-precision neutrino-argon cross section measurements.

\begin{figure}[h!]
	\centering
	\includegraphics[width=0.70\linewidth]{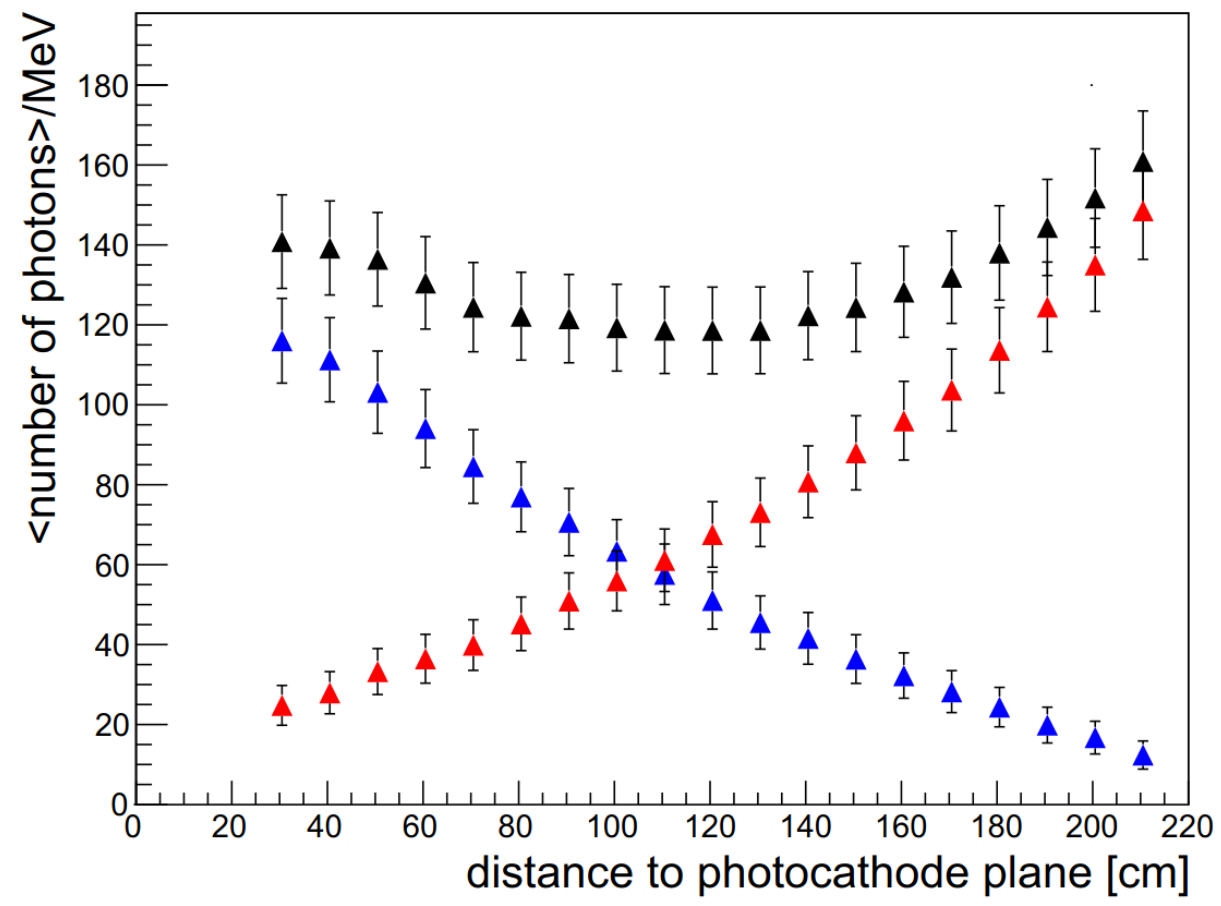}
	\caption{Average number of detected photons per MeV as a function of the distance to the photocathode plane (PDS at 0) The direct UV light detection (in blue) decreases with the distance of the source, while the reflected light (in red) increases. The total light (black) is more uniform along the source distance from the PDS system~\cite{sbnd_foils}.}
	\label{fig:sbnd_foils}
\end{figure}

The SBND will count with 200~\xara\ double-cell and 120~PMTs of 8~inches. The light detection system in SBND will allow a more precise vertex localization and will help to handle the background induced by cosmic rays (as it is a ground level experiment). To improve the light collection, the SBND cathode plane has reflective foils coated with TPB. This effect is shown in Figure~\ref{fig:sbnd_foils}~\cite{sbnd_foils}, where the average number of detected photons per MeV for different distances to photocathode plane (equivalent to distance to the PDS) is displayed. 
Direct scintillation light is displayed in blue and decreases the farther the light source is from the PDS. On the other hand, the reflected light, in red, will increase. The total light is displayed in black and shows a good uniformity along the drift direction.
 
In order to detect the 430~nm light, 20\% of the PMTs are sensitive to visible light while the other 80\% are coated with TPB to detect UV light. Besides, 100 \xara s will be sensitive to UV light and 100 \xara s will be sensitive to visible light.

\section{Silicon Photomultipliers (SiPMs)}
\label{sec:sipms}

\subsection{Principle of working}
The \ara\ device will use Silicon Photomultipliers (SiPMs) or Multi-Pixel Photon Counters (MPPCs) as active photon sensors. Silicon (Si) has an intrinsic high resistance, however, doping with atoms of 5 or 3 electrons in the outermost electron shell (such as Phosphorus or Aluminium) can turn the doped Si into an electron or hole (positive) donor, respectively. An electron donor region is named N region and the hole donor is named P region. The N region is formed by the donation of 4 electrons from the electron donor to complete the Si valence band and an extra free electron in the conduction band, this free electron becomes the charge carrier. In a similar way, the P region is formed by donation of 3 electrons from the donor to the Si valence band, leaving a free empty space (hole) for an electron, this hole can pass from atom to atom in the crystal lattice and becomes the positive charge carrier. 

\begin{figure}[h!]
	\centering
	\begin{subfigure}{0.443\textwidth}
		\includegraphics[width=0.99\linewidth]{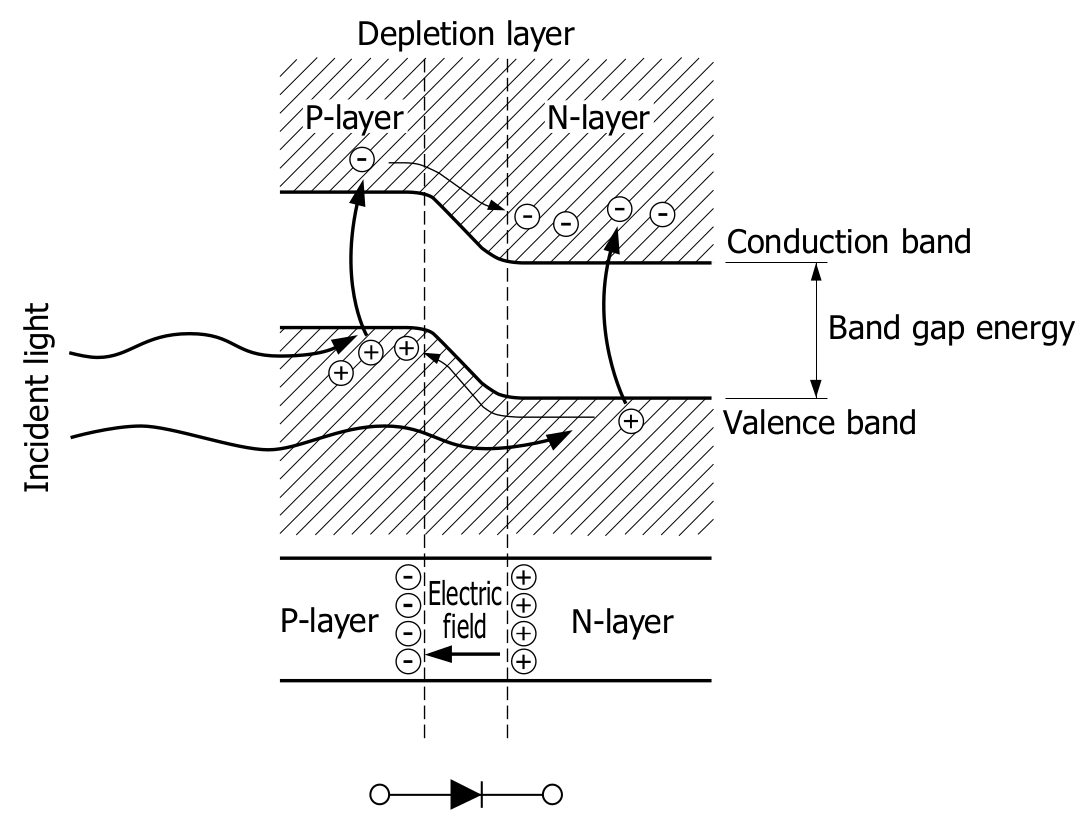}
		\caption{}
		\label{fig:sipm_band_gap}
	\end{subfigure}
	\begin{subfigure}{0.545\textwidth}
		\includegraphics[width=0.99\linewidth]{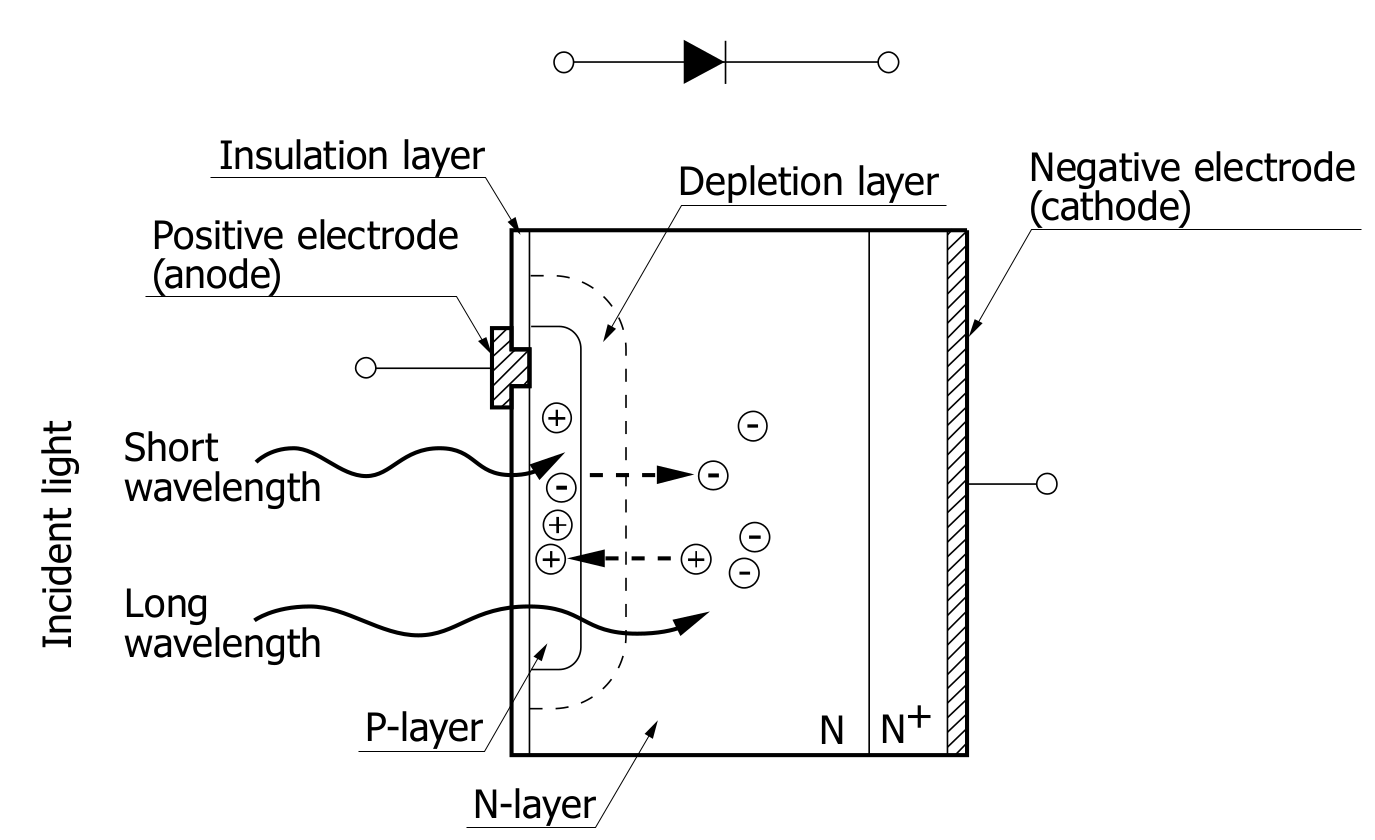}
		\caption{}
		\label{fig:sipm_structure}
	\end{subfigure}
	\caption{\textbf{(a)}~Doped silicon energy gab between the valence and conduction bands~\cite{hmmt_manual}. \textbf{(b)}~Absorption of light by a Silicon PN photodiode~\cite{hmmt_manual}.}
	\label{fig:sipm_explanation}
\end{figure}

A PN junction is formed by the contact of the two regions: electrons from the N region will drift towards the P junction and occupy empty holes, while holes in the P region will drift towards the N junction for free electrons. The accumulation of negative charges and positive charges in the P and N region will produce an electric field that opposes to the transition of charge carries to other regions. The junction enters in equilibrium, where the electron does not have enough energy to cross the junction layer, called \textit{depletion layer}. This is the standard configuration of a diode, where the P layer is the Anode and the N layer is the Cathode as shown in bottom of Figure~\ref{fig:sipm_band_gap}~\cite{hmmt_manual}. Just above, a diagram shows the positive and negative charge accumulation forming the electric field. 

The energy gap between the valence and conduction bands from the layer is kept the same (typically of 1.14~eV for Si~\cite{hmmt_manual}), however, P layer bands will increase energy while N layers will decrease\footnote{The energy bands of P layers is naturally higher than N layers, because the impurities in the P layer can exert lower force in the electrons, resulting in a slightly later orbit with higher energy.}, creating a potential barrier that the electrons cannot surpass. One can break this balance by introducing an opposite electric field, which will lower the potential and let the electrons freely drift back. This is namely forward bias, where positive voltage is applied to the anode (P layer) and negative to the cathode (N layer). In the case of SiPM, a reverse bias is used, increasing the intensity of the electric field and the gap between the conduction bands of P and N layers.

The SiPMs light detection will happen when photoelectric effect happens in the P layer or N layer passing an electron from the valence band to the conduction band, forming an electron-hole pair. The electron drifts towards the cathode N (positively biased) and holes towards the anode. The electric field gives enough energy to the charge carriers to produce new ionization while traveling, creating an avalanche effect. 

Thermal agitation can cause a surplus of charge carriers resulting in the same process, this is defined as ``Dark Current''. In SiPMs at room temperature the Dark Current dominates, typically making unfeasible to see single photo-electrons.

\begin{figure}[h!]
	\centering
	\hspace{70pt}\includegraphics[width=0.55\linewidth]{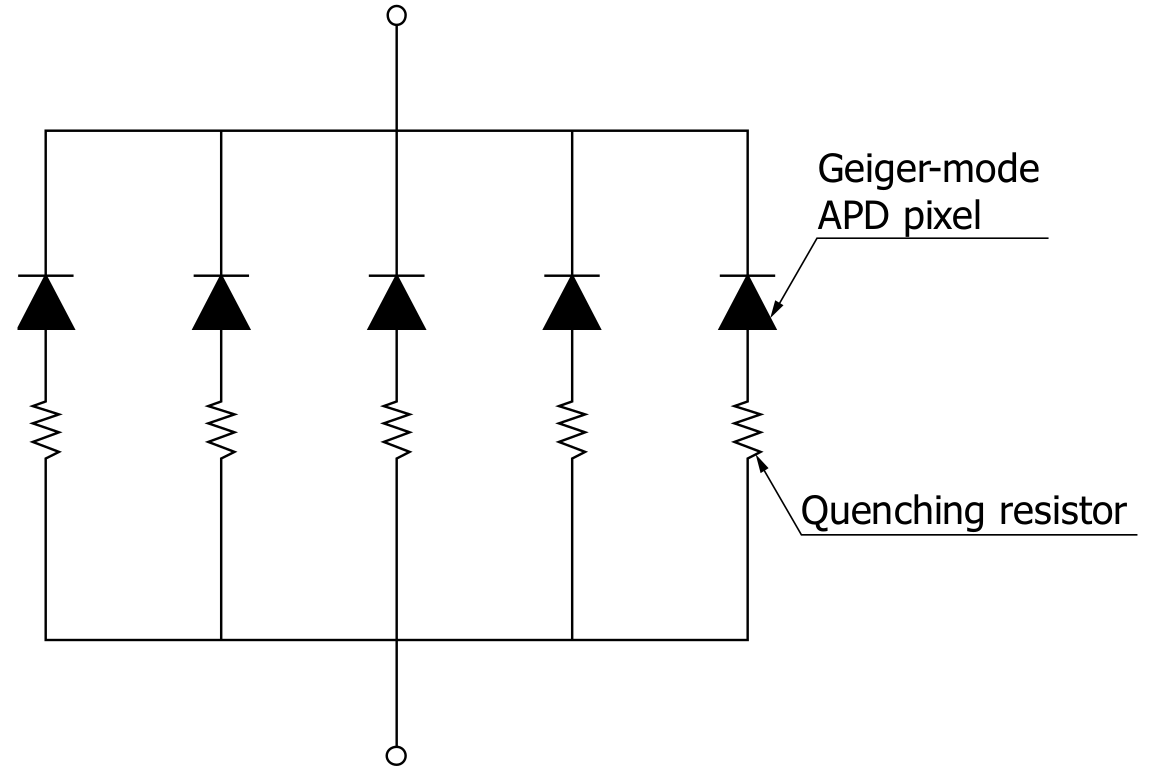}
	\caption{Concept of the MPPC matrix of Geiger-mode Avalanche Photodiode (GAPD) pixels connected in parallel~\cite{hmmt_manual}.}
	\label{fig:sipm_parallel_gang}
\end{figure}

The diagram of Fig~\ref{fig:sipm_structure} shows one microcell or pixel, for a P-to-N pixel\footnote{P-to-N are more sensitive to Ultra-Violet/blue, because shorter wavelength photons have a low penetration depth, being more easily absorbed in the P-layer. N-to-P pixels are, therefore, more sensitive to red/Near-infrared wavelengths.}.
A resistor (namely quenching resistor) is introduced in the anode to quench the current driven by the avalanche. The quenching resistor allows to operate the photodiode in Geiger-mode, that is above the breakdown voltage where the gain stops to be linear and increases exponentially (in theory, infinity gain). The microcells are ganged together in parallel as shown in Figure~\ref{fig:sipm_parallel_gang} forming the SiPM matrix, biased positively in the cathode (top connector) and signal read on the anode\footnote{The read-out at the negatively biased anode make the SiPM signal positive.} (bottom conector). The center-to-center distance between the microcells, defined as pitch, can change for different SiPMs. The S13360-6050VE MPPC used in the liquid argon tests in Brazil (Chapter~\ref{chap:lar_test}) has a pixel pitch of 50~$\mu$m and a 6$\times6$~mm$^2$ unity is composed by 14336 pixels.  

A SiPM will have a gain around $10^5$ to $10^6$ operating in the Geiger-mode. The Avalanche Photodiode (APD), utilized in the Monochromator tests (Sec.~\ref{sec:monochromator}), consist of a single pixel operated in the linear region gain of the reverse bias, having a lower gain from one to a few hundreds. The APD high sensitivity compensate the low gain, being suitable to detect higher flux of photons with respect to a SiPM. 

\subsection{Afterpulses and Cross-talk}

\begin{figure}[h!]
	\centering
	\begin{subfigure}{0.49\textwidth}
		\includegraphics[width=0.99\linewidth]{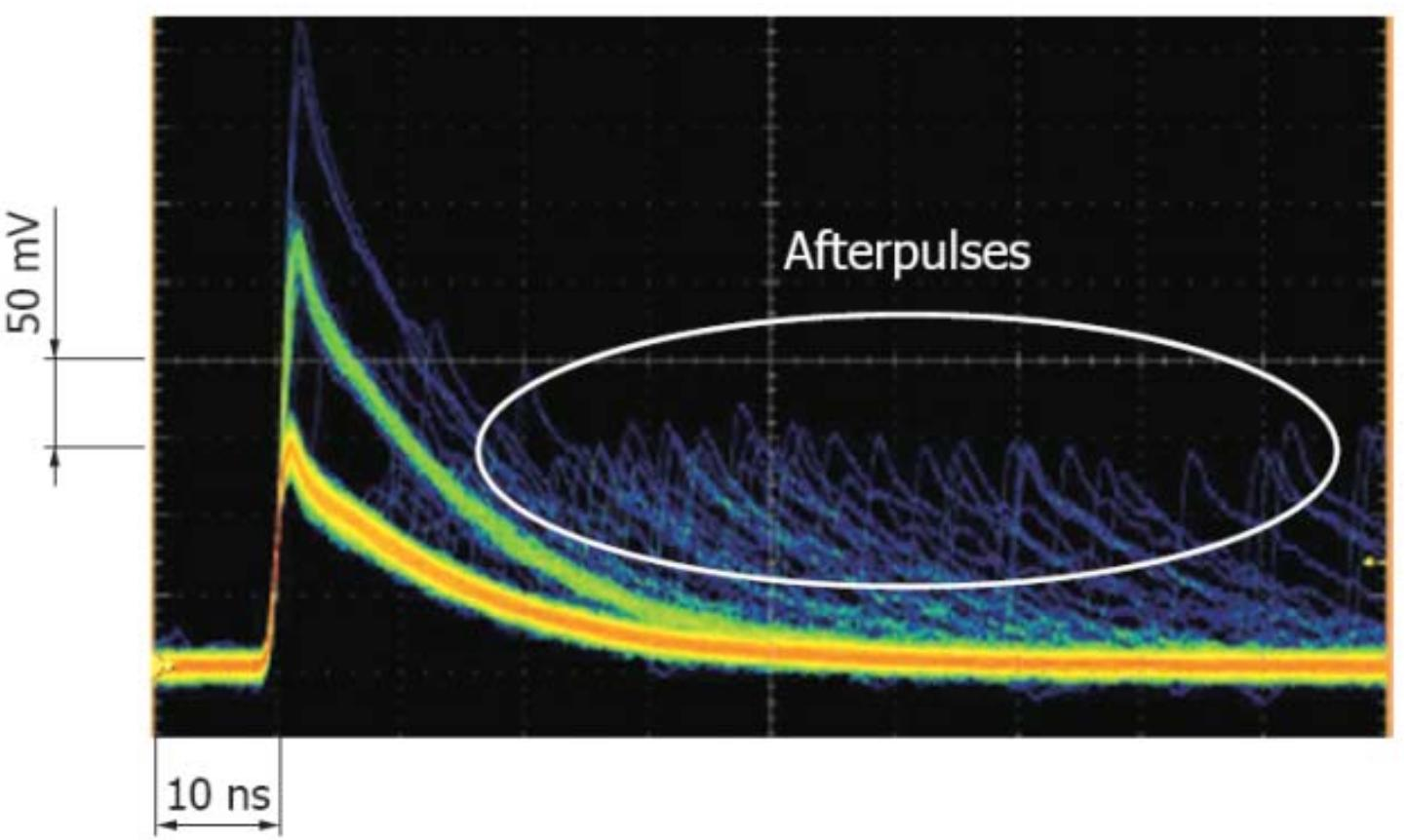}
		\caption{}
		\label{fig:sipm_afterpulse_example}
	\end{subfigure}
	\begin{subfigure}{0.49\textwidth}
		\includegraphics[width=0.99\linewidth]{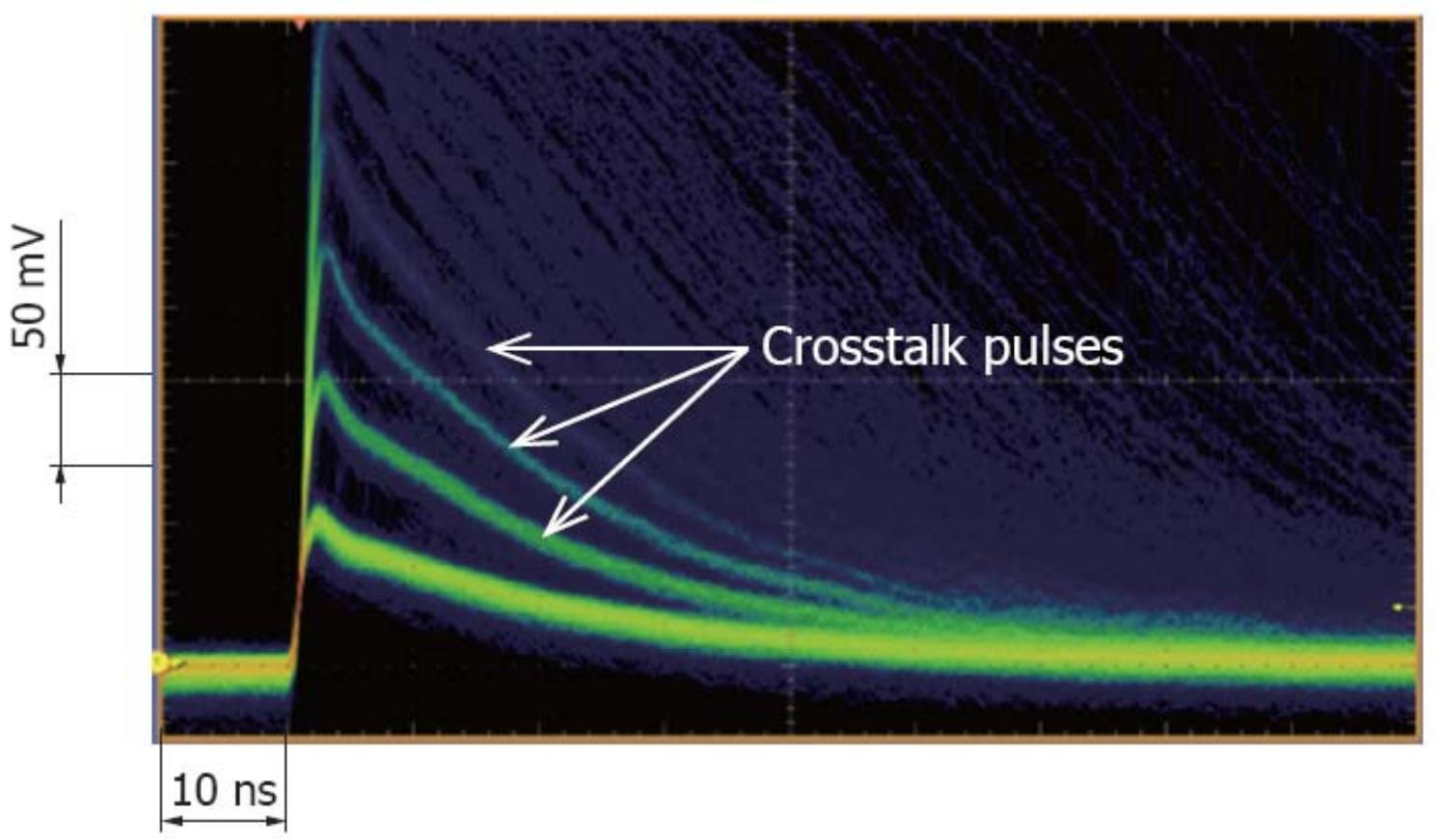}
		\caption{}
		\label{fig:sipm_crosstalk_example}
	\end{subfigure}
	\caption{Waveforms showing the occurrence of afterpulses \textbf{(a)}~and cross-talk \textbf{(b)}~\cite{hmmt_manual}. One can notice that afterpulses have amplitude lower than 1~\phe, while the cross-talk increases the events with more than 1~\phe.}
	\label{fig:sipm_afterpulses_crosstalk}
\end{figure}

Afterpulses are caused by impurities that trap a portion of the avalanche carriers for a short time during the avalanche. A new avalanche will start in their release causing a secondary late signal. If the afterpulse is released during the SiPM recovery time (the characteristic discharge time), its amplitude will be shorter than of a regular pulse of one photo-electron (\phe) as shown in Figure~\ref{fig:sipm_afterpulse_example}. If the release is done after the recovery time of the SiPM, the afterpulse will have a full 1~\phe.

Cross-talk, or optical crosstalk, is caused due to energy loss of the avalanche carriers into emission of photons with Near-infrared (NIR) wavelength~\cite{sipm_sensL}. These secondary photons can reach a neighboring pixel and initiate avalanches in them. The output is a signal with amplitude and charge higher than the original signal as shown in Figure~\ref{fig:sipm_afterpulses_crosstalk}, where the number of events with amplitude greater than 1~\phe\ has increased due to cross-talk.

Both effects, afterpulses and cross-talk, can be triggered by photons or thermal carriers (Dark Current).

The DUNE FD module will have 48 SiPMs ganged together with the requirement to operate with a $S/N>4$ and a dark noise rate, or Dark Count Rate (DCR), lower than 1~kHz. 


\section{Wavelength shifter}
\label{sec:ptp_coating}

\begin{figure}[h!]
	\centering
	\includegraphics[width=0.55\linewidth]{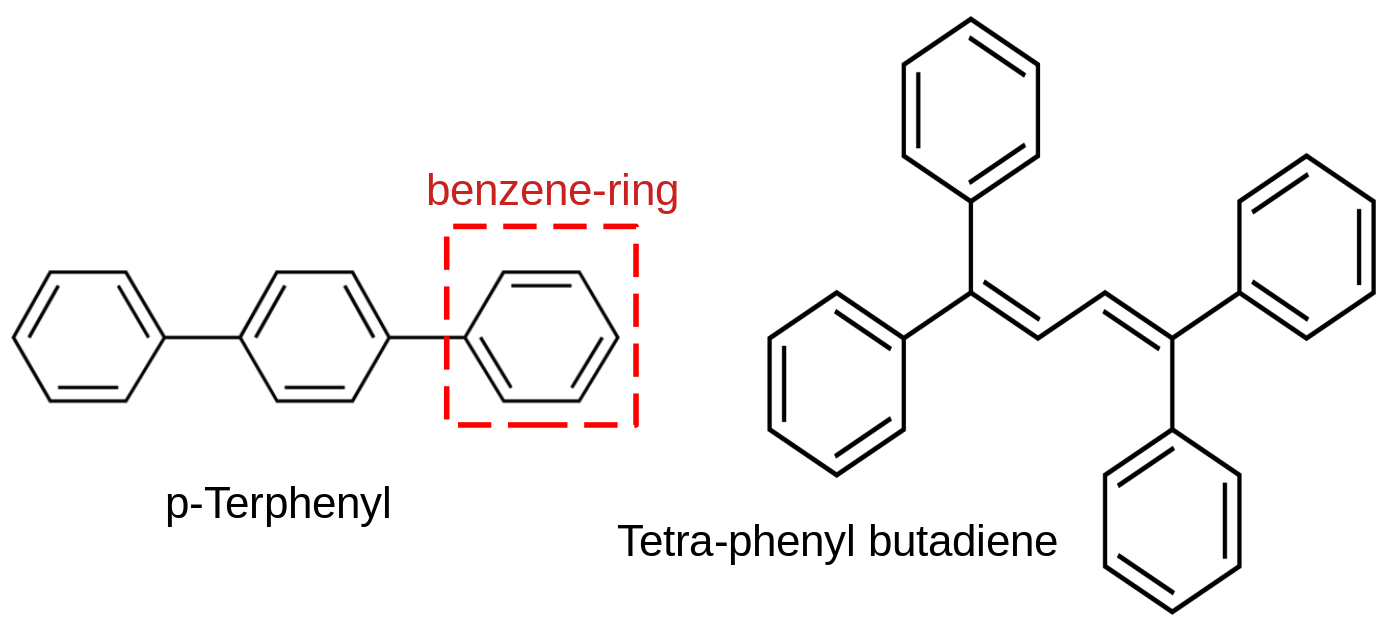}
	\caption{Organic wavelength shifters para-Terphenyl (\ptp) and tetra-phenyl butadiene (TPB).}
	\label{fig:ptp_and_tpb_rings}
\end{figure}

The outer Wavelength shifter (WLS) of the S-\ara\ and the \xara\ is the para-Terphenyl (\ptp)~\cite{pTP}. For the S-\ara\, tetra-phenyl butadiene (TPB)~\cite{TPB_LAr} was used as the inner wavelength shifter. Organic WLS, such \ptp\ and TPB are aromatic hydrocarbons compounds that contain linked benzene-ring structures, as shown in Figure~\ref{fig:ptp_and_tpb_rings}. 

\begin{figure}[h!]
	\centering
	\begin{subfigure}{0.44\textwidth}
		\vspace{14pt}
		\includegraphics[width=0.99\linewidth]{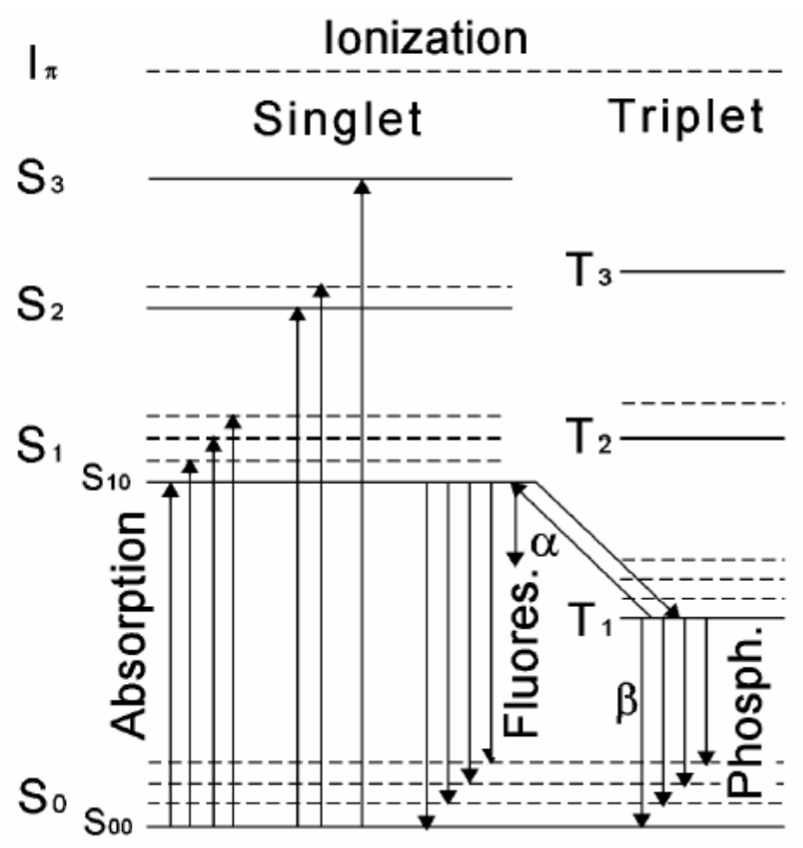}
		\vspace*{20pt}
		\caption{}
		\label{fig:organic_scint_energy_bands}
	\end{subfigure}
	\begin{subfigure}{0.55\textwidth}
		\includegraphics[width=0.99\linewidth]{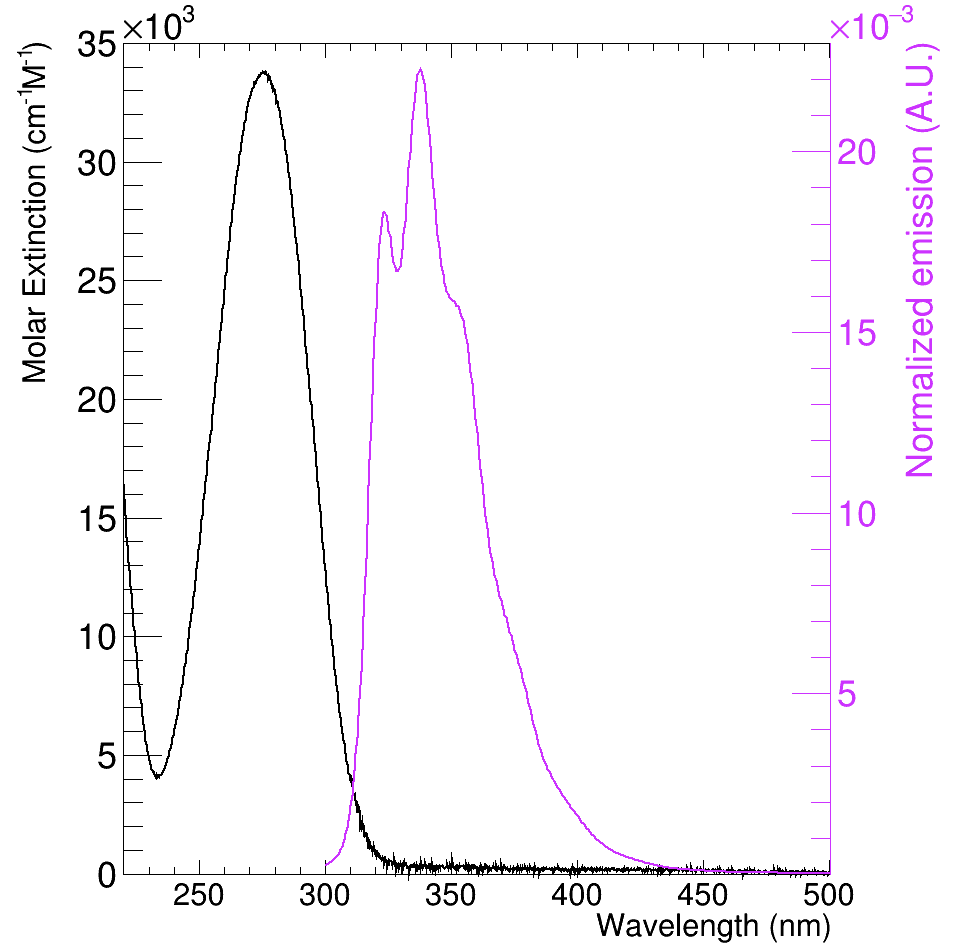}
		\caption{}
		\label{fig:absorption_emission_ptp}
	\end{subfigure}
	\caption{\textbf{(a)}~Jablonski diagram showing the energy levels of an organic wavelength shifter~\cite{organic_scintillator}. \textbf{(b)}~Absorption and emission spectra of \ptp (taken from Ref.~\cite{ptp_spetrum}). The absorption is giving in cm$^{-1}\;M^{-1}$ where $M$ = mol/m$^3$ is the molar concentration. }
	\label{fig:organic_wls}
\end{figure}

The wave shifting is a process where higher energy photons transfer their energy to the WLS which re-emits photons with a lower energy. Free valence electrons (occupying the so called $\pi$-molecular orbital~\cite{Leo2012}) absorb the photons and make a transition from the singlet ground state $S_0$ to an excited singlet states ($S_1, S_2, S_3$, etc.) as shown in Figure~\ref{fig:organic_scint_energy_bands} (thinner lines are vibrational states).

After excitation, electrons can immediately decay ($\le10$~ps) to the $S_1$ state without radiation emission, a process named internal degradation. Molecules also go through vibration relaxation to lose any excess vibrational energy, transferring the electrons from the vibrational states to the lowest excited state (tick lines). There is now a high probability of a de-excitation from $S_1$ state to one of the ground singlet vibrational states within a few nanoseconds. This process is called fluorescence, with a prompt light emission. The fact that $S_1$ states can decay to $S_0$ vibrational states with emission of photons with energy lower than that required for the transition $S_0\rightarrow S_1$ also explains why the material is typically transparent to its own light~\cite{Leo2012}. 

Alternatively, electrons can also jump to triplet states ($T_1, T_2$, etc.) by inter-system crossing\footnote{Photons will, ideally, not transfer to the triplet states from the ground singlet state because it is a forbidden transition.}. The decay from triplet states to ground state is a highly forbidden transition by selection rule, therefore, this process (called phosphorescence) takes place with a ``long time'' emission, it can take from milliseconds to several minutes or even hours. The contribution of light from phosphorescence is negligible for most organic crystals.  

Figure~\ref{fig:absorption_emission_ptp} shows the \ptp\ absorption and emission spectra~\cite{ptp_spetrum}. It is very common to have the absorption and emission spectrum intercepting each other as seen around 300~nm. A good WLS needs to have a high shifting efficiency, emit mostly prompt light and needs to be transparent to its own light. 

For the WLS slab, the principle of scintillation is the same. The main difference is that they are a solution of organic scintillators in a solid plastic solvent. For instance, the EJ-286 slab uses polivyniltoluene (PVT) as plastic solvent (the organic scintillator is a company secret). The WLS developed by the G2P co.\ and tested at Milano Bicocca (see Sec.~\ref{sec:warm_tests_bicocca}) uses Poly(methyl methacrylate)~(PMMA) doped  with 2,5-Bis(5-tert-butyl-benzoxazol-2-yl)thiophene~(BBT).

\section{Dichroic filter}
\label{sec:dichroic}

The dichroic filter plays a major role in the trapping effect of the \ara. The dichroic filter is a thin interference film, like tantalum or silicon oxides, that selectively let a specific band of light pass through the filter while being highly reflective to other colors. The principle of work is the same as thin oil film in water: a thin layer of oil will reflect an incident light on its surface and in the contact face with water. Depending on the wavelength and length traveled by the reflected light with water surface it can create a constructive or destructive interference with the incident reflected light. Dichroic filters use this property by coating layers of thin materials over an optical surface.

\begin{figure}[h!]
	\centering
	\begin{subfigure}{0.475\textwidth}
		\includegraphics[width=0.99\linewidth]{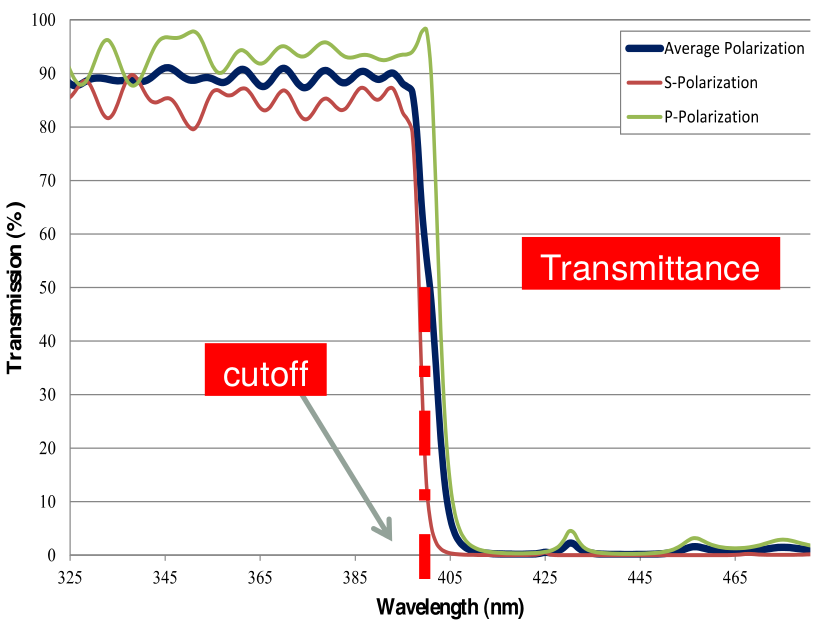}
		\caption{}
		\label{fig:transmitancia}
	\end{subfigure}
	\begin{subfigure}{0.505\textwidth}
		\includegraphics[width=0.99\linewidth]{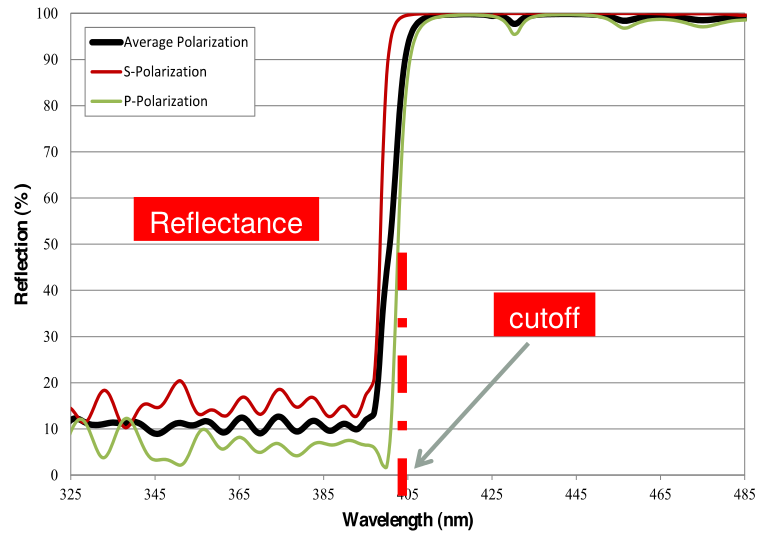}
		\caption{}
		\label{fig:refletancia}
	\end{subfigure}
	\caption{Dichroic filter transmittance \textbf{(a)}~and reflectance \textbf{(b)}~measured for two different light polarities and the average~\cite{propostaARA}.} 
	\label{fig:transmitance_and_reflectance}
\end{figure}

Figures~\ref{fig:transmitancia} and~\ref{fig:refletancia} show the transmittance and reflectance of the (Edmund optics~\cite{edmunds}) dichroic filter. Transmittance and reflectance of the filter will also depend on the angle and polarization of incident light, therefore an average is given. If compared with Fig.~\ref{fig:ptp_ej286_spectrum} with the \ptp\ and EJ-286 emission spectrum, one can notice that \ptp\ light is transmitted through the filter with $\sim$90\% efficiency and the EJ-286 re-emitted light will be reflected (and therefore trapped) with a $\sim$98\% efficiency.   

As the \ptp\ is coated onto the surface of the dichroic filter, it is of great importance that it has good adhesion to the \ptp\ and is able to survive to several cryogenic cycles without losing the quality of the coated film or filter. In total, 48,000 dichroic filters will be deployed in the DUNE FD module. Dichroic filters from four different companies where investigated, Edmunds optics~\cite{edmunds}, Opto Electronica S.A.~\cite{opto_br}, Ashai~\cite{ashai} and Omega Optical Inc.~\cite{omega}. Edmunds dichroic filters did not present a good performance for the S-\ara\ tests. The high cost of the Omega filters is impractical for the usage in DUNE. Opto filters were chosen as the baseline option by the Collaboration, since they showed performances competitive with the other filters and are much cheaper.

%% file: r_n_d.tex
\chapter{Research and Development}
\label{chap:RnD}
\thispagestyle{myheadings}

In this chapter the research and development (R\&D) of the S-\ara\ and \xara\ carried out during the period of this Ph.D thesis will be discussed. In the first section, the Dark Box set-up is presented, in which the trapping effect of the \ara\ could be tested. In Section~\ref{sec:monochromator} and~\ref{sec:grid_transparency} the measurements with Monochromator are presented. Section~\ref{sec:warm_tests_bicocca} show the room and cryogenic temperature tests performed in Italy for a new wavelength shifter to enhance the \xara\ light collection efficiency. 

\section{Dark Box Arapuca prototype}
\label{sec:dark_box}

The liquid argon tests performed for the \xara\ and presented in Chapter~\ref{chap:lar_test} are not easy to be performed. The prototyping of the device involves the coating of \ptp\ on a dichroic filter and, for the S-\ara, the coating of TPB on \viku. After assembled, the \ara\ needs to go under liquid argon, a complex procedure that needs time and investment. In order to speed up the optimization process, several tests of \ara\ prototypes were performed in a specifically designed Dark Box. 

\begin{figure}[h!]
	\centering
	\begin{subfigure}{0.44\textwidth}
		\includegraphics[width=0.99\linewidth]{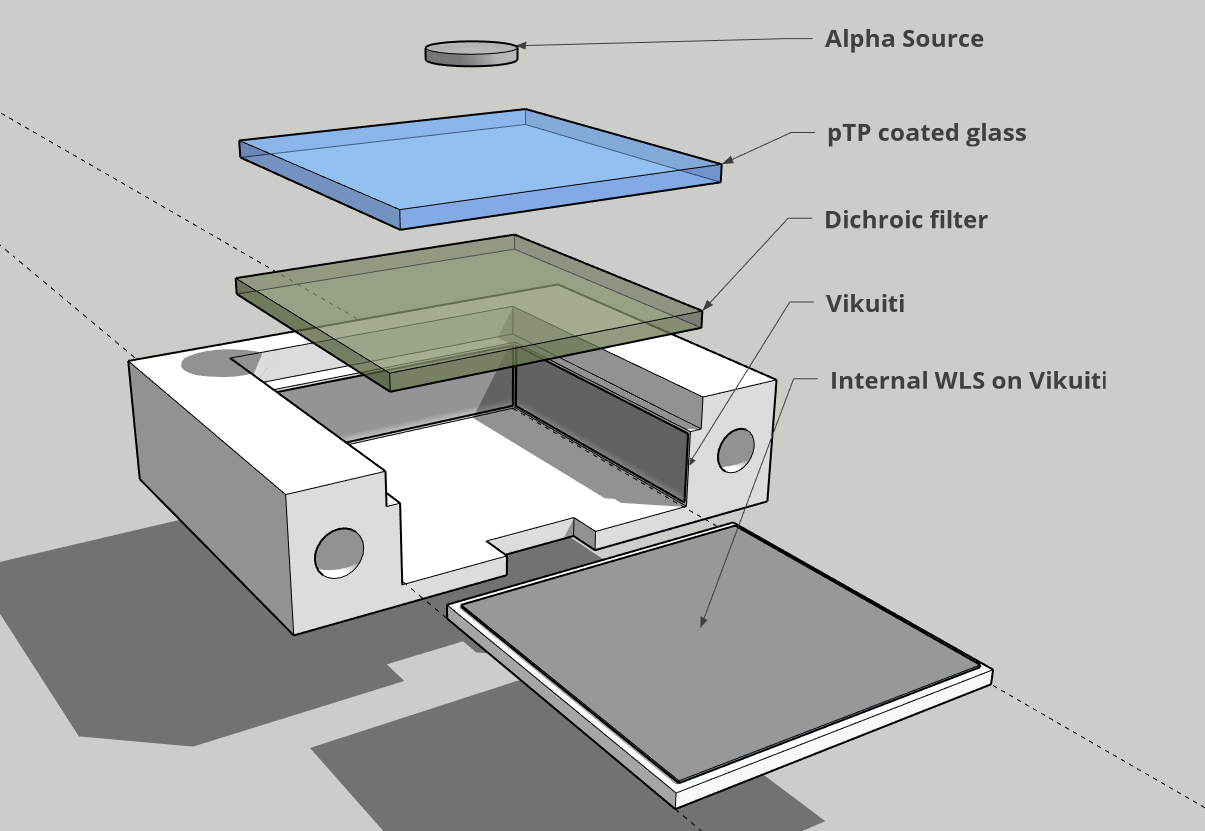}
		\caption{}
		\label{fig:darkbox_schematic}
	\end{subfigure}
	\begin{subfigure}{0.54\textwidth}
		\includegraphics[width=0.99\textwidth]{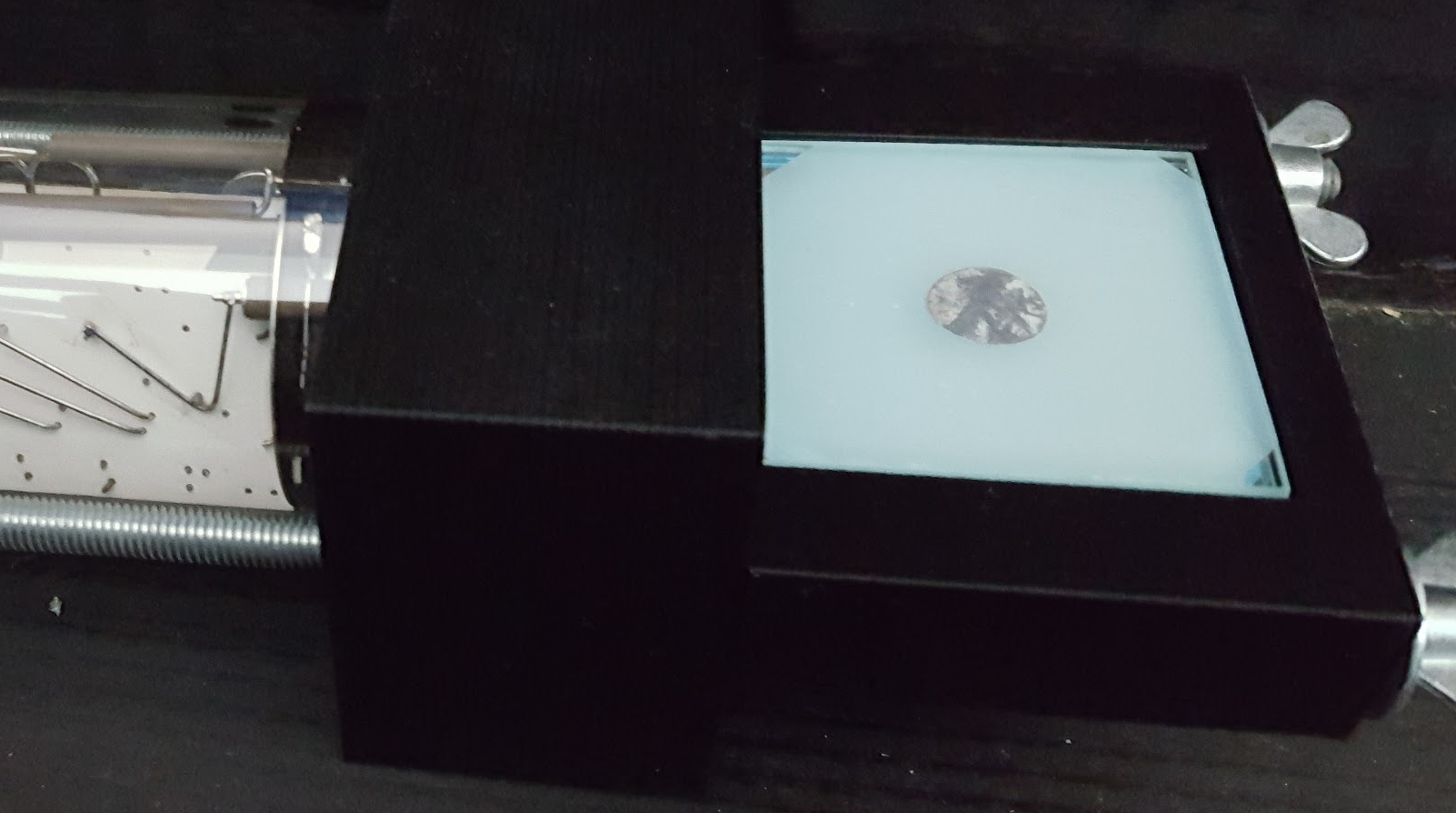}
		\caption{}
		\label{fig:setup}
	\end{subfigure}
	\caption{\textbf{(a)}~S-\ara\ 3D model used in the Dark Box test. \textbf{(b)}~Photo of the experimental setup showing the PMT ETL D750 attached to the \ara.}
	\label{fig:darkbox}
\end{figure}

The Dark Box set-up consists in an 3D printed S-\ara\ coupled with a Photomultiplier tube (PMT) as shown in Figure~\ref{fig:darkbox}. The S-\ara\ was a 5~$\times$~5~cm$^2$ area and 1~cm height cavity with \viku~\cite{vikuiti} on the internal surfaces. A drawer-like compartment holds the internal wavelength shifter coated on \viku. The \ptp\ was coated on optical glass and could be placed over the dichroic filter as seen in Fig.~\ref{fig:darkbox_schematic}. A natural Uranium alpha source (see~Sec.~\ref{sec:xara_single_cell}) is placed over the \ptp\ which will scintillate, producing light around 350~nm. The light detection is then performed by a Photomultiplier ETL D750 (see Appendix~\ref{chap:pmt}) in the open cavity of the device, with a 3D mask to cover any external light (Fig~\ref{fig:setup}). Assembling the S-\ara\ as described makes it much easier to change between different components of the device, enabling the study of different WLS molecules type and thicknesses and dichroic filters.

The setup was read-out by a CAEN Digitizer DT5720B with 12~bit resolution and 250~MSamples/s.

\subsection{Calibration}
\begin{figure}[h!]
	\centering
	\includegraphics[width=0.8\linewidth]{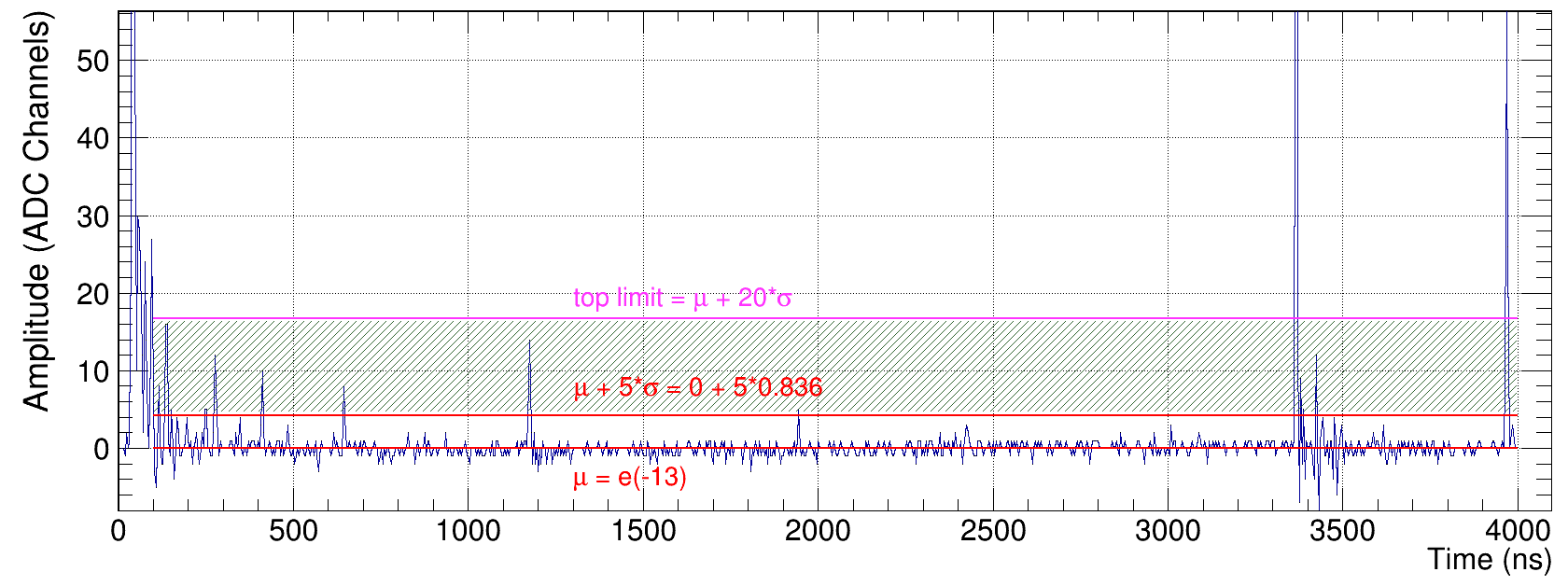}
	\caption{Single photo-electron selection, events inside 5 and 20$\sigma$ from the baseline are integrated for 36~ns.}
	\label{fig:event3_lines_area_tex}
\end{figure}

The system was calibrated by searching for single photo-electrons in the tail of the signal as shown in Figure~\ref{fig:event3_lines_area_tex}, a method similar to the one described at Sec.~\ref{sec:calibration_double_cell}. Peaks of photo-electrons produced by thermionic emission are searched between 5$\sigma$ and~20$\sigma$ from the baseline and integrated for 36~ns. In this illustrative waveform, three signals from the scintillation of $\alpha$ particles into \ptp\ can be seen, six other signals are integrated. 

\begin{figure}[h!]
	\centering
	\includegraphics[width=0.72\linewidth]{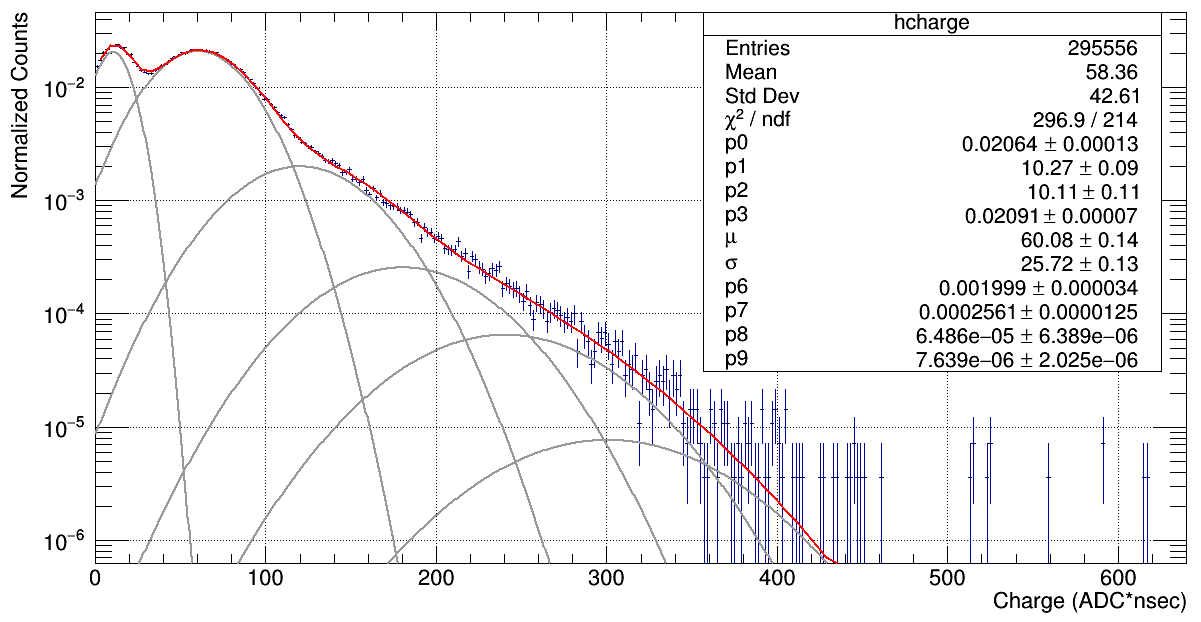}
	\caption{Single photo-electron charge spectrum obtained, fit described in the text.}
	\label{fig:sphe_500pts_3sigma}
\end{figure}

Figure~\ref{fig:sphe_500pts_3sigma} shows the single photo-electron (\sphe) charge spectrum obtained. The fit is performed with $N+1$ Gaussian distributions, where the first one corresponds to the baseline noise and the others correspond to one or more photo-electrons response. The noise and \sphe\ Gaussian distributions have 3 free parameters each (mean value, standard deviation and normalization constant). The other Gaussians have mean and standard deviation constrained to be $N\times\mu_1$ and $\sqrt{N}\times\sigma_1$, where $\mu_1$ and $\sigma_1$ are the mean and standard deviation of the \sphe. 

The charge of alpha signals was always divided by value of the \sphe\ charge (fitted as 60.1\error0.1~ADC$\cdot$ns in Fig.~\ref{fig:sphe_500pts_3sigma}) found for each run to give the resulting photo-electrons detected. 

\subsection{Trapping effect}
To verify the consistence of the Dark Box tests, the trapping effect of the \ara\ was tested. First, the alpha spectrum was measured with the S-\ara. Later, the dichroic filter was replaced by a normal optical glass and the measure was repeated. The waveforms are acquired for 4~$\mu$s, for the baseline and \sphe\ calibration, and integrated for 100~ns.

\begin{figure}[h!]
	\centering
	\begin{subfigure}{0.436\textwidth}
		\includegraphics[width=0.99\linewidth]{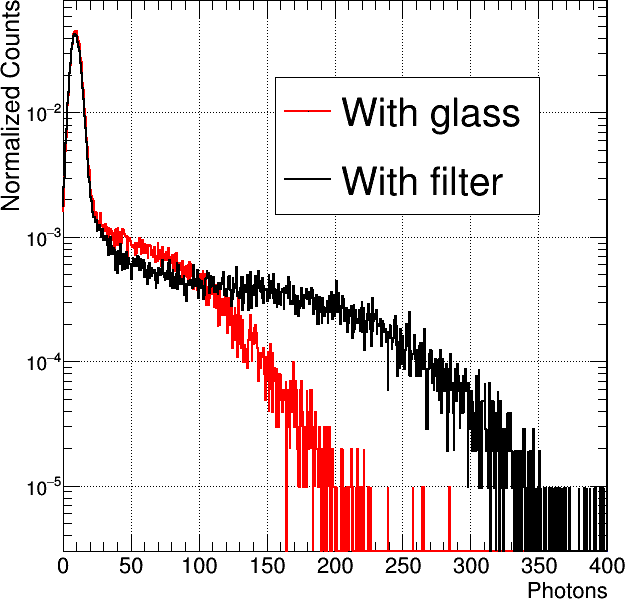}
		\caption{}
		\label{fig:dark_box_photons}
	\end{subfigure}
	\begin{subfigure}{0.43\textwidth}
		\includegraphics[width=0.99\textwidth]{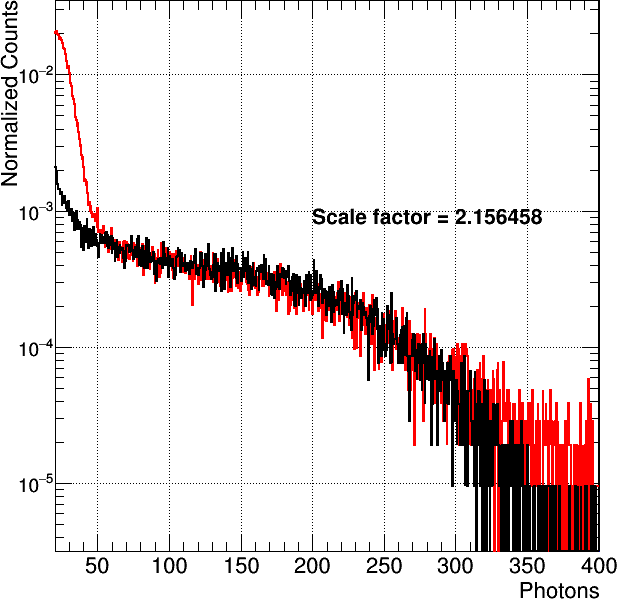}
		\caption{}
		\label{fig:dark_box_factor}
	\end{subfigure}
	\caption{\textbf{(a)}~Alpha source spectrum measured with the dichroic filter (black) and with the optical glass replacing it (red). \textbf{(b)}~Factor of $\sim$2.16 found between the two histograms through $\chi^2$ minimization.}
	\label{fig:darkbox_concept}
\end{figure}

Figure~\ref{fig:dark_box_photons} shows the resulting spectra found for each configuration. It is noticeable that the trapping effect of the \ara\ device significantly increases the amount of light detected. 

To quantify the increase in light collection, a $\chi^2$ minimization was performed between the two histograms using the TMinuit tool from Root CERN~\cite{root_cern}. Figure~\ref{fig:dark_box_factor} shows the result, two free parameters are used: a normalization constant and a scaling factor that is directly related to the collection efficiency. A scaling factor of $\sim$2.16 was found, meaning that the trapping effect of the \ara\ increases the efficiency by $\sim$116\%.
\subsection{Results}
\begin{figure}[h!]
	\centering
	\begin{subfigure}{0.44\textwidth}
		\includegraphics[width=0.99\linewidth]{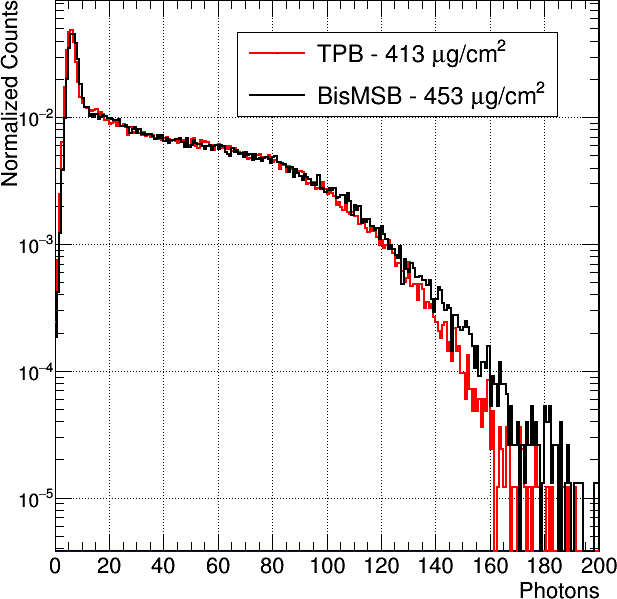}
		\caption{}
		\label{fig:result1}
	\end{subfigure}
	\begin{subfigure}{0.44\textwidth}
		\includegraphics[width=0.99\textwidth]{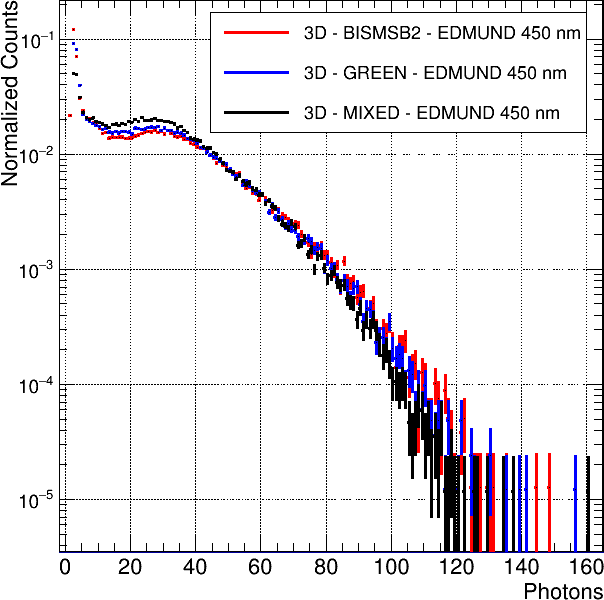}
		\caption{}
		\label{fig:result2}
	\end{subfigure}
	\caption{\textbf{(a)}~Comparison between TPB and BisMSB as internal WLS for the S-\ara\, an increase of $\sim$5\% was noticed for BisMSB. \textbf{(b)}~Liquid WLS compared with BisMSB giving comparable spectra. GREEN and MIXED refer to emission in green and mixture of green/blue.}
	\label{fig:results}
\end{figure}

The S-\ara\ has a lower efficiency than the \xara~\cite{LAr_arapuca_test,x_arapuca_article}, therefore only preliminary results are given here as no further investigation was taken. Two interesting results point to two alternatives to TPB: (1) BisMSB (1,4-Bis(2-methylstyryl)benzol), where preliminary results (Fig.~\ref{fig:result1}) show an efficiency $\sim$5\% greater for BisMSB and (2) liquid WLS, that, as shown in Fig.~\ref{fig:result2} has almost the same performance as the BisMSB.  

However, BisMSB may degrade when exposed to UV light and it has not been widely tested and the liquid wavelength shifters deposition did not surpass a liquid nitrogen test, making necessary to develop a better deposition technique. 

\section{Monochromator tests}
\label{sec:monochromator}

\begin{figure}[h!]
	\centering
	\includegraphics[width=0.9\linewidth]{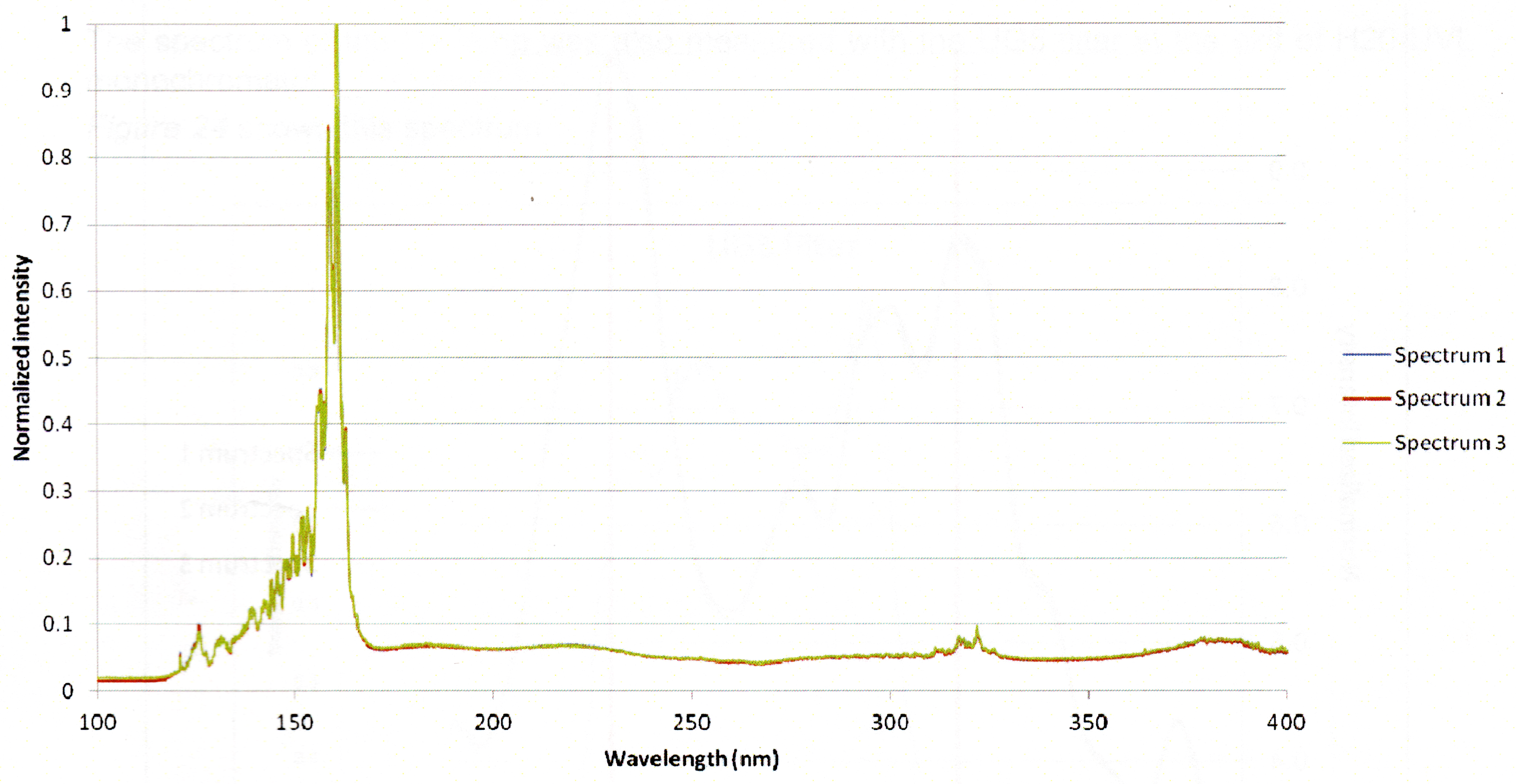}
	\caption{Monochromator emission spectrum ranging from 120 to 400~nm.}
	\label{fig:spectrum_monochromator}
\end{figure}

Prototypes and components were also exposed to monochromatic light from a H20-UVL Horiba~\cite{horiba} Monochromator installed at the ``\textit{Laboratório de Léptons}'' in Campinas. The Monochromator uses a granting diffracted light from a  Deuterium light source (150~W). Figure~\ref{fig:spectrum_monochromator} shows the emission spectra of the Monochromator that goes from 120 to 400~nm ($\pm$0.1~nm). The Monochromator must be operated in vacuum, specially bellow 185~nm due to air (O$_2$) absorption of the light~\cite{vuv_o2_absorption}. Therefore, it was a more ``complicate'' system than the Dark Box but with the same idea of speeding up the optimization process, limiting the use of liquid argon.

\begin{figure}[h!]
	\centering
	\includegraphics[width=0.95\linewidth]{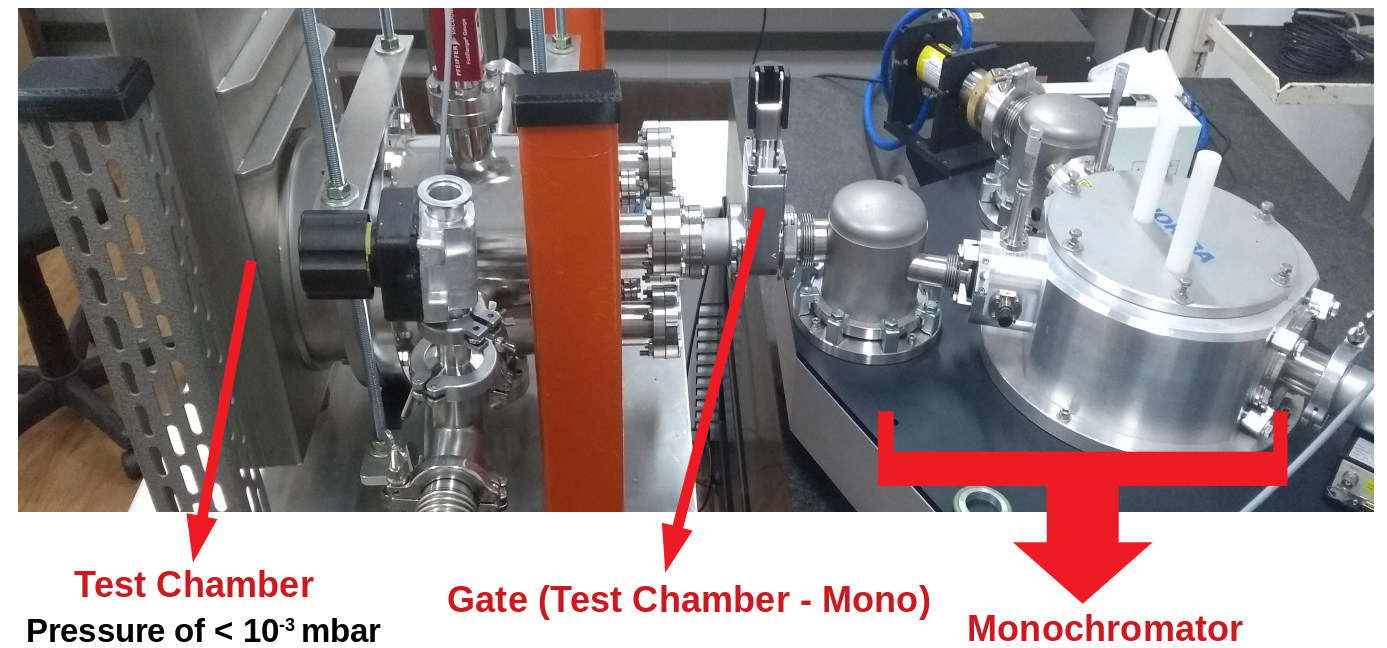}
	\caption{Experimental setup, the test chamber is vacuum isolated from the Monochromator through the Gate.}
	\label{fig:photo_monochromator}
\end{figure}

Figure~\ref{fig:photo_monochromator} shows the experimental setup: the Monochromator kept in high vacuum \mbox{$P\sim10^{-5}$~mbar} isolated by the gate valve in the middle. The test chamber can be operated separately and the gate valve is opened when the pressure is below $10^{-3}$~mbar. With this, one measure can be done in about 1~hour and half, including vacuum breaking and pumping. 

\begin{figure}[h!]
	\centering
	\begin{subfigure}{0.35\textwidth}
		\includegraphics[width=0.99\linewidth]{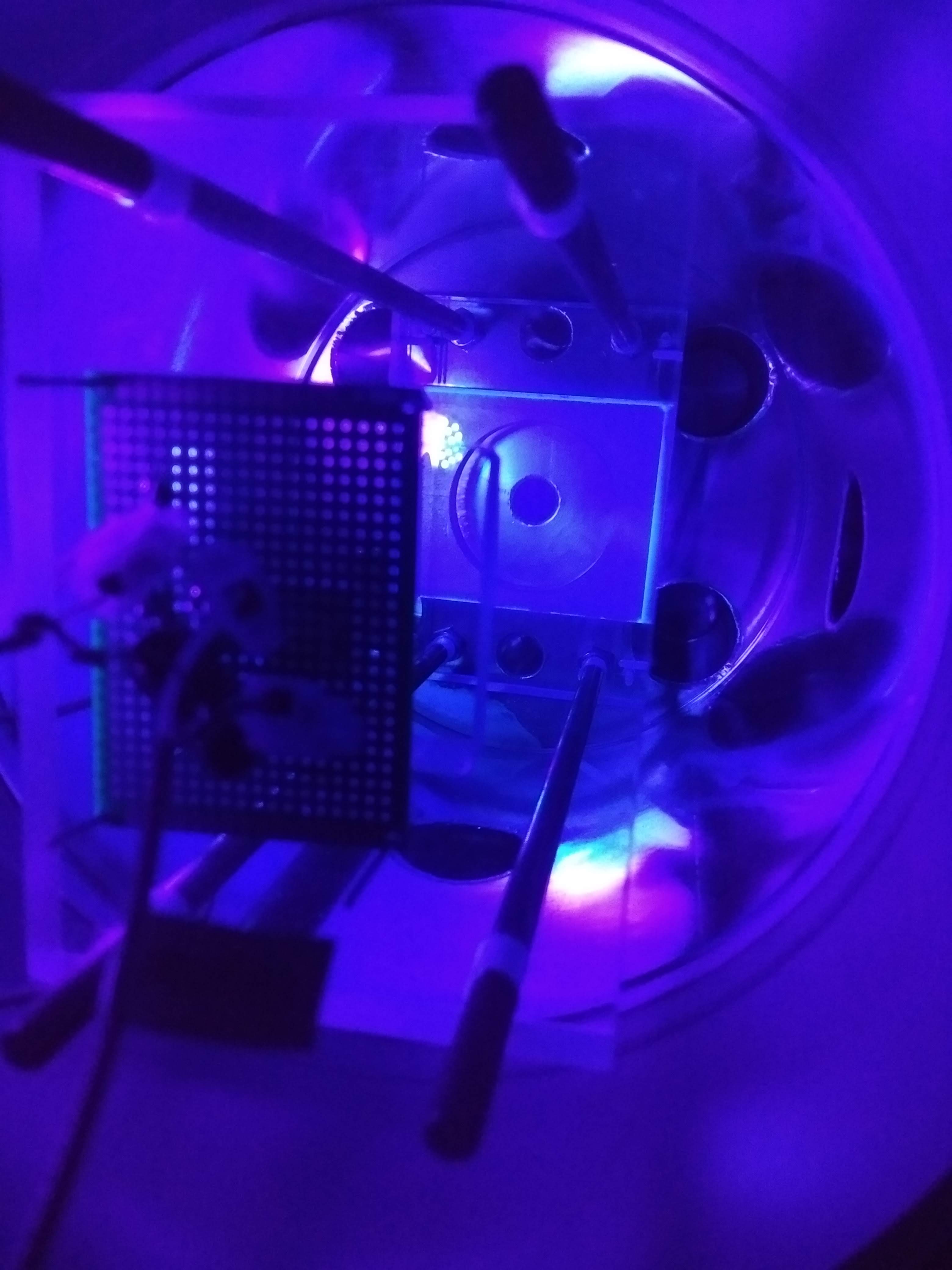}
		\caption{}
		\label{fig:photo_apd_sample}
	\end{subfigure}
	\begin{subfigure}{0.64\textwidth}
		\vspace{23pt}
		\includegraphics[width=0.99\linewidth]{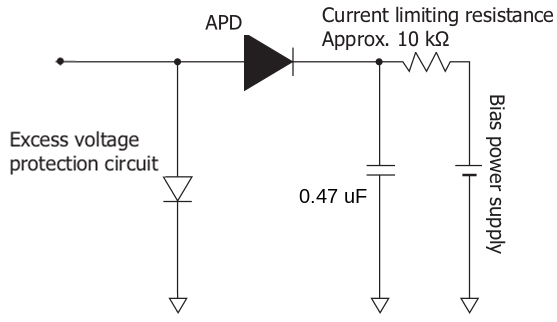}
		\vspace{22pt}
		\caption{}
		\label{fig:apd_circuit}
	\end{subfigure}
	\caption{\textbf{(a)}~Photo inside the testing chamber, a wavelength shifter exposed to UV light and in the back of it (at the left side of the photo) the APD to detect the light. \textbf{(b)}~Diagram of the APD readout~\cite{hmmt_manual}.}
	\label{fig:apd_at_mono}
\end{figure}

The light detection was done with an APD (see Sec~\ref{sec:sipms}) S8664-1010 from Hamamatsu~\cite{hmmt_s13360} placed behind the sample as shown in Figure~\ref{fig:photo_apd_sample}. The APD has a breakdown voltage of 400~V and it was operated at 350~V with a gain $G=50$ and quantum efficiency of about 70\% for 400~nm. The producer gives a spectral response ranging from 320~to~1000~nm, however, detection of light was possible in all the spectral range of the Monochromator.

Figure~\ref{fig:apd_circuit} shows the circuit to readout the APD~\cite{hmmt_manual}, the RC circuit close to the Bias power supply serves as a low pass filter and current limiting. The resistance and capacitance were chosen to efficiently cut frequencies above the $f_C \approx 33$~Hz, significantly reducing fast oscillation noise and the 60$\sim$70~Hz from the electric wires. A diode is placed on the APD anode exit as an excess voltage protection. The voltage amplitude was read-out with an oscilloscope, always retrieving the baseline by measuring in dark. The Deuterium light bulb requires approximately ten minutes to stabilize after turning on.  

In order to make  a relative comparison between two different samples, for instance two types of WLS, there is no need to correct for the emission spectrum of Fig~\ref{fig:spectrum_monochromator}. However, for absolute measurements this characterization is essential. In this case, a reference measurement was taken but, ideally, the reference measurement of the beam must be taken online, using a beam splitter with another light detector. These new feature are being implemented.

\subsection{Results}

\subsubsection{Fused silica versus B270}
\label{sec:fused_silica}

Fused silica is typically used as dichroic filter substrate due to its optical properties~\cite{fused_silica}, with excellent light transmittance to wavelengths of 195~nm up to 2.1~$\mu$m. Due to the high cost of fused silica, other alternatives of dichroic filter substrates were investigated, seeking good transparency, good  \ptp\ adhesion and low cost. Among the candidates, the optical glass B270 showed to have good \ptp\ adhesion and a higher robustness of the film at liquid nitrogen temperatures, in addition to being easier to cut and cheaper than fused silica.

\begin{figure}[h!]
	\centering
	\includegraphics[width=0.7\linewidth]{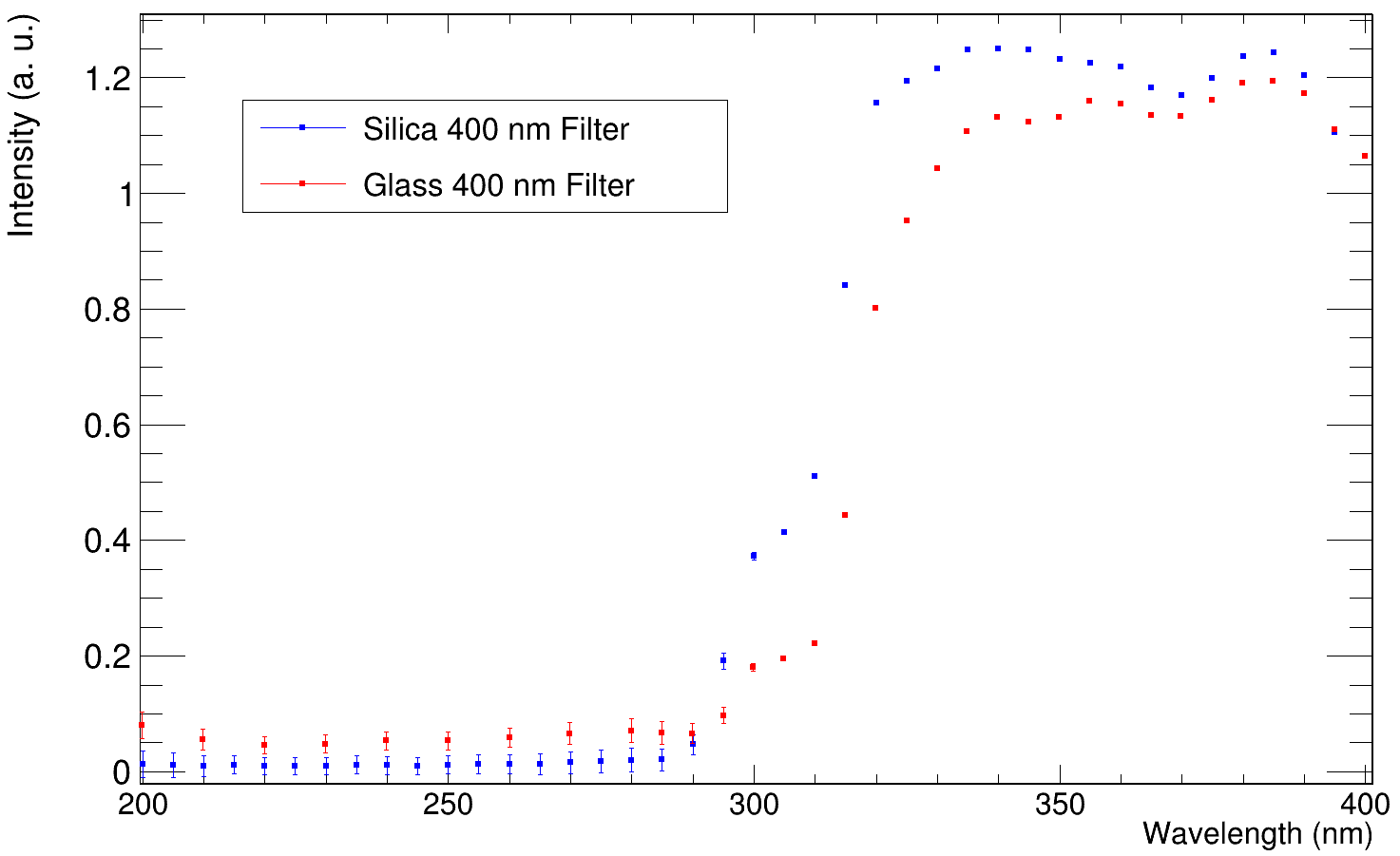}
	\caption{Transmission from  Opto dichroic filter with thin film deposition in Fused Silica and optical glass (B270).}
	\label{fig:filters_B270_fs}
\end{figure}

A comparison between B270 and fused silica was then performed. Figure~\ref{fig:filters_B270_fs} shows the transmittance for two dichroic filters produced by Opto Co., one with thin film deposition on B270 and another one on fused silica. The spectra are simply normalized by the direct light of the Monochromator. It was noticed that both substrates have similar performance around the \ptp\ emission and, therefore, B270 was chosen due its adhesion, lower cost and because it is easier to cut.  

\begin{figure}[h!]
	\centering
	\includegraphics[width=0.9\linewidth]{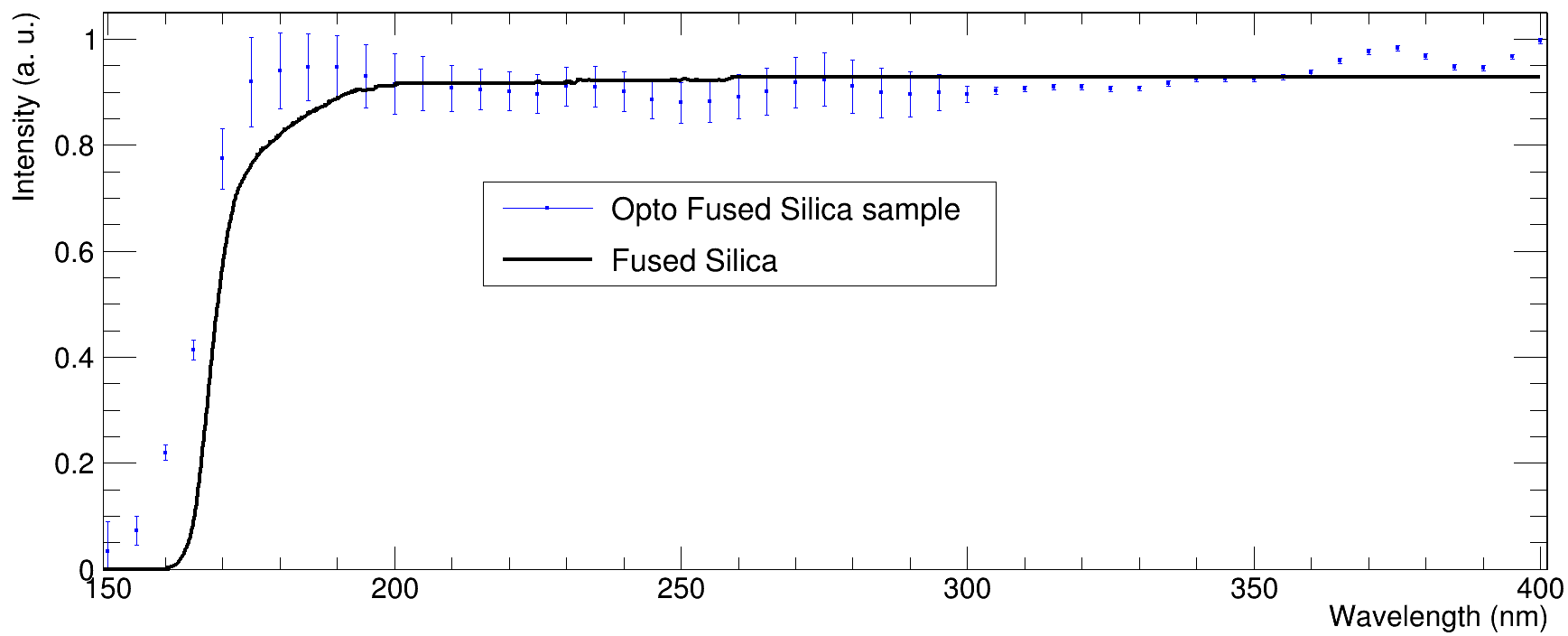}
	\caption{Transmission measurements (data points) for sample of the Opto Co.\ fused silica. The fused silica data was normalized to match the given transmittance~\cite{fused_silica} of the fused silica (thick black line).}
	\label{fig:compare_b270_silica}
\end{figure}

The fused silica substrate, without the thin layers that characterize the dichroic filter (Sec.~\ref{sec:dichroic}), was tested. Figure~\ref{fig:compare_b270_silica} shows the normalized transmittance found, the Opto fused silica sample data (blue) was normalize to match the given transmission~\cite{fused_silica} of fused silica~(thick black line).  

This test was originally performed to better understand the transparency of the materials. However, it came in hand later to define the fused silica as the ideal material to block LAr scintillation light without significantly attenuate Xenon light (175~nm) for the Xenon Doping tests performed at ProtoDUNE-SP (see Sec.~\ref{sec:xe_dop}).

\section{Grounding mesh transparency}
\label{sec:grid_transparency}

\begin{figure}[h!]
	\centering
	\begin{subfigure}{0.428\textwidth}
		\includegraphics[width=0.98\linewidth]{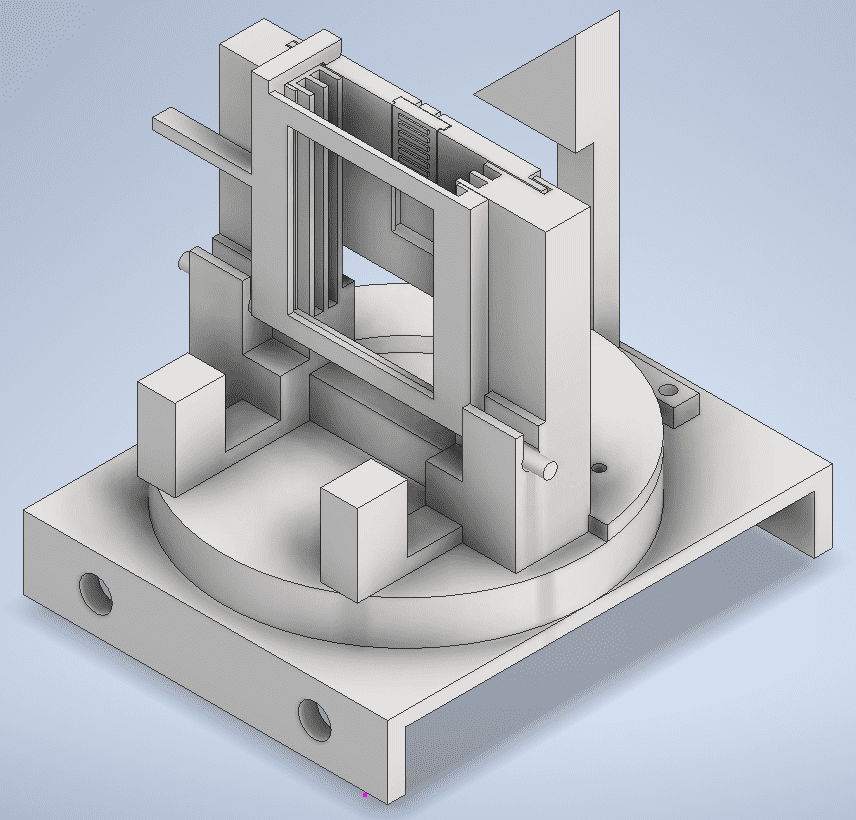}
		\caption{}
		\label{fig:grelha_montagem}
	\end{subfigure}
	\hspace{2pt}
	\begin{subfigure}{0.304\textwidth}
		\includegraphics[width=0.99\textwidth]{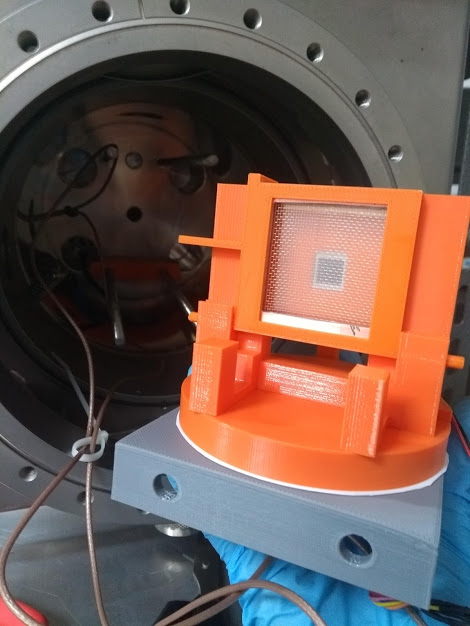}
		\caption{}
		\label{fig:grelha_foto}
	\end{subfigure}
	\caption{\textbf{(a)}~3D model of the experimental setup. \textbf{(b)}~Photo of the 3D printed setup. In the background the testing chamber can be seen with the hole in which the beam comes from. The system is designed so the mesh falls down when it reached 80$^\circ$.}
	\label{fig:grelha}
\end{figure}

Transparency measurements of the grounding mesh that surrounds the APA, so ionizing radiation inside the APA does not produce signals, were performed in the Monochromator. This is an important feature to be implemented in light simulations to correctly perform the energy reconstruction, for example. Figure~\ref{fig:grelha_montagem} shows the 3D model of the experimental setup while Figure~\ref{fig:grelha_foto} shows a photo of the 3D-printed model in front of the Monochromator.

\begin{figure}[h!]
	\centering
	\begin{subfigure}{0.2\textwidth}
		\includegraphics[width=0.93\linewidth]{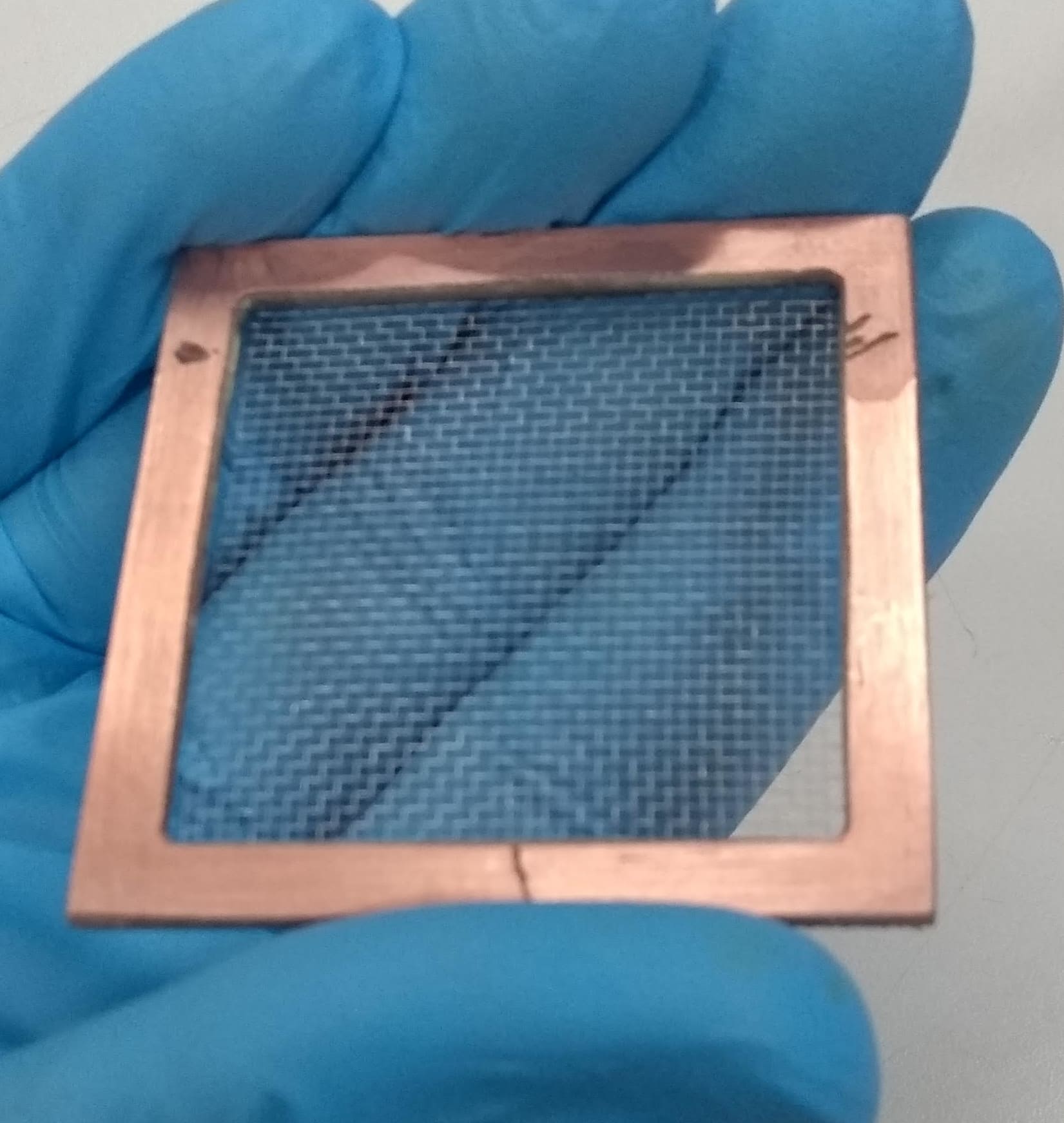}
		\vspace{5pt}
		
		\includegraphics[width=0.93\linewidth]{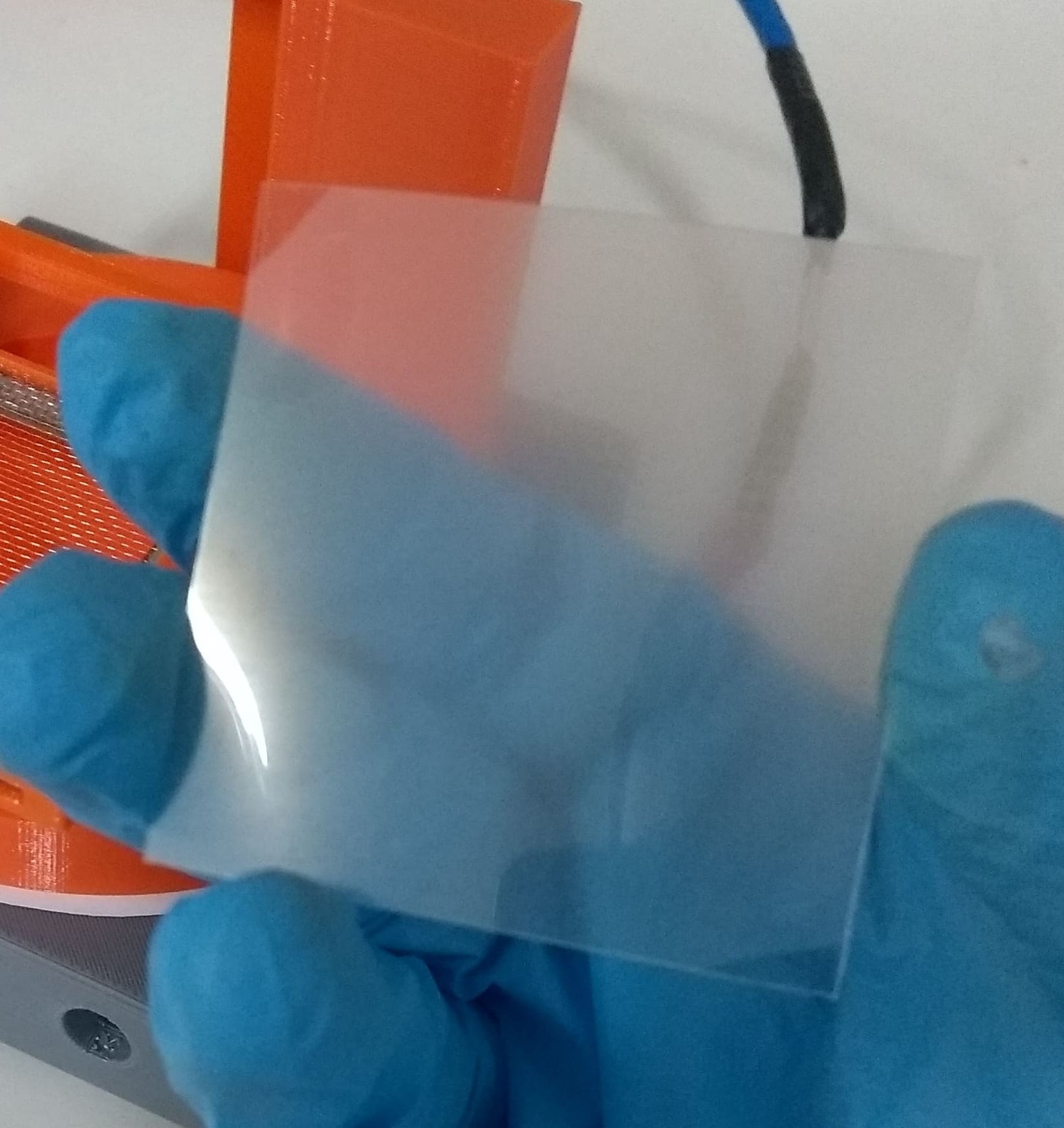}
		\caption{}
		\label{fig:photo_grid_and_ITO}
	\end{subfigure}
	\begin{subfigure}{0.334\textwidth}
		\includegraphics[width=0.99\linewidth]{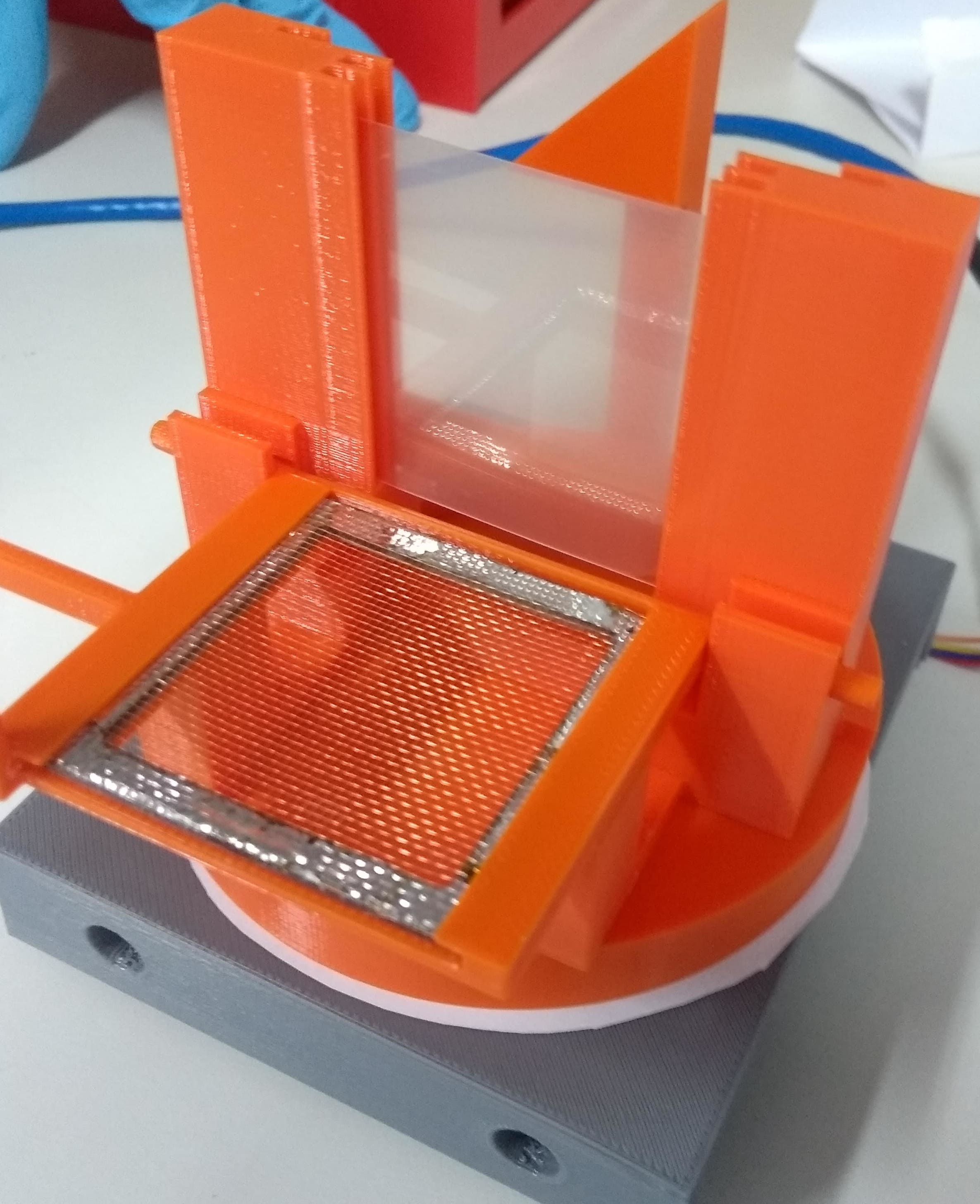}
		\caption{}
		\label{fig:photo_grid_down}
	\end{subfigure}
	\caption{\textbf{(a)}~(top) Photo of the grounding mesh (bottom) Indium tin oxide (ITO) thin layer coated with \ptp. \textbf{(b)}~Photo of the experimental setup with the grounding mesh down, for direct light data taking.}
	\label{fig:grid_grid_ito_down}
\end{figure}

The APD is placed in the back of the setup. The grounding mesh to be tested is installed in the front of the beam and, in between the mesh and the APD, there is a indium tin oxide\footnote{Indium tin oxide is widely used in research and industry. It is a transparent conducting oxide, which is electrical conductor and optical transparent. It was chosen due to its transparency to \ptp\ emission light and thickness (another set of measurements were planned with electric field applied to the ITO).} (ITO) thin film coated with \ptp.  Figure~\ref{fig:photo_grid_and_ITO} shows the grounding mesh and the ITO coated with \ptp\ (top and bottom, respectively).

A step motor controlled by an Arduino rotates the upper part (orange in the Fig.~\ref{fig:grelha_foto}) over the mesh vertical axis to study the angular dependence of the transparency. The setup is built so the grounding mesh will fall down when it reaches 80$^\circ$ (Fig.~\ref{fig:photo_grid_down}). In this way, it is possible to measure the direct light, as a reference for the transparency, without changing anything in the system (such as vacuum level, setup position, light intensity, etc.).


\begin{figure}[h!]
	\centering
	\begin{subfigure}{0.42\textwidth}
		\vspace{20pt}
		\includegraphics[width=0.99\linewidth]{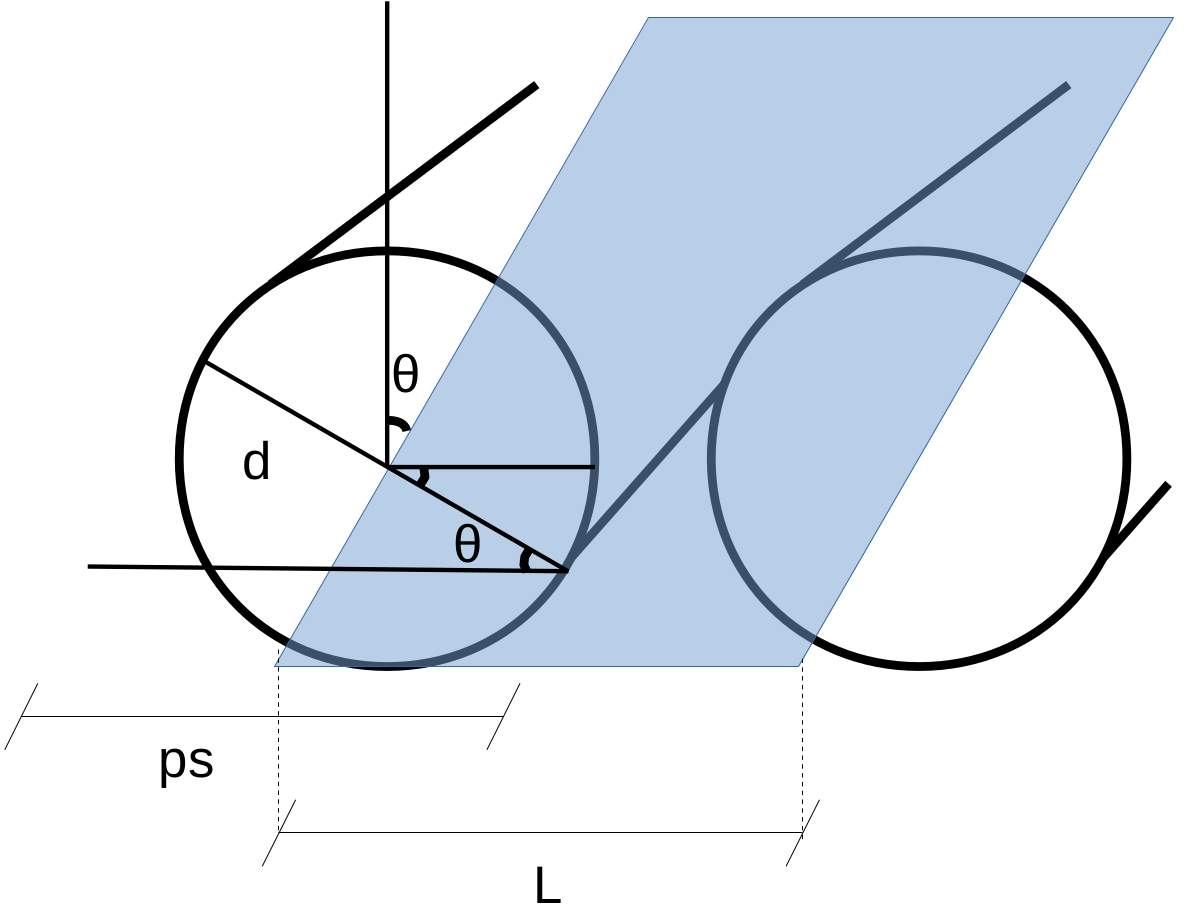}
		\vspace{25.5pt}
		\caption{}
		\label{fig:geometric_arguments}
	\end{subfigure}
	\begin{subfigure}{0.42\textwidth}
		\includegraphics[width=0.99\textwidth]{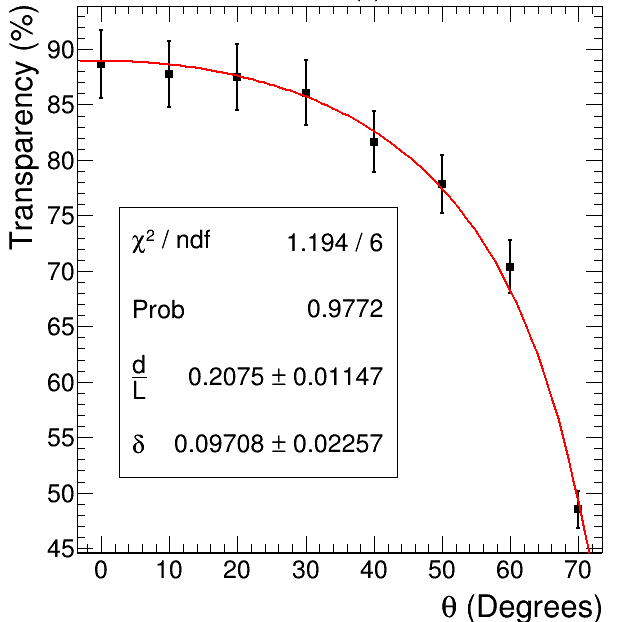}
		\caption{}
		\label{fig:results_mesh}
	\end{subfigure}
	\caption{\textbf{(a)}~Illustrative light (blue) with width L arriving with an angle $\theta$ at the mesh wires of diameter $d$ and pitch $L$. \textbf{(b)}~Grounding mesh transparency as function of the light incident angle $\theta$. The maximum angle is 70$^\circ$ due to the mechanical constrains of the project.}
	\label{fig:geometric_arg_and_results_mesh}
\end{figure}

A simplified model can be used to predict and interpret the data. In Figure~\ref{fig:geometric_arguments} a representation of the incident light (in blue) is made arriving at the grounding mesh with angle~$\theta$. A section of the mesh is displayed (the other one is rotated by 90$^\circ$ and not displayed for simplicity), the wires with diameter $d$ and pitch $L$ will produce a projected shadow (ps). By using the angle relations:
\begin{equation}
	\cos(\theta) = \frac{d}{\text{ps}}.
\end{equation}

Therefore, for a light strip with width $L$, the proportional shadow ($S_P$) created by the two wires can be written as:
\begin{equation}
	S_P = \frac{\text{ps}}{L} = \frac{d}{L\;\cos(\theta)},
\end{equation}

and, the transparency $T = 1-S_P$ is:
\begin{equation}
	\label{eq:mesh_transparency}
	T = 1 - \frac{d}{L\;\cos(\theta)}.
\end{equation}

Figure~\ref{fig:results_mesh} shows the transparency of the mesh as function of the incidence angle $\theta$. The transparency is defined as the ratio between the voltage of the APD with the mesh in front by the voltage with direct incidence. An average was made among the three different wavelengths\footnote{The choice of the wavelength is due to the emission amplitude of the Monochromator (see Fig.~\ref{fig:spectrum_monochromator}) and to have a good reference in the Xenon emission wavelength (see Chap.~\ref{chap:protoDUNE}).} used: 165, 170 and 175~nm. 

The data were fitted with Eq.~\ref{eq:mesh_transparency} plus a constant factor~$\delta$, as  the model does not take into account the second section of wires that, most probably, has a constant relation with the incidence angle. Therefore, the transparency as a function of $\theta$ can be written as:
\begin{equation}
		\label{eq:mesh_transparency_numerical}
		T(\theta) = 1-\frac{0.2075}{\cos(\theta)}+0.09708
\end{equation}
and the total average transparency ($T_U$) for an isotropic light source ca be estimated by integrating over the solid angle:
\begin{equation}
	T_U = \frac{1}{2\pi} \int_0^{\phi}\diff\phi \int_0^{\theta'} T(\theta)\sin(\theta)\diff\theta \;\; \cong \;\;54.4\%,
\end{equation}
where $\theta'\approx79.1^\circ$ was set so $T(\theta')=0$ to avoid negative values. 

This is not a realistic approach as the light does not arrive uniformly. Muons will have a long track along the TPC, while electrons will produce showers with length of 1$\sim$2~meters long. As informed in a private communication with professor Franciole Marino, the angular distribution of the photons coming from a punctual light source 1~meter away form the detector can be described by the polynomial:
\begin{equation}
	\label{eq:franciole_light_dist}
	D(\theta) = 0.0111797+0.000675671\cdot \theta-1.84617\cdot 10^{-5}\cdot \theta^2 + 1.06234\cdot10^{-7}\cdot \theta^3,
\end{equation}
that gives:
\begin{equation}
	T_L = \int_0^{\theta'} T(\theta)D(\theta)\diff\theta = 80.9\%.
\end{equation}

Therefore, a transparency of 80.9\% can be expected from the grounding mesh, close to the value of 85\% given by Ref.~\cite{DUNE_vol4}.

The study of the grounding mesh transparency was very useful to understand the data from the Xenon doping tests in ProtoDUNE-SP, as will be described in Chapter~\ref{chap:protoDUNE}. Equations~\ref{eq:mesh_transparency} and~\ref{eq:mesh_transparency_numerical} were used to better simulate photons arriving in the \xara\ deployed for the tests. 

\section{Glass-to-Power Wavelength Shifter}
\label{sec:warm_tests_bicocca}

A new wavelength shifter was designed and developed by the team of University of Milano Bicocca in collaboration with Glass-to-Power (G2P) company~\cite{G2P} a spin-off from Milano-Bicocca. The WLS slab is produced with Poly(methyl methacrylate) (PMMA), acrylic glass, doped with 2,5-Bis(5-tert-butyl-benzoxazol-2-yl)thiophene (BBT)~\cite{enhancement_xara}. The prototype of this WLS are named here as FB118. In this section the preliminary results that led to the evaluation of the \xara\ double-cell efficiency in LAr will be discussed. The final result is shown in Chapter~\ref{chap:lar_test}.

\begin{figure}[h!]
	\centering
	\includegraphics[width=0.75\linewidth]{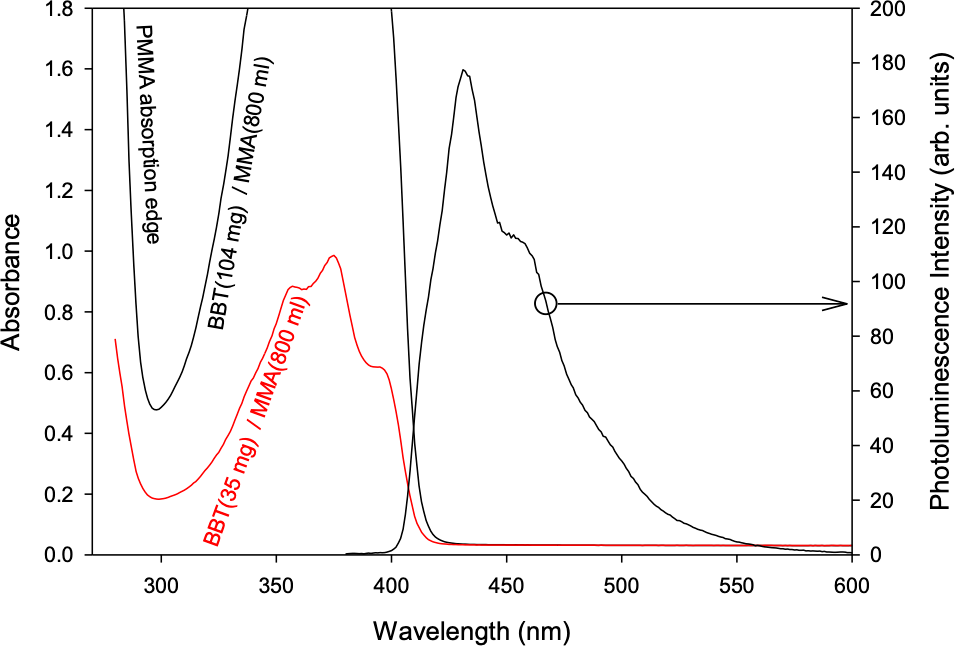}
	\caption{The secondary wavelength shifter absorption and emission spectra together with the PMMA absorption~\cite{enhancement_xara}.}
	\label{fig:SWLS_abs_em}
\end{figure}

The PMMA, without doping, has a good transparency for the \ptp\ emission spectrum as shown in Figure~\ref{fig:SWLS_abs_em} where the absorption edge of PMMA is around 300~nm. However, the absorption spectra peaks around 370~nm when doped with BBT, as can be seen in the red line. Two BBT concentrations are displayed, 104~mg and 35~mg, so its possible to see that increasing the BBT concentration in the 800~ml of PMMA the absorbance also increases\footnote{Note that absorbance here is defined as $A=-\log_{10}(T)$, giving the corresponding transmittance. Therefore, an absorbance of $A=1$ means a transmittance of 10\%, while $A=1.8$ means $T\approx1.6\%$.}. During the tests performed at Milano-Bicocca three different concentrations were used. The initial WLS slab tested at warm had good results, therefore, double and half concentrations were tested. For the cryogenic tests and liquid argon tests (Sec.~\ref{sec:cryo_tests} and Sec.~\ref{sec:xara_double_cell}) the double concentration one was used. 

The emission spectra is also shown in Fig.~\ref{fig:SWLS_abs_em} with the label on the right side. The peak emission around 430~nm and the spectrum starting edge around 400~nm allow the use of the same dichroic filters with cutoff of 400~nm. 

A set of measurements in warm and cold were done in order to measure the enhancement in efficiency with respect to EJ-286 slab from Eljen~\cite{eljen_286}. It was also possible to measure the increase in collection efficiency by placing \viku\ reflector foils on the WLS slab edges, leaving gaps for the SiPM devices.  

\subsection{Warm measurements}
\label{sec:warm_measurements}

\begin{figure}[htbp]
	\centering
	\includegraphics[width=.26\textwidth]{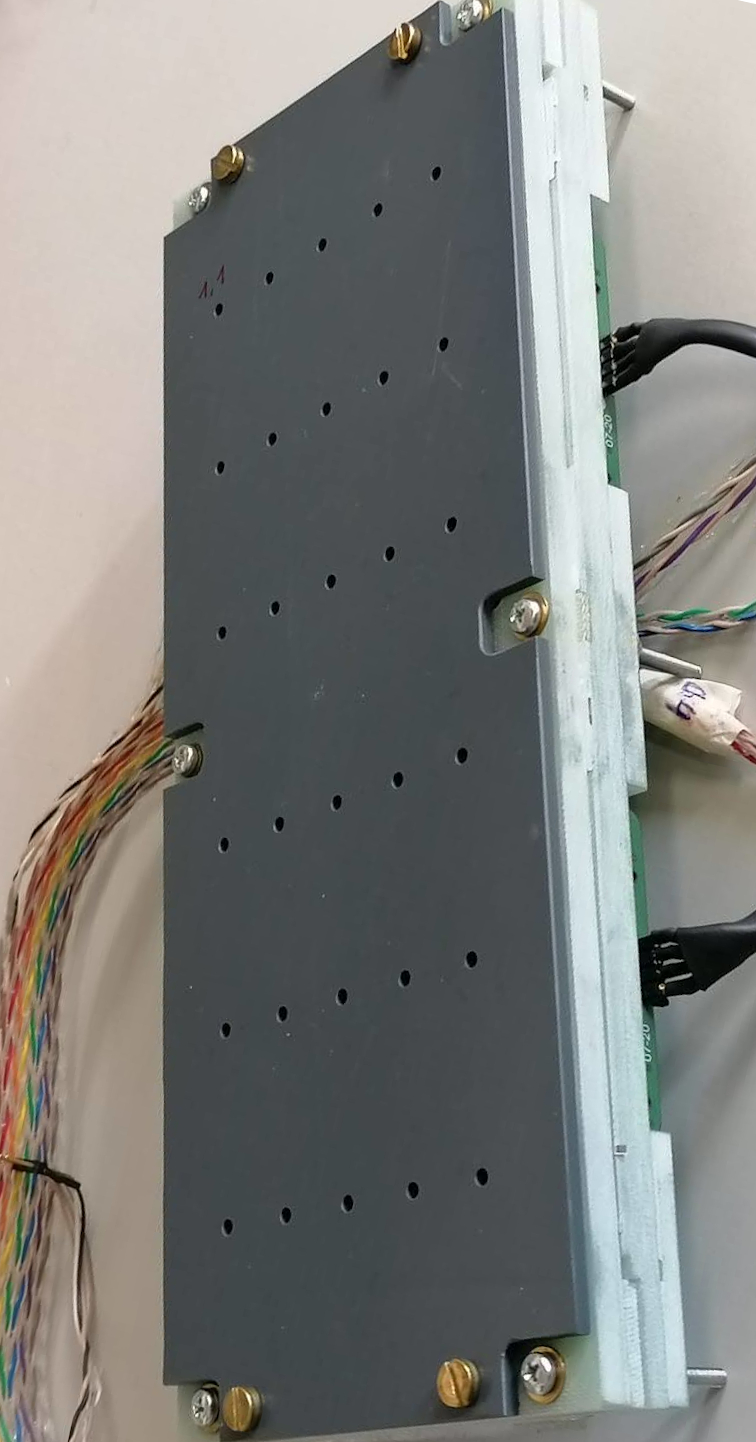}
	\includegraphics[width=.32\textwidth]{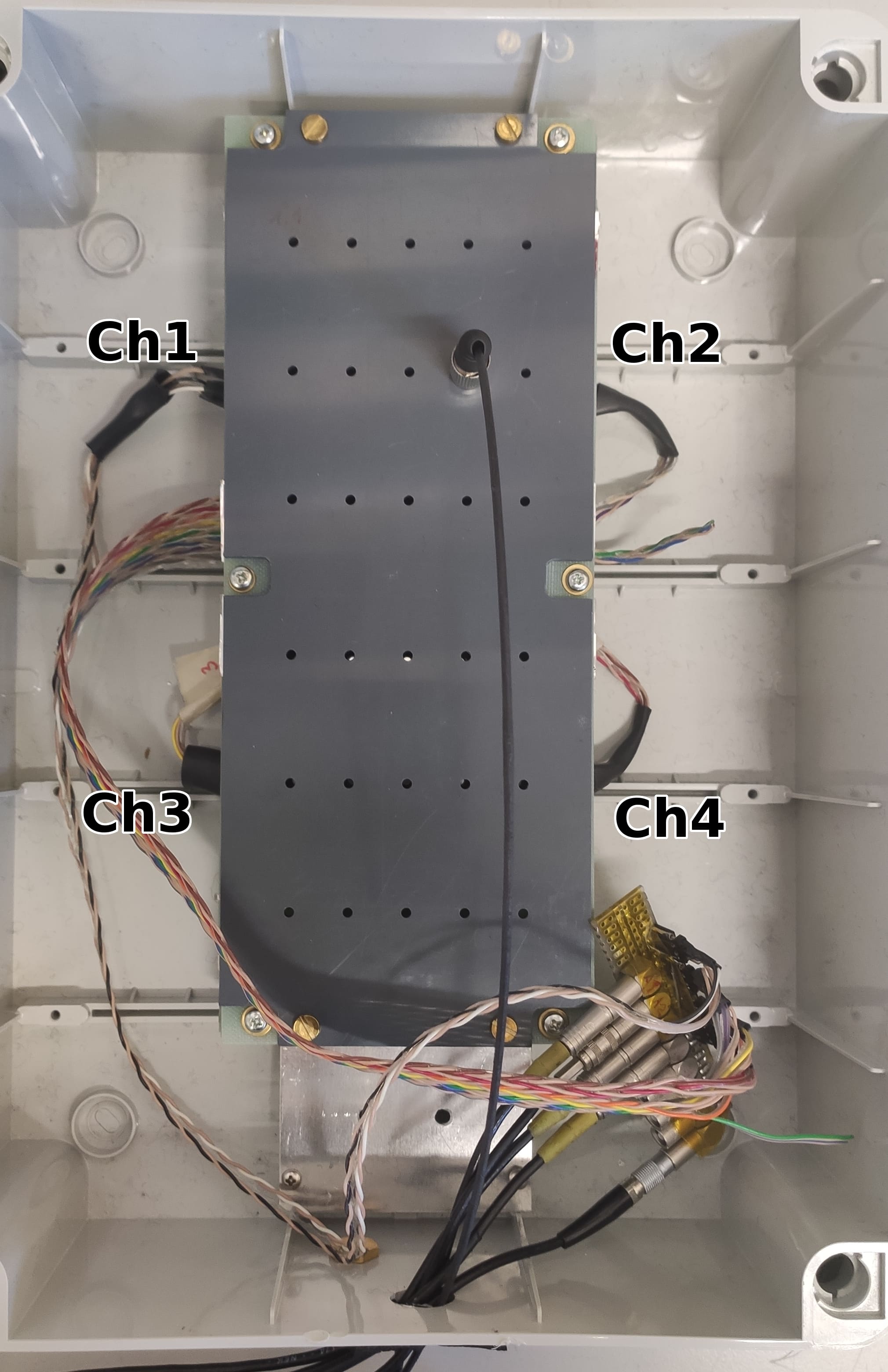}
	\caption{The X-ARAPUCA with the plastic mask and the optical fiber plugged, a total of 30~deployment points could be used to scan the WLS uniformity~\cite{enhancement_xara}.}
	\label{fig:xara_mask}
\end{figure}

Tests at room temperature were initially performed to search for hints of the photon detection efficiency (PDE) and to understand the photon detection uniformity (PDU) of each WLS. For these measurements, a CAEN SP5601 LED with peak emission $\sim$405~nm was used and the dichroic filter was replaced by a plastic mask with equally spaced points where the optical fiber could be connected, as shown in Fig~\ref{fig:xara_mask}. Four channels, each with an array of four Hamamatsu S14160-6050HS SiPMs, read out the LED pulses.

As these are room temperature measurements, single photo-electron calibration to evaluate the absolute number of photo-electrons could not be performed due to the high dark current. Therefore, LED pulses signals where measured in area and amplitude, always discounting the baseline area with the LED turned off three times during the measurement, to check the stability. To be able to compare the two WLS response, the LED intensity should be the same in between the tests, as much as dark current, SiPM bias voltage, etc.

For both, amplitude and area, the PDU matrix of each WLS slab is determined by the measurement $V_{ij}$ at each position coordinate $i$ and $j$ as:
\begin{equation}
	\label{eq:RT_uniformity}
	V_{ij} = \frac{V^{\mbox{ch1}}_{ij} + V^{\mbox{ch2}}_{ij} + V^{\mbox{ch3}}_{ij} + V^{\mbox{ch4}}_{ij}}{\mbox{Average}(V^{\mbox{ch1}} + V^{\mbox{ch2}} + V^{\mbox{ch3}} + V^{\mbox{ch4}})},
\end{equation}
where $V^{\mbox{ch1}}_{ij}$ corresponds to the readout of channel 1 (see Fig.~\ref{fig:xara_mask}) at position $ij$ and $V$ denotes the readout area or amplitude. This was done for both slabs with and without applying \viku\ on the edges. To evaluate the dispersion, the RMS of the PDU matrix is computed.

\begin{figure}[htbp]
	\centering
	\includegraphics[width=.39\textwidth]{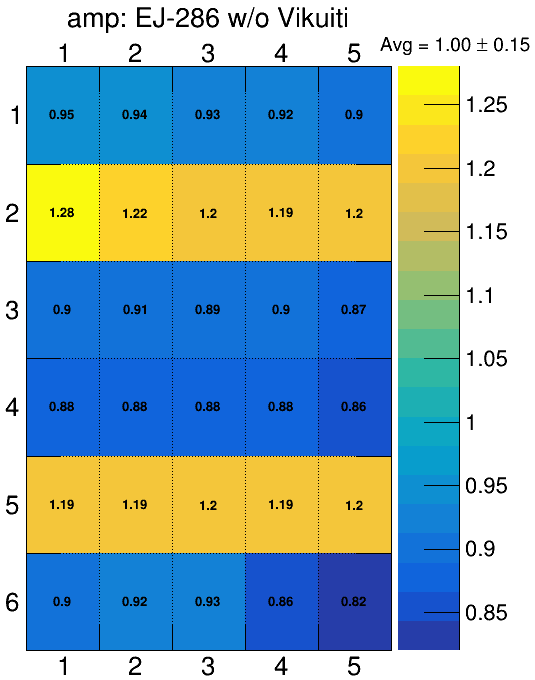}
	\includegraphics[width=.39\textwidth]{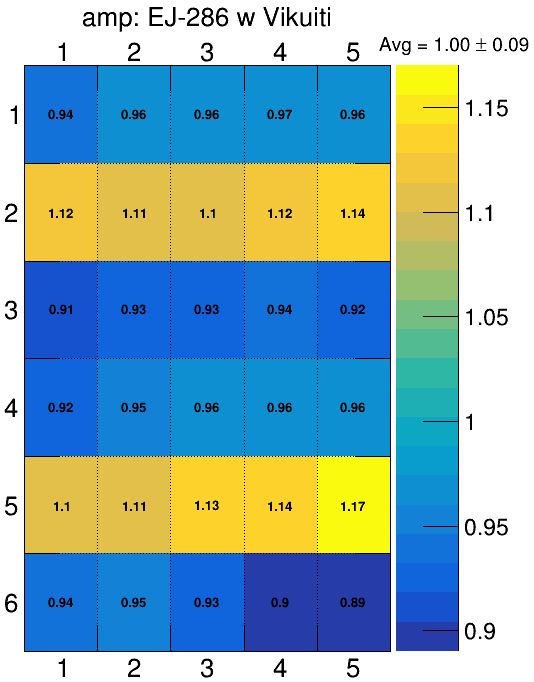}
	\caption{The PDU matrix of the EJ-286 slab, without (left) and with (right) \viku.}
	\label{fig:ej_RT}
\end{figure}

Figure~\ref{fig:ej_RT} shows the PDU matrix in terms of amplitude for the EJ-286 slab, without (Left) and with (Right) the \viku\ applied. One can see that applying \viku\ reflector directly on the WLS slab edges the RMS decreases from 15\% to 9\%, that is, the uniformity response of the device increases. For the FB118 slab the RMS decreased from 17\% to 9\%. It is noticeable that the second and fifth rows have a higher value, this is caused by the fact that the SiPM array is positioned directly in front of the optical fiber.

\begin{figure}[htbp]
	\centering
	\includegraphics[width=.385\textwidth]{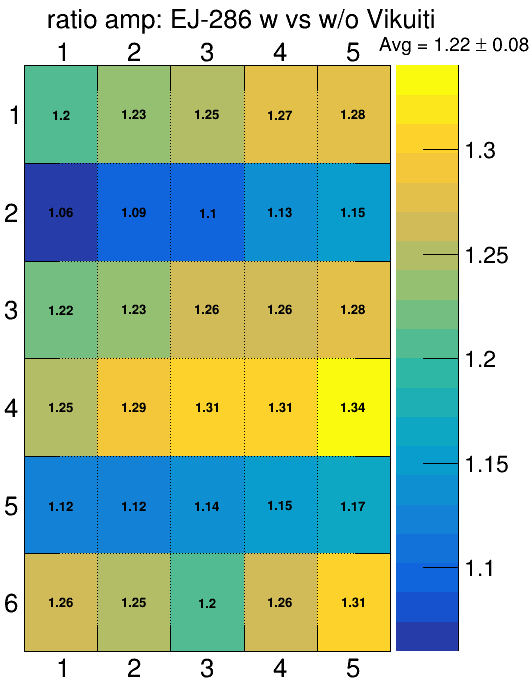}
	\includegraphics[width=.385\textwidth]{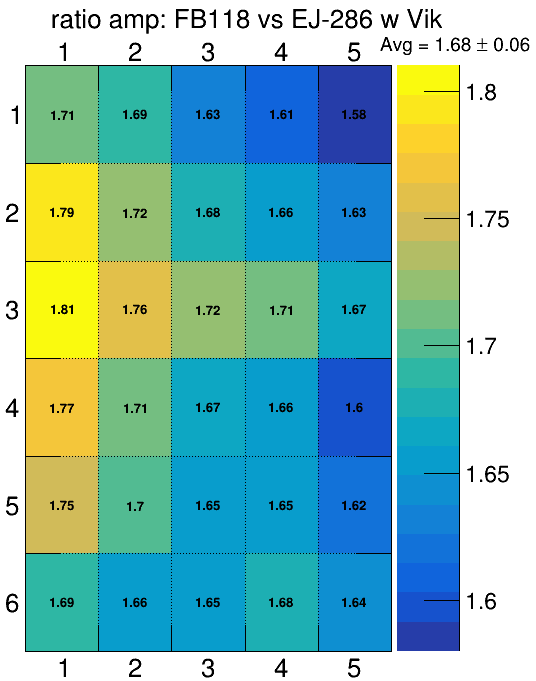}
	\caption{The PDE ratio matrix for the slab with versus without \viku\ (Left) and the comparison of FB118 versus EJ-286, both with \viku\ (Right)~\cite{enhancement_xara}.}
	\label{fig:ej_FB_RT}
\end{figure}

Another relevant measurement is the comparison between two samples $A$ versus $B$. The relative photon detection efficiency (PDE) variation between the two samples can be computed by the ratio matrix $R_{ij}$:
\begin{equation}
	\label{eq:RT_ratio}
	R_{ij} = \frac{A^{\mbox{ch1}}_{ij} + A^{\mbox{ch2}}_{ij} + A^{\mbox{ch3}}_{ij} + A^{\mbox{ch4}}_{ij}}{B^{\mbox{ch1}}_{ij} + B^{\mbox{ch2}}_{ij} + B^{\mbox{ch3}}_{ij} + B^{\mbox{ch4}}_{ij}},
\end{equation}
where $A^{\mbox{ch1}}_{ij}$ is the signal collected in the $ij$-th position by channel 1 for sample A. Figure~\ref{fig:ej_FB_RT} (Left) shows the relative measurement between the EJ-286 slab with versus without \viku\ applied. An increase of $\sim$22\% is noticeable, while the FB118 had an increase of $\sim$33\%. On the other hand, Figure~\ref{fig:ej_FB_RT} (Right) shows the relative measurement in terms of amplitude between the FB118 versus the EJ-286 slabs (both with \viku). By computing the average, the relative gain of FB118 over EJ-286 was found to be ($\sim$50\%) $\sim$65\% when (not) applying the \viku. 

\begin{figure}[h!]
	\centering
	\includegraphics[width=0.9\linewidth]{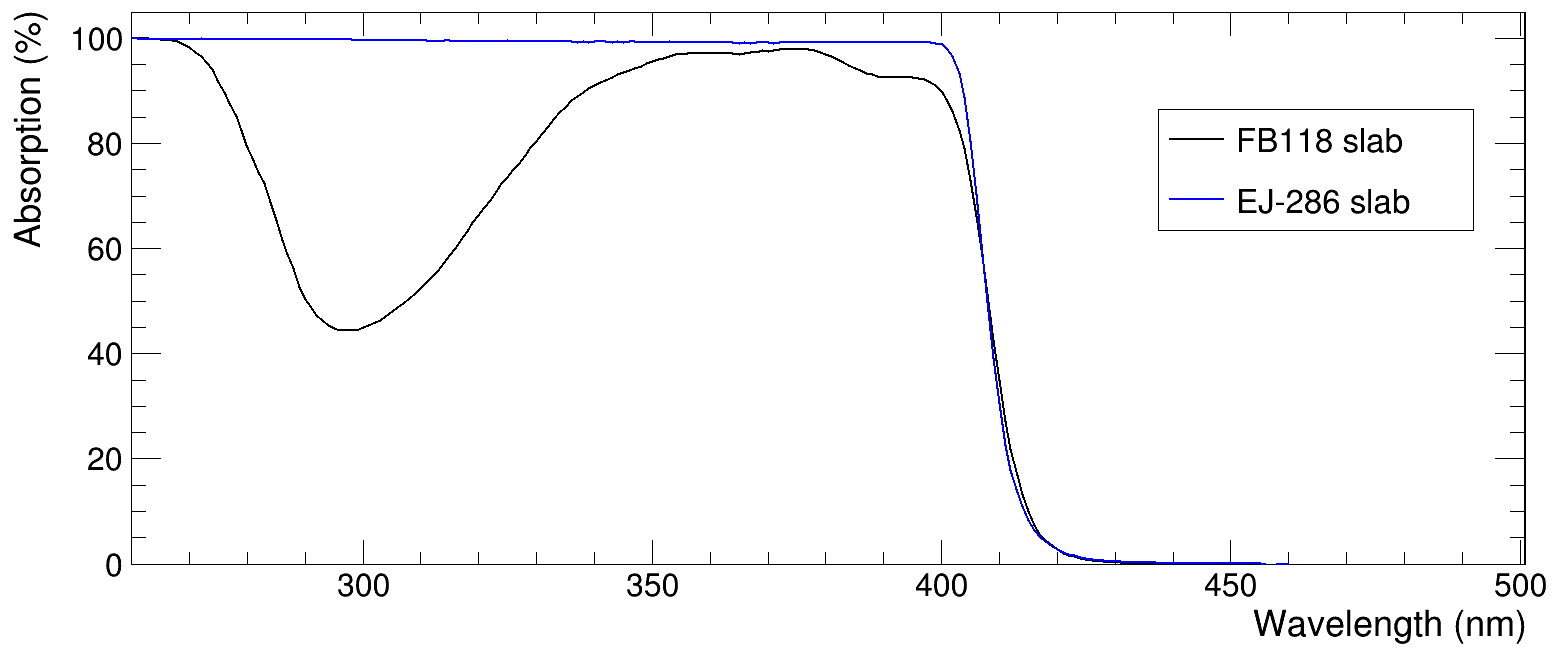}
	\caption{Absorption in percentage for different wavelengths for EJ-286 and FB118 slab. Data taken from Eljen data-sheet~\cite{eljen_286} and provided by G2P Co.~\cite{G2P}.}
	\label{fig:absorption_ej_fb}
\end{figure}

The measurements with signal amplitude and integrals were consistent among them. The gain in PDE found at room temperature gave a great hint about the relative measurements of the two slabs; however, there are some drawbacks: (1) The detection is performed without \sphe\ calibration, (2) the trapping effect of the \xara\ is not taking place and (3) the wavelength of the LED peaks at 405~nm, where the absorbance of the slabs are different, as can be seen in Figure~\ref{fig:absorption_ej_fb} where EJ-286 have an absorption of $\sim$80.2\% and the FB118 slab $\sim$72.9\%.

\subsection{Cryogenic measurements}
\label{sec:cryo_tests}

Prior to moving to the tests of the \xara\ in liquid argon (Chapter~\ref{chap:lar_test}), similar tests were performed at cryogenic temperature.

\begin{figure}[h!]
	\centering
	\hspace*{70pt}
	\includegraphics[width=0.55\linewidth]{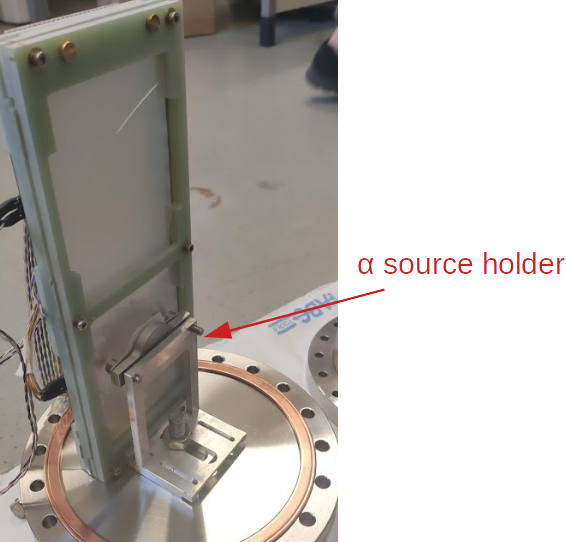}
	\caption{Photo of the \xara\ with the $\alpha$ source holder a few millimeters away from the dichroic filter.}
	\label{fig:xara_cryo_meas}
\end{figure}

In these tests, the cryogenic stainless steel chamber sizing 250~mm in diameter by 310~mm in height (see~Sec.~\ref{sec:x_ara_double_cell_setup}) was pumped down to a pressure of about 10$^{-3}$~mbar. Inside the chamber the \xara\ was mounted (with the dichroic filter deployed) and an americium alpha source ($^{241}$Am) was placed a few millimeters from the \ptp\ coated in the dichroic filter as can be seen in Figure~\ref{fig:xara_cryo_meas}. This configuration was inspired by the Dark Box tests (Sec.~\ref{sec:dark_box}): $\alpha$-particles can travel through the vacuum and stimulate the \ptp\ scintillation.

The testing chamber was submerged in a thermal bath of liquid nitrogen (LN$_2$) and a temperature sensor PT100 was placed in contact with the \xara\ so the measurements could be performed at a known temperature\footnote{As the chamber is in vacuum, the thermal contact between the \xara\ and the external is through the bottom flange to which the device is attached. Therefore, LN$_2$ baths were done overnight.}. Performing a cryogenic test in such way allows a better measurement of the wavelength downshifting properties of the slabs, as the \ptp\ emitted light follows the normal emission spectrum (Fig.~\ref{fig:absorption_emission_ptp}).

\begin{figure}[h!]
	\centering
	\includegraphics[width=0.8\linewidth]{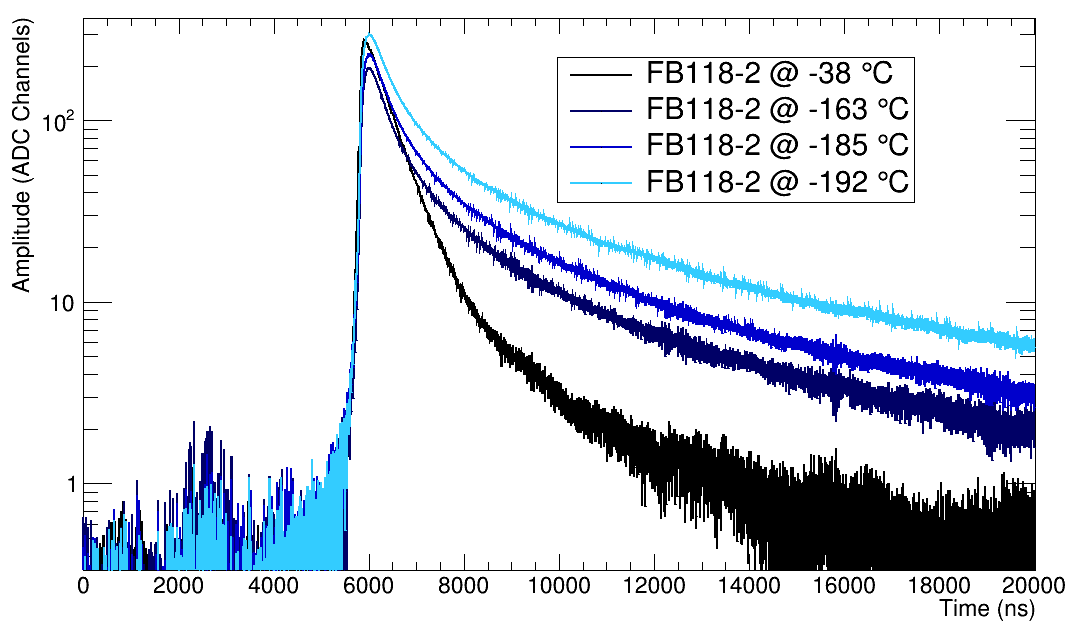}
	\caption{Averaged waveforms for four different temperatures.}
	\label{fig:temperatures_ch3_amp_signal_WLSFB118}
\end{figure}

The waveforms were acquired for 20~$\mu$s with a CAEN Digitizer DT5725 and signals where monitored during the cooling down. Figure~\ref{fig:temperatures_ch3_amp_signal_WLSFB118} shows the averaged waveforms for one of the channels. It is noticeable that the decay constant time of the waveforms increases with the drop of temperature. This is mostly probably related to the \ptp\ intrinsic scintillation. Unfortunately, the cold trans-impedance amplifier was not available at the time, so measurements had low signal-to-noise due to warm amplification and the \sphe\ calibration was rudimentary and no deconvolution of the signal was possible (see Sec.~\ref{sec:calibration_double_cell}) 

\begin{figure}[htbp]
	\centering
	\includegraphics[width=0.85\textwidth]{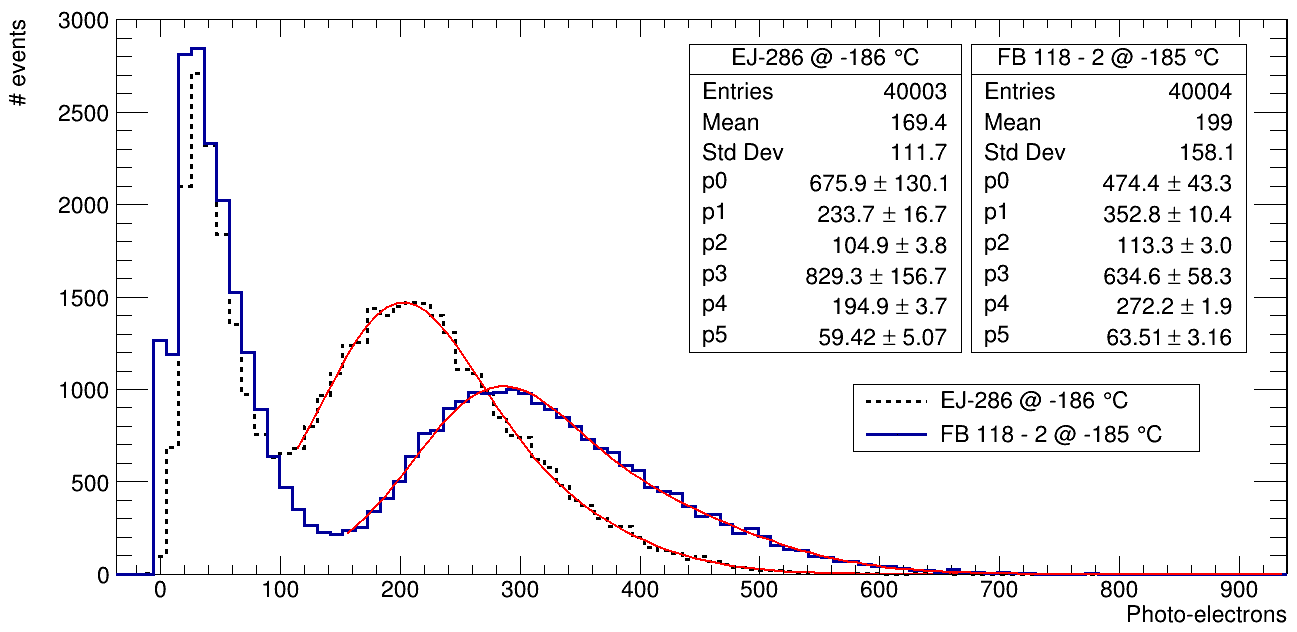}
	\caption{Alpha spectrum in photo-electrons detected by the \xara\ with EJ-286 (black dashed) and FB118 (blue solid) slabs. The fit was performed with the sum of two Gaussians~\cite{enhancement_xara}.}
	\label{fig:result_vacuum}
\end{figure}

As there was a different output depending on the temperature, the comparison between EJ-286 and FB118 slabs was performed with the closest temperature possible. Figure~\ref{fig:result_vacuum} shows the $\alpha$ spectrum in terms of photo-electrons detected for the \xara\ deployed with EJ-286 (black dashed) or FB118 (blue solid) WLS slabs. An empirical fit was performed with two Gaussians, with the intention of modeling and also taking into account the fraction of light from \ptp\ that is emitted back towards the vacuum and reflected back by the $\alpha$ source holder. Figure~\ref{fig:result_vacuum} reports the set of parameters from two Gaussian functions as:
\begin{equation}
	f(x) = p0\cdot\exp(-\frac{1}{2}\left(\frac{x-p1}{p2}\right)^2) +  p3\cdot\exp(-\frac{1}{2}\left(\frac{x-p4}{p5}\right)^2),
\end{equation}
where $p0$ and $p3$ correspond to the amplitudes, $p1$ and $p4$ to the means and $p2$ and $p5$ to the standard deviations of the two Gaussians respectively~\cite{enhancement_xara}.

By comparing the peak positions of the two spectra, a light collection increase of (42\error13)\% can be estimated. The large uncertainty is mainly due to the \sphe\ calibration and to the weighted average of the two Gaussian peaks. 

Besides this measurement, the intrinsic scintillation of the secondary wavelength shifters was studied. To do so, the alpha source was removed so the \xara\ was exposed only to cosmic muons. In this case, muons crossing the EJ-286 slab generated scintillation light signals well above 100 \phe. It happens that, EJ-286 slab is based on polivyniltoluene (PVT), that is known to be a scintillator~\cite{plasticsci14} with light yield of $\mathcal{O}(10^3)$ photons/MeV~\cite{enhancement_xara}. the Eljen producer, in the EJ-286 data sheet~\cite{eljen_286}, mentions the attempt to reduce the direct scintillation component. There was no noticeable scintillation in the FB118 slab tested in vacuum, as PMMA is not known to scintillate. 

The results achieved at room and cryogenic temperature motivated the comparison of the two wavelength shifters in liquid argon.

%% file: lar_tests.tex
\chapter{\xara\ Efficiency test}
\label{chap:lar_test}
\thispagestyle{myheadings}

Two different sets of experiments were performed in Brazil and Italy to evaluate the light collection efficiency of the \xara\ device. The experiments in Brazil were performed from October 2019 to March 2020 and consisted in exposing the device to $\alpha$ and $\gamma$ sources and to cosmic muons~\cite{x_arapuca_article}. A GEANT4 dedicated simulation was performed to retrieve the device efficiency with the three different ionizing radiations. In Italy, experiments to retrieve the efficiency and enhancement of the light collection were done in 2020/2021 by exposing the device to an $^{241}$Am source. Different positions along the longitudinal axis of the cell were tested and a Monte Carlo toy-model evaluated the number of expected photons arriving on the \xara\ acceptance window~\cite{enhancement_xara}. In this chapter, these two experiments are presented. 

\section{\xara\ single-cell tests}
\label{sec:xara_single_cell}
\begin{figure}[!h]
	\centering
	\begin{subfigure}{0.36\textwidth}
		\includegraphics[width=0.99\textwidth]{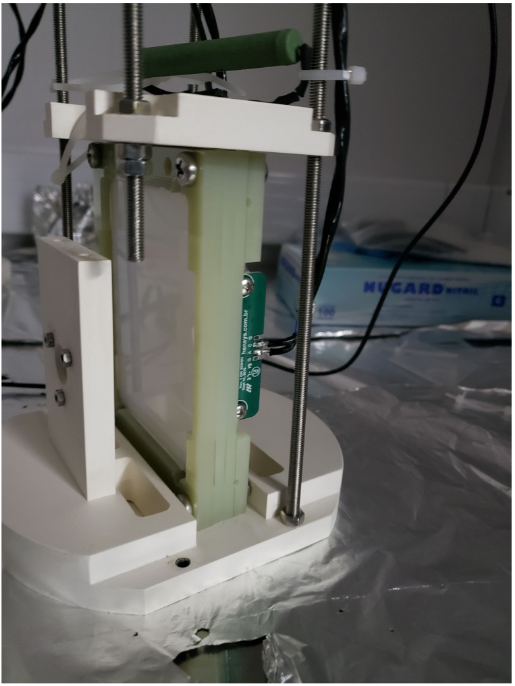}
		\caption{ }
		\label{fig:photo_arapuca_frame}
	\end{subfigure}
	\begin{subfigure}{0.36\textwidth}
		\includegraphics[width=0.99\textwidth]{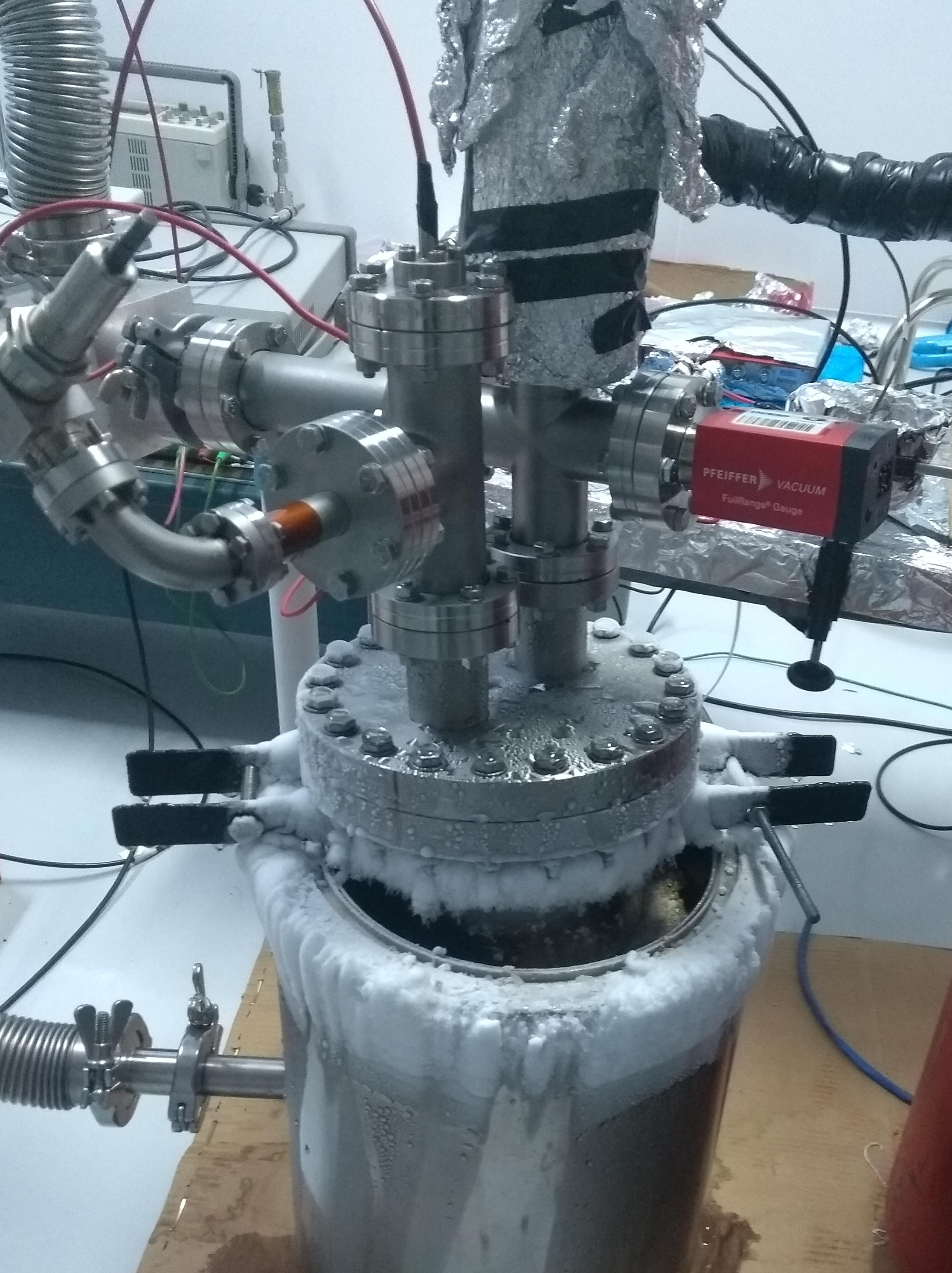}
		\caption{ }
		\label{fig:cryostat}
	\end{subfigure}
	\caption{\textbf{(a)}~the prototype when installed in its sustaining mechanical structure together with the holder of the $\alpha$-source. Level sensor can be seen above the prototype with a green color. \textbf{(b)}~Picture of the cryogenic setup during $\alpha$-source run~\cite{x_arapuca_article}.}
\end{figure}

The \textit{device under test} (DUT) was a \xara\ prototype with external dimensions of 9.6~cm$\times$12.5~cm hosting one single dichroic filter with an area of 8~cm$\times$10~cm as shown in Fig.~\ref{fig:photo_arapuca_frame}. The dichroic filter, from OPTO~\cite{opto_br} (see Sec.~\ref{sec:monochromator}), is designed to have a cut-off wavelength of 400~nm, a transmittance band which goes from 300~nm to 400~nm (transmissivity $\simeq$~90\%) and a reflective band ranging from 400~nm to 500~nm (reflectivity $\simeq$~98\%). The internal wavelength shifter (WLS) selected was a 3.5~mm thick Eljen EJ-286~\cite{eljen_286}. The absorption band spectrum of the WLS plate is well matched with the emission spectrum of the \ptp\ with maximum absorbance around 350~nm, while its emission is centered around 430~nm, inside the reflective band of the filter. The filter was coated with $\sim$400~$\mu$g/cm$^2$ of \ptp\ (see Sec.~\ref{sec:ptp_coating}). Two arrays of 4 SiPM each are installed on the 10~cm long lateral sides of the cavity. The SiPMs are Hamamatsu S13360-6050VE with active area of 6$\times6$~mm$^2$~\cite{hmmt_s13360}. The internal surfaces of the cavity are lined with 3M ESR \viku\ reflective foils, as well as the lateral sides of the WLS plate, with the exception of the regions where the SiPMs are installed.

\subsection{Experimental setup}
\label{sec:xara_single_cell_setup}

\begin{figure}[htbp]
	\centering
	\includegraphics[width=0.99\textwidth]{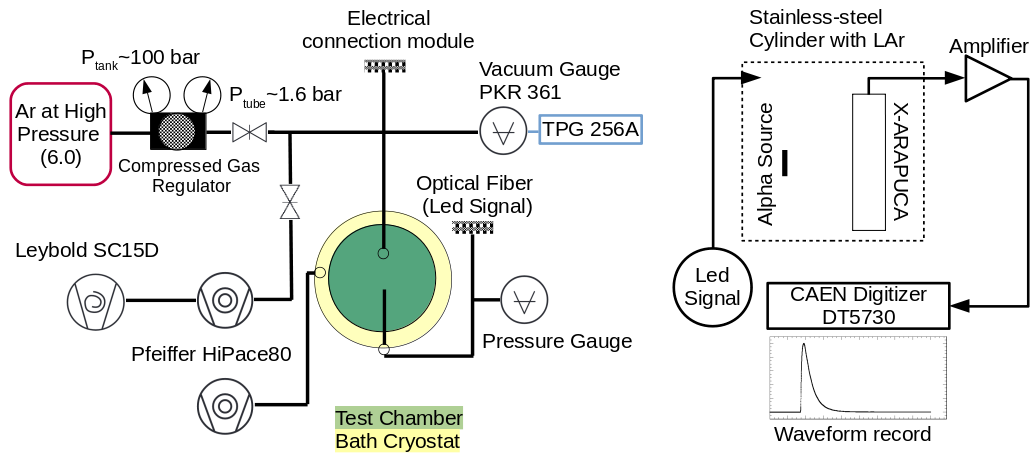}
	\caption{\label{fig:schmatics} (Left)~Schematic diagram of the cryogenic setup. (Right)~Schematic diagram of the DAQ~\cite{x_arapuca_article}.}
\end{figure}

The cryogenic setup of Figure~\ref{fig:cryostat} was assembled to carry out the \xara\ photon collection efficiency tests. The schematic diagram of the set-up is shown in Fig.~\ref{fig:schmatics}.  The inner stainless-steel cylinder (green) is pumped down to a pressure $\sim$$10^{-4}$~mbar at room temperature. Then, a cryogenic pump down to a pressure of $\sim$$10^{-6}$~mbar is performed by filling the external open thermal bath (yellow) with liquid argon (LAr)\footnote{In Italy, we discovered that a more successful method is to not make a cryogenic pump down. Instead, slowing filling with argon gas and stopping the pumps before starting the thermal bath will reduce the chance of contamination.}. The pump's gate valve is closed and Gas Argon~6.0\footnote{Gas Argon 6.0 refers to a contaminants level equal or below 1~ppm.} is injected in the cylinder, keeping the inner pressure $1<P_{\text{work}}<1.4$~bar. Argon gas (GAr) is liquefied thanks to the heat exchange with the thermal bath and a level sensor\footnote{The level sensor consists of a resistor sensitive to temperature variation. By monitoring the increase of the current going through the circuit one can determine when the resistor is fully covered by LAr.} installed inside the stainless-steel cylinder ensures that the \xara\ is completely submerged in LAr (the level sensor can be seen in Figure~\ref{fig:photo_arapuca_frame}, as a green resistor over the DUT). A vacuum tight optical feed-through was installed inside the stainless-steel cylinder to generate pulsed LED flashes in order to properly calibrate the system (see Sec.~\ref{sec:calibration_single_cell}).

The \xara\ is installed facing an $\alpha$-source 3~cm away from the center of the dichroic filter, as shown in the diagram of Figure~\ref{fig:schmatics} (Right) and the picture of Figure~\ref{fig:photo_arapuca_frame}. The DUT is held vertically by the Polyvinyl chloride (PVC) support that hosts the $\alpha$ source. For the $\gamma$-rays and cosmic-$\mu$ tests, the $\alpha$-source was removed but the holder was kept in place to maintain the same geometrical configuration and not to bias the measurements with different amount of environmental reflected light. The signals from the two SiPM arrays are amplified with a gain of 32~dB by a custom made, low-noise, two channels, pre-amplifier produced  by the Brazilian Company Hensys and read-out with a 14~bit CAEN DT5730 digitizer with sampling frequency of 500~MHz.

The $\alpha$-source is made of an alloy of aluminium and natural uranium in the form of a thin disk with thickness of 0.14~mm and diameter of 1~cm. An aluminium cover was placed in front of the alloy to reduce the count rate of the device, this is important to retrieve a clear spectrum. This source was previously studied by the Chronology group of University of Campinas~\cite{LAr_arapuca_test}. The energy, relative abundance and parent nucleus of the $\alpha$ source are shown in Table~\ref{tab:alpha_spectrum}~\cite{alpha_spectrum_tab}. For the $\gamma$-rays test, the $\alpha$-source was removed and a $^{60}$Co source was placed outside the cryostat in a position where the emission of the source faces the \xara\ acceptance window.
\begin{table}[htbp]
	\centering
	\caption{\label{tab:alpha_spectrum}Energy, relative intensity and parent nucleus of the $\alpha$ particles emitted by the natural uranium~\cite{alpha_spectrum_tab}.}
	\begin{tabular}{|c|c|c|}
		\hline
		$\alpha$ energy (MeV) & relative intensity & parent nucleus               \\ \hline
		4.187                 & 48.9\%             & $^{238}$U \\
		4.464                 & 2.2\%              & $^{235}$U \\
		4.759                 & 48.9\%             & $^{234}$U \\ \hline
	\end{tabular}
\end{table}

\subsection{Data Acquisition}
\label{sec:xara_single_cell_daq}
The data acquisition (DAQ) was performed with a customized version of the \textit{wavedump} software from CAEN. The ADC read-out was performed every time that one of the two arrays exceeded a fixed threshold, which corresponded to approximately 5~photo-electrons. However, a fine tuning of the threshold was done manually for each run searching for a good balance between trigger rate and spectrum resolution. For each triggered event a waveform with 18~$\mu$s record length, which corresponds to 6,000 samples, with 3.6~$\mu$s of pre-trigger was saved for both channels. A total of 40,000 trigger events were acquire for $\alpha$ and $\mu$ runs. The $^{60}$Co $\gamma$-source has a low activity being necessary to subtract the background. For that, the data were taken during a fixed period of time of about 30 minutes with and without the source with a fixed trigger.

Dedicated calibration runs were performed each time the SiPMs overvoltage was changed. A blue LED was flashed using pulses of approximately 50~ns, frequency of 1~kHz and amplitude set in order to produce signals of few photo-electrons. The acquisition of the signals was performed with the external trigger of the Pulse Generator \textit{Agilent} with the same record length of 18~$\mu$s used for source signals.

\begin{figure}[htbp]
	\centering
	\includegraphics[width=0.74\textwidth]{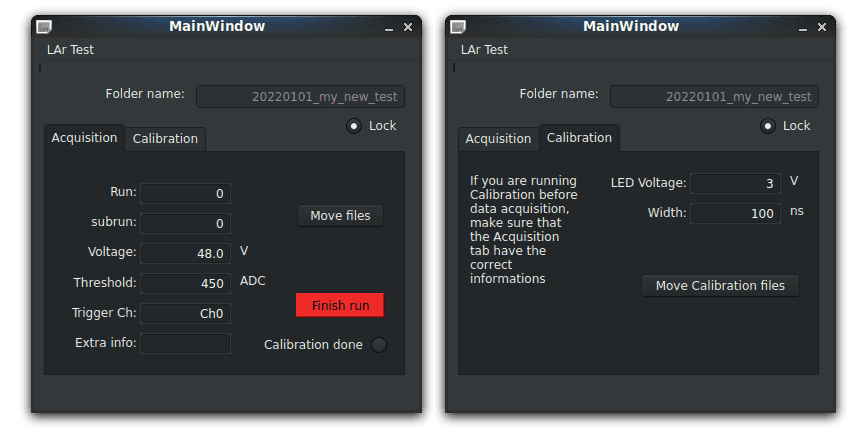}
	\caption{\label{fig:gui_daq} (Left)~Tab for the main acquisition. (Right)~Tab for calibration runs.}
\end{figure}

A graphical user interface (GUI) was build in C++ with QtCreator software tool to automatically save the data. Figure~\ref{fig:gui_daq} shows the two taps of the GUI. The data were saved in folders with the specific name given by the user, for example, ``\textit{20220101\_my\_new\_test}''. The run and sub-run\footnote{Sub-runs were necessary because data were saved in packs of 10,000 events.} numbers are changed automatically by the GUI, while the bias voltage, threshold, trigger channel and extra information are inserted by the user. The pulse width and voltage amplitude of the LED flashes can be set for the calibration runs. In the example of Fig.~\ref{fig:gui_daq}, the run file for channel 0 would be named as ``\textit{0\_wave0\_48V0\_450ADC\_Ch0.dat}'' in a folder named ``\textit{run0\_48V0\_450ADC\_Ch0}''. Calibration runs are saved in an internal folder ``\textit{Calibration}'' as ``\textit{wave0\_48V00\_3V00\_100ns.dat}''. The data was collected initially as ASCII format and later changed to binary format due to its better compression.

The collected waveforms are pre-processed and stored in root files. Each channel has a class associated with stored information such as: event ID, maximum amplitude value and position, charge, $F_{\text{prompt}}$ (see Sec~\ref{sec:lar_purity_double_cell}) and selection cuts. The root files have the waveforms stored with the baseline already calculated and filtered with the denoise algorithm for 1-D signals~\cite{denoising}. The filter preserves the signal rise-time and integral~\cite{protoDUNE_first_results} and the C code is available on the authors website. The baseline is calculated in the first 3.6~$\mu$s, the pre-trigger region.

\subsection{Calibration}
\label{sec:calibration_single_cell}

\subsubsection{Single photo-electron (\sphe) calibration}

The SiPMs were operated in reverse bias with a few volts over the breakdown voltage of 43.0~$\pm$~0.2~V. Starting from +3.5~V overvoltage (O.V.) to +5.5~V with steps of 0.5~V. Between the different overvoltages (O.V.) used, the values of +5.0 and +5.5~V~($\pm 0.2$~V)~O.V.\ were selected due to the quality of the calibration and spectrum.

In order to properly evaluate the amount of photo-electrons (\phe) detected by the device, the calibration data were used to retrieve the single photo-electron (\sphe) response. The waveforms, acquired with the LED as external trigger, are integrated for 900~ns (decay time of \sphe\ signal $\sim$300~ns) starting from the trigger point (3.6~$\mu$s)~\cite{SiPM_better}.

\begin{figure}[ht]
	\centering
	\includegraphics[width=.45\textwidth]{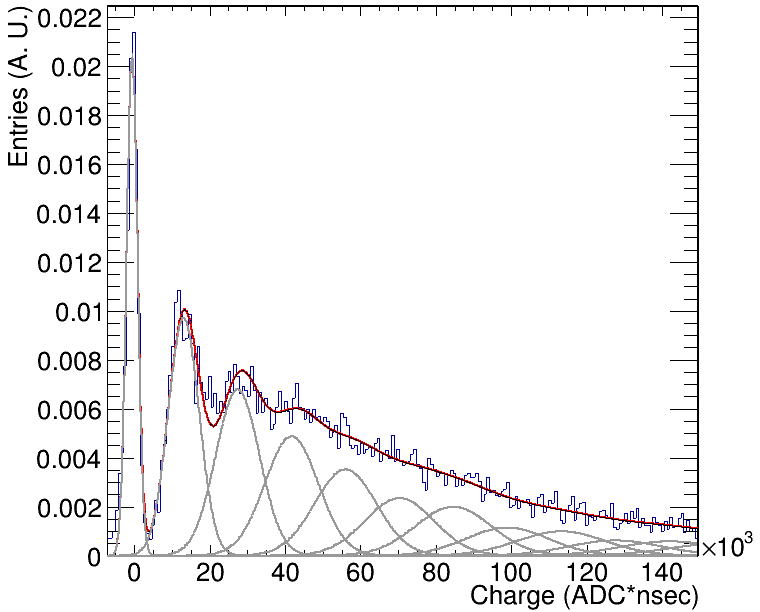}
	\includegraphics[width=.44\textwidth]{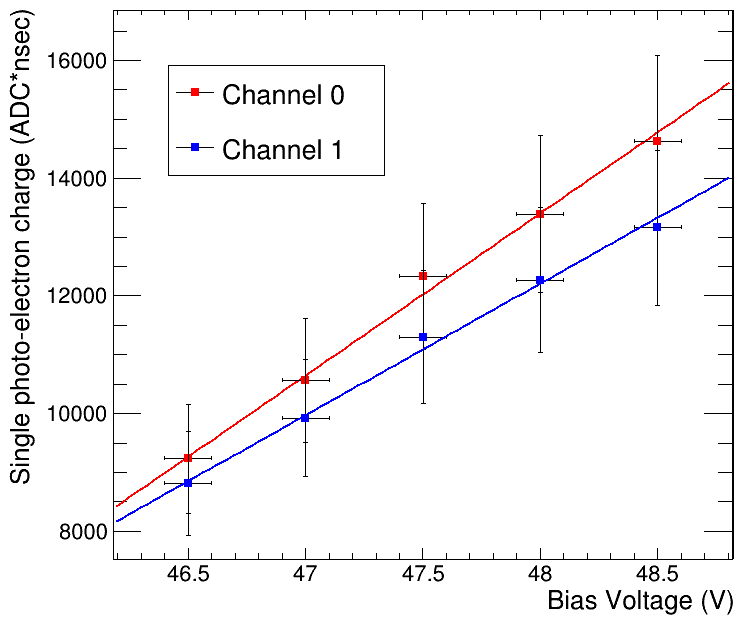}
	\caption{\label{fig:sphe} (Left)~Single photo-electron response obtained during calibration. The distance between the first and second peaks after the noise gives the gain of the SiPM. (Right)~Gain versus bias voltage for the two channels of the \xara. Red and blue lines represent linear fits of the experimental points. The breakdown voltage was found to be 43.0~$\pm$~0.2~V~\cite{x_arapuca_article}.}
\end{figure}

A spectrum obtained with a bias voltage of 48.0~V (+ 5.0~$\pm$~0.2 O.V.) is shown in Figure~\ref{fig:sphe}. The fit is performed with $N+1$ Gaussian distributions, where the first one corresponds to the electronic noise and the others correspond to first, second and $N$-th photo-electron responses. The noise and \sphe\ Gaussian distributions have 3 free parameters each (mean value, standard deviation and normalization constant). The mean values and standard deviations of the other peaks are constrained by the following conditions:

\begin{itemize}
	\item the mean value of the \textit{N-th} peak is given by $N~\times~\mu_1$, where $\mu_1$ is the mean value of the single photo-electron peak;
	\item the standard deviation of the \textit{N-th} peak is given by $\sqrt{N}~\times~\sigma_1$, where $\sigma_1$ is the standard deviation of the single photo-electron peak;
\end{itemize}

The gain was considered to be the distance between the first and second photo-electron. The result for different bias voltages is shown in Figure~\ref{fig:sphe} for the two channels of the \xara. A linear fit is shown together with the data. 

\subsubsection{Matching filter}

Unfortunately, during the run with cosmic muons and $\gamma$-rays there was the appearance of a periodic external noise that prevented a proper calibration for low voltages. For +5.0 and +5.5 O.V. a noise suppression algorithm was made with a \textit{Matching filter} or \textit{Optimum filter}.

The filter consists of a convolution between the reference signal $r(t)$ and the waveform signal $s(t)$. The conventional convolution inverts, for instance, the reference signal during the integration. However, in the \textit{Matching filter} the signal is unchanged, that is:
\begin{equation}
\text{matching}(s, r)[n] = \sum_{i=0}^{k-1} s[n+i]\times r[i],
\end{equation}
where $n$ is the number of points in the complete event $s$ and $k$ is the number of points in the reference $r$. Once the convolution results in the area under the two curves, it is necessary to normalize the signals to compensate the difference in amplitude. In this way, it is necessary to perform the following operation:
\begin{equation}
\label{eq:matching}
\text{matching}_{\text{norm}}(s,r)[n]=\dfrac{\sum_{i=0}^{k-1} s[n+i]\times r[i]}{\sqrt{\sum_{i=0}^{k-1} s[n+i]^2 \times \sum_{i=0}^{k-1} r[i]^2}}.
\end{equation}

\begin{figure}[h!]
	\centering
	\includegraphics[width=0.7\linewidth]{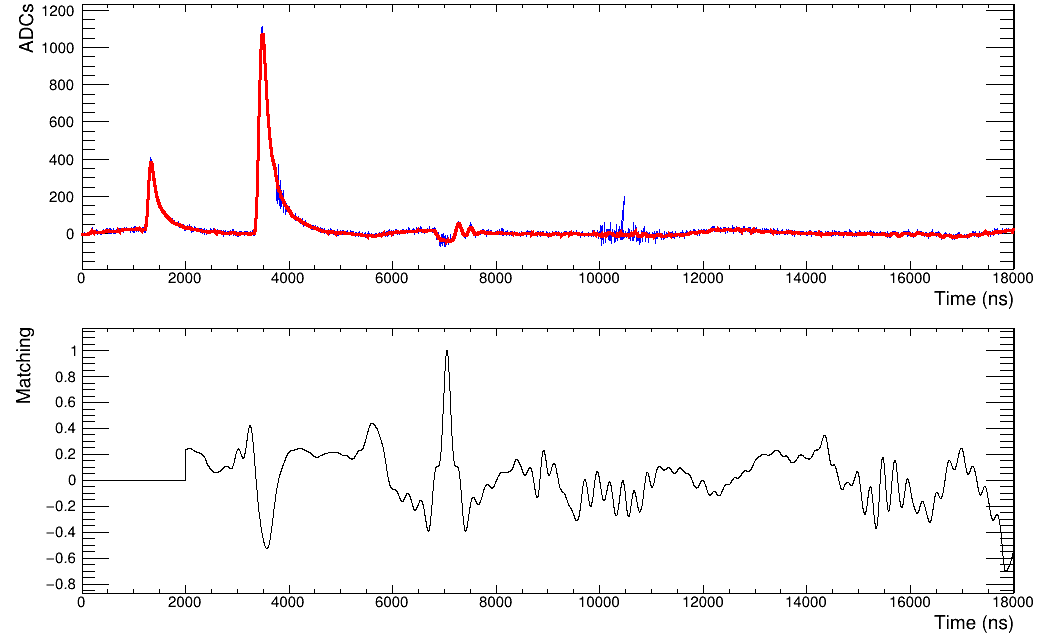}
	\caption{(Top) Example of a waveform in blue with a moving average of 35 points in red. (Bottom) Result obtained by applying the \textit{Matching filter} of Eq.~\ref{eq:matching}, using as reference the points between 6000 and 8100~ns. A Matching value equal to 1.0 means a perfect match. The filter was applied only for $t>2000$~ns.}
	\label{fig:sample_matchin}
\end{figure}

Lets take, for example, the signal $r$ extracted from $s$ as in the Figure~\ref{fig:sample_matchin} (top), where the red line (moving average) represents the signal $s$. Taking $r$ as the signal between 6000~ns and 8100~ns, which corresponds to the periodic noise (the blue line is the normal signal, without any processing). In this way, one can imagine that in a certain point $n'$, one has $s[n'+k] = r[k]$ for all $k$ in the sum (precisely between 6000 and 8100~ns). It is clear, from Eq.~\ref{eq:matching} that the result is equal to 1. For every other point, where the matching is not 100\%, the factor $\text{matching}_{\text{norm}}$ will be always lower than 1 ($\text{matching}_{\text{norm}} \le 1$). In the bottom part of Fig.~\ref{fig:sample_matchin} the matching for the red signal from the top part using $r$ as described above is presented. One can notice that the maximum matching is achieved around 7,000~ns.

Therefore, in the calibration analysis of $\gamma$ and $\mu$ runs any pulse with a matching $>0.55$ was discarded. This value was chosen manually trying to minimize the amount of false positive events lost. As lower voltages decrease the amplitude of the SiPMs signals but does not affect the amplitude of the noise, the matching filter was not enough to discard noisy signals and the calibration was not successfully performed. 

This filter can also be used to search \sphe\ peaks instead of noise. This, however, was not necessary due to the dedicated calibration procedure.

\subsubsection{Cross-talk evaluation}
\label{sec:cross_talk_single_cell}
The cross-talk probability strongly depends on the temperature of the sensor. Operating them at LAr temperature, makes the cross-talk probability quoted by the vendor not usable, since it is referred to room temperature. To estimate the cross-talk probability and find a correction factor to better estimate the number of photo-electrons the method described in~\cite{Cross_talk_vino} was applied. The method assumes that:
\begin{enumerate}[label=(\roman*)]
	\item The number of ``primary'' pulses has the Poisson distribution with $\lambda = L$.
	\item Each ``primary'' pulse can generate one or several duplicated pulses independently of preceding events.
	
In this case, the number of duplicated pulses for each primary pulse has the geometrical distribution:
	\begin{enumerate}[label=(\alph*)]
		\item The first additional pulse is generated with probability $p$.
		\item In the presence of the first additional pulse, the second additional pulse is generated with the same probability $p$, and so on;
		\item The chain of duplicated pulses is unlimited.
	\end{enumerate}
	\item Primary pulses and all the chains of secondary pulses generated by them are detected.
\end{enumerate} 

The recursive function to calculate the compound Poisson distribution for $k$ photo-electrons was found to be:
\begin{equation}
\label{eq:ct_vino}
\begin{split}
P_k(L,p) &= \exp(-L)\sum_{i=0}^{k} B_{i,k} \cdot [L(1-p)]^i \cdot p^{k-i}
\\
\text{where,}
\\
B_{i,k} &= \left\{
\begin{array}{lr}
\qquad 1 \quad \qquad \text{if } i = 0 \text{ and } k = 0 \\
\qquad 0 \quad \qquad \text{if } i = 0 \text{ and } k > 0 \\
\frac{(k-1)!}{i!(i-1)!(k-i)!} \quad \text{otherwise}
\end{array}
\right.
\end{split}
\end{equation}
where $L$ is the mean value of the Poisson distribution and $p$ is the cross-talk probability~\cite{Cross_talk_vino}.

To apply this method into the calibration data, we considered that the probability of detecting $k$ photo-electron is given by the total number of events containing $k$ photo-electrons, divided by the total number of events. The integration of the $k$-th Gaussian distribution from Figure~\ref{fig:sphe} gives the number of events with $k$ \phe. 

\begin{figure}[!h]
	\centering
	\begin{subfigure}{0.49\textwidth}
		\includegraphics[width=0.99\textwidth]{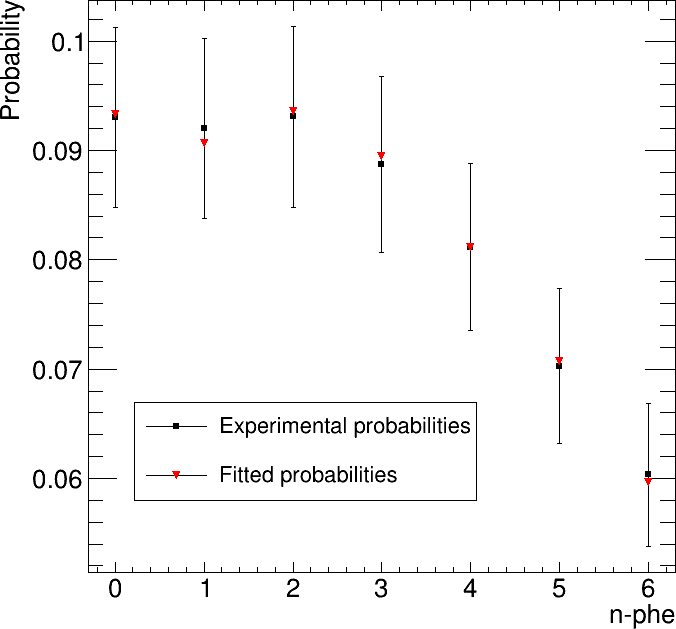}
		\caption{ }
		\label{fig:CT_vinos}
	\end{subfigure}
	\begin{subfigure}{0.49\textwidth}
		\includegraphics[width=0.99\textwidth]{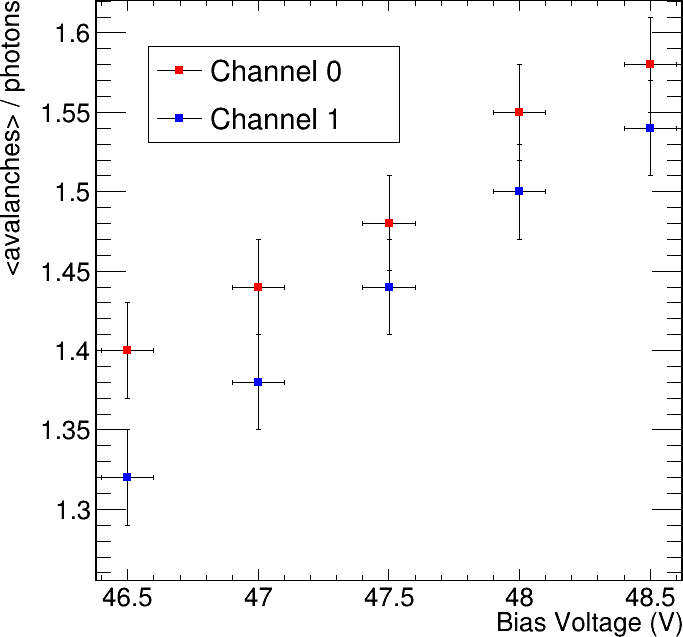}
		\caption{ }
		\label{fig:CT_vinos_bias}
	\end{subfigure}
	\caption{\textbf{(a)}~Probability of detecting $n$ photo-electrons (black squares). A $\chi^2$ minimization of the experimental points with equation~\ref{eq:ct_vino}~\cite{Cross_talk_vino} was performed and the result is shown as red triangles. \textbf{(b)}~Avalanches per photons versus bias voltage for each channel~\cite{x_arapuca_article}.}
\end{figure}

Equation~\ref{eq:ct_vino} was used to minimize the normalized calibration spectra, where $L$ and $p$ were considered as free parameters together with a normalization constant.
One of the results of the fitting procedures is shown in Figure~\ref{fig:CT_vinos}.
The plot of the number of SiPM avalanches per photon ($1+p$) vs.\ the bias voltage is shown in Figure~\ref{fig:CT_vinos_bias} for each channel. These values are used as correction factors when we convert the number of detected avalanches into photo-electrons. It ranges between 1.32 and 1.58 avalanches per photon and is consistent with what was found in the first run of the  protoDUNE detector~\cite{protoDUNE_first_results}. The calibration to convert the charge read-out as ADC$\times$ns to single photo-electrons is performed by dividing the charge of the signal by the gain of the $\sphe$ times the number of avalanches per photon.

The cross-talk probability is quite high if compared with the $\sim$7\% reported in the SiPM's datasheet. The reason for that may be explained by the cryogenic temperatures along side with the SiPM pulse resolution. However, during the measurements performed at Milano-Bicocca, dedicated tests were performed in the S13360-6050VE SiPMs and it was noticed small after-pulses in the single photo-electron pulses that degraded the resolution of the spectrum. In fact, one can see that the spectrum of Fig.~\ref{fig:sphe} is very broad and the Gaussian distributions overlap significantly compromising the resolution of single peaks. 

The two main sources of uncertainties are the single photo-electron calibration and the cross-talk probability. An uncertainty of 15\% due to the \sphe\ was taken based in the quality of the fits of Fig.~\ref{fig:sphe}. The cross-talk calibration had a maximum variation of $\sim$10\% which was conservatively chosen. The combined uncertainty was taken as 20\%.

\subsubsection{Liquid Argon purity}
\label{sec:lar_purity_single_cell}
\begin{figure}[htbp]
	\centering
	\begin{subfigure}{0.49\textwidth}
		\includegraphics[width=.99\textwidth]{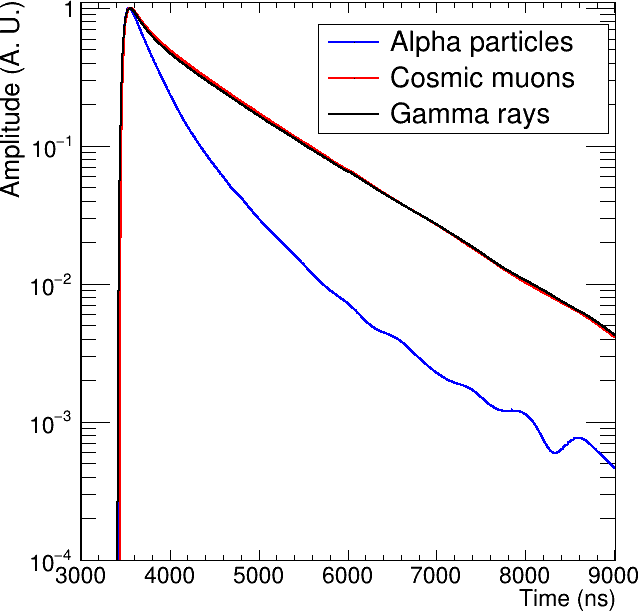}
		\caption{ }
		\label{fig:mean_signals}
	\end{subfigure}
	\begin{subfigure}{0.49\textwidth}
	\includegraphics[width=.99\textwidth]{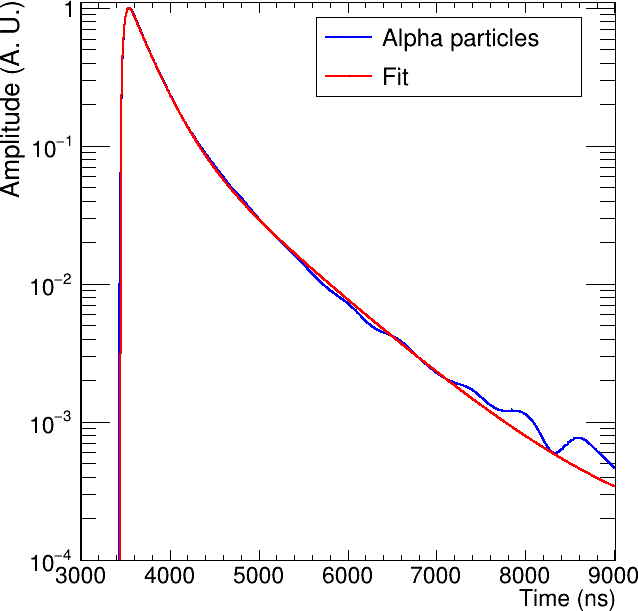}
		\caption{ }
		\label{fig:mean_signals_fit}
	\end{subfigure}
	\caption{\textbf{(a)}~Average signals for the three different particles. \textbf{(b)}~Average signal of $\alpha$-particles fitted with equation~\ref{eq:signal_fit}~\cite{x_arapuca_article}.}
\end{figure}

The liquid argon purity was monitored by the triplet time constant (see Sec.~\ref{sec:lar_scintillation}) for all measurements. The information of the triplet time constant is important to validate the tests and correct the total amount of photo-electrons detected. To perform this analysis an average signal was retrieved as shown in Figure~\ref{fig:mean_signals} for $\alpha$-particles (red), $\gamma$-rays (black) and cosmic muons (blue). The difference in the ratio between singlet and triplet contributions is evident by the plots, where the $\alpha$'s have more singlet contribution and electrons and muons have a greater triplet contribution~\cite{LAr_fund_properties,ICARUS,abudance_dependence}. It was shown  that N$_2$ and O$_2$ contamination, at a few ppm level, can significantly quench the triplet scintillation emission, thus reducing the total photon yield~\cite{nitrogen_contamination_roberto}. In order to correct for this effect, an overall fit of the average scintillation waveforms is performed and the value of the triplet scintillation decay time is extracted. This value is used to extrapolate the number of photo-electrons which would be detected in the case of perfectly pure LAr.

Equation~\ref{eq:signal_fit} is used to fit the waveform as presented in Figure~\ref{fig:mean_signals_fit}, where $\tau_S$ and $\tau_T$ are the singlet and triplet components and $A_S + A_T = 1$ are the relative amplitudes~\cite{Segreto_2021}. The third term of the sum is a component related to a possible \ptp\ delayed emission~\cite{Ettore_tpb}, with $N$ and $A$ as constants depending on the nature of the scintillator and $t_a$ is the relaxation time. $A$ and $t_a$ where constrained to vary between $0.22 - 0.55$ and $20 - 100$~ns, respectively. The fit is performed with a $\chi^2$ minimization of the signal with the light pulses of eq.~\ref{eq:signal_fit}, convoluted with a Gaussian, which takes into account the effect of the readout electronics.
\begin{equation}
\label{eq:signal_fit}
L(t) = \frac{A_f}{\tau_f} e^{-t/\tau_f} + \frac{A_s}{\tau_s} e^{-t/\tau_s} + \frac{N}{[1+A\;ln(1+t/t_a)]^2 (1+t/t_a)}
\end{equation}

The fit was performed with the TMinuit tool from Root Cern~\cite{root_cern}. The discrete convolution is performed considering that each point from Eq.~\ref{eq:signal_fit} has a corresponding normalized Gaussian distribution with a certain fixed standard deviation $\sigma$. The convolution of a certain point at $t=t'$ is performed from left to right, where only points with $t\leq t'$ are considered. The convoluted point at $t'$, that is $C(t')$, will be the sum of every point within\footnote{This value is chosen to take only Gaussian that are effectively contributing to the convolution and to minimize the time consumption.} 5$\sigma$ from $L(t)$ times the corresponding Gaussian amplitude of that specific point, that is:
\begin{equation}
	C(t') = \sum_{t=0}^{t'} \text{Gaus}(t,t',\sigma) \times L(t),
\end{equation}
where, $\text{Gaus}(t,t',\sigma)$ means a Gaussian with mean value $t'$, standard deviation $\sigma$ evaluated at the point $t$. This operation is performed on the waveform at every step of the minimization, therefore, it is very time consuming. The approach done in the \xara\ double-cell was to deconvolve the waveform as shown in Sec.~\ref{sec:xara_double_cell}.

The triplet time constant was found to be 780\error20, 1100\error20 and 1050\error20~ns for $\alpha$'s, $\gamma$'s and muons, respectively.  

Comparing the values of the triplet decay times, found with this fitting procedure, with the expected values~\cite{nitrogen_contamination_roberto}, we determined the correction factors of 1.16, 1.43 and 1.47 for $\alpha$-particles, $\gamma$-rays and cosmic muons, respectively.

\subsection{Experimental setup Monte Carlo Simulation}
\label{sec:simulation_single_cell}
	
A dedicated GEANT4 Monte Carlo simulation was developed~\cite{simulation_rafa} for each type of ionizing radiation to properly estimate the device photon collection efficiency. All the dimensions of the experimental setup were carefully measured and implemented: the stainless-steal cylinder and cryostat, the \xara\ and $\alpha$-source PVC support (Figure~\ref{fig:photo_arapuca_frame}) and the \xara\ prototype.

The liquid argon scintillation light yield was set at 51,000 photons per~MeV~($\gamma/$MeV)~\cite{LAr_fund_properties,abudance_dependence} times the quenching factor, which was set at 0.71 for the $\alpha$-particles and at 0.78 for electrons and muons~\cite{LAr_fund_properties,model_nuclear_recoil_nl,x_arapuca,x_arapuca_article}. 

The emission of the $\alpha$-source was set uniformly and isotropically inside the aluminium disc with the energy distribution given in Table~\ref{tab:alpha_spectrum}. The $\alpha$-particles lose a fraction of their initial energy inside the aluminium disc (which goes undetected) and leave the aluminium alloy with a continuous residual energy distribution~\cite{alpha_equation}.

\begin{figure}[htbp]
	\centering
	\includegraphics[width=0.60\textwidth]{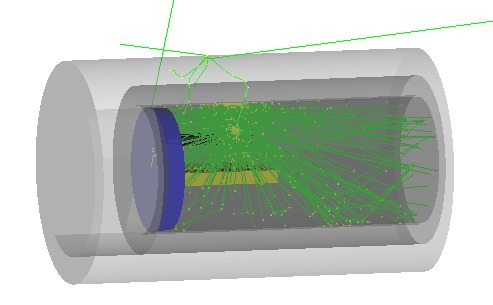}
	\caption{\label{fig:ex_simu} Example of the experimental setup with simulated gamma source outside the cryostat and the optical photons (green lines) reaching the \xara~\cite{x_arapuca_article}.}
\end{figure}

To simulate the detection of $\gamma$-rays, the $^{60}$Co source was positioned outside the cryostat. The two emission lines, one at 1.17~MeV and the other at 1.33~MeV~\cite{cobalt_60}, were chosen to be isotropic. Figure~\ref{fig:ex_simu} shows an example, with the gamma source placed outside the cryostat and optical photons (green) generated in the LAr reaching the \xara\ (yellow).

\begin{figure}[tbph!]
	\centering
	\includegraphics[width=0.85\linewidth]{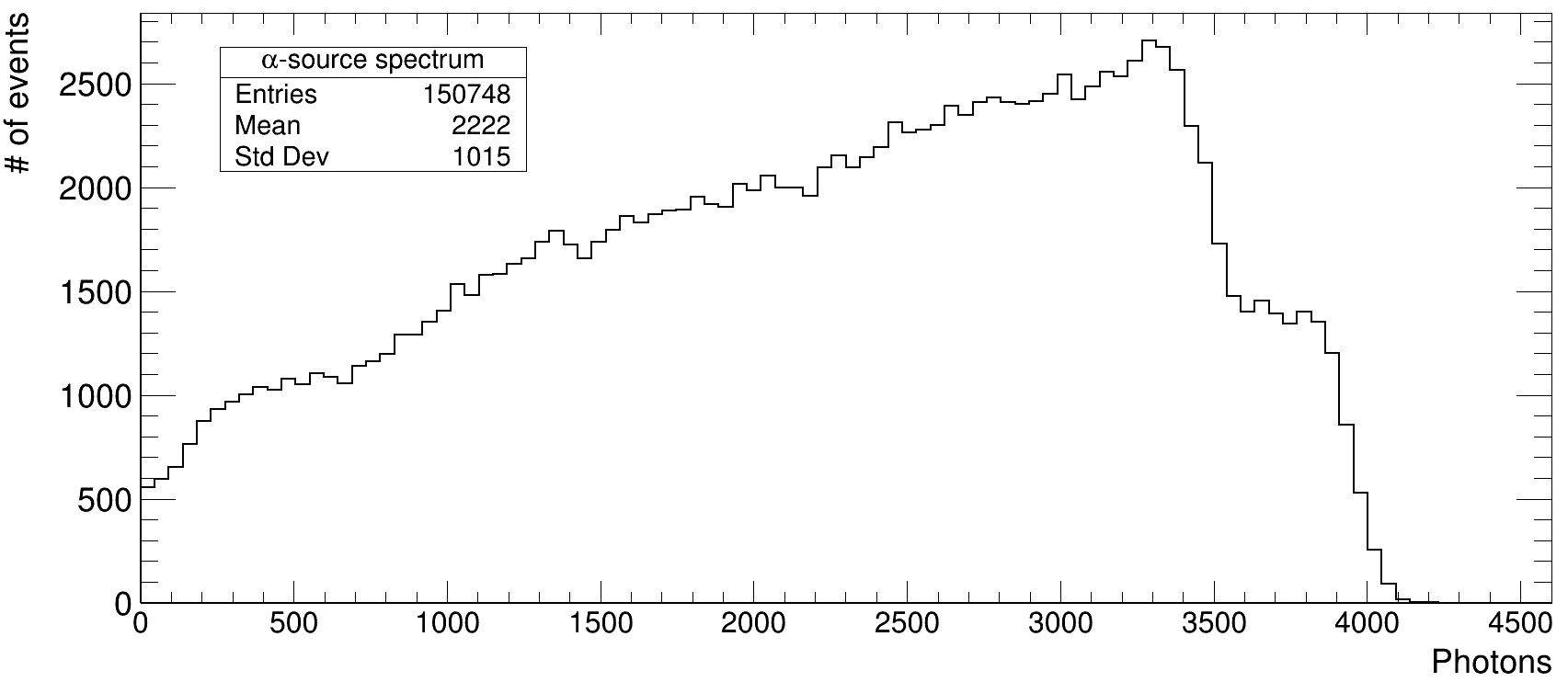}
	\caption{Monte Carlo simulation output, the number of expected photons compose the $\alpha$-source spectrum.}
	\label{fig:simulationalphaspectrumrafa}
\end{figure}

Cosmic muons were generated uniformly in a 4 meters diameter disc positioned 50~cm above the cryostat, with a $\cos^2{\theta}$ angular distribution (where $\theta$ is the zenith angle), with an uniform azimuth angle distribution and a fixed energy of 4~GeV, which corresponds to the average energy of muons at ground level~\cite{pdg,MuonsCecchini2012}. The reflectivity of the internal stainless-steel surfaces was set at 20\% for the scintillation photons. The final Monte Carlo output is the number of photons that reaches the \xara\ acceptance window per particle generated. The simulated spectrum is shown at Figure~\ref{fig:simulationalphaspectrumrafa}.

\subsection{Efficiency analysis}

The final light collection efficiency of the \xara\ single-cell was measured comparing the simulated spectrum with the experimentally measured one. The digitized waveforms were integrated over 15.3~$\mu$s starting from 300~ns before the onset. The integral value is first multiplied by the correction factor due to the LAr purity and secondly divided by the gain of the SiPMs times the number of avalanches per photon found. The output is the total amount of photo-electrons detected.

\begin{figure}[tbph!]
	\centering
	\includegraphics[width=0.8\linewidth]{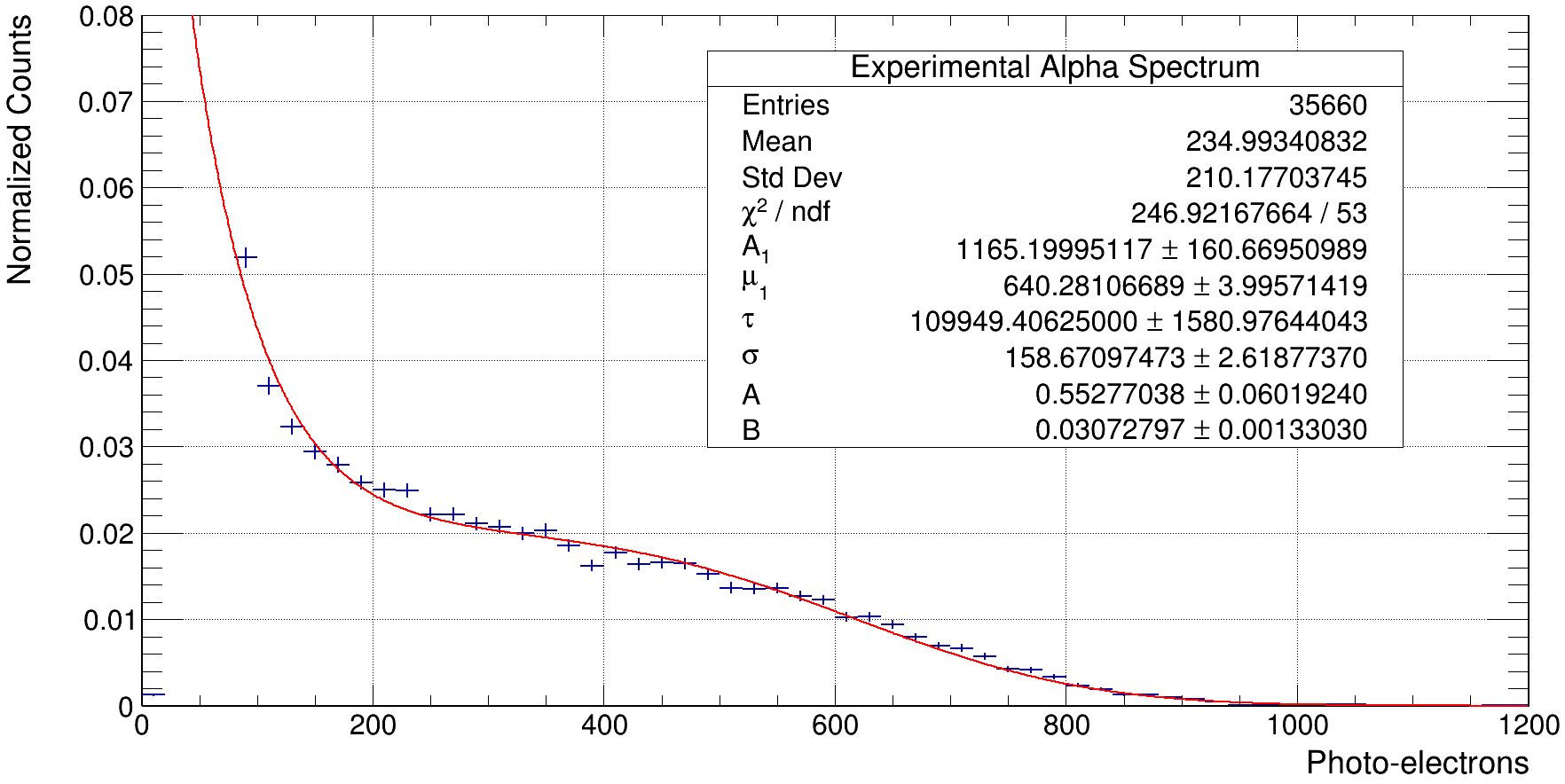}
	\caption{Fit of the $\alpha$-spectrum following Eq.~\ref{eq:alpha_spectrum_campinas} with an exponential noise added.}
	\label{fig:alphaspectrumfit}
\end{figure}

The simulated spectrum must go through a Gaussian smearing, which simulates the electronic response of the detector. 
The $\alpha$-source spectrum can be represented by the convolution of a Gaussian and three exponentials, one for each line of the spectrum~\cite{alpha_equation}:
\begin{equation}
\label{eq:alpha_spectrum_campinas}
F(E) = \sum_{i=1}^{3} \frac{A_i}{2\tau}\exp(\frac{E-\mu_i}{\tau} + \frac{\sigma^2}{2\tau^2})\;\erf{\left(\frac{1}{\sqrt{2}}\left(\frac{E-\mu_i}{\sigma} +\frac{\sigma}{\tau}\right)\right)},
\end{equation}
where $\mu_i$ is the peak position for the energy line $i$, $A_i$ is the relative abundance of each spectral line, $\sigma$ is the width of the alpha-peaks and $\tau$ is the slope of the tail on the peak low-energy side. The fit is performed by setting the two lower energy lines of Tab.~\ref{tab:alpha_spectrum} as function of the higher energy one. An example of the fit is shown in Figure~\ref{fig:alphaspectrumfit}, an exponential decaying term is added to MC spectra to take into account the very low energy background entering the stainless-steel vessel and which is not included in the MC simulation. 

The simulated spectrum of Fig.~\ref{fig:simulationalphaspectrumrafa} is smeared with the Gaussian's standard deviation $\sigma$ taken from the fit shown in Fig.~\ref{fig:alphaspectrumfit} and the exponential noise is inserted based on the fit parameters. The fitting procedure of the experimental spectrum on top of the MC ones foresees the use of four free parameters: two parameters for the low energy exponential noise, an overall normalization constant and a scale factor, which directly represents the total detection efficiency of the \xara.

\begin{figure}[tbph!]
	\centering
	\begin{subfigure}{0.9\textwidth}
		\includegraphics[width=0.99\linewidth]{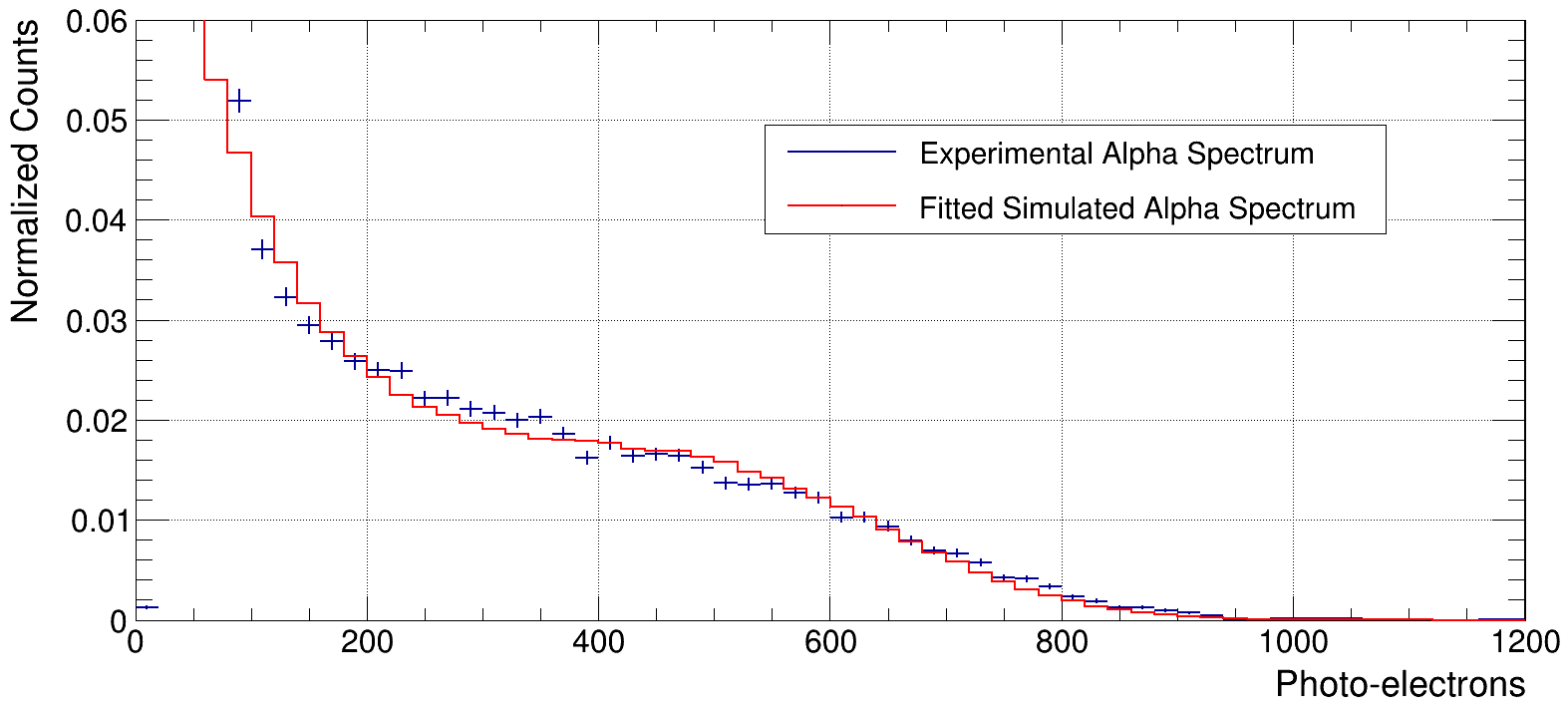}
		\caption{ }
		\label{fig:alpha3}
	\end{subfigure}

	\begin{subfigure}{0.9\textwidth}
		\includegraphics[width=0.99\linewidth]{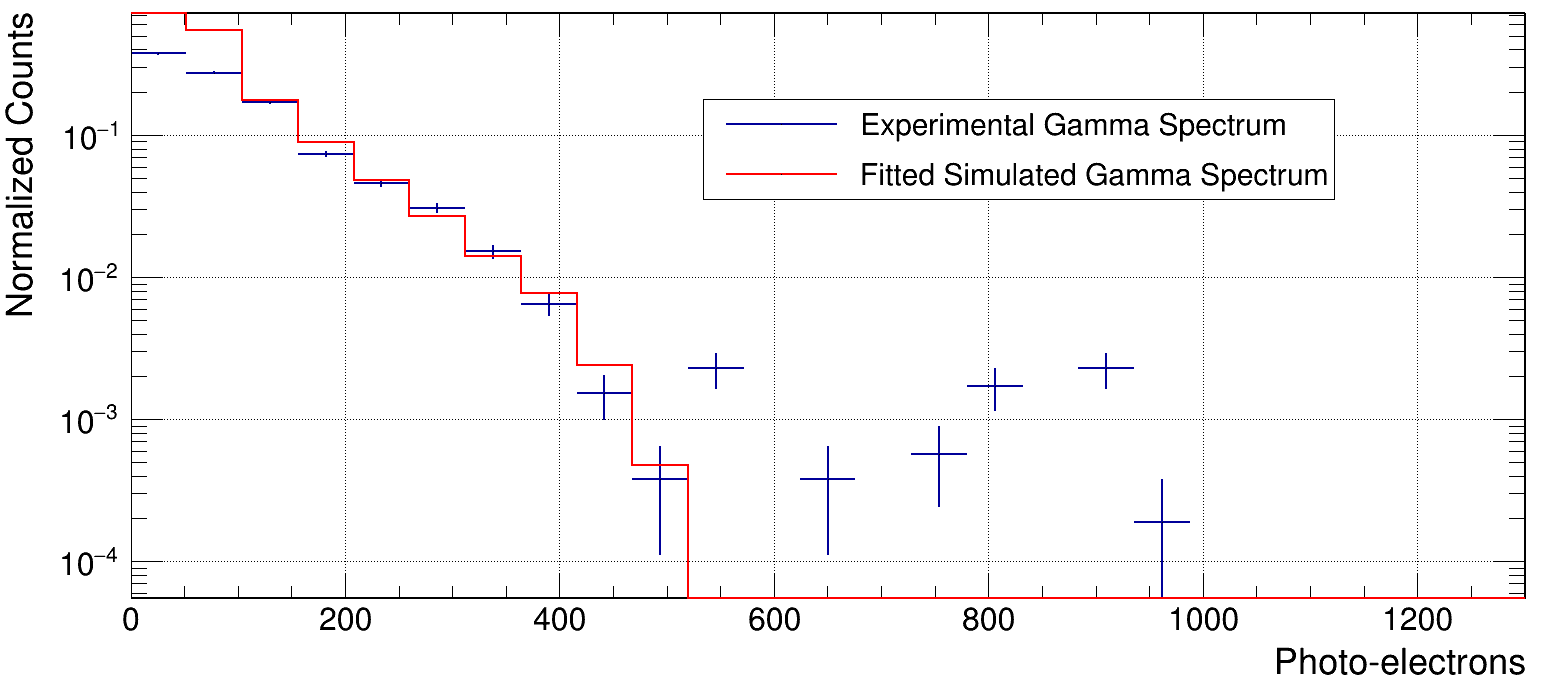}
		\caption{ }
		\label{fig:gammas3}
	\end{subfigure}

	\begin{subfigure}{0.9\textwidth}
		\includegraphics[width=0.99\linewidth]{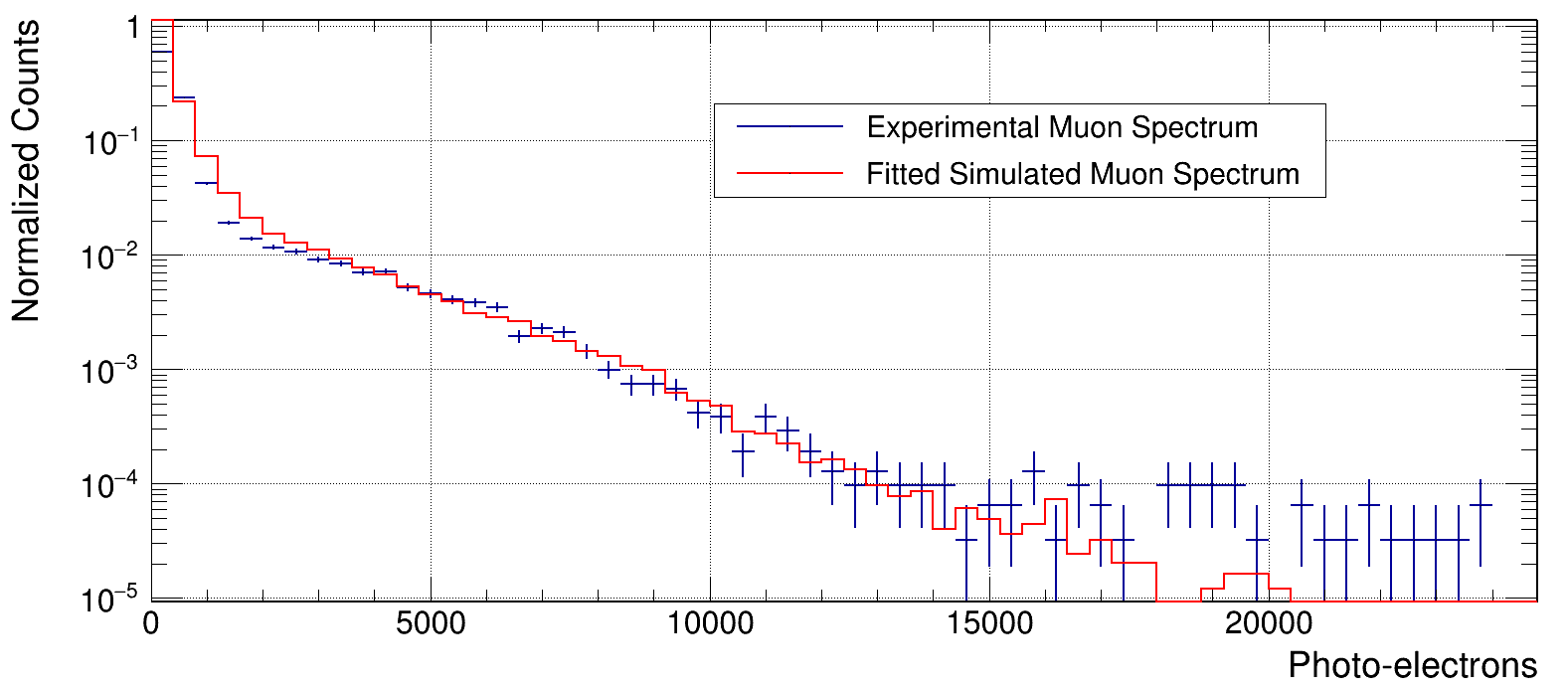}
		\caption{ }
		\label{fig:muons3}
	\end{subfigure}
	\caption{\label{fig:fitted_spectrums}Monte Carlo simulated spectrum (red) fitted to the experimental (blue) spectrum of \textbf{(a)}~$\alpha$-particles, \textbf{(b)}~$^{60}$Co $\gamma$-rays and \textbf{(c)} cosmic muons~\cite{x_arapuca_article}.}
\end{figure}

There was no need to apply a Gaussian smearing to $\gamma$-rays and cosmic muons MC outputs, as data were already well fitted without smearing. The experimental $\gamma$-source spectrum is taken by subtracting the histogram done without source from the histogram with the source. Figure~\ref{fig:fitted_spectrums} shows the $\chi^2$ minimization for all the ionizing radiations for a bias voltage of 48.0~V (+ 5.0~$\pm$~0.2 O.V.). The $\alpha$ spectrum in Fig.~\ref{fig:alpha3} was fitted starting with 75~\phe, the $\gamma$-ray spectrum in Fig.~\ref{fig:gammas3} was fitted between 200 and 500~\phe~and the $\mu$ spectrum in Fig.~\ref{fig:muons3} was fitted starting with 3400~\phe. 

\begin{figure}[tbph!]
	\centering
	\begin{subfigure}{0.49\textwidth}
		\includegraphics[width=0.99\linewidth]{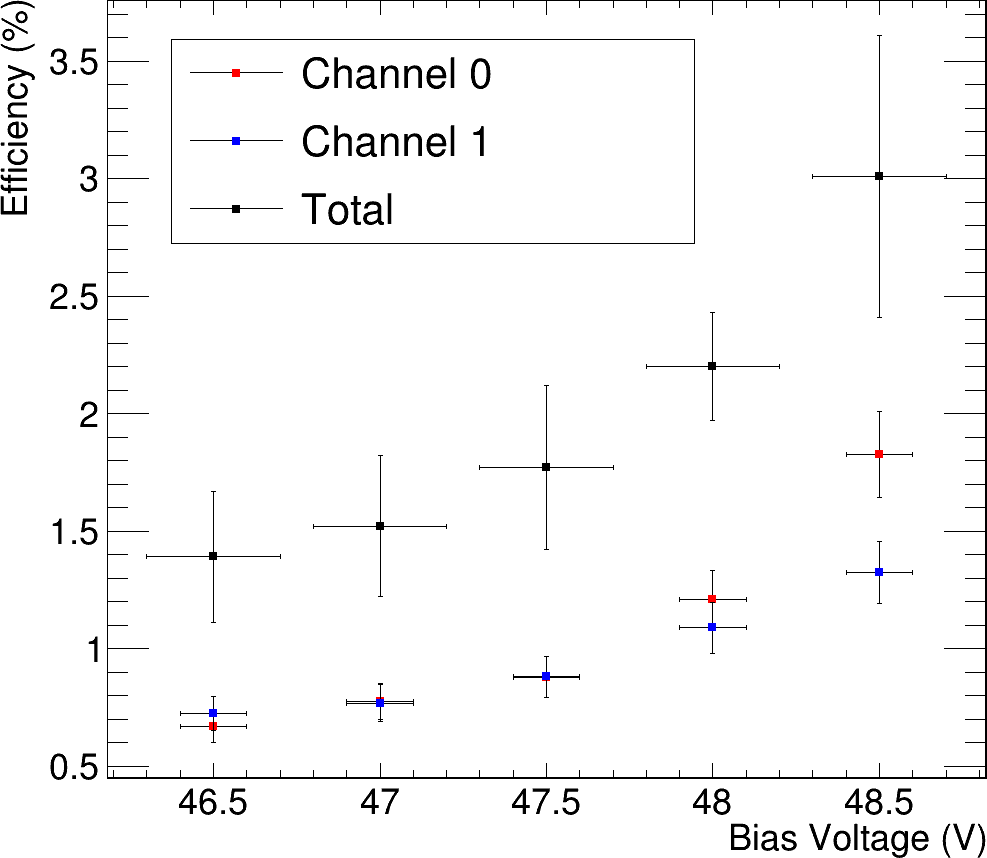}
		\caption{ }
		\label{fig:alphaefficienciesct}
	\end{subfigure}
	\begin{subfigure}{0.49\textwidth}
		\includegraphics[width=0.99\linewidth]{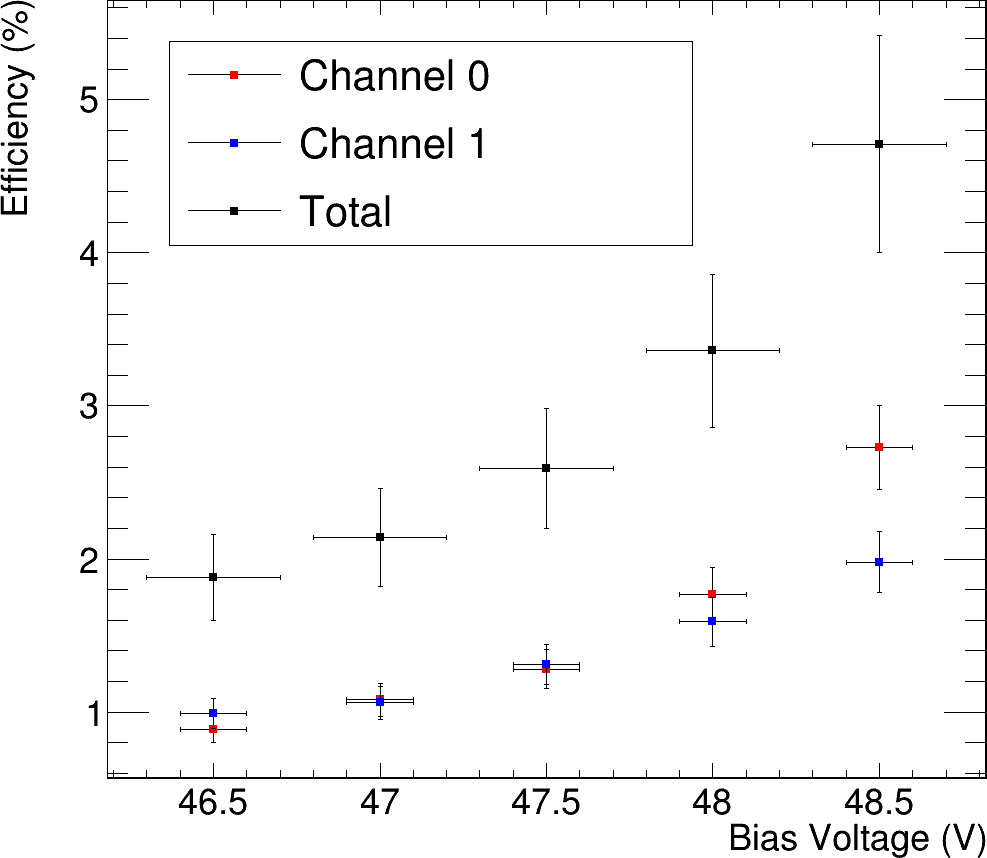}
		\caption{ }
		\label{fig:alphaefficiencies}
	\end{subfigure}
	\caption{Photon collection efficiency versus bian voltage without correcting \textbf{(a)}~and correcting \textbf{(b)}~for the cross-talk factor.}
\end{figure}

As previously mentioned, the calibration failed for an overvoltage below 5~V in the $\gamma$-ray and cosmic muons runs. Figure~\ref{fig:alphaefficiencies} and Fig.~\ref{fig:alphaefficienciesct} show the \xara\ photon collection efficiency, taken with the $\alpha$-source, versus the bias voltage without and with cross-talk correction, respectively. In order to compare these results with the ones achieved in Italy~\cite{enhancement_xara} (described in Sec.~\ref{sec:xara_double_cell}), the voltage of 48~V, which corresponds to 5.0~$\pm$~0.2 O.V., should be taken. This is due to the fact that the SiPM reaches 50\% quantum efficiency at this O.V.~\cite{hmmt_s13360}, the same quantum efficiency at which the SiPMs were operated in Italy. The uncertainties were taken as described in the calibration (Sec.~\ref{sec:calibration_single_cell}), i.e., 15\% without cross-talk correction and 20\% with correction. 

Finally, the results for the \xara\ single-cell efficiency are summarized in table~\ref{tab:results} comparing between the three ionizing radiations. The cases where the cross-talk is considered or not are both shown. One can notice that the results are compatible considering the uncertainties between $\alpha$'s, $\gamma$-rays and cosmic muons, which shows that the estimation of the \xara\ efficiency does not depend on the particle type nor on the topology of the energy deposition.
\begin{table}[htbp]
	\centering
	\caption{\label{tab:results}Efficiency obtained not considering and considering the cross-talk correction.}
	\smallskip
	\begin{tabular}{c|c|c|c|cc}
		\cline{2-4}
		& \begin{tabular}[c]{@{}c@{}}$\mu$ eff. (\%)\end{tabular} & \begin{tabular}[c]{@{}c@{}}$\gamma$ eff. (\%)\end{tabular} & \begin{tabular}[c]{@{}c@{}}$\alpha$ eff. (\%)\end{tabular} &  &  \\ \cline{2-4}
		& \multicolumn{3}{c|}{Without Cross-talk}                                                                                                                                           &  &  \\ \cline{1-4}
		\multicolumn{1}{|c|}{48.0 V} & 3.47~$\pm$~0.52                                          & 3.43~$\pm$~0.51                                           & 3.36~$\pm$~0.50                                       &  &  \\ \cline{1-4}
		\multicolumn{1}{|c|}{48.5 V} & 5.00~$\pm$~0.75                                          & 4.20~$\pm$~0.63                                          & 4.71~$\pm$~0.71                                          &  &  \\ \cline{1-4}
		\multicolumn{1}{|c|}{}       & \multicolumn{3}{c|}{With Cross-talk}                                                                                                                                              &  &  \\ \cline{1-4}
		\multicolumn{1}{|c|}{48.0 V} & 2.33~$\pm$~0.47                                           & 2.32~$\pm$~0.46                                           & 2.20~$\pm$~0.44                                                       &  &  \\ \cline{1-4}
		\multicolumn{1}{|c|}{48.5 V} & 3.13~$\pm$~0.63                                           & 2.65~$\pm$~0.53                                           & 3.01~$\pm$~0.60                                          &  &  \\ \cline{1-4}
	\end{tabular}
\end{table}

\begin{figure}[h!]
	\centering
	\includegraphics[width=0.6\linewidth]{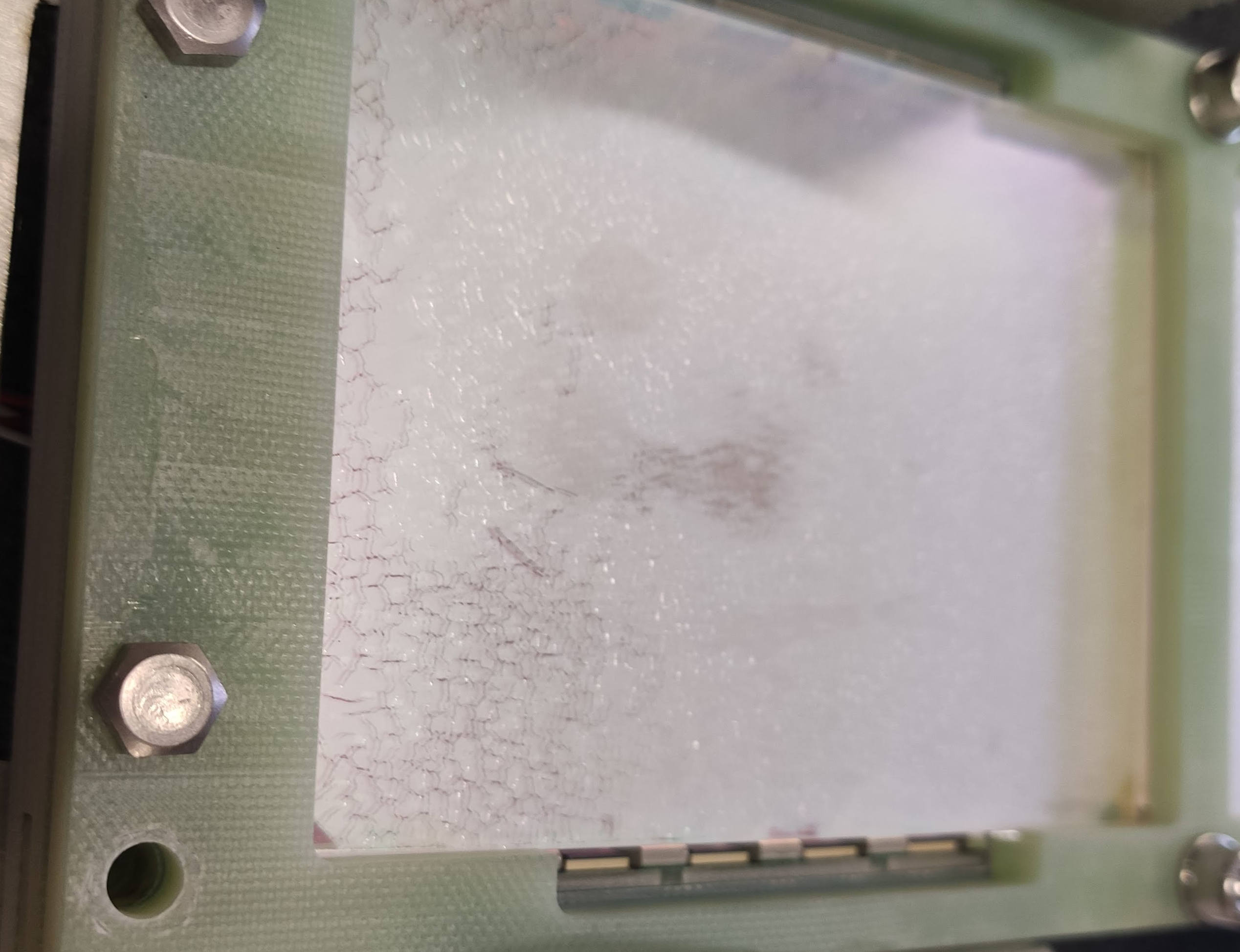}
	\caption{Micro-cracks in the \ptp\ deposition after several thermal cycles.}
	\label{fig:ptp_cracks}
\end{figure}

Before these three tests in LAr, other runs with $\alpha$-source to evaluate the DUT efficiency were performed with a reasonable agreement~\cite{first_lar_test}. However, a decrease in efficiency was noticed between the first and last tests, which gave efficiencies of (3.5~$\pm$~0.5)\% and (3.01~$\pm$~0.60)\% respectively. We believe this could be due to the several thermal stresses along these tests. In fact we also noticed micro-cracks formation in the internal WLS slab and in the \ptp\ coating, as shown in Figure~\ref{fig:ptp_cracks}. The cracks formation in \ptp\ can be reduced by avoiding exposure of the \xara\ to moister while performing the thermal cycle. 

\section{\xara\ double-cell tests}
\label{sec:xara_double_cell}

\begin{figure}[htbp]
	\center
	\includegraphics[width=0.28\textwidth]{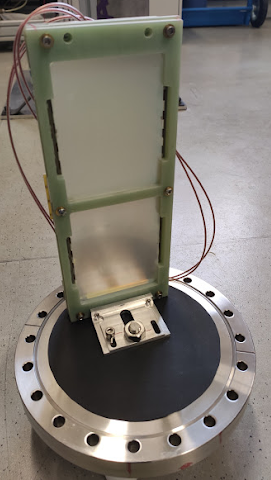}
	\includegraphics[width=0.30\textwidth]{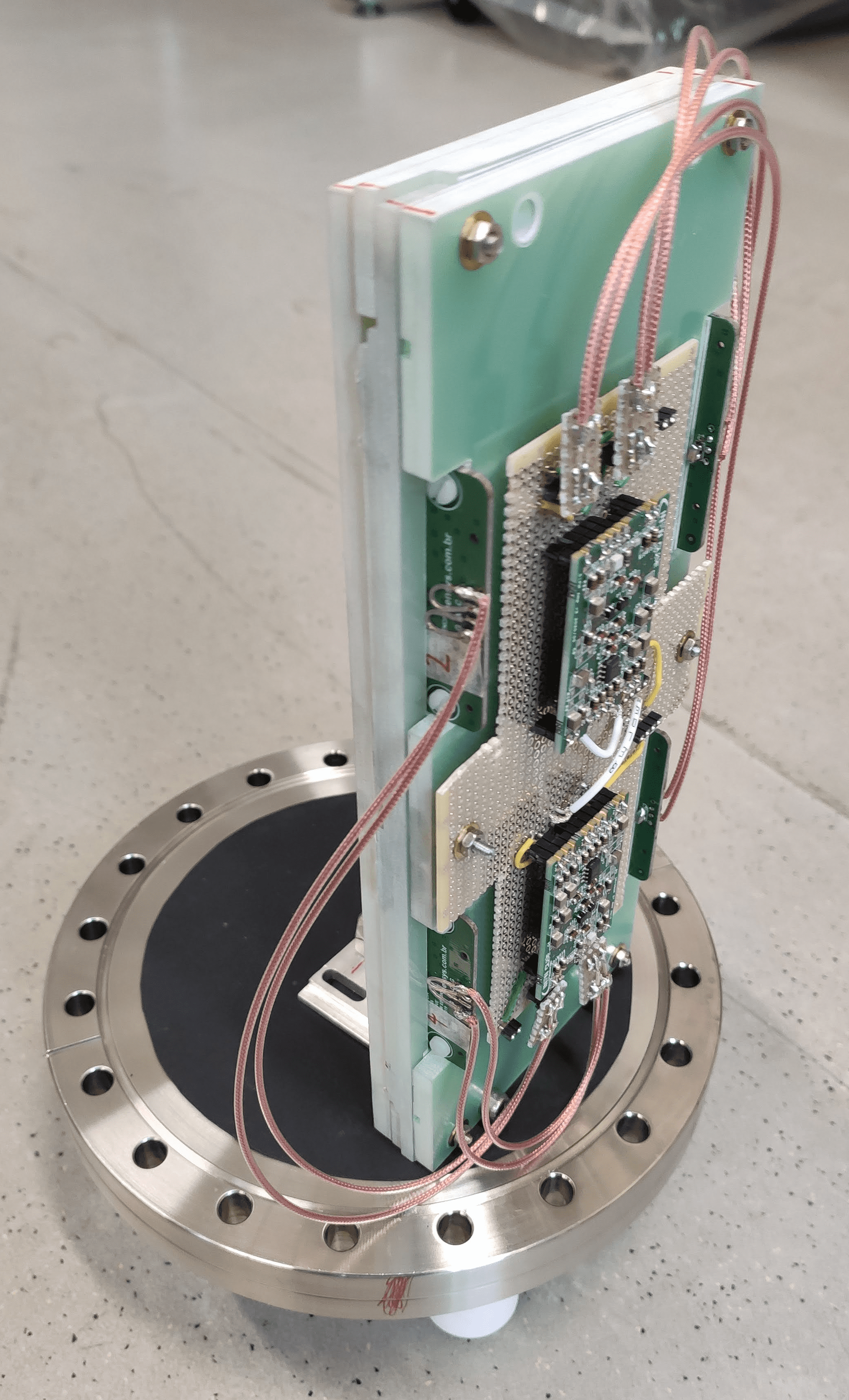}
	\caption{Front and back views of the \xara\ device mounted on the bottom flange of the cryogenic chamber. On the back two readout trans-impedance amplifier circuits are mounted on a service board providing the input/output electrical connections and contacts to the SiPMs. Eight SiPMs are ganged in input to each circuit~\cite{enhancement_xara}.}
	\label{fig:XAonFlange}
\end{figure}

In Italy, the DUT was the \xara\ double-cell sizing (200~$\times$~75)~mm$^2$ in Fig.~\ref{fig:XAonFlange}~\cite{enhancement_xara}, the same model that will be deployed in SBND~\cite{sbnd}. The two dichroic filters were from the same company (OPTO), but the \ptp\ coating was slightly higher at $\sim$500~$\mu$g/cm$^2$. The SiPMs were replaced by 4 boards with 4 S14160-6050HS with active area of (6$\times$6)~mm$^2$ produced by Hamamatsu~\cite{hmmt_s13360}, maintaining the same coverage area of the device. The goal of the experiment was not only to measure the photon detection efficiency of the \xara, but to verify the light collection enhancement due to the WLS developed at Milano-Bicocca (see Sec.~\ref{sec:warm_tests_bicocca}). Therefore, the standard WLS Eljen EJ-286~\cite{eljen_286} and the Glass to Power (G2P)~\cite{G2P}, denoted FB118, were tested. Besides, the enhancement of the light collection by inserting \viku\ on the edges of the WLS slab was also tested with the EJ-286 slab. The setup also featured two cold trans-impedance amplifiers, ganging two boards of SiPMs in parallel.

\subsection{Experimental setup and DAQ}
\label{sec:x_ara_double_cell_setup}

\begin{figure}[htbp]
	\center
	\includegraphics[width=0.2\textwidth]{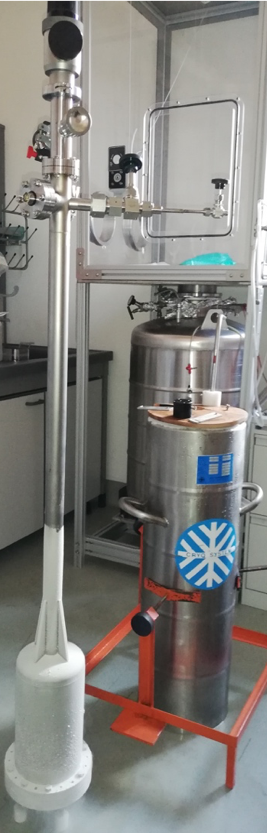}
	\includegraphics[width=0.318\textwidth]{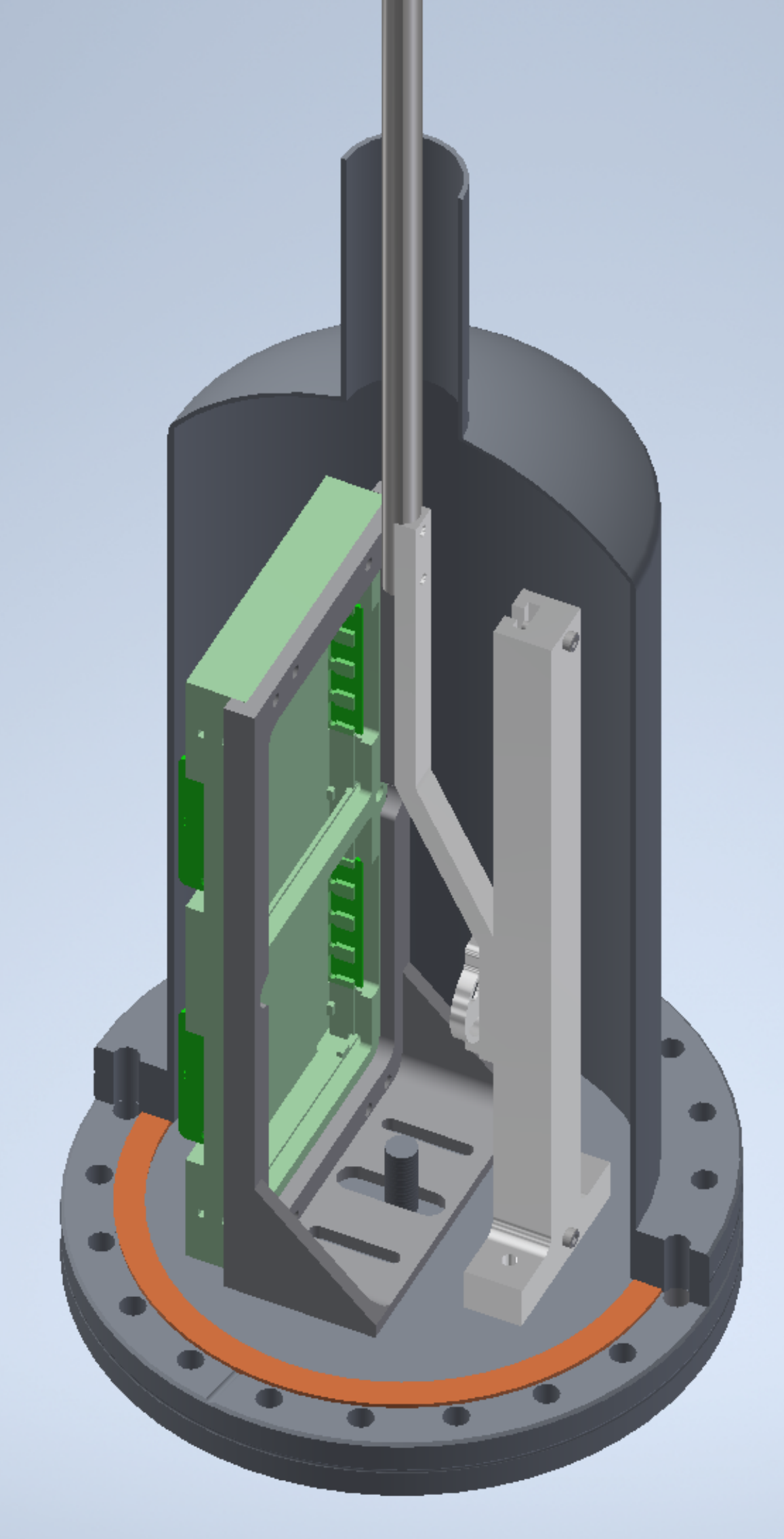}
	\includegraphics[width=0.34\textwidth]{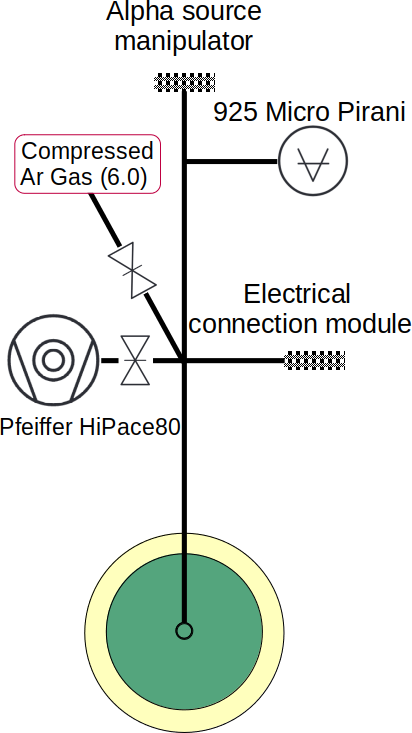}
	\caption{The test chamber after removal from the LAr bath (left), 3D drawing of the chamber with the X-Arapuca and the source sliding rail (center) and schematic diagram of the cryogenic setup (right).}
	\label{fig:cryosetup}
\end{figure}

The experimental setup, shown in Fig.~\ref{fig:cryosetup}, uses the same thermal bath of the \xara\ single-cell test at Campinas but with a few differences. An exposed 5.49~MeV $^{241}$Am source was used. An alpha source manipulator was installed, which allowed to change the height of the alpha source along the DUT longitudinal axis and eventually to completely remove it. A longer pipe to connect the external warm electronics to the testing chamber minimized the heat transfer, allowing measurements overnight and reduced LAr losses. There was no optical fiber for the \sphe\ calibration.

The method to liquefy the GAr was improved and a better LAr purity was reached. After pumping down the chamber to a pressure of $\sim$10$^{-3}$~mbar, gas argon 6.0 was carefully injected into the system, instead of making the cryogenic pump down as done in the single-cell test. Keeping the GAr at a pressure of $\sim$1.4~bar inside the chamber after making the vacuum prevent impurities to get into the stainless steel wall when cooling down.

\begin{figure}[tbph!]
	\centering
	\includegraphics[width=0.25\linewidth]{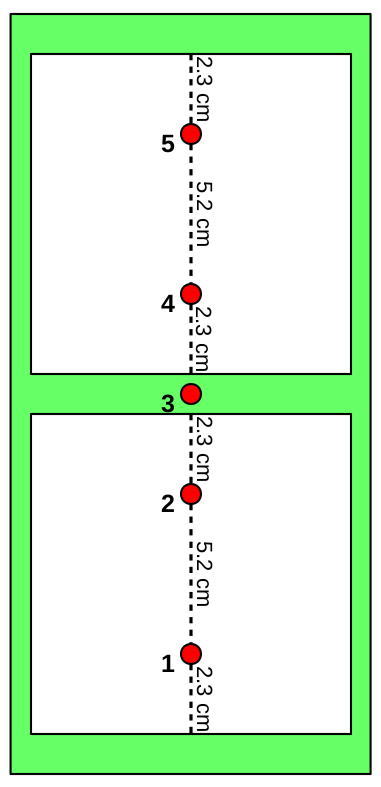}
	\caption{The five positions at fixed distance from the X-Arapuca used to characterize the WLS bars~\cite{enhancement_xara}.}
	\label{fig:points_scheme}
\end{figure}

The $\alpha$-source was fixed to the rail at a distance of 5.5\error0.2~cm from the dichroic filter and five different positions were tested, as shown in Fig.~\ref{fig:points_scheme}, one at the center of the \xara\ and the other four symmetrically at $\pm$2.3 and $\pm$7.5~cm away from the center. The source-to-DUT distance was chosen to not saturate the cold amplifiers as the dynamic range was about 500~\phe. The exposed source was covered with a Teflon mask to reduce the trigger rate.

The readout was performed by a CAEN Digitizer DT5725 with 250~MSamples/s and 14~bits resolution. As the digitizer was similar, the same DAQ presented in Sec.~\ref{sec:xara_single_cell_daq} was used.  The ADC readout was done whenever one of the two channels exceeded a fixed threshold of about 10 to 15 \phe. The threshold was tuned to reach a count rate around 300~Hz and 60~Hz for $\alpha$ and muon runs, respectively. Triggered events were recorded for 20~$\mu$s (5,000 points) for each channels with 5.6~$\mu$s of pre-trigger.  

For each position a total of 40,000 events were recorded. The same root files format adopted in the \xara\ single-cell test, with the denoise algorithm~\cite{denoising} and the baseline (calculated in the first 5~$\mu$s of the pretrigger) applied, was used.

\subsection{Calibration}
\label{sec:calibration_double_cell}

\subsubsection{Single photo-electron (\sphe) calibration}

\begin{figure}[!h]
	\centering
	\begin{subfigure}{0.49\textwidth}
		\includegraphics[width=0.99\textwidth]{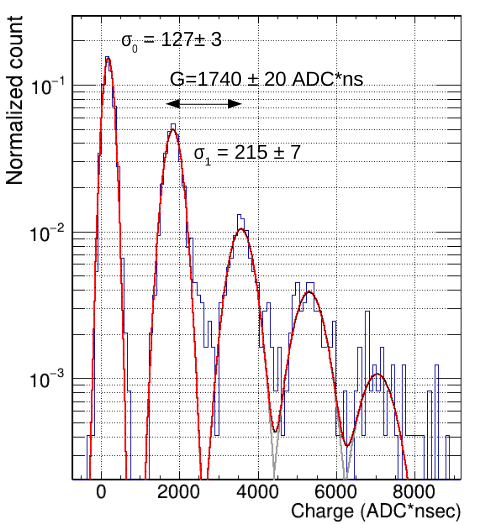}
		\caption{ }
		\label{fig:sphespectrum}
	\end{subfigure}
	\begin{subfigure}{0.49\textwidth}
		\includegraphics[width=0.99\textwidth]{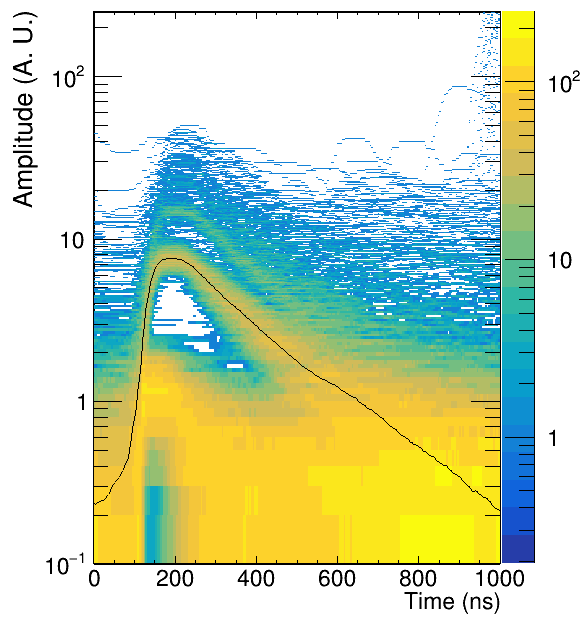}
		\caption{ }
		\label{fig:persistancelogsphe}
	\end{subfigure}
	\caption{\textbf{(a)}~Charge histogram of the selected peaks, the gain is defined by the distance between the first and second peak. The fit is described in the text. \textbf{(b)}~Persistence histogram of the selected waveforms together with the average waveform of one single photo-electron~\cite{enhancement_xara}.}
\end{figure}

The \sphe\ calibration had to be performed searching for dark counts in the pre-trigger, as the experimental setup did not featured an optical fiber. The \sphe\ candidates are searched in the first 5~$\mu$s of the pre-trigger. All pulses above 3.5~r.m.s of the baseline are integrated for 0.46~$\mu$s and and their trace stored for $\sim$1~$\mu$s from the onset. Figure~\ref{fig:sphespectrum} shows the obtained spectrum  with five well defined peaks, corresponding to the baseline and one to four \phe. By comparing with the \sphe\ spectra obtained at Unicamp (see Fig.~\ref{fig:sphe}), one can notice that the spectrum of Fig.~\ref{fig:sphespectrum} has a higher resolution. This is a benefit of the use of the cold amplification, that reduces the electrical noise, together with the SiPM model, that has a better shaped \sphe\ compared to the S13360 series. 

The fit performed to obtain the gain (G) of the SiPMs is the same as described in Sec.~\ref{sec:calibration_single_cell}. The recorded traces are displayed at Figure~\ref{fig:persistancelogsphe}, which contains also the averaged waveform (black line) of \sphe\ obtained by selecting waveforms inside plus or minus one sigma interval of the \sphe\ ($\sigma_1$) in Fig.~\ref{fig:sphespectrum}.

\subsubsection{Cross-talk evaluation}

\begin{figure}[tbph!]
	\centering
	\includegraphics[width=0.99\linewidth]{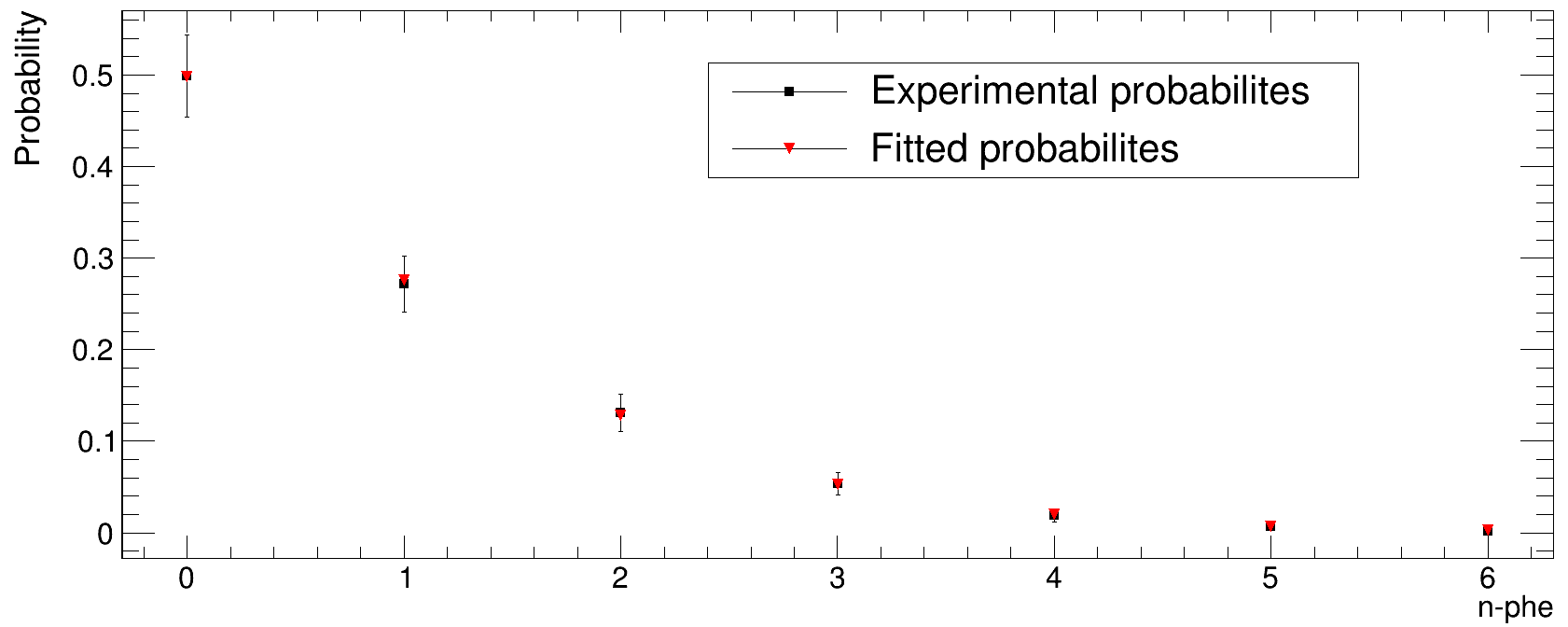}
	\caption{Probability of detecting $n$ photo-electrons (black squares) for the SiPM S14160-6050HS. A $\chi^2$ minimization of the experimental points with equation~\ref{eq:ct_vino}~\cite{Cross_talk_vino} was performed and the result is shown as red triangles.}
	\label{fig:crosstalkdoublecell}
\end{figure}

The cross-talk calibration was performed offline, by submerging the array of SiPMs in liquid nitrogen and flashing a blue LED through an optical fiber. The same method described in Sec.~\ref{sec:cross_talk_single_cell} was applied on the data and Figure~\ref{fig:crosstalkdoublecell} shows the result of the fit. The cross-talk probability is evaluated to be (18.7\error0.4)\%.

A more precise measurement, that was adopted by the community to characterize the SiPMs candidates for DUNE, was performed to calculate the cross-talk probability. In this measurement the SiPM was kept in dark and events triggered above half \sphe\ were recorded. The only assumption in this method is that the dark count should be (basically) composed of \sphe\ and that any signal above one photo-electron is caused by cross-talk. Therefore, the cross-talk probability is defined as the ratio between the number of events above 1 \phe\ and the total number of the event registered~\cite{SiPM_better}. By choosing this method, a cross-talk probability ranging from 20~to~24\% was measured,  corresponding to  1.22~$\pm$~0.02 avalanches/photon. Afterpulses, i.e. the number of secondary pulses within 10 $\mu$s, contribute for 11.7\%~\cite{enhancement_xara}. 

\subsubsection{Liquid Argon purity}
\label{sec:lar_purity_double_cell}

\begin{figure}[!h]
	\centering
	\begin{subfigure}{0.49\textwidth}
		\includegraphics[width=0.99\textwidth]{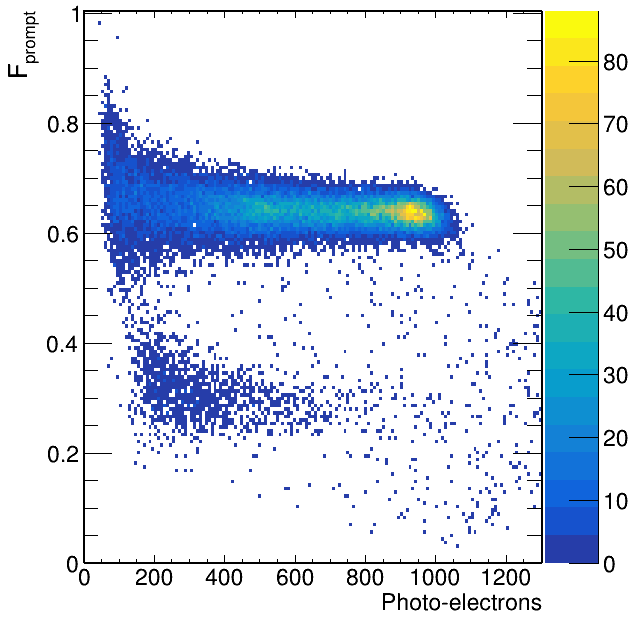}
		\caption{ }
		\label{fig:fprompt}
	\end{subfigure}
	\begin{subfigure}{0.49\textwidth}
		\includegraphics[width=0.99\textwidth]{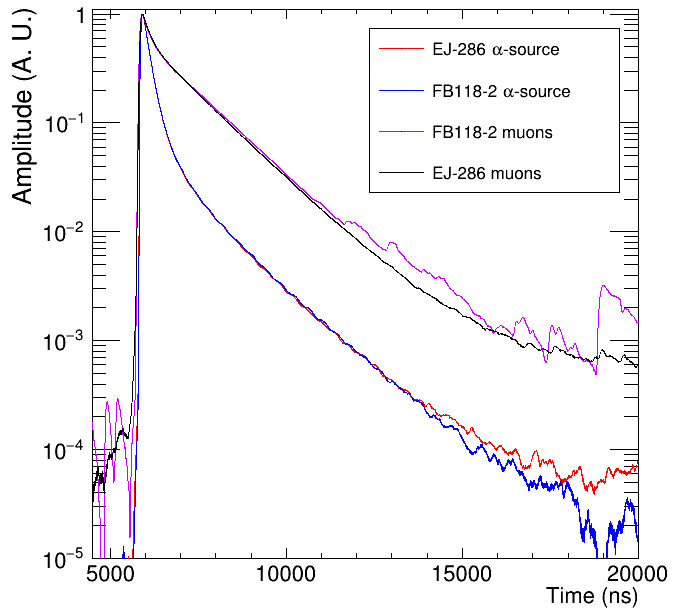}
		\caption{ }
		\label{fig:averaged_waveforms_double_cell}
	\end{subfigure}
	\caption{\textbf{(a)}~F$_{\text{prompt}}$ versus number of \phe: alphas and muons have F$_{\text{prompt}}$>~0.5  and <~0.5 respectively. \textbf{(b)}~The normalized average waveforms of the events, selected on the F$_{\text{prompt}}$ classifier~\cite{enhancement_xara}.}
\end{figure}

The purity of the LAr was monitored through the triplet time constant as in the \xara\ single-cell test (Sec.~\ref{sec:lar_purity_single_cell}). In the tests performed at Italy, there was the possibility to make muon runs during the $\alpha$-source measurements by rising the source on the top part of the testing chamber. This reduces the amount of light reaching the DUT originated from $\alpha$-particles. Rising the threshold will, therefore, increase the number of events from muons. It was not possible to surpass the amplitude of signals from $\alpha$-particles due to the saturation of the cold amplifier. The particle identification was performed via \fprompt\ classifier, defined as the ratio of the prompt ($<$600~ns) to the total charge~\cite{LAr_arapuca_test,pulse_shape_analysis,fprompt_psd} as:

\begin{equation}
\label{eq:fprompt}
F_\text{prompt} = \frac{\int_{t_0}^{t_0 + 0.6~\mu\text{s}}I(t)\diff t}{\int_{t_0}^{t_0 + 15.3~\mu\text{s}}I(t)\diff t},
\end{equation}

where $I(t)$ is the waveform, $t_0$ is the starting point of the integration chosen as 300~ns before the onset and the total charge is evaluated by integrating up to 15.3~$\mu$s. Figure~\ref{fig:fprompt} shows the \fprompt\ versus the number of \phe. As $\alpha$ particles have a higher contribution of the singlet component and muons will have a higher triplet component (Sec.~\ref{sec:lar_scintillation}), one can choose \fprompt$>0.5$ for $\alpha$ particles and \fprompt$< 0.5$ for muons. This selection allows to obtain the averaged waveforms for each type of particle in the two different tests, with EJ-286 and FB118. Figure~\ref{fig:averaged_waveforms_double_cell} presents the waveforms, where one can immediately notice that the LAr purity was the same for the two different tests, as the triplet time constant is basically the same.  

The high signal-to-noise resolution noticed in Fig.~\ref{fig:persistancelogsphe} due to the cold amplification allows to deconvolve the \sphe\ response in the averaged waveform. In this scenario, deconvolving was a more precise method than the convolution described in Sec.~\ref{sec:lar_purity_single_cell}. The ROOT Cern class TSpectrum~\cite{root_cern} was used, with the Gold deconvolution algorithm~\cite{deconvolve} already implemented. Details about the method are described in Appendix~\ref{chap:deconvolution}. The \sphe\ averaged waveform from Fig.~\ref{fig:persistancelogsphe} is deconvolved from the averaged waveforms of Fig.~\ref{fig:averaged_waveforms_double_cell}. 

\begin{figure}[tbph!]
	\centering
	\includegraphics[width=.9\textwidth]{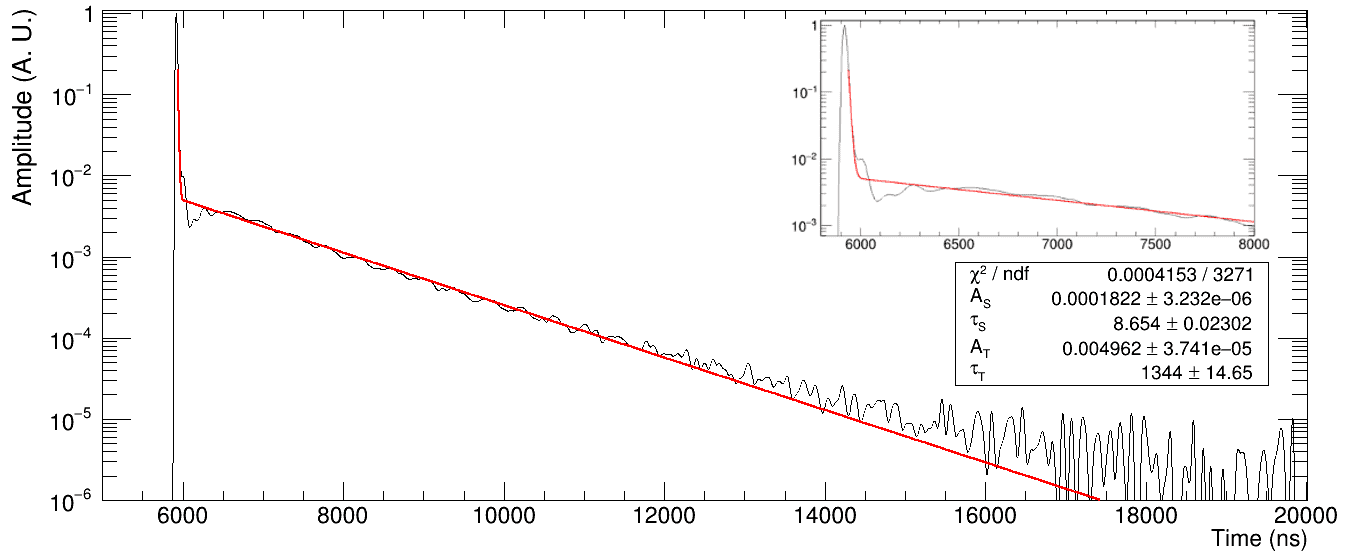}
	\includegraphics[width=.9\textwidth]{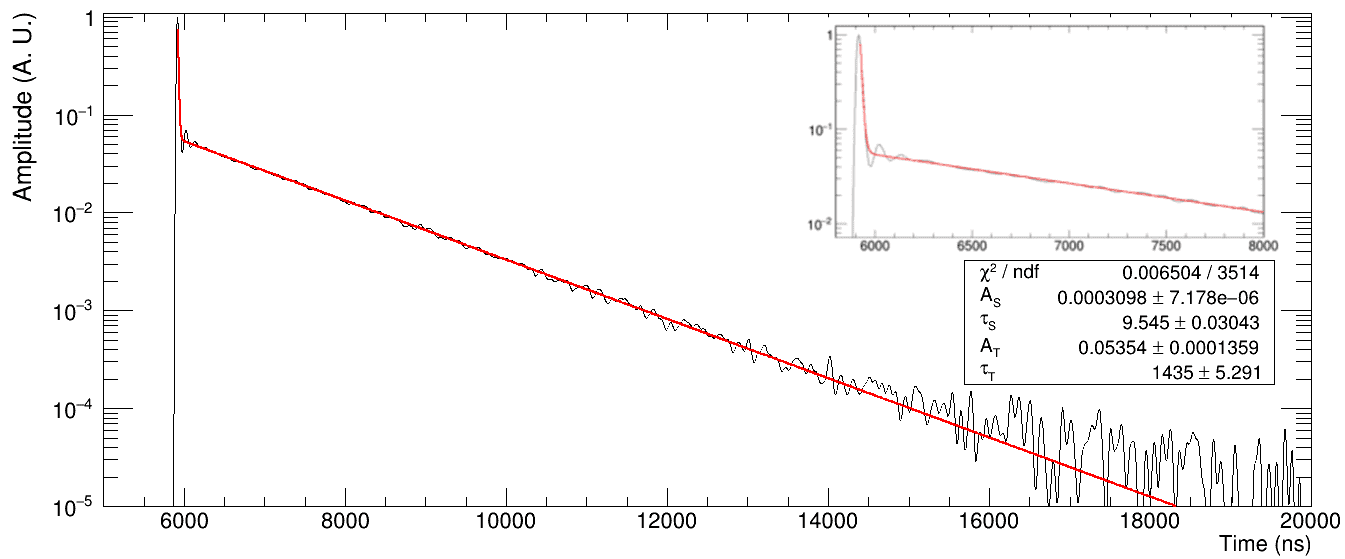}
	\caption{The deconvolved averaged waveforms of $\alpha$-particles (top) and muons (bottom). A fit of two exponential is made to retrieve the triplet time constant ($\tau_T$). The same fit is displayed with a zoom on the top right corner~\cite{enhancement_xara}.}
	\label{fig:deconvolved_waveforms}
\end{figure}

Figure~\ref{fig:deconvolved_waveforms} shows the deconvolved waveforms for $\alpha$ particles (top) and cosmic muons (bottom). A zoom is displayed on the top right corner, so the reader can notice the oscillation on the peak of the waveform which is, most probably, caused by the divergence between the fit and the true \sphe\ waveform (see Fig.~\ref{fig:sphe_fitted}). The waveform is fitted by the function $I(t)$ (from Sec.~\ref{sec:lar_scintillation}):
\begin{equation}
	\label{eq:fast_and_slow_fit}
	I(t) = A_S \exp(-\frac{t}{\tau_S}) + A_T \exp(-\frac{t}{\tau_T}),
\end{equation}
where $A_S$ and $A_T$ are the relative amplitudes and $\tau_S$ and $\tau_T$ are the time constants of the singlet and of the triplet dimer states respectively.

The $\tau_S$ and $\tau_T$ were computed by the fit of Eq.~\ref{eq:fast_and_slow_fit}. $A_S$ was taken by integrating the deconvolved waveform of Fig.~\ref{fig:deconvolved_waveforms} from the rising edge up to 4 times $\tau_S$ and the $A_T$ by integrating the triplet component fitted from the rising edge up to 20~$\mu$s. 
Three different measurements where taken: (1) DUT with EJ-286 slab without \viku\ on the edges, (2) DUT with EJ-286 slab with \viku\ on the edges and (3) DUT with FB118 slab with \viku. Table~\ref{tab:lar_parameters} reports, for each measurement (row),  the average values among all the available runs. The uncertainties are the combination of the individual standard deviations with the uncertainty of the average. The ratio $\langle A_S/A_T \rangle$ was computed from all the runs within each measurement, the ratio weighted average is then computed, properly propagating the uncertainties~\cite{enhancement_xara}.

Consistent values were found among the three measurements, both for the singlet/triplet decay times and for the relative amplitudes, allowing to evaluate the average. The uncertainties include both the statistical and the systematics which are dominated by the single \phe\ calibration. The average triplet time constant of $1414\pm21$~ns and $1294\pm35$~ns for muons and alphas respectively are in agreement with the literature~\cite{first_lar_test,nitrogen_contamination_roberto,abudance_dependence} as it is the singlet decay time constant. However, one should notice that the fast component is the most difficult to extract due to the poor fitting of the \sphe\ waveform shown in Fig.~\ref{fig:sphe_fitted}. The ratio $\langle A_S/A_T \rangle$ also agrees, even though there are some divergences such as the one found in Ref.~\cite{abudance_dependence}.

\begin{table}[tbph!]
	\centering
	\caption {The singlet and triplet decay time constants ($\tau_S$ and $\tau_T$) and their relative amplitudes ($A_S$ and $A_T$) from the waveforms fit for each liquefaction and their averaged values~\cite{enhancement_xara}.}
	\label{tab:lar_parameters}
	\begin{tabular}{cc|c|c|c|c|c|}
		\cline{3-7}
		& & $\tau_S$ (ns) & $\tau_T$ (ns) & $A_S$ & $A_T$ & $A_S/A_T$       \\ \hline
		\multicolumn{1}{|l|}{\multirow{3}{*}{$\alpha$}} & EJ-286 w.o.\ \viku\ & 9.3~$\pm$~0.9 & 1287 $\pm$ 35 & 0.76~$\pm$~0.06 & 0.24~$\pm$~0.03 & 3.10 $\pm$ 0.42 \\ 
		\multicolumn{1}{|l|}{}& EJ-286 w.\ \viku\ &  11.9~$\pm$~1.8 & 1286~$\pm$~37 & 0.77~$\pm$~0.06 & 0.23~$\pm$~0.03 & 3.41 $\pm$ 0.51 \\ 
		\multicolumn{1}{|l|}{}& FB118 w.\ \viku\ & 7.8~$\pm$~1.8 & 1331~$\pm$~22 & 0.79~$\pm$~0.07 & 0.21~$\pm$~0.03 & 3.67 $\pm$ 0.58 \\  \hline
		\multicolumn{1}{l|}{} & {\color[HTML]{FE0000} Average} & {\color[HTML]{FE0000} 9.7~$\pm$~3.0} & {\color[HTML]{FE0000} 1294 $\pm$ 35} & {\color[HTML]{FE0000} 0.77~$\pm$~0.04} & {\color[HTML]{FE0000} 0.23~$\pm$~0.02} & {\color[HTML]{FE0000} 3.35 $\pm$ 0.28} \\ \cline{2-7}\cline{1-7}
		\multicolumn{1}{|l|}{\multirow{3}{*}{$\mu$}} & EJ-286 w.o.\ \viku\ & 10.6~$\pm$~0.5 & 1371 $\pm$ 18 & 0.22~$\pm$~0.04 & 0.78~$\pm$~0.10 & 0.28 $\pm$ 0.07 \\
		\multicolumn{1}{|l|}{}& EJ-286 w.\ \viku\ & 9.6~$\pm$~0.1 & 1411 $\pm$ 33 & 0.23~$\pm$~0.05 & 0.77~$\pm$~0.11 & 0.29~$\pm$~0.08 \\
		\multicolumn{1}{|l|}{}& FB118 w.\ \viku\ & 10.6~$\pm$~4.4 & 1459 $\pm$ 35 & 0.23~$\pm$~0.05 & 0.77~$\pm$~0.11 & 0.30~$\pm$~0.08 \\ \hline
		\multicolumn{1}{l|}{} & {\color[HTML]{FE0000} Average} & {\color[HTML]{FE0000} 10.2~$\pm$~5.1}  & {\color[HTML]{FE0000} 1414 $\pm$ 21} & {\color[HTML]{FE0000} 0.18~$\pm$~0.03} & {\color[HTML]{FE0000} 0.82~$\pm$~0.04} & {\color[HTML]{FE0000} 0.29 $\pm$ 0.03} \\ \cline{2-7}
	\end{tabular}
\end{table}

The higher triplet time constant of $1294\pm35$~ns in the \xara\ double-cell test (compared with the 780\error20 in Sec.~\ref{sec:lar_purity_single_cell}) implied a correction of only +1.4 to 2.6\% more light. However, the precise evaluation of the waveforms showed that all the three measurements reached the same LAr purity level. Therefore, it is possible to evaluate the difference in light collection among the three measurements.

\subsection{Monte Carlo ``Toy-model''}
\label{sec:simulation_double_cell}

A Monte Carlo numerical evaluation was performed to estimate the amount of light that would reach the \xara\ acceptance window. The name ``toy-model'' comes from the fact that this is not a complete Geant4 simulation of the experimental setup, as the one described at Sec.~\ref{sec:simulation_single_cell}, but a geometrical estimation using Monte Carlo through the ROOT Cern libraries~\cite{root_cern}.

The $\alpha$ particles emission light was considered to be point-like from the center of the disk 5.5\error0.2~cm away from the center of the DUT. The point like approach is quite enough, because the open surface of the source was about 6~mm diameter and the maximum range for the $\alpha$ particles in LAr will be around 50~$\mu$m~\cite{alpha_range}. The light is emitted isotropically, assuming a light yield of~51,000 $\gamma$/MeV times the quenching factor of 0.71~\cite{LAr_fund_properties,abudance_dependence,model_nuclear_recoil_nl} (the same values used in Sec.~\ref{sec:simulation_single_cell}). To speed up the simulation, a factor of 0.01 was applied to the photons production. Reducing the amount of photons by this factor is equivalent to having a collection efficiency of 1\%. 

\begin{figure}[tbph!]
 	\centering
 	\includegraphics[width=0.9\linewidth]{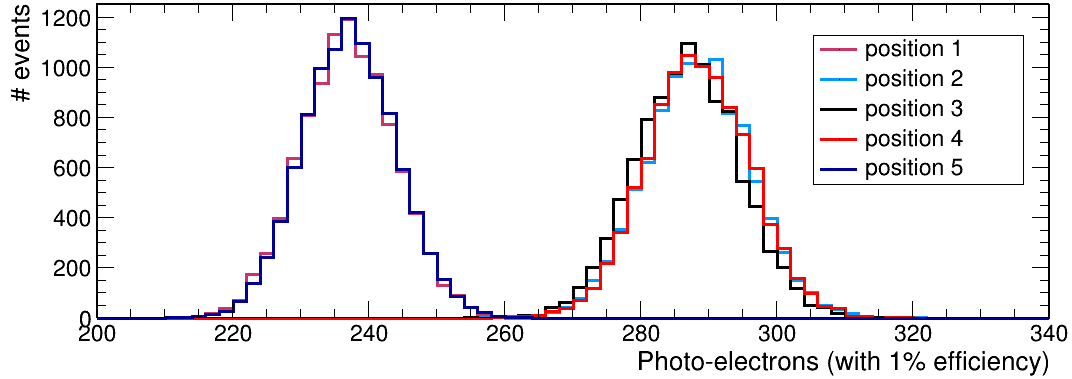}
 	\caption{Expected photo-electron for each position considering an efficiency of 1\%.}
 	\label{fig:toymodeloutput}
 \end{figure}

The photons are tracked and the amount of photons reaching the DUT dichroic filter is computed. Figure~\ref{fig:toymodeloutput} shows the expected number of photons for each position of the $\alpha$ source (as shown in Fig.~\ref{fig:points_scheme}). One can notice positions 1 and 5 are indeed symmetric and for the central positions 2, 3 and 4 the output is basically the same. This is quite expected as the solid angle for the central positions is basically the same. 

The expected number of photo-electrons is retrieved by fitting the spectra with a Gaussian distribution and the uncertainty set as 2\% by taking the deviation due the alpha-to-device distance uncertainty. The results are presented in Tab.~\ref{tab:expected_values_mc} for the analysis of the \xara\ efficiency.

\subsection{Efficiency and enhancement analysis}
\label{sec:efficiency_double_cell}

\begin{figure}[tbph!]
	\centering
	\includegraphics[width=0.456\textwidth]{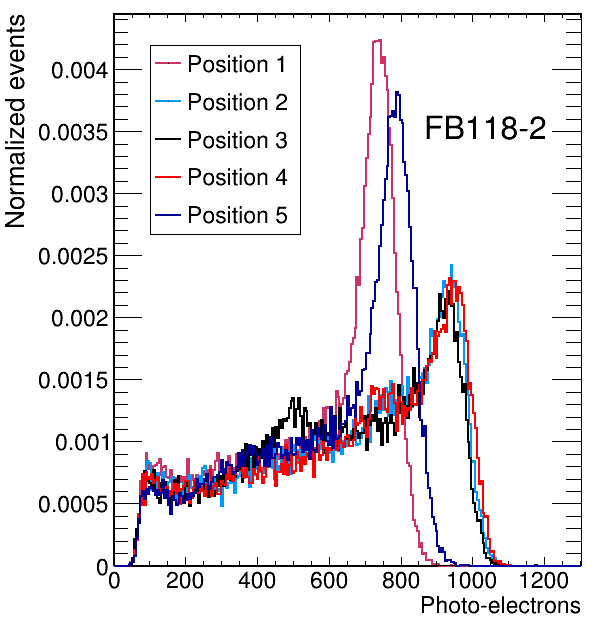}
	\includegraphics[width=0.45\textwidth]{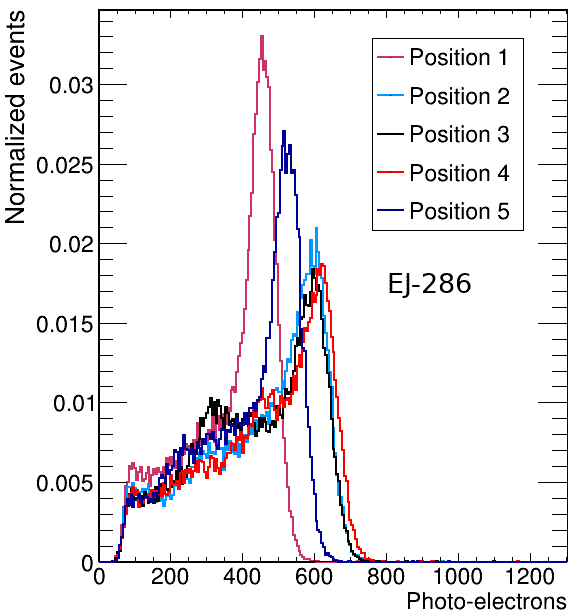}
	\caption{The $\alpha$ spectrum in number of detected photo-electrons  for each of the five source positions: the X-Arapuca is equipped with FB118 (Left)~and EJ-286 (Right)~respectively~\cite{enhancement_xara}.}
	\label{fig:alpha_spectrums_lar}
\end{figure}

The waveforms were integrated from 5.6 to 11~$\mu$s and the \sphe\ calibration applied. Figure~\ref{fig:alpha_spectrums_lar} shows the calibrated $\alpha$ spectrum, for both the FB118 (left) and EJ-286 (right). As expected from the geometry and the ``toy-model'', the spectra for position 2, 3 and 4 overlap. However, a significant difference can be noticed between position 1 and 5, which are expected to be identical. This is attributed to the shadowing of the $\alpha$-source holder of Fig.~\ref{fig:cryosetup} (central figure), which blocks light on the top part of the DUT when the source is at the bottom. Because of that, the position 1 was not used for the evaluation of the \xara\ light collection efficiency, but was still used to estimate the enhancement in the light collection

The spectra of Fig.~\ref{fig:alpha_spectrums_lar} are presented with the same range, so it is straightforward to notice the sizable increase of the collected photons when the optical cell is equipped with the FB118 instead of the EJ-286.

\begin{figure}[htbp]
	\centering
	\includegraphics[width=0.9\textwidth]{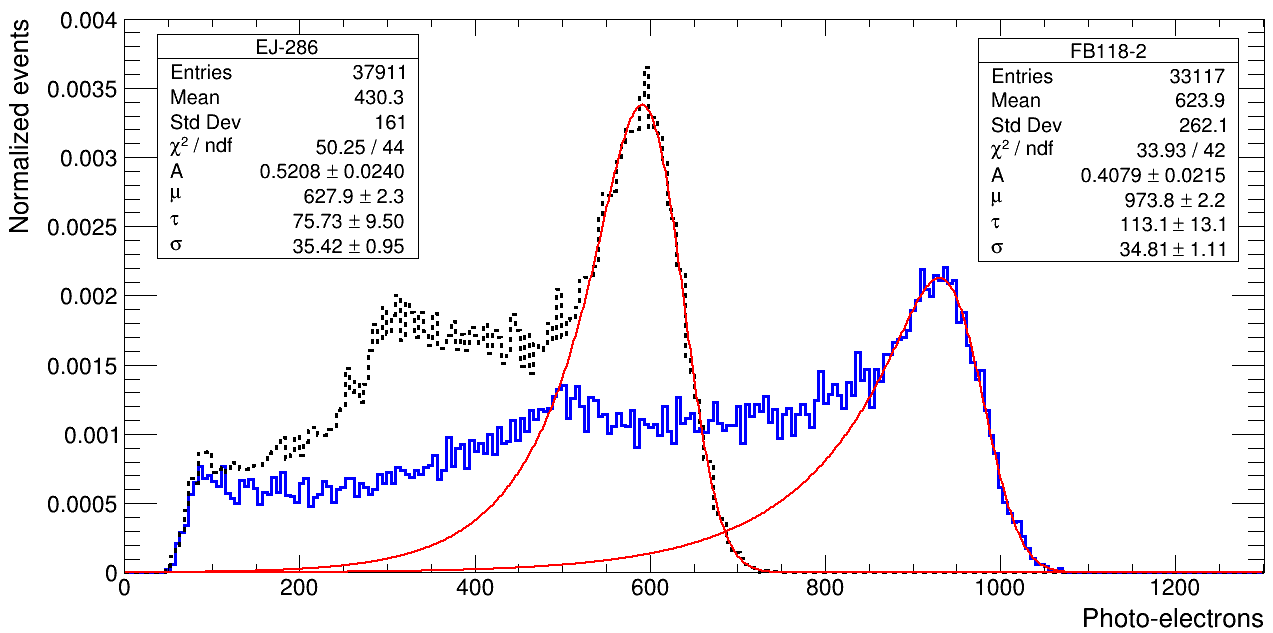}
	\caption{The $\alpha$ spectrum fit with Eq.~\ref{eq:alpha_spectrum} for the source in the central position (\#3)  for EJ-286 (black, dashed line) and FB118 (blue, solid line)~\cite{enhancement_xara}.}
	\label{fig:sample_fits_spectrum_lar}
\end{figure}

To determine the alpha peak position, the same function as in Eq.~\ref{eq:alpha_spectrum_campinas} was used, however with only one emission line, that is~\cite{alpha_equation}:
\begin{equation}
\label{eq:alpha_spectrum}
F(E) = \frac{A}{2\tau}\exp(\frac{E-\mu}{\tau} + \frac{\sigma^2}{2\tau^2})\;\erf{\left(\frac{1}{\sqrt{2}}\left(\frac{E-\mu}{\sigma} +\frac{\sigma}{\tau}\right)\right)},
\end{equation}
where $\mu$ is the peak position, A is the area, $\sigma$ is the spread of the alpha-peak and $\tau$ is the slope of the tail on the peak low-energy side. Figure~\ref{fig:sample_fits_spectrum_lar} shows the spectrum  when the source is in the central position (\#3), and superimposed the fits for both the EJ-286 and FB118.

Table~\ref{tab:expected_values_mc} shows the expected number of photo-electrons for a 1\% efficiency \xara, and the detected ones for the three measurements performed. The uncertainty of the peak was assumed as 2.5\% due to the \sphe\ calibration and the average of position 2, 3 and 4 was taken to make a better comparison. These values can be directly compared in order to retrieve the light collection enhancement of the DUT with the FB118 slab or by the application of the \viku. For the efficiency of the device, however, the cross-talk and liquid argon purity must be applied. 

\begin{table}[tbph!]
	\centering
	\caption{Expected value of photons reaching the X-Arapuca acceptance window given by numerical evaluation described in Sec.~\ref{sec:simulation_double_cell} and number of photo-electrons detected in the three different configurations. An average of position 2, 3 and 4 was taken~\cite{enhancement_xara}.}
	\label{tab:expected_values_mc}
	\begin{tabular}{|c|c|c|c|c|}
		\hline
		Positions & Expected (MC)               & EJ-286 w/o \viku\ & EJ-286 w/ \viku\ & FB118      \\ 
		& \# p.e.\ & \# p.e.\ & \# p.e.\ & \# p.e.\ \\ \hline
		2,3,4     & 288 $\pm$ 6  & 607 $\pm$ 9           & 637 $\pm$ 9          & 986 $\pm$ 14 \\ \hline
		5         & 236 $\pm$ 5  & 515 $\pm$ 13          & 548 $\pm$ 14         & 825 $\pm$ 20 \\ \hline
		1         & 236 $\pm$ 5 & 443 $\pm$ 11          & 473 $\pm$ 12         & 770 $\pm$ 18 \\ \hline
	\end{tabular}
\end{table}

From the ratio of the alpha-peak means, for each source position, the photo detection efficiency variation $G_\epsilon$ was computed when the \xara\ had the FB118 as secondary wavelength shifter instead of the EJ-286, the result is presented at Table~\ref{tab:lar_results_relative}, defining $G_\epsilon$ as:
\begin{equation}
G_\epsilon = \left(\frac{\mu_{\text{FB118}}}{\mu_{\text{EJ-286}}}-1\right)\times 100 \; [\%],
\end{equation}
where $\mu$ is the alpha-peak mean from Eq.~\ref{eq:alpha_spectrum} already presented in Table \ref{tab:expected_values_mc}.
\begin{table}[htbp]
	\centering
	\caption{\label{tab:lar_results_relative} The PDE increase ($G_\epsilon$) when replacing the EJ-286 in the X-Arapuca with the FB118, for the five source positions as in Table~\ref{tab:expected_values_mc}~\cite{enhancement_xara}.}
	\smallskip
	\begin{tabular}{|c|c|}
		\hline
		Positions & $G_\epsilon$  \\ \hline
		2,3,4     & 55 $\pm$ 5 \%  \\ \hline
		5         & 50 $\pm$ 5 \%  \\ \hline
		1         & 63 $\pm$ 6 \%   \\ \hline
	\end{tabular}
\end{table}

Finally, the absolute efficiencies of the X-Arapuca with EJ-286 (with and without \viku) and FB118 with \viku\ are displayed in Table~\ref{tab:lar_results_efficiency}. The efficiency uncertainty includes those from LAr purity correction, the single photo-electron calibration and the solid angle~\cite{enhancement_xara}. The signal-to-noise ratio is reported and defined as:
\begin{equation}
\label{eq:signal_to_noise}
\text{S/N} = \frac{\mu_1}{\sqrt{\sigma_0^2 + \sigma_1^2}},
\end{equation}
where $\mu_1$ is the mean charge of a single \phe, $\sigma_0$ and $\sigma_1$ are the standard deviation of the noise and \sphe, respectively, in Fig.~\ref{fig:sphespectrum}. This is an important feature for the trans-impedance amplifier, as one of the DUNE requirements is a signal-to-noise $>4$~\cite{DUNE_vol4}. It is important to notice that here, however, only eight SiPMs were ganged together and for the supercell module there will be 48 SiPMs ganged. The resolution ($\sigma/\mu$) of the alpha peak ($\sigma \text{ and } \mu$ from the fit function~\ref{eq:alpha_spectrum}) is also reported and, as expected, it scales with the number of detected photons. 

\begin{table}[htbp]
	\centering
	\caption{Summary of the relevant facts for the X-Arapuca efficiency measurements by $^{241}$Am $\alpha$ source radiation  and the three considered configurations: Energy Resolution of the $\alpha$ peak, S/N: signal-to-noise, $\epsilon_{\text{raw}}$, $\epsilon$; Efficiency prior (raw) and post corrections  respectively~\cite{enhancement_xara}.}
	\label{tab:lar_results_efficiency}
	\smallskip
	\begin{tabular}{|c|c|c|c|}
		\hline
		& EJ-286 w/o \viku\ & EJ-286 w/ \viku\ & FB118           \\ \hline
		En. res. ($\sigma/\mu$)     & 6.3 $\pm$ 0.2 \%     & 6.0 $\pm$ 0.2 \%
		& 3.6 $\pm$ 0.1 \%\\ \hline
		S/N                 & 6.8~$\pm$~0.3                   & 7.3~$\pm$~0.3                  & 7.3~$\pm$~0.3               \\ \hline
		$\epsilon_\text{raw}$         & 2.1 $\pm$ 0.1 \%      & 2.3 $\pm$ 0.1 \%     & 3.5 $\pm$ 0.1 \%  \\ \hline
		$\tau_T$           & \multicolumn{3}{c|}{1294~$\pm$~35~ns}                                   \\ \hline
		LAr purity correction          & \multicolumn{3}{c|}{+ (1.4 to 2.6)~\%}                                   \\ \hline
		Cross-talk correction           & \multicolumn{3}{c|}{- (18~$\pm$~1)~\%}                                   \\ \hline
		$\epsilon$         &  1.8 $\pm$ 0.1\%      &  1.9 $\pm$ 0.1\%     &  2.9 $\pm$ 0.1\%  \\ \hline
	\end{tabular}
\end{table}

The application of \viku\ on the EJ-286 slab edges resulted in an increase from 3 to 11\% where positions (1 and 5) show the largest increase as measured at room temperature (see Sec.~\ref{sec:warm_tests_bicocca}). However, the increase in the PDE was lower in LAr than in air/vacuum. We believe this is caused by the critical angle in LAr of about 55.4$^\circ$ and 51.4$^\circ$ for FB118 and EJ-286, respectively, compared with 41.8$^\circ$ and 39.2$^\circ$ in air, therefore less photons are trapped in LAr reducing the effectiveness of applying \viku\ on the edges.

The enhancement of about 50\% with the Glass-to-Power WLS with respect to the Eljen EJ-286 WLS was a very important result: increasing the efficiency of the \xara\ helps to approach the DUNE specifications of average light yield of \phe/MeV (see Sec.~\ref{sec:dune_spec}). The research developed in Milano-Bicocca has set a more direct contact between the DUNE collaboration and the vendor of the WLS. New batches of both WLS were sent for different groups of the collaboration for the tests of the \xara\ supercell, Eljen Co. has provided EJ-286 slabs with two times and four times more concentration than the commercial batches. In ProtoDUNE run II, both WLS will be deployed which will help the selection for the DUNE first far detector.

The efficiency of (1.9\error0.1)\% for the \xara\ double-cell with the EJ-286 slab agrees with the efficiency of (2.20\error0.44)\% found at Sec.~\ref{sec:xara_single_cell} for the single-cell, as both SiPMs operated at 50\% PDE~\cite{hmmt_s13360}. This efficiency satisfies the minimum requirement of the DUNE FD for an efficiency $>1.3$\%~\cite{DUNE_vol4}. The \xara\ equipped with the FB118 slab, however, presented an efficiency of (2.9\error0.1)\% which satisfies the average efficiency requirement.

%% file: protodune.tex
\chapter{The ProtoDUNE-SP and Xenon Doping}
\label{chap:protoDUNE}
\thispagestyle{myheadings}

\section{The ProtoDUNE-SP experiment}

Even though the LArTPC technology has been quite explored and proved to meet the requirements for excellent neutrino detection and physics, the DUNE experiment gather around many new technologies and techniques with major challenges. The biggest LArTPC built up to now is the ICARUS T600~\cite{ICARUS}, with a LAr fiducial volume of about 500~t divided into two drift regions, the detector is basically 20 times smaller than one DUNE FD module. Besides, the ICARUS T600 maximum drift distance of 1.5~m requires a high voltage of $-75$~kV, while the DUNE FD will have a drift distance of 3.5~m and high voltage of $-180$~kV. 

\begin{figure}[h!]
	\centering
	\includegraphics[width=0.55\linewidth]{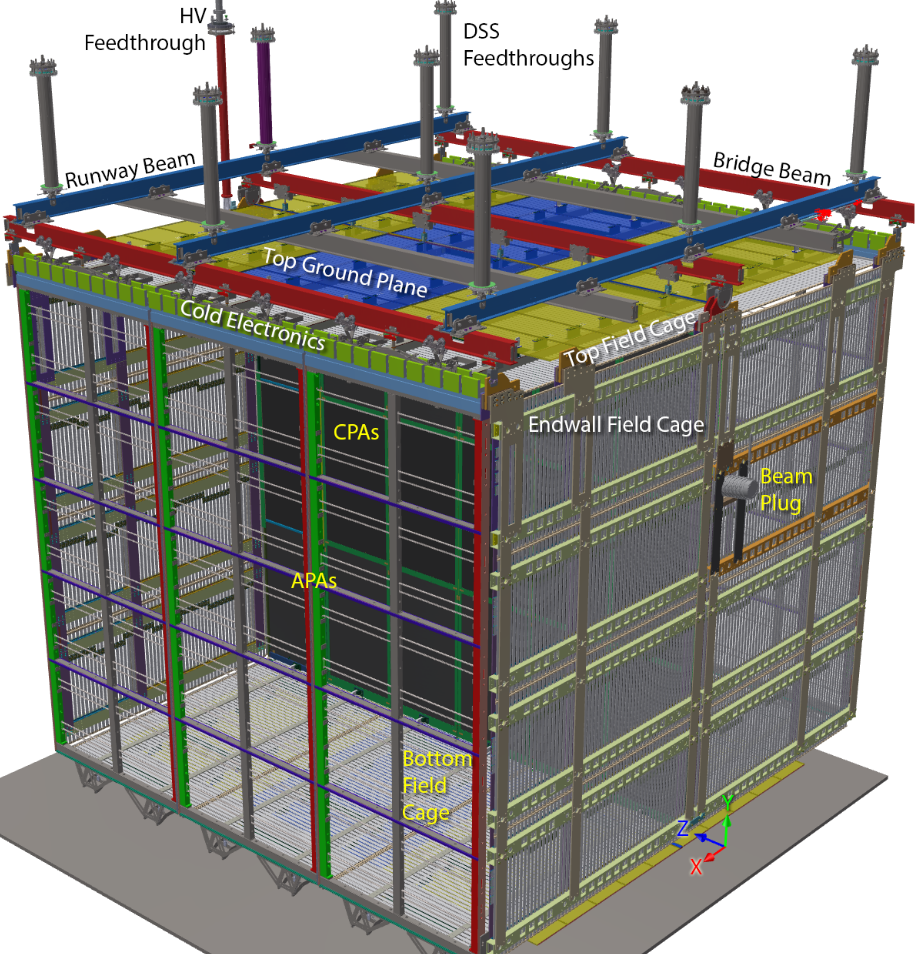}
	\caption{View of the ProtoDUNE TPC, the APAs area placed in the side of the detector and the CPA in the middle. The beam arrives in one right-side TPC~\cite{protoDUNE_first_results}.}
	\label{fig:protoDUNE_schematic}
\end{figure}

The ProtoDUNE-SP experiment~\cite{protoDUNE_TDR} is a prototype built at CERN with 419~t of instrumented LAr mass. The two TPC volumes with  6.086~(h)~×~3.597~(w)~×~7.045~(l)~m$^3$  are shown in Figure~\ref{fig:protoDUNE_schematic} with total active volume of 2$\times$154~m$^3$, each with drift distance of 3.572~m and $-180$~kV high voltage bias in the APA. 
One of the instrumented TPCs receives a beam that can deliver pions, kaons, protons, muons and electrons with momenta ranging from 0.3~GeV/c to 7~GeV/c. The wires pitch, distance and voltage in the APAs have the same construction proposed for DUNE (described at Sec.~\ref{sec:far_detector_dune}).

The prototype operated in the first run from September 2018 to January 2020 and had as primary goals to: (1)~stress-test the production and quality assurance of the detector components, (2)~validate the installation procedures, (3) validate the detector performance with cosmic rays and (4) measure the physics response of the detector with beam data~\cite{DUNE_Vol1_TDR}.

All the details about the performance of the TPC and the light detectors are very well described in the performance paper~\cite{protoDUNE_first_results}. It is worth mentioning the light collection efficiency found for the \ara\ devices installed in the prototype. 

A total of 60 light collector modules were installed: two of them are based on the S-\ara\ concept (see Sec.~\ref{sec:s_arapuca}) and the other 58 were equally divided into dip-coated and double-shift light guides~\cite{protoDUNE_first_results}. The dip-coated light guide consist of 207.4~$\times$~8.2~cm$^2$ wavelength shifter bars with TPB dip-coated to downshift the light to 430~nm. The double-shift light guides have TPB evaporated in a wavelength shifter that downshifts the 430~nm light to 490~nm. Both devices detect the total internal reflected light in the shorter edge of the bar with 12~SiPMs. 

One S-\ara\ module consisted of 12 cells, eight with optical area of 9.8~$\times$~7.9~cm$^2$ and 4 with 19.6~$\times$~7.9~cm$^2$. Each cell with its own readout of the 12 Hamamatsu S13360-6050CQ SiPMs. The ratio of photon sensors area to the light collectors surface area was 0.25\% for the light-guides and 5.6\% and 2.8\% for the two S-\ara s modules with ``normal'' and double area, respectively. 

It was found an efficiency of (2.00\error0.25)\% and (1.06\error0.09)\% for the S-\ara\ cells with normal and double area, respectively. An efficiency of (0.21\error0.03)\% and (0.08\error0.02)\% was found for the double-shift and dip-coated light guides~\cite{protoDUNE_first_results}. It is worth mentioning that the LAr tests performed in Brazil (Sec.~\ref{sec:xara_single_cell}) had the same type of SiPM installed in the \xara\ with an area ratio of 3.6\% and an efficiency of (2.2\error0.44)\% with $\alpha$-particles. This shows the improvement of the \xara.

\begin{figure}[h!]
	\centering
	\begin{subfigure}{0.47\textwidth}
		\includegraphics[width=0.99\linewidth]{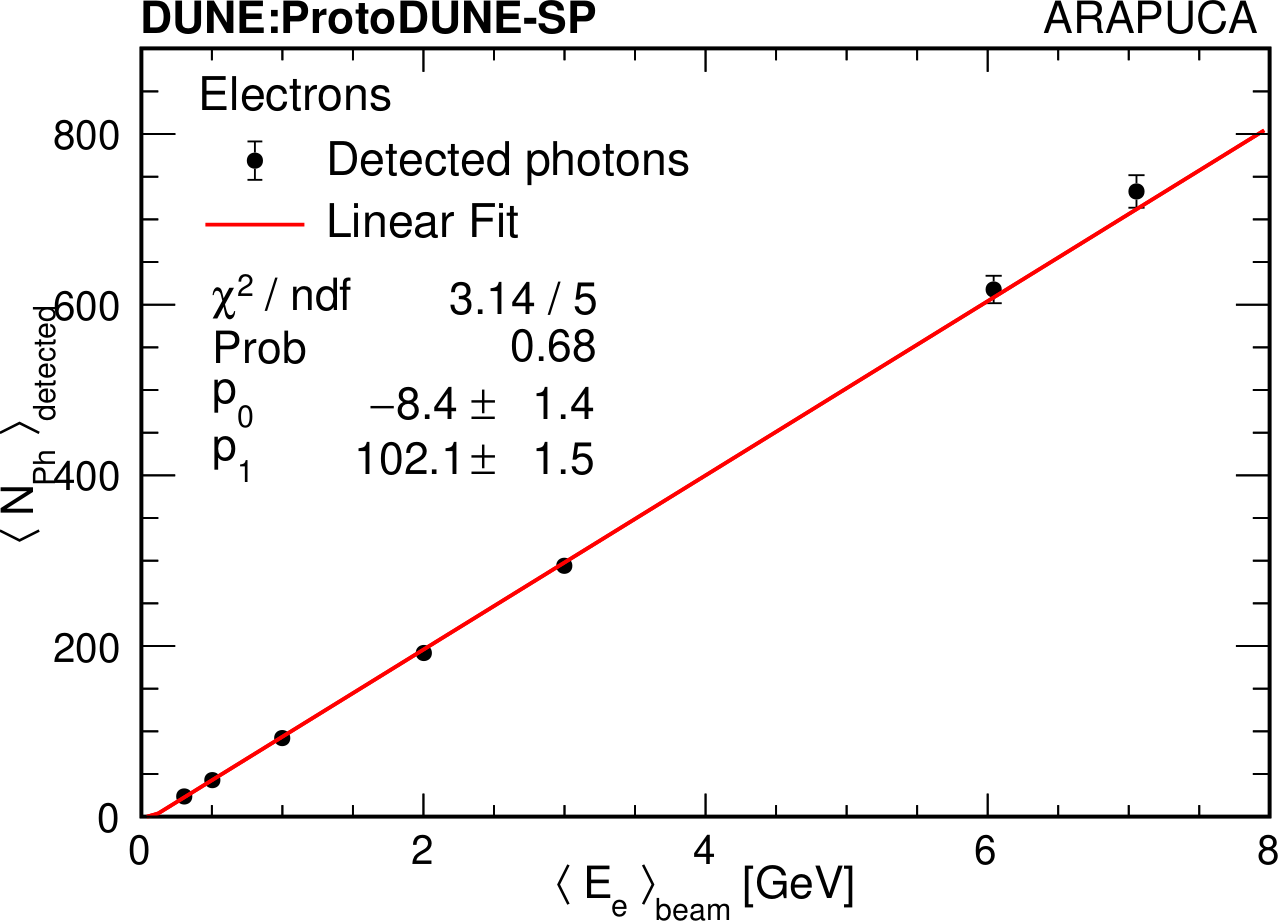}
		\caption{}
		\label{fig:protoDUNE_light_yield}
	\end{subfigure}
	\begin{subfigure}{0.51\textwidth}
		\includegraphics[width=0.99\textwidth]{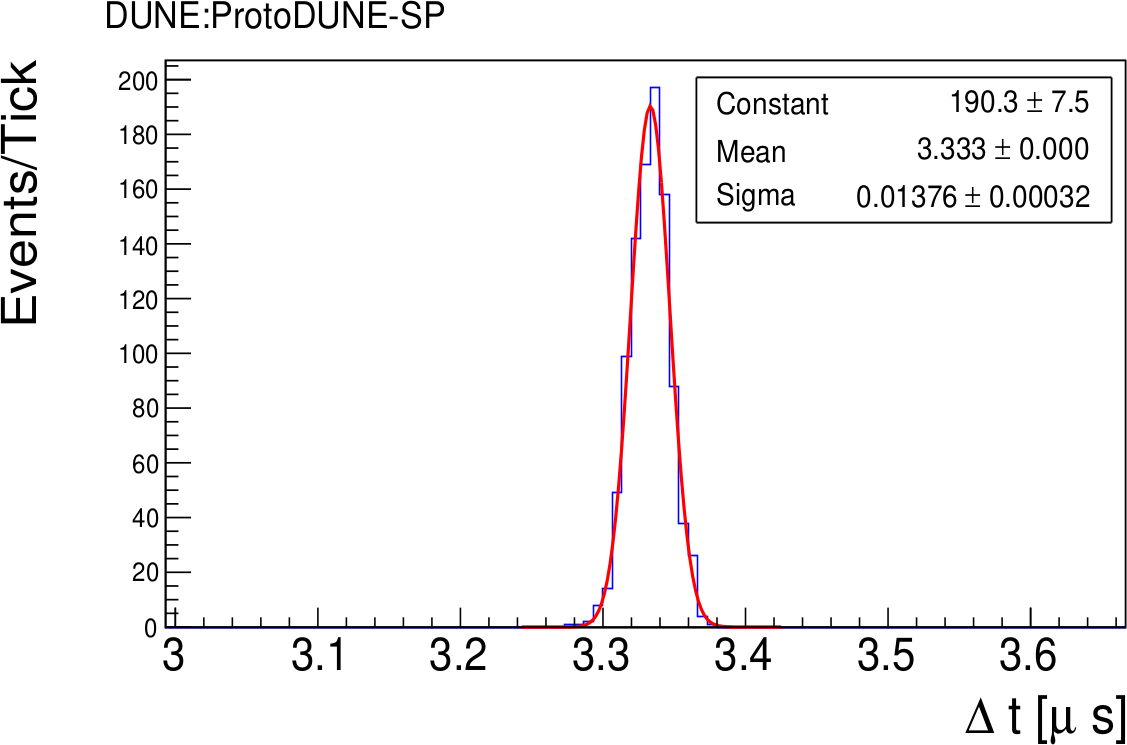}
		\caption{}
		\label{fig:protoDUNE_light_flash}
	\end{subfigure}
	\caption{\textbf{(a)}~Average detected photons versus as function of incident electron energy for the \ara\ module. \textbf{(b)} PDS time resolution (using the time difference between two LED signals)~\cite{protoDUNE_first_results}.}
	\label{fig:protoDUNE_plots}
\end{figure}

The PDS light yield was evaluated by measuring the average light detected by the \ara\ module for different electron energies from the beam, as shown in Figure~\ref{fig:protoDUNE_light_yield}~\cite{protoDUNE_first_results}. By comparing the number of detected photons with the expected number of incident photons simulated and by extrapolating to a PDS composed only of \ara s, a light yield of 1.9~\phe/MeV was found. A value greater than the 0.5~\phe/MeV is required by DUNE~\cite{DUNE_vol4}.

The time resolution of the PDS was taken by measuring the time difference between two LED light signals, as shown in Figure~\ref{fig:protoDUNE_light_flash}. The value of 14~ns fulfills the DUNE requirement of a time resolution lower than 100~ns~\cite{protoDUNE_first_results,DUNE_vol4}. Finally, it was reported an electron lifetime greater than 20~ms, while it is required $e$-lifetime~>~3~ms. 
\section{Xenon Doping tests}
\label{sec:xe_dop}

At the end of the ProtoDUNE-SP run I there was a failure in the gas re-circulation pump, which resulted in a certain amount of air inside the detector. The LArTPC purification system could remove the O$_2$, CO$_2$ and H$_2$O: however, the system was not able to remove $N_2$. It was estimated that about 5~ppm of N$_2$ were leaked into the detector.

As described in Sec.~\ref{sec:quenching_nitrogen}, nitrogen contaminant decreases the light yield of the detector by quenching the triplet state light emission. 

Xenon emits scintillation light at 175~nm~\cite{xe_pulse_shape_discrim,Xe_dopping_fast_comp,Xe_dopping_first}, that can be more efficiently collected, and have a longer Rayleigh scattering length~\cite{rayleigh_ar_xe,rayleigh_ar_new}, which can improve the uniformity of the light response across a large detector such as DUNE. Besides, the xenon light emission is faster than argon, which can improve the detector time resolution. Because of that, the DUNE Collaboration was already studying the possibility of doping LAr with a few tens of ppm level of xenon (Xe).


The nitrogen contamination event was a great opportunity to test this technology in a large scale experiment. In the case where the TPC works properly and the PDS is restored, Xe doping may be considered as a solution for the DUNE experiment in a case of contamination leakage such the one just described.

\subsection{Xe doping effect in LAr} 

Doping LAr with small amounts of xenon (at few ppm levels) causes the transfer of the excitation energy from argon to xenon dimers through the following series of reactions~\cite{xe_pulse_shape_discrim,pulse_shape_analysis}:
\begin{align}
	\text{Ar}^*_2 + \text{Xe} + \text{migration} \rightarrow (\text{ArXe})^* + \text{Ar} \\
	(\text{ArXe})^* + \text{Xe} + \text{migration} \rightarrow \text{Xe}^*_2 + \text{Ar},
\end{align}
where the excimer singlet and triplet states will decay with a characteristic time of 4,2~ns and 22~ns, respectively, by emitting light around 175~nm. As for the quenching effect (see Sec.~\ref{sec:quenching_nitrogen}), the Xe energy transfer will affect mostly the argon triplet state as it is the longest living one. The light production rate can be taken as the sum of the Ar and Xe excimer distributions, that is:
\begin{equation}
	r = \lambda_{\text{Ar},1}N_{\text{Ar},1} + \lambda_{\text{Ar},3}N_{\text{Ar},3} + \lambda_{\text{Xe},1}N_{\text{Xe},1} + \lambda_{\text{Xe},3}N_{\text{Xe},3}, 
\end{equation}
where $N$ corresponds to the number of singlet and triplets (depending on the index 1 or 3) of Ar or Xe, and $\lambda$ is the decay constant of each state. 

Defining $\lambda_m$ as the transition rate to the excimer ArXe, $\lambda_d$ as the transition rate to the Xe excimer, $q$ as the fraction of Ar excimer in the singlet state and $p$ as the fraction of Xe excimer in the singlet state, one can obtain the transition differential equations for different states~\cite{xe_pulse_shape_discrim}:
\begin{align}
	\frac{dN_{Ar,1}}{dt} &= - \lambda_{Ar,1} N_{Ar,1} - \lambda_m N_{Ar,1} \\
	\frac{dN_{Ar,3}}{dt} &= - \lambda_{Ar,3} N_{Ar,3} - \lambda_m N_{Ar,3} \\
	\frac{dN_{ArXe}}{dt} &= + \lambda_{m} N_{Ar,1} + \lambda_m N_{Ar,3} - \lambda_d N_{ArXe} \\
	\frac{dN_{Xe,1}}{dt} &= + \lambda_{d} p N_{ArXe} - \lambda_{Xe,1} N_{Xe,1} \\
	\frac{dN_{Xe,3}}{dt} &= + \lambda_{d} (1-p) N_{ArXe} - \lambda_{Xe,3} N_{Xe,3}.
\end{align}

The argon light output will have the same usual two exponential decays. By defining:
\begin{align}
	\lambda_{1m} &\equiv \lambda_{Ar,1} + \lambda_{m} \\
	\lambda_{3m} &\equiv \lambda_{Ar,3} + \lambda_{m},
\end{align}
the solved equation can be simplified by disregarding the exponential terms with $\lambda_{Xe,1}$ and $\lambda_{Xe,3}$ as~\cite{pulse_shape_analysis}:
\begin{equation}
	\label{eq:shape}
	r = A_1 e^{-\lambda_{1m}t} + A_2 e^{-\lambda_{3m}t} - A_3 e^{-\lambda_{d}t},
\end{equation}
where the decay time constants fast, slow and delayed can be defined as:
\begin{align}
	\label{eq:taus_def}
	\tau_\text{fast} &= \frac{1}{\lambda_{1m}}\\
	\tau_\text{slow} &= \frac{1}{\lambda_{3m}}\\
	\tau_d &= \frac{1}{\lambda_{d}}.
\end{align}

Equation~\ref{eq:shape} translates the competition between LAr scintillation emission and Xe shifting. The light output becomes faster with the increasing of Xe concentration: as the transfer rates $\lambda_m$ and $\lambda_d$ increase, the argon triplet contribution decreases. Therefore, the time constants $\tau_\text{slow}$ and $\tau_d$ are expected to decrease at higher concentrations. The assumption that a few ppm of N$_2$ does not affect the singlet or triplet excimer de-excitation is reasonable due the time scales. Consequently, it is expected that increasing the Xe concentration in the case of $N_2$ contamination, the Xe shift process happens before the quenching, restoring the light yield.

\subsection{Experimental setup}

The xenon doping test consisted of doping the LAr five times, the first had Xe injected so the proportion of 0.8~ppm was achieved. A proportion of 3.3~ppm, 11~ppm, 15~ppm and 20~ppm where achieved in the following Xe injections (dopings).

\begin{figure}[h!]
	\centering
	\begin{subfigure}{0.47\textwidth}
		\includegraphics[width=0.99\textwidth]{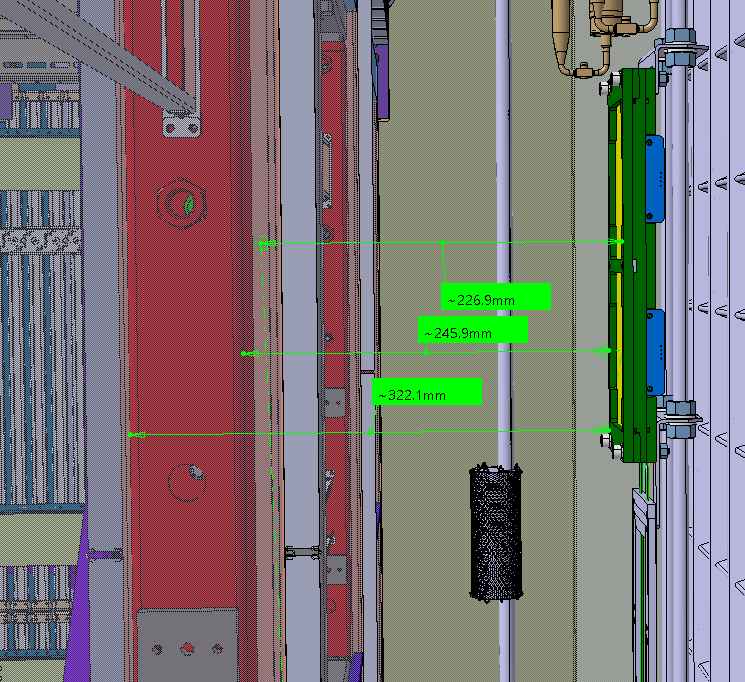}
		\caption{}
		\label{fig:arapuca_in_protodune}
	\end{subfigure}
	\begin{subfigure}{0.52\textwidth}
		\includegraphics[width=0.99\linewidth]{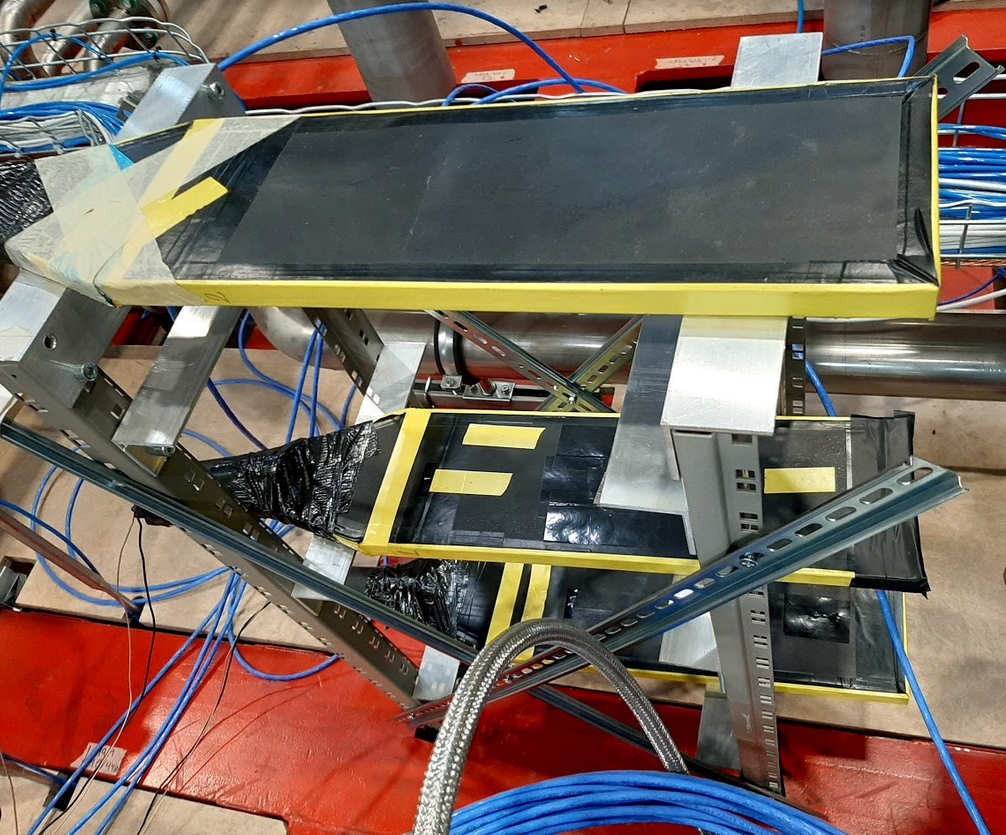}
		\caption{}
		\label{fig:muonscope}
	\end{subfigure}
	\caption{\textbf{(a)}~Schematic assembly of the \xara\ behind the APA. \textbf{(b)}~Muon telescope composed with three plastic scintillator detector above the ProtoDUNE-SP cryostat.}
	\label{fig:explanation_setup_protodune}
\end{figure}

It was decided to deploy two \xara\ modules of the SBND type in the cryostat. The \xara s were installed behind the APA, as can be seen in Figure~\ref{fig:arapuca_in_protodune}. The APA (and the PDS modules) can be seen 22.69~cm away form the \xara\ on the top. The second \xara\ is 36~cm bellow. The double-cell was of the same size and construction as the one described in Sec.~\ref{sec:xara_double_cell}, with EJ-286 WLS slab and \ptp\ coated in the Opto Co. dichroic filter, but with SiPMs Hamamatsu S13360-6050VE, the same used in the single-cell test. The tests performed in the Monochromator (Sec.~\ref{sec:fused_silica}) showed that fused silica (quartz) has a transparency of 80\%~to~90\% for the 175~nm while completely blocking the 128~nm LAr scintillation light. Therefore, the bottom \xara\ was deployed with a quartz window in the front, to detect only Xe emitted light. 

An external trigger was used with three detectors (denoted \textit{paddles}) of plastic scintillator sizing 44~$\times$~15.5~cm$^2$ coupled to PMTs and vertically aligned, as shown in Figure~\ref{fig:muonscope} with the paddles over the LArTPC cryostat. The top paddle is 34~cm away from the center one, while the bottom paddle is distanced 29~cm. The triple coincidence of the detectors selects mostly vertical muons and was used as external trigger for the readout of the \xara s. The readout was performed by an ADC of 150~MSamples/s with a total of 2000 ``ticks'' each with 6.67~nm. 

\subsection{Waveform analysis}
\label{sec:protodune_waveforms}

A dedicate team performed the analysis of the data collected with xenon doped LAr. Here is described the waveform analysis performed by the author and Niccoló Gallice (Università degli Studi di Milano). The waveform behavior along the increasing Xe doping concentration was analyzed.

The averaged waveforms for each of the four channels of the \xara\ was taken. As each channel has four SiPMs ganged passively in parallel, the time response of the detector is not enough to retrieve the shape described by Eq.~\ref{eq:shape}. Two different approaches were taken to retrieve the time constants. In the first one, the single photo-electron (\sphe) response was deconvolved from the averaged signal so a fit with Eq.~\ref{eq:shape} could be performed, the same approach used for the \xara\ double-cell (Sec.~\ref{sec:xara_double_cell}). The second approach consisted in convolving the \sphe\ response with the theoretical signal and performing a $\chi^2$ minimization of the experimental waveform, a method similar to the one used for the \xara\ single-cell (Sec.~\ref{sec:xara_single_cell}).

\begin{figure}[h!]
	\centering
	\begin{subfigure}{0.505\textwidth}
		\includegraphics[width=0.99\textwidth]{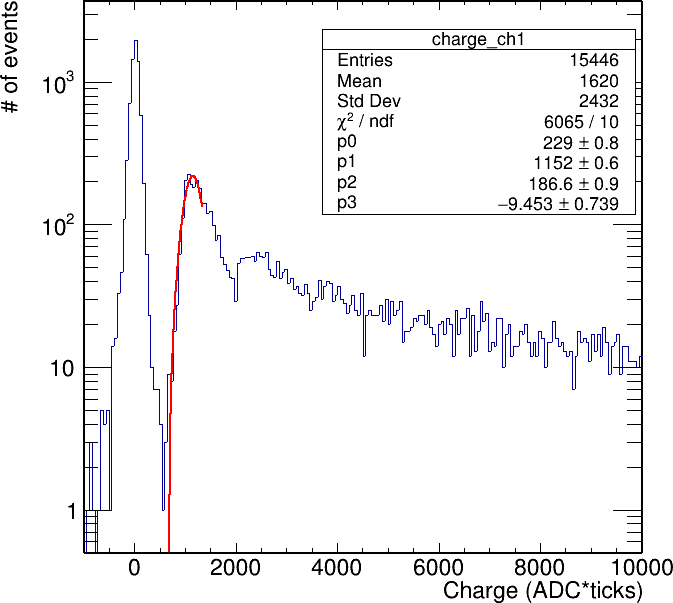}
		\caption{}
		\label{fig:sphe_spectrum}
	\end{subfigure}
	\begin{subfigure}{0.481\textwidth}
		\includegraphics[width=0.99\linewidth]{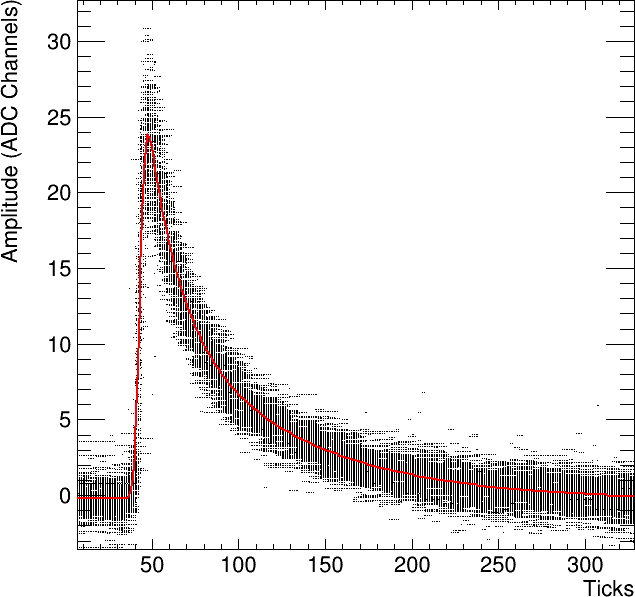}
		\caption{}
		\label{fig:sphe_waveforms}
	\end{subfigure}
	\caption{\textbf{(a)}~Single photo-electron spectrum obtained, a Gaussian was adjusted in the data (red) to retrieve the corresponding \sphe\ events. \textbf{(b)}~Waveforms (black points) selected inside 1$\sigma$ of Gaussian fit and the averaged waveform fit (red line).}
	\label{fig:two_sphe_stuff}
\end{figure}

For the deconvolution, the \sphe\ responses where searched and selected in the pretrigger (the first 400~ticks), the \sphe\ spectrum of Figure~\ref{fig:sphe_spectrum} was taken by integrating the pulses for about 200~ticks. The averaged \sphe\ waveforms is taken by selecting events with charge inside 1$\sigma$ from the adjusted Gaussian in Figure~\ref{fig:sphe_spectrum}. The resulting waveform, displayed in Figure~\ref{fig:sphe_waveforms}, is than fitted with two exponentials convoluted with a Gaussian response (see Appendix~\ref{chap:deconvolution}), represented in red. 

\begin{figure}[h!]
	\centering
	\begin{subfigure}{0.485\textwidth}
		\includegraphics[width=0.99\textwidth]{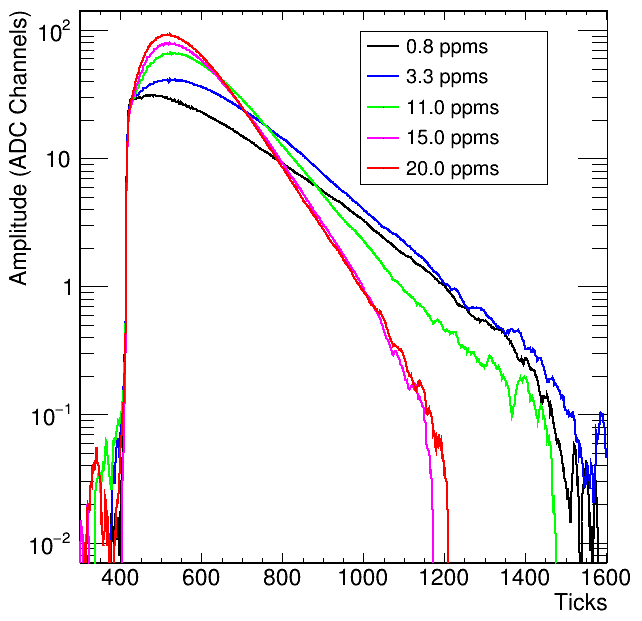}
		\caption{}
		\label{fig:waveforms_dopings}
	\end{subfigure}
	\begin{subfigure}{0.49\textwidth}
		\includegraphics[width=0.99\linewidth]{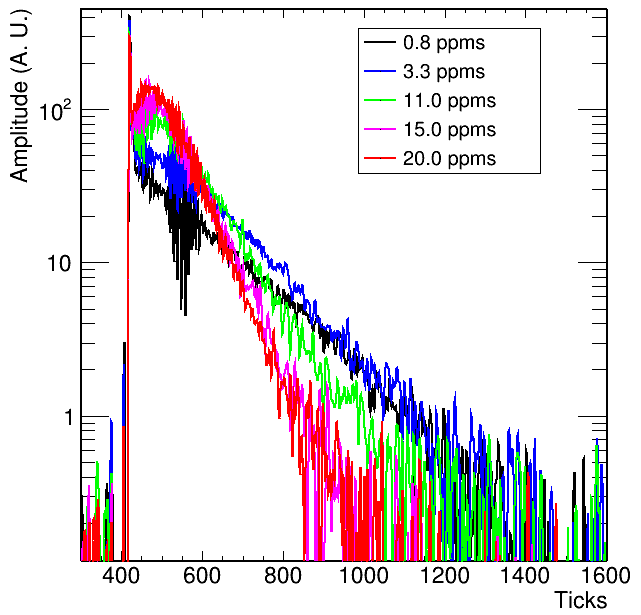}
		\caption{}
		\label{fig:waveforms_dopings_deconvolved}
	\end{subfigure}
	\caption{\textbf{(a)}~Averaged waveforms obtained for each xenon doping concentration, each one contain 10 thousand events. \textbf{(b)}~Resulting waveforms after deconvolving the \sphe\ response.}
	\label{fig:two_waveforms}
\end{figure}

Figure~\ref{fig:waveforms_dopings} shows the averaged waveforms of the five Xe concentrations for the \xara\ without quartz window.  The  ``Gold deconvolution algorithm'' implemented in the TSpectrum class of Root Cern~\cite{root_cern} was used (see Appendix~\ref{chap:deconvolution}) and the results are presented in Figure~\ref{fig:waveforms_dopings_deconvolved}. The fast, slow and delayed components are more evident in the deconvolved waveforms and can be more accurately retrieved. One can verify that the slow component is restored after the first doping (0.8~ppm) and decreases with the increasing concentration while the delayed light increases. Reaching stability after 15~ppm. 

\begin{figure}[h!]
	\centering
	\begin{subfigure}{0.49\textwidth}
		\includegraphics[width=0.99\linewidth]{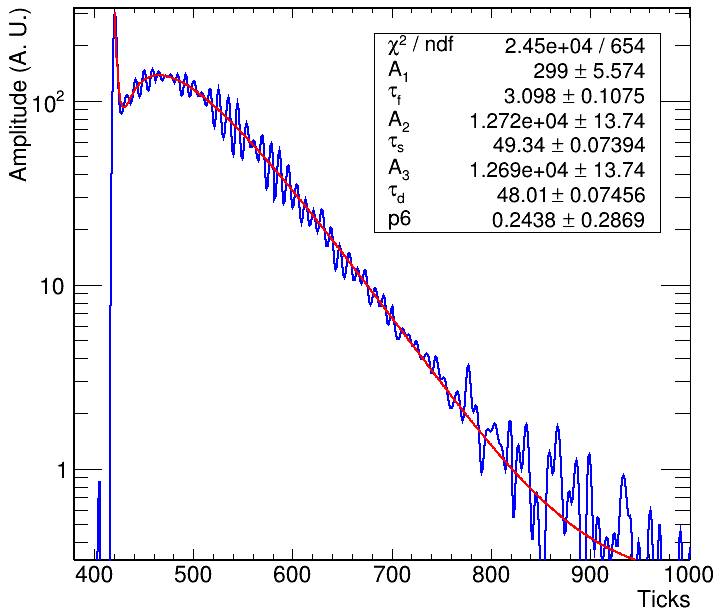}
		\caption{}
		\label{fig:fit_deconvolve}
	\end{subfigure}
	\begin{subfigure}{0.49\textwidth}
		\includegraphics[width=0.99\textwidth]{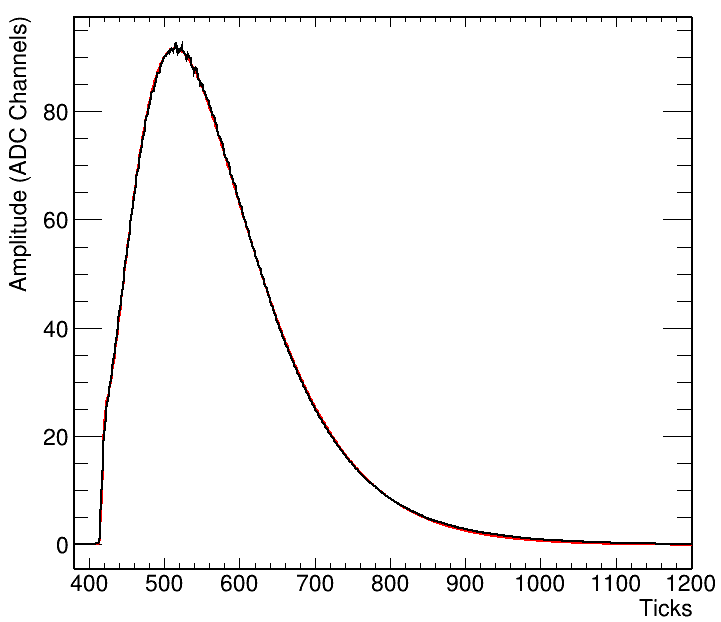}
		\caption{}
		\label{fig:fit_convolve}
	\end{subfigure}
	\caption{\textbf{(a)}~Deconvolved waveform (blue) adjusted by Eq.~\ref{eq:shape} (red) with the addition of a baseline constant (fitted values are in ticks). \textbf{(b)}~Averaged waveform (black) fitted by the expected signal Eq.~\ref{eq:shape} convoluted with the \sphe\ response (red).}
	\label{fig:two_fits}
\end{figure}

Finally, Figure~\ref{fig:two_fits} shows the fits performed with the two methods, where Fig.~\ref{fig:fit_deconvolve} shows the fit of Eq.~\ref{eq:shape} in the deconvolved waveform for the last doping. On the other hand, Figure~\ref{fig:fit_convolve} shows the $\chi^2$ minimization with TMinuit~\cite{root_cern} of the averaged waveforms (black) with the exponential terms of Eq.~\ref{eq:shape} convoluted with the \sphe\ response (red).

\begin{figure}[h!]
	\centering
	\begin{subfigure}{0.99\textwidth}
		\includegraphics[width=0.99\textwidth]{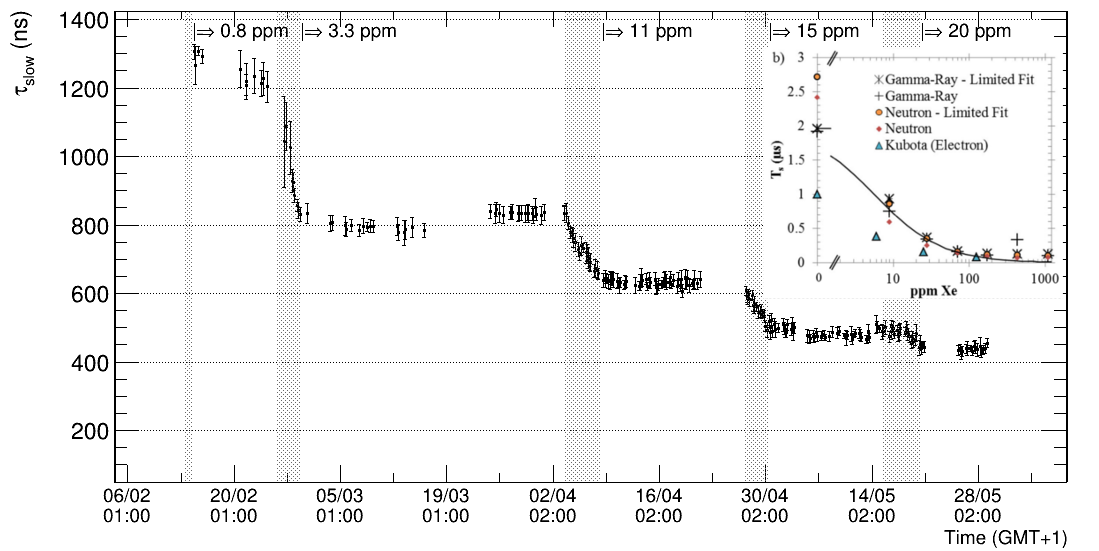}
		\caption{}
		\label{fig:tau_slow}
	\end{subfigure}
	\begin{subfigure}{0.99\textwidth}
		\includegraphics[width=0.99\linewidth]{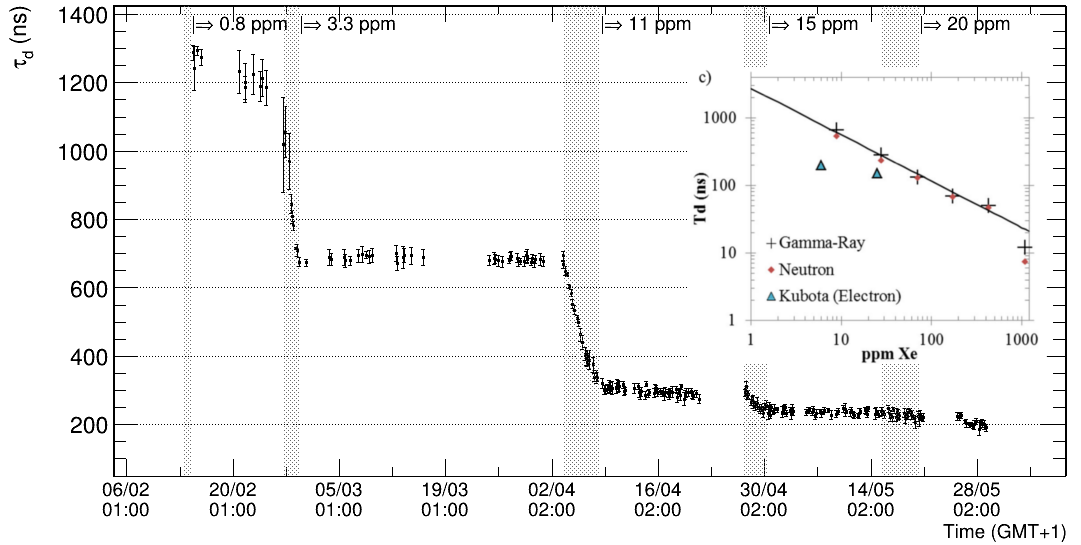}
		\caption{}
		\label{fig:tau_d}
	\end{subfigure}
	\caption{\textbf{(a)}~Slow time constant and \textbf{(b)}~delayed dimer constant versus time. The internal plots are the results found by Ref.~\cite{pulse_shape_analysis} for different Xe concentrations.} 
	\label{fig:two_taus}
\end{figure}

The data were collected for several days, each day had about four runs with 10 thousand events each. The methods just described were applied for each run. Both \xara s deployed presented problems in one of the channels, so the analysis was done in six channels. Figure~\ref{fig:tau_slow} shows the slow time constant ($\tau_\text{slow}$) as function of time, while Figure~\ref{fig:tau_d} shows the result for the delayed time constant ($\tau_d$). The doping periods are labeled with a crosshatched area. One can note the decreasing behavior of both time constants, including during doping periods, where the Xe shifting takes place. Inside the plots, the time constants found by Ref.~\cite{pulse_shape_analysis} are displayed, in which a good correspondence was found. 

The presented values where taken by averaging the response of the six channels, where the uncertainty was calculated with the combination of the fit uncertainty with the average uncertainty calculated with the difference of the extreme values divided by root square of six.

\subsection{Light yield recovery}

\begin{figure}[h!]
	\centering
	\begin{subfigure}{0.99\textwidth}
		\includegraphics[width=0.99\textwidth]{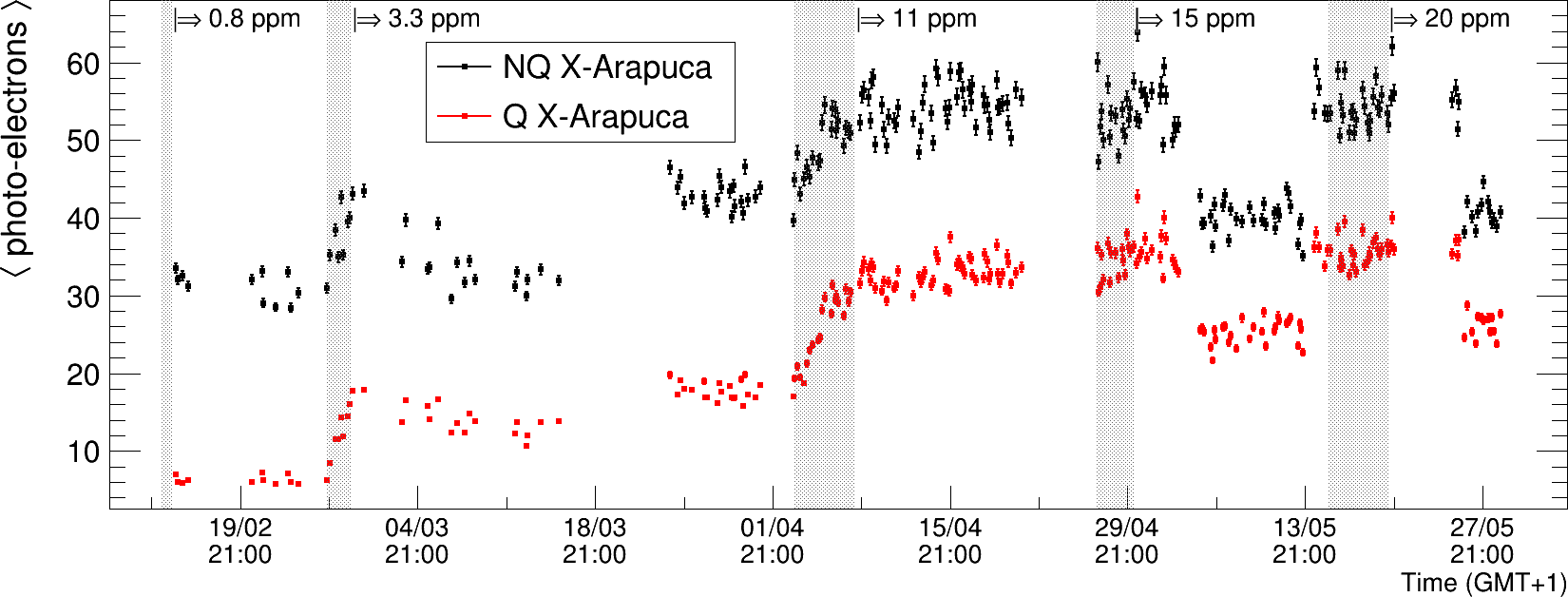}
		\caption{}
		\label{fig:photons}
	\end{subfigure}
	\begin{subfigure}{0.99\textwidth}
		\includegraphics[width=0.99\linewidth]{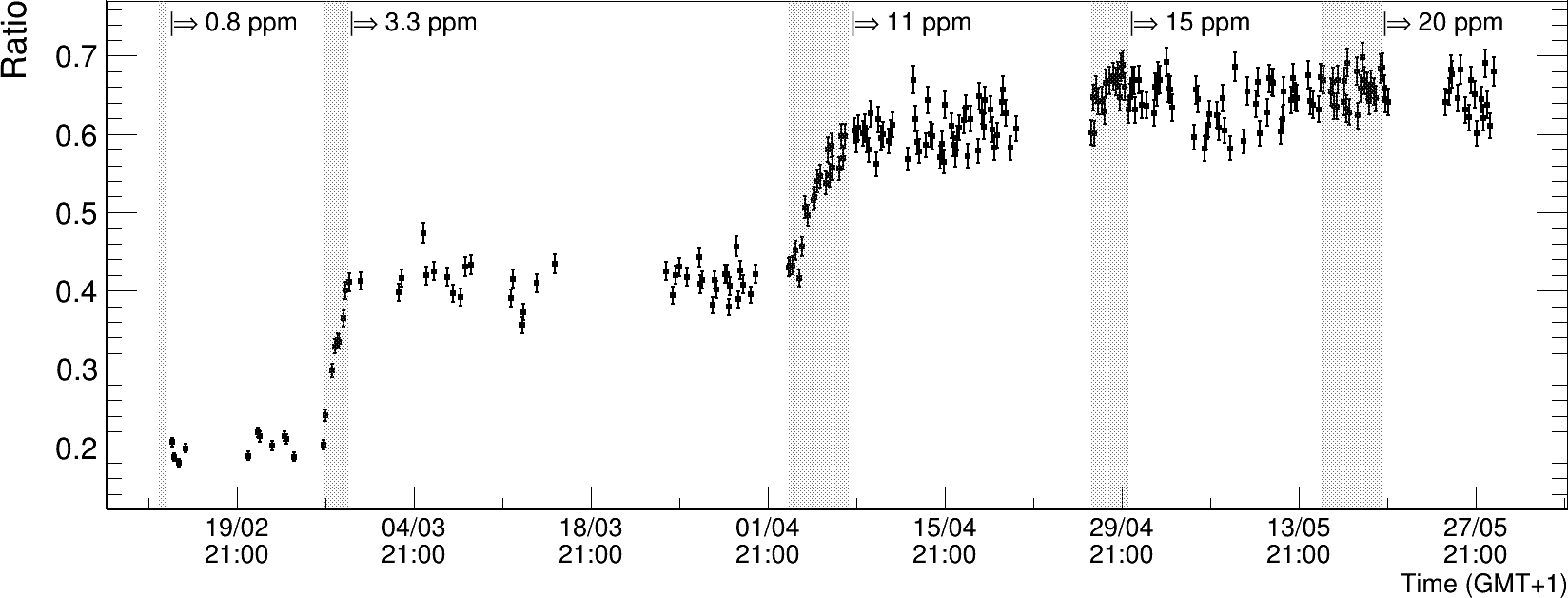}
		\caption{}
		\label{fig:ratio}
	\end{subfigure}
	\caption{\textbf{(a)}~Average number of photons detected versus times for the \xara\ without quartz (NQ) and with quartz (Q) window. \textbf{(b)}~Ratio between Xe light and total light (Xe+Ar) versus time.} 
	\label{fig:photons_and_ratio}
\end{figure}

To verify the recovery in the light yield, the average number of photo-electrons (\phe) detected was computed for each run. For each channel, the amount of \phe\ was evaluated by dividing the averaged charge by the \sphe\ charge in the respective run. The response of the three channels was summed up, resulting in the average number of photo-electrons of the run, as can be seen in Figure~\ref{fig:photons}. The NQ~\xara\ is the device deployed without the quartz window that collects LAr scintillation light and Xe shifted light. On the other hand, the Q~\xara\ is sensitive only to Xe shifted light (175~nm) as it was deployed with the quartz window. The drops in the average number of \phe\ is caused by the electric field applied in the LArTPC (see Sec.~\ref{sec:recombination}).

The increase in light yield is evident in Figure~\ref{fig:photons}, showing the recovery expected by doping with xenon. Another interesting feature is the the ratio between \phe\ of the Q by the NQ \xara, that can be translated as:
\begin{equation}
	\label{eq:ratio_q_nq}
	\text{Ratio} = \frac{\varepsilon\times\text{Xe light}}{\text{Ar light + Xe light}},
\end{equation}
where $\varepsilon$ is the quartz window transparency to Xe light. Figure~\ref{fig:ratio} shows this result, were the ratio seems to stabilize after the fourth doping, that is, at 15~ppm of Xe. Taking into account that the measured transparency of the quartz window was between 80\%~and~90\%, the true ratio between the light can be computed as 0.72~to~0.81, which corresponds to the relative intensity of triplet and singlet for LAr (see Table~\ref{tab:lar_properties}). This means that, at this concentration, the argon triplet component is completely eliminated by the Xe. 

By analyzing the data from the \xara s deployed and from the S-\ara s, which compose the ProtoDUNE PDS, it was concluded that 95\% of the light lost due to the nitrogen contamination was recovered with the xenon doping.  

The recovery in the light yield plus the fact that the LArTPC operated normally during the tests leave the xenon doping as an alternative for the DUNE FD module. The results found by the collaboration are yet unpublished and they may significantly help to understand the operation of DUNE if the xenon is used. 

\subsection{Toy model Monte Carlo}
\label{sec:toy_model_protodune}

\begin{figure}[h!]
	\centering
	\includegraphics[width=0.95\linewidth]{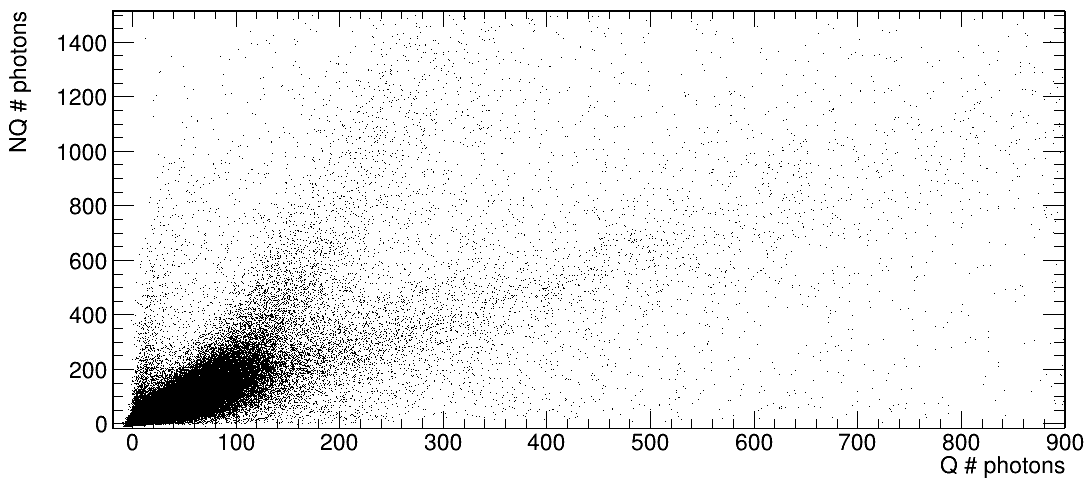}
	\caption{Number of \phe\ detected by the \xara\ without quartz window (NQ) versus quartz (Q).}
	\label{fig:lobesdop3}
\end{figure}

During the period of analysis, it was not easy to explain the plot of the number of photo-electrons detected by the NQ versus Q~\xara\ for each external trigger as shown in Figure~\ref{fig:lobesdop3}. The number of \phe\ was computed by the sum of the three channels during the third doping. It is noticeable that there are three distinct populated regions (lobes) with different angular coefficients.

In order to investigate this behavior, a ``Toy Model'' Monte Carlo simulation was developed to verify if it was a geometrical effect of the detector. The simulation was performed using Root Cern~\cite{root_cern}, and it was made trying to bring plausible and simple approaches to the problem. The geometry of the \xara s was implemented, the 3.6~$\times$~6~$\times$~7~m$^3$ cryostat and the two structures that could interfere (drastically) with the light collection: the PDS modules and the APA division (in red in Fig.~\ref{fig:arapuca_in_protodune}). 

The muons are generated in the first \textit{paddle} of the muon telescope with uniform distribution along the azimuth angle and $\cos[2](\theta)$ for the zenith~\cite{cecchini2012atmospheric}. The energy distribution of the muons was taken from the vertical muon flux at sea level~\cite{energy_spectrum_muons}:
\begin{equation}
	\label{eq:muonflux}
	D_{\mu}(p, h = 1030~\text{g/cm}^2, \theta = 0^{\circ}) = C p^{-(\gamma_0+\gamma_1 \log p + \gamma_2 \log^2 p + \gamma_3 \log^3 p)},
\end{equation}
where $p$ is the momentum, $h$ is the atmospheric depth adjusted to the sea level, $\theta$ is the zenith angle, and the parameters C, $\gamma_0$, $\gamma_1$, $\gamma_2$, $\gamma_3$ are given at Table~\ref{tab:parametersflux}.

\begin{table}[!h]
	\centering
	\caption{Parameters used at Eq.~\ref{eq:muonflux}~\cite{energy_spectrum_muons}}.
	\label{tab:parametersflux}
	\begin{tabular}{cccccc}
		\hline
		\multicolumn{1}{|c}{Momentum Range (GeV/c)} & C (cm$^{-2}$s$^{-1}$sr$^{-1}$GeV$^{-1}$) & $\gamma_0$ & $\gamma_1$ & $\gamma_2$  & \multicolumn{1}{c|}{$\gamma_3$} \\ \hline
		\multicolumn{1}{|c}{1 - 10}                 & $2.95 \times 10^{-3}$                                & 0.3061 & 1.2743 & -0.2630 & \multicolumn{1}{c|}{0.0252} \\ \hline                      
	\end{tabular}
\end{table}

\begin{figure}[h!]
	\centering
	\includegraphics[width=0.9\linewidth]{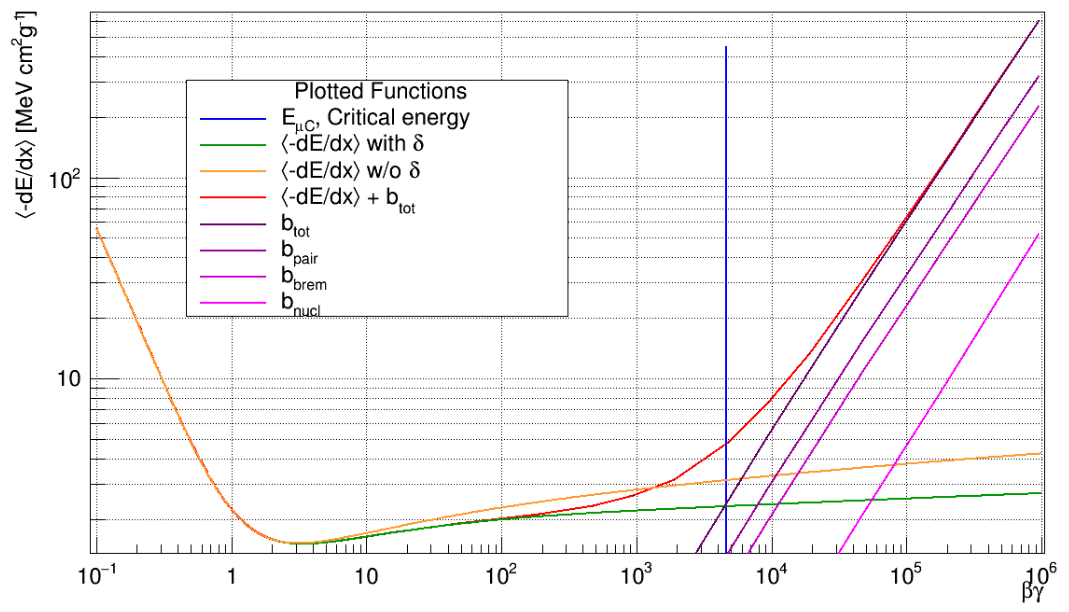}
	\caption{Expected muons energy loss in liquid argon~\cite{muon_energy_loss}.}
	\label{fig:bethebloch}
\end{figure}

The external trigger of the experiment was simulated by selecting only muons that cross all the three detectors. A few images can be found in Appendix~\ref{chap:images_toy_model}. The muon selection can be verified in Fig.~\ref{fig:exampletpcmc2}. A step of 1~cm was defined for the muon interactions after entering the TPC. In this case, for each 1~cm traveled by the muon, an energy loss is defined by the Bethe-Block curve of Figure~\ref{fig:bethebloch}  and converted to 400 photons per MeV emitted isotropically. The number of photons is set by the quenched factor of 0.8 for muons (see Sec.~\ref{sec:lar_scintillation}) divided by 100, for a faster simulation.

To implement statistical fluctuations, the Landau distribution was used in conjunction with the energy loss. The most probable energy loss is given by the Bethe-Block curve and the full width at half maximum (FWHM) set to 4$\varepsilon$, defined as~\cite{pdg}:
\begin{equation}
	\varepsilon = (K/2)(Z/A)(x/\beta^2)\;\; \text{MeV},
\end{equation}
where $K = 4\pi N_Ar_e^2 m_e c^2$ ($N_A$ is the Avogadro constant, $r_e$ is the classic electron radius and $m_e c^2 = 0.511$~MeV is the electron mass), $Z=18$ and $A=40$ are, respectively, the atomic number and mass of the absorber (LAr), $\beta = v/c$ is the relative velocity of the muon and $x$ is the thickness of LAr in g/cm$^2$. 

An absorption length of 20~meters was set for the light and Rayleigh scattering length was set to 1~meter~\cite{lar_atten_length,rayleigh_ar_xe,rayleigh_ar_new}. The simulation does not treat photons with the wavelength of LAr scintillation (127~nm) or xenon shifted (175~nm) any differently. To optimize the run time\footnote{No difference was noticed when changing the selection from the production to the detection.}, the ``type'' of the photons are selected if they are detected by the \xara. A proportion of 30\% LAr light and 70\% Xe shifted light was set, corresponding to a doping greater than 11~ppm. 

The APA wires and the grounding mesh were not simulated. The angle of incidence of the photons arriving in the \xara\ acceptance window was recorded and, in the post production, a random selection based on the grid transparency (see Sec.~\ref{sec:grid_transparency}) was performed. A transparency of 80\% was used for the quartz window for the Xe light (see Sec.~\ref{sec:fused_silica}).

\begin{figure}[h!]
 	\centering
 	\includegraphics[width=0.95\linewidth]{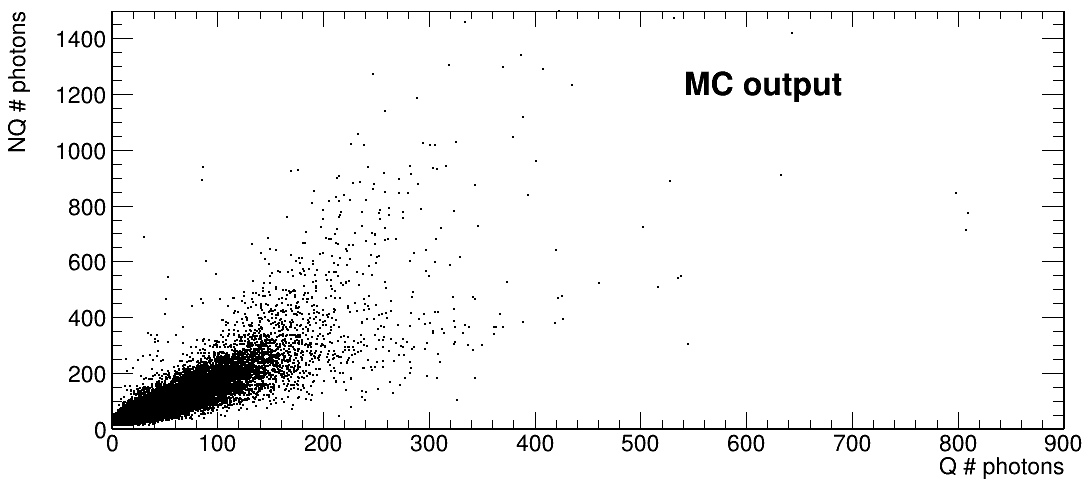}
 	\caption{Results of the Monte Carlo simulation for the number of photons detected by the \xara\ without quartz (NQ) versus quartz (Q).}
 	\label{fig:lobesmc}
 \end{figure}

The results of the simulation are shown in Figure~\ref{fig:lobesmc}, with number of photo-electrons detected by the \xara\ without quartz window (NQ) versus quartz (Q). An efficiency of 2.1\% was set for the \xara s. Although the simulation does not retrieve the expected results with precision, the goal was reached by reproducing the two main lobes and, eventually, the third one when comparing with Fig.~\ref{fig:lobesdop3}. The populated regions appeared when the Rayleigh scattering was implemented and accentuated when the TPC column and PDS modules were implemented.  

%% file: Conclusion.tex
\chapter*{Conclusion}
\addcontentsline{toc}{chapter}{Conclusion}
\thispagestyle{myheadings}
\label{chap:conclusao}

The notable CP-violation measurement proposed by the Deep Underground Neutrino Experiment, together with a rich primary and ancillary science program were discussed in Chapter~\ref{chap:dune_nu}. The DUNE experiment will be an essential tool to advance in the current knowledge and understanding of neutrino physics. In Chapter~\ref{chap:dune}, the experiment was described in details together with the Liquid Argon Time Projection Chamber principle of work. An essential aspect of the experiment is the detection of liquid argon scintillation light. The scintillation light was explored in Chapter~\ref{chap:pd_system} with the introduction of the \ara\ as the DUNE photon detection system.

The research and development of the S-\ara\ and \xara\ performed in the period of this thesis were presented in Chapter~\ref{chap:RnD}. The trapping effect of the device was tested in 2018 in a dark box at room temperature. The Dark Box test aided the search for improvements in the S-\ara\ device in a simple experimental setup and its concept was later implemented for the room and cryogenic temperature tests performed in Italy. The tests with the Monochromator have successfully characterized different dichroic filters, substrates and wavelength shifters. Moreover, the transparency measurements of the grounding mesh were used as input for Monte Carlo simulations, helping to model better the light signals in the LArTPC. The experimental setup preparation for these tests have significantly helped to establish the Leptons Laboratory at the University of Campinas as an outstanding research facility. 

In Chapter~\ref{chap:lar_test}, the tests in liquid argon performed with a \xara\ single-cell and double-cell in Brazil and Italy, respectively, were described. The author of this thesis was responsible for the data analysis of the experiments and the results are published as Refs.~\cite{x_arapuca_article,enhancement_xara}.

The \xara\ single-cell was characterized with three different ionizing radiations: alpha particles, gamma rays and cosmic muons. The comparable efficiencies found completely characterize the device, showing that the efficiency does not depend on the particle type nor on the energy deposition topology. The efficiency of (2.2\error0.4)\% and (3.0\error0.6)\% were found with the SiPM biased with 5.0~V and 5.5~V overvoltage, respectively. The analysis were performed with a direct comparison between Geant4 Monte Carlo simulation.

The \xara\ double-cell was characterized with an $^{241}$Am source in five different positions along the device longitudinal axis. Measurements comparing the enhancement in light collection showed an efficiency increase of about 50\% when using the new wavelength shifter developed at Milano- Bicocca. An efficiency of (1.9\error0.1)\% was found for the \xara, a result in agreement with the efficiency found for the single-cell. Moreover, by deploying the new wavelength shifter, an efficiency of (2.9\error0.1)\% was found. These results probe the use of the \xara\ as the standard photon detection system (PDS) of DUNE, fulfilling the requirements of 1.3\% and 2.6\% efficiency for proton decay search and supernova neutrino burst detection.

The deployment of two \xara s in ProtoDUNE-SP for the xenon doping experiments was explored in Chapter~\ref{chap:protoDUNE}. The joint analysis of the \xara s and S-\ara s (part of the ProtoDUNE PDS), together with data from the TPC, showed a recovery of 95\% of the light lost due to nitrogen contamination. The carried out research has established the xenon doping as an alternative for DUNE in the case of a contamination with nitrogen. 

The \xara\ super-cells are going to be tested in dedicated setups in Brazil, Italy and Spain and are going to be deployed in ProtoDUNE for the second run. The validation of the super-cell is crucial for the DUNE collaboration. Currently, different versions of \xara\ are under study for the vertical drift modules, where there \xara\ will be placed in the cathode plane and on the field cage of the LArTPC. The first module PDS assembly is foreseen to finish in beginning of 2023 and the first module operation should start in mid 2026, with a lot of work to be done.

\newpage
\def\thispagestyle#1{} 
\bibliographystyle{naturemag}
\bibliography{bibliography_henrique}


\begin{appendices}
	

\chapter{Deconvolution method}
\thispagestyle{myheadings}
\label{chap:deconvolution}

The ROOT Cern class TSpectrum~\cite{root_cern} has the Gold deconvolution~\cite{deconvolve} algorithm already implemented. The one dimensional deconvolution wants to find the original discrete impulse response $x(i)$  from the convoluted waveform $y(i)$ in the form:
\begin{equation}
y(i) = \sum_{k=0}^{N-1} h(i-k)x(k),\qquad i = 0,1,2,...,N-1,
\end{equation}
where, in the case of this analysis, $y(i)$ is the output averaged waveform from $\alpha$'s or muons, $h(i)$ is the input average waveform of a single photo-electron and $x(i)$ is the original LAr light response. In matrix form we have\footnote{Here, to match the dimensions, we are using the indexes in loop. That is, the $N-1$ vector y on the left hand is composed as $y(0) = y(0) + y(N)$, $y(1) = y(1)+y(N+1)$, ..., $y(N-1) = y(N-1)+y(2N-1)$ }:
\begin{equation}
\begin{bmatrix}
y(0)\\
y(1)\\
... \\
y(2N-2)\\
y(2N-1)
\end{bmatrix}
=
\begin{bmatrix}
h(0) & 0 & 0 & ... & 0\\
h(1) & h(0) & 0 & ... & 0\\
h(2)  & h(1) & h(0) & ... & ...\\
... & ... & h(1) & ... & ... \\
... & ... & ... & ... & ...\\
h(N-1) & h(N-2) & ... & ... & 0\\
0 & h(N-1) & h(N-2) & ... & h(0)\\
0 & 0 & h(N-1) & ... & h(1)\\
... & ... & ... & ... & ...\\
0 & 0 & 0 & ... & h(N-1)
\end{bmatrix}
\begin{bmatrix}
x(0)\\
x(1)\\
... \\
x(N-2)\\
x(N-1)
\end{bmatrix}.
\end{equation}

The proposed solution utilizes the transposed matrix $H$ to find a solution to $x$ as:
\begin{align}
y &= Hx \nonumber\\
H^T y &= H^T H x \nonumber\\
y' &= H'x,
\end{align}
and the Gold deconvolution algorithm solution is:
\begin{equation}
x^{k+1}(i) = \frac{y'(i)}{\sum_{m=0}^{N-1}H'_{im}x^k(m)} \cdot x^k(i),
\end{equation}
where k is the number of iterations and $x^0 = [1,1,...,1]^T$ is the initial guess~\cite{root_cern,deconvolve}.

The algorithm has a drawback here,  it is suitable to process positive definite data, such as histograms. Therefore, an offset needs to be applied in the waveforms to avoid negative values. Moreover, the \sphe\ average waveform of Fig.~\ref{fig:persistancelogsphe} is still too noise to extract a decent output from the deconvolution, specially because of the small oscillation present at 600~ns.

Because of that, a fit to retrieve $h(t)$ is performed with three exponential and a Heaviside step function convoluted with a Gaussian as:
\begin{equation}
\label{eq:sphe_convolution}
h(t) = \frac{1}{\sqrt{2\pi\sigma^2}}\; \int_{-\infty}^{\infty} \; \exp(-\frac{(t-x)^2}{2\sigma^2})H(x)\sum_{i=1}^{3}\exp(-\frac{x}{\tau_i}) \diff x,
\end{equation}
where $H(x)$ is the Heaviside step function and the Gaussian was set as common for all the exponential. Solving this equation is quite simple, one can notice that the equation:
\begin{equation*}
\exp(-\frac{(t-x)^2}{2\sigma^2})\exp(-\frac{x}{\tau}) = \exp(-\frac{\left(x-\left(t-\frac{\sigma^2}{\tau}\right)\right)^2}{2\sigma^2})\exp(-\frac{t}{\tau})\exp(\frac{2\sigma^2}{\tau})
\end{equation*}
can be split in the integral. Because of the Heaviside step function, only the positive area will contribute to the integral. Therefore, we need to solve:
\begin{equation}
I(t) = \frac{1}{\sqrt{2\pi\sigma^2}} \exp(-\frac{t}{\tau})\exp(\frac{\sigma^2}{2\tau}) \int_{0}^{\infty} \exp(-\left[\frac{x-\left(t-\frac{\sigma^2}{\tau}\right)}{\sqrt{2}\sigma}\right]^2) \diff x.
\end{equation}

Taking the variable substitution:
\begin{equation*}
u = \frac{x-\left(t-\frac{\sigma^2}{\tau}\right)}{\sqrt{2}\sigma}\;\; \xrightarrow{\qquad} \;\; \diff x = \sqrt{2}\sigma \diff u
\end{equation*}
the integral becomes:
\begin{equation}
I(t) = \frac{1}{\sqrt{\pi}} \exp(-\frac{t}{\tau})\exp(\frac{\sigma^2}{2\tau}) \int_{-\frac{t}{\sqrt{2}\sigma}+\frac{\sigma}{\sqrt{2}\tau}}^{\infty} \me^{-u^2}\diff u,
\end{equation}
which is solved by using the definition of the $\erf(x)$ function:
\begin{equation*}
I(t) =  \frac{1}{\cancelto{}{\sqrt{\pi}}}\; \exp(-\frac{t}{\tau})\exp(\frac{\sigma^2}{2\tau}) \left[\frac{\cancelto{}{\sqrt{\pi}}}{2} \erf(x)\right]^\infty_{-\frac{t}{\sqrt{2}\sigma}+\frac{\sigma}{\sqrt{2}\tau}}.
\end{equation*}

Finally, using the fact that $\mathrm{erfc}(x)= 1-\erf(x)$, the solution is given by:
\begin{equation}
\label{eq:sphe_function_integral}
I(t) = \frac{1}{2} 	\exp(-\frac{t}{\tau})\exp(\frac{\sigma^2}{2\tau}) \mathrm{erfc}\left(-\frac{t}{\sqrt{2}\sigma}+\frac{\sigma}{\sqrt{2}\tau} \right).
\end{equation}

\begin{figure}[tbph!]
	\centering
	\includegraphics[width=0.95\linewidth]{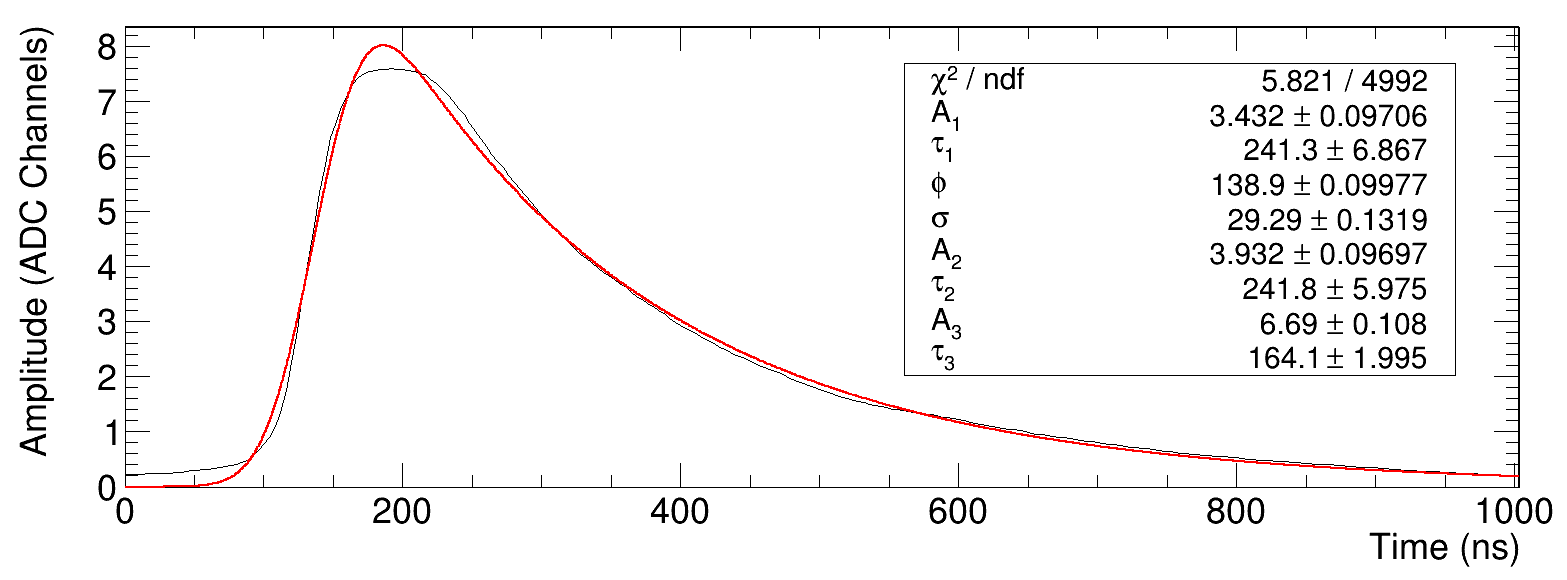}
	\caption{Single photo-electron averaged waveform form Fig.~\ref{fig:persistancelogsphe} fitted with Eq.~\ref{eq:sphe_fit_function}. }
	\label{fig:sphe_fitted}
\end{figure}

Using the result of Eq.~\ref{eq:sphe_function_integral} into Eq.~\ref{eq:sphe_convolution}, adding a phase $\phi$ for the onset of the waveform and a amplitude $A_i$ for each exponential, the final fit function is:
\begin{equation}
\label{eq:sphe_fit_function}
h(t) = \sum_{i=1}^{3} A_i \exp(-\frac{t-\phi}{\tau_i})\exp(\frac{\sigma^2}{2\tau_i}) \mathrm{erfc}\left(\frac{\phi-t}{\sqrt{2}\sigma}+\frac{\sigma}{\sqrt{2}\tau_i} \right),
\end{equation}
with eight free parameters. The method described above is the best estimation we could find to avoid the oscillation,  the result is shown in Fig.~\ref{fig:sphe_fitted} where one can notice the peak of the waveform is not well fitted. Unfortunately, this could not be solved in the fit and the consequence is a not well deconvolved peak.  

\chapter{Total internal reflection}
\thispagestyle{myheadings}
\label{chap:total_internal_reflection}

Total internal reflection happens when an incident light crosses the interface of two different index of refraction mediums ($n_{\text{in}}>n_{\text{out}}$) with a high enough oblique angle. The \textit{Snell's law} relates the angle of incidence the incident light $\theta_{\text{in}}$ with the refracted light $\theta_{\text{out}}$ as:
\begin{equation}
\label{eq:snells_law}
n_{\text{out}}\cdot\sin{\theta_{\text{out}}} = n_{\text{in}}\cdot\sin{\theta_{\text{in}}},
\end{equation}
that can be extended to the total internal reflection considering that $\theta_{\text{out}} = 90^\circ$. In this case, the critical angle $(\theta_C)$ in which the effect starts can be calculated by:
\begin{equation}
	\label{eq:snells_law_index}
	\theta_{\text{C}} = \sin[-1](\frac{n_\text{out}}{n_{\text{in}}}),
\end{equation}
with solution only if $n_{\text{in}}>n_{\text{out}}$. In the specific case of the \xara, the outer medium can be liquid argon or vacuum/air, and the inner medium is the wavelength shifter slab. For instance, the EJ-286 slab has an index of refraction $n_\text{EJ-286} = 1.58$, this means that in vacuum the critical angle would be $\theta_C \sim 39.2^\circ$ and in LAr ($n_\text{LAr}(\lambda=430$~nm$) = 1.2344$) $\theta_C \sim 51.4^\circ$.

\begin{figure}[h!]
	\centering
	\includegraphics[width=0.64\linewidth]{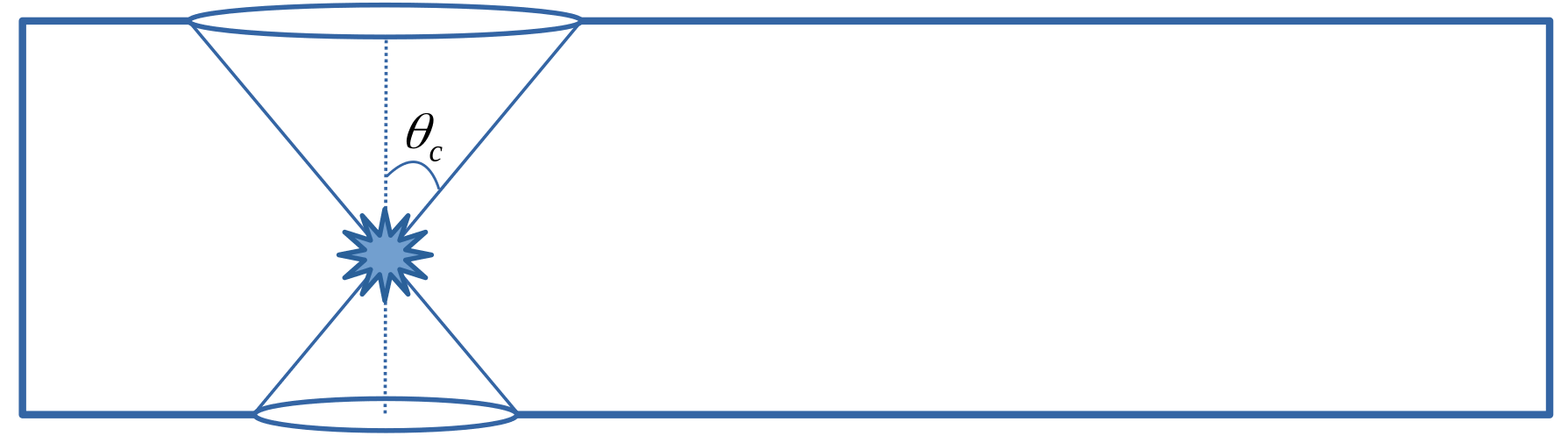}
	\caption{Solid angle of the photons emitted in the critical angle $\theta_C$, illustration based on the image found in the MUONCA manual by Prof. Anderson Campos Fauth.}
	\label{fig:totalinternalreflection}
\end{figure}

To evaluate the amount of photons trapped by total internal reflection, one can take solid angle of emission of the photons as illustrated in Figure~\ref{fig:totalinternalreflection}. The fraction of light exiting the WLS ($F_E$) is related to the solid angle as:
\begin{equation}
	\label{eq:critical_angle}
	F_E = \frac{2\Omega}{4\pi} = \frac{1}{2\pi}\int_0^{\theta_C}\int_0^{2\pi}\diff\phi\sin\theta\diff\theta = 1-\cos\theta_C = 1-\sqrt{1-\sin[2](\theta_C)}
\end{equation}

Using Eq.~\ref{eq:critical_angle} together with Eq.~\ref{eq:snells_law_index}, one can take the fraction of light trapped in the slab due to total internal reflection:
\begin{equation}
	\label{eq:fraction_light_trapped}
	F_T = 1-F_E = \sqrt{1-\left(\frac{n_\text{out}}{n_{\text{in}}}\right)^2}.
\end{equation}

About 77.4\% and 62.4\% of the light is trapped in vacuum and LAr, respectively.

\chapter{Photomultiplier tube (PMT)}
\thispagestyle{myheadings}
\label{chap:pmt}

\begin{figure}[!h]
	\center
	\includegraphics[width=0.9\textwidth]{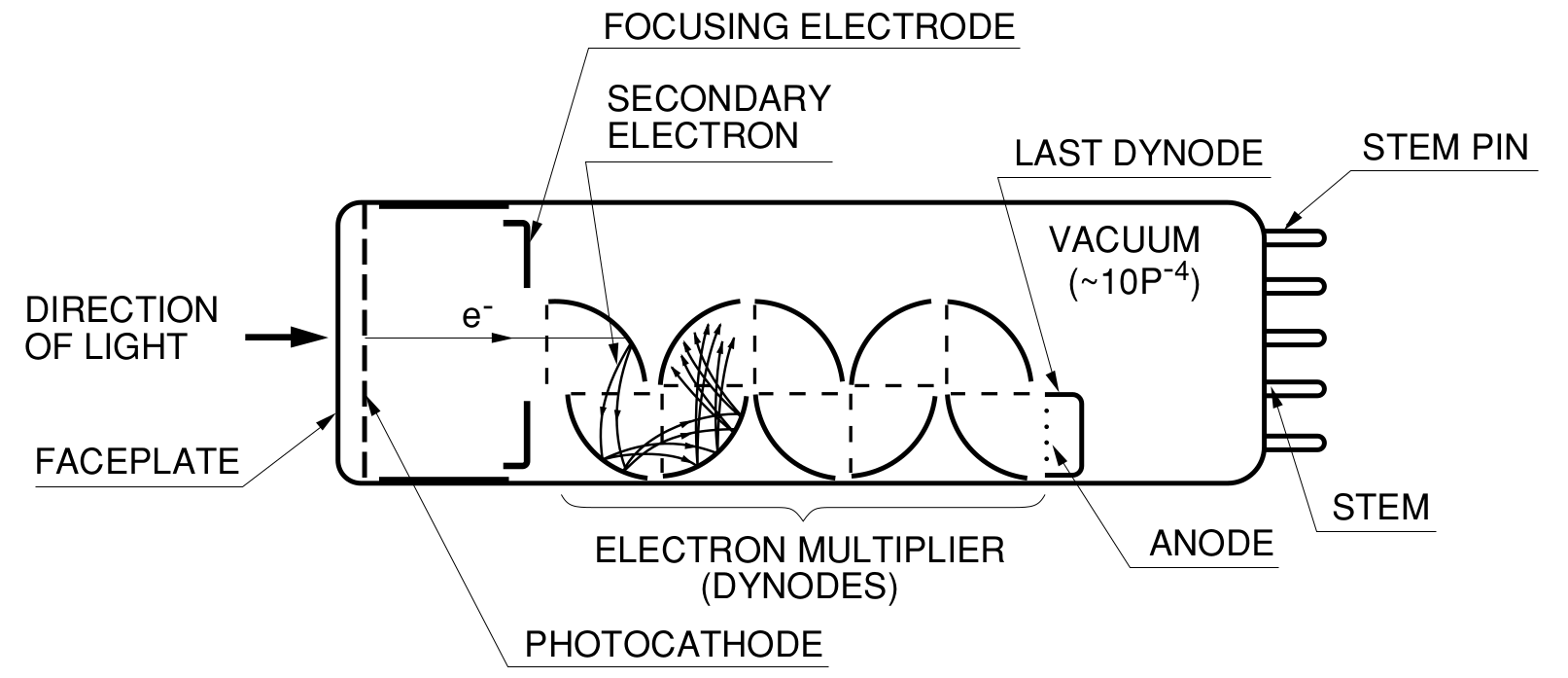}
	\caption{Schematic operation of a photomultiplier tube. Image from Hamamatsu~\cite{hmmt_s13360}.}
	\label{fig:ilustracaoPMT}
\end{figure}

A Photomultiplier tube (PMT) works through the photoelectric effect, in which photons colliding with a metallic material may eject electrons from it. The electron has the energy of the photon subtracting the energy necessary for the ejection~\cite{relatividade}:
\begin{equation}
	K = h\nu -\phi,
	\label{eq:efeitofe}
\end{equation}
where $K$ is the kinetic energy of the electron, $h\nu$ is the photon energy and $\phi$ is the material work function, that is, the energy necessary to eject the bounded electron. The electrons, called photo-electrons, are ejected according to the quantum efficiency of the photo-cathode. A high voltage accelerate the photo-electrons towards dynodes stages that releases more electrons up collision. The multiplied electrons are than collected by the anode. The process is illustrated in Figure~\ref{fig:ilustracaoPMT}.

The gain of the PMT depends on the number of dynodes stages and high voltage applied between the anode and photo-cathode, but its typically of 10$^6$ up to 10$^8$. A negative or positive high voltage power supply of 1~kV up to 2~kV is required.

\chapter{Toy Model images}
\thispagestyle{myheadings}
\label{chap:images_toy_model}

\begin{figure}[h!]
	\centering
	\includegraphics[width=0.99\linewidth]{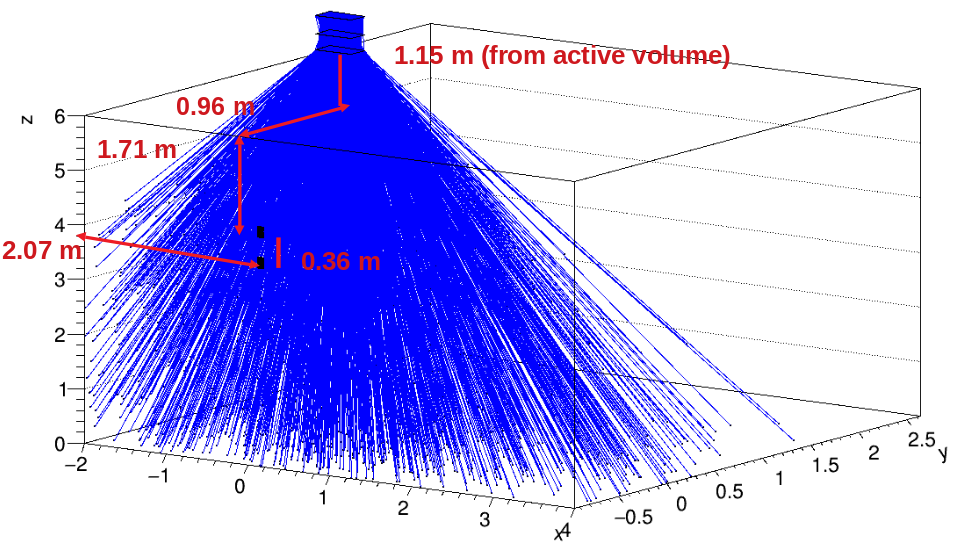}
	\caption{Selected muons (blue lines) by the telescope composed by three detectors of plastic scintillator above the LArTPC. The two \xara s can be seen in the plane $y=-1$~m as black boxes.}
	\label{fig:exampletpcmc2}
\end{figure}

\begin{figure}[h!]
	\centering
	\includegraphics[width=0.99\linewidth]{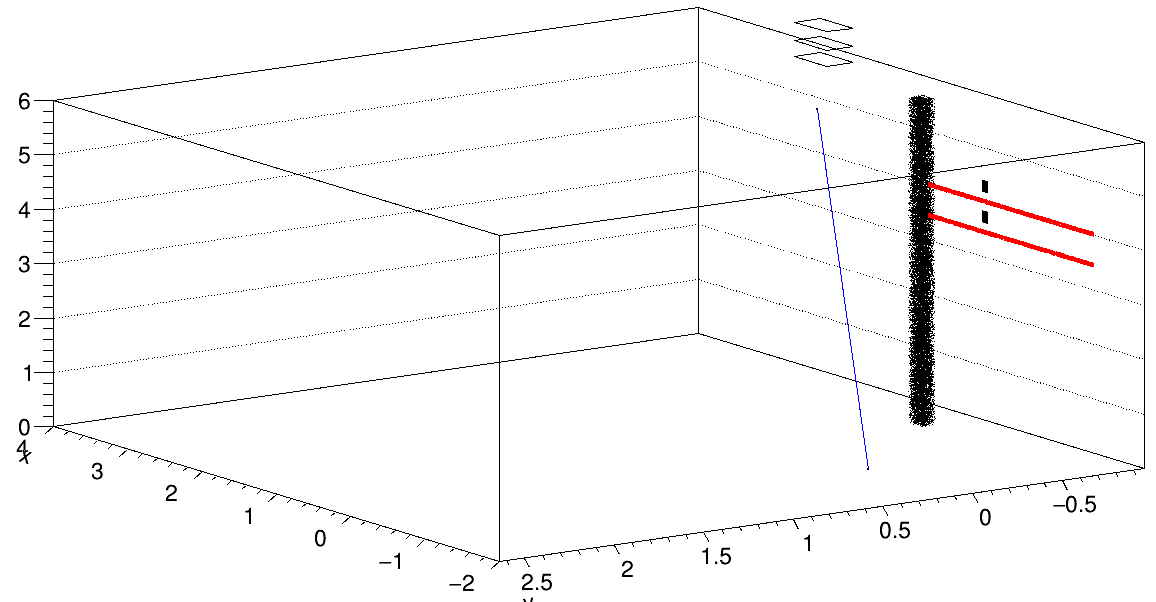}
	\caption{A muon (blue line) crossing the LArTPC will emit light only after 1~m inside the TPC. The APA column and the two \xara s are illustrated in black. The PDS modules are being show in red.}
	\label{fig:exampletpcmc}
\end{figure}

\begin{figure}[h!]
	\centering
	\includegraphics[width=0.99\linewidth]{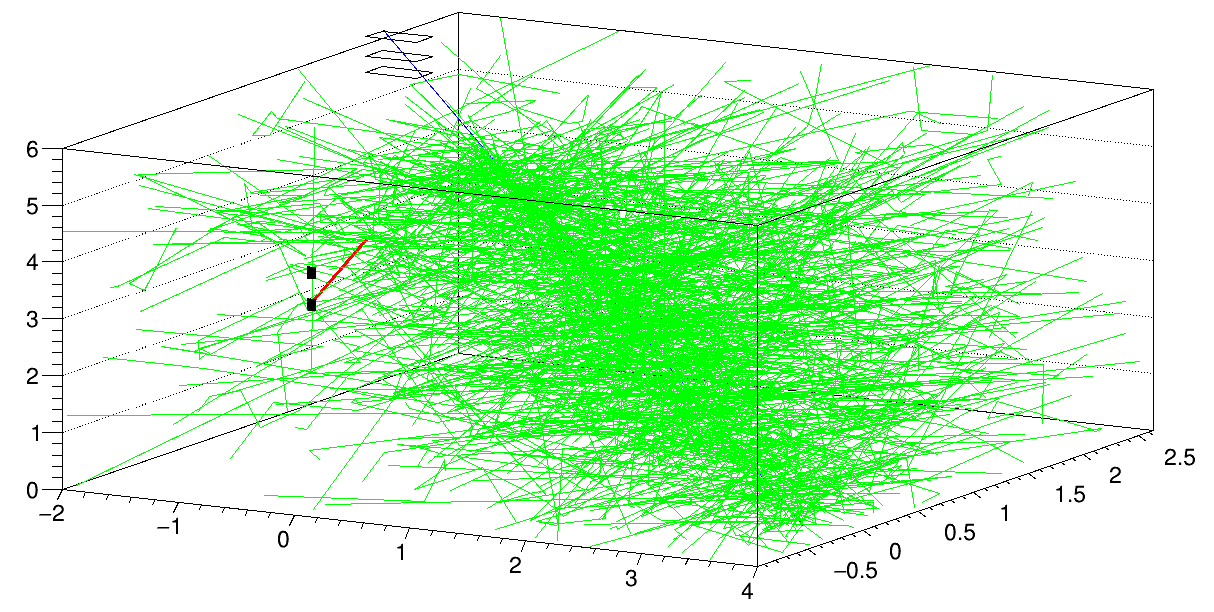}
	\caption{Example of a muon (blue line) crossing the LAr and emitting scintillation photons (green lines) isotropically. The red line represents a photon that was successfully collected after begin Rayleigh scattered. The light yield was significantly reduced to produce the image.}
	\label{fig:exampletpcmc3}
\end{figure}

\end{appendices}